\documentclass[twocolumn]{aa}
\usepackage[multidot]{grffile}
\usepackage{graphicx}
\usepackage{txfonts}
\usepackage{natbib}
\usepackage[colorlinks,allcolors=blue,bookmarks=false,hypertexnames=true]{hyperref} 
\pdfoutput=1
\bibpunct{(}{)}{;}{a}{}{,}
\begin{document}

\title
{
Multicomponent, multiwavelength benchmarks \\ for source- and filament-extraction methods
}


\author
{
A.~Men'shchikov 
}


\institute
{
AIM, IRFU, CEA, CNRS, Universit{\'e} Paris-Saclay, Universit{\'e} Paris Diderot, Sorbonne Paris Cit{\'e}, F-91191 Gif-sur-Yvette, 
France\\\email{alexander.menshchikov@cea.fr}
}

\date{Received 14 June 2021 / Accepted 11 August 2021}

\offprints{Alexander Men'shchikov}
\titlerunning{Benchmarks for source- and filament-extraction methods}
\authorrunning{A.~Men'shchikov}


\abstract
{ 
Modern multiwavelength observations of star-forming regions that reveal complex, highly structured molecular clouds require
adequate extraction methods that provide both complete detection of the structures and their accurate measurements. The
omnipresence of filamentary structures and their physical connection to prestellar cores make it necessary to use methods that are
able to disentangle and extract both sources and filaments. It is fundamentally important to test all extraction methods on
standard benchmarks to compare their detection and measurement qualities and fully understand their capabilities before their
scientific applications. A recent publication described \textsl{getsf}, the new method for source and filament extraction that
employs the separation of the structural components, a successor to \textsl{getsources}, \textsl{getfilaments}, and
\textsl{getimages} (collectively referred to as \textsl{getold}). That paper describes a detailed benchmarking of both
\textsl{getsf} and \textsl{getold} using two multicomponent, multiwavelength benchmarks resembling the \emph{Herschel} observations
of the nearby star-forming regions. Each benchmark consists of simulated images at six \emph{Herschel} wavelengths and one
additional derived surface density image with a $13${\arcsec} resolution. The structural components of the benchmark images include
a background cloud, a dense filament, hundreds of starless and protostellar cores, and instrumental noise. Five variants of
benchmark images of different complexity are derived from the two benchmarks and are used to perform the source and filament
extractions with \textsl{getsf} and \textsl{getold}. A formalism for evaluating source detection and measurement qualities is
presented, allowing quantitative comparisons of different extraction methods in terms of their completeness, reliability, and
goodness, as well as the detection and measurement accuracies and the overall quality. A detailed analysis of the benchmarking
results shows that \textsl{getsf} has better source and filament extraction qualities than \textsl{getold} and that the best choice
of the images for source detection with \textsl{getsf} is the high-resolution surface density, alone or with the other available
\emph{Herschel} images. The benchmarks explored in this paper provide the standard tests for calibrating existing and future
source- and filament-extraction methods to choose the best tool for astrophysical studies.
} 
\maketitle


\section{Introduction}
\label{introduction}

Extraction methods are critically important research tools, interfacing the astronomical imaging observations with their analyses
and physical interpretations. Many different methods have been applied in various studies of star formation in the recent decades
to extract sources and filaments and derive their physical properties. The launch of the \emph{Herschel} Space Observatory
stimulated the development of a number of new source-extraction methods, for example, \textsl{cutex} \citep{Molinari_etal2011},
\textsl{getsources} \citep{Men'shchikov_etal2012}, \textsl{csar} \citep{Kirk_etal2013}, and \textsl{fellwalker} \citep{Berry2015}.
Ubiquitous filamentary structures observed with \emph{Herschel} prompted the creation of several filament-extraction methods, for
example, \textsl{disperse} \citep{Sousbie2011}, \textsl{getfilaments} \citep{Men'shchikov2013}, a Hessian matrix-based method
\citep{Schisano_etal2014}, \textsl{rht} \citep{Clark_etal2014}, \textsl{filfinder} \citep{KochRosolowsky2015}, and \textsl{tm}
\citep{Juvela2016}. Most of the methods provide solutions to the problem of detecting sources or filaments, whereas a complete
extraction entails their accurate measurements, for which knowledge of their backgrounds is necessary. However, the backgrounds of
sources and filaments embedded in the complex, filamentary molecular clouds that strongly fluctuate on all spatial scales are
highly uncertain, which induces increasingly larger measurement errors for fainter structures.

The methods employ very different approaches, and it is quite reasonable to expect the qualities of their results obtained for the
same observed image to be dissimilar. Experience shows that various methods perform differently on increasingly complex images,
although they tend to show more comparable results when tested on the simplest images. It seems unlikely that various independent
tools would provide the same or consistent results in terms of detection completeness, number of false positive (spurious)
detections, and measurement accuracy. The various uncalibrated tools applied in different studies have the potential to bring about
contradictory results and wrong conclusions and to create serious long-term confusion in our understanding of the observed
astrophysical reality.

It is highly important to benchmark the extraction methods before their astrophysical applications. Although new extraction methods
are usually validated before publication on either observed or simulated images, the test images are different for each method,
have dissimilar components and complexity levels, and are not always available for independent evaluations and future comparisons
with other tools. The validation images, used to test older methods at the time of their publication, are unlikely to resemble the
higher complexity observed with new telescopes that have improved angular resolution, sensitivity, and dynamic range. Before the
use of the older tools for such improved generations of images, their performance must be reevaluated and compared with other
methods on newer images that resemble the new observations. New methods must also be tested on the same set of images to
demonstrate their advantages over the older methods.

Comparisons of extraction methods using observed images cannot be conclusive. Only the proper benchmarks would be able to reveal
the true qualities and capabilities of the extraction tools. In this paper, the term ``benchmark'' refers to a standard
multiwavelength set of simulated images with fully known properties of all their components, resembling a certain type of observed
image in their components and complexity. To benchmark extraction methods means to run them on the simulated images without any
knowledge of the model parameters, as if such images were the true observed images. Subsequent comparisons of the resulting
extraction catalogs with the truth catalogs using a reasonable set of quality estimators would determine their detection and
measurement qualities, inaccuracies, and biases. It would be highly desirable, if various studies used the extraction tool that
shows the best performance in benchmarks, to exclude any discrepancies caused by different methods. Notwithstanding that such an
approach is sometimes practiced within research consortia, it does not solve the problem entirely, because the results and
conclusions derived for the same images by independent groups with completely different tools would still likely be incompatible.

Systematic benchmarking of different extraction methods to guide researchers in their selection of the most appropriate tool for
their star-formation studies are hard to find in the literature. A quantitative benchmarking of eight source-extraction methods,
referred to by \cite{Men'shchikov_etal2012}, was instrumental in the selection of the best tool to apply for the \emph{Herschel}
Gould Belt Survey \citep[HGBS,][]{Andre_etal2010} and \emph{Herschel} Imaging Survey of OB Young Stellar Objects
\citep[HOBYS,][]{Motte_etal2010}, but that work remains unpublished. It would not make sense to publish the old results now,
because some of the methods have been improved over the years, while the others have become outdated and are not used for the
modern, complex images. Any publication of benchmarking results for a selection of extraction tools might quickly lose its value,
because it cannot include any improved and newly developed methods. In this work, a completely different approach was taken.

A recent publication \citep[][hereafter referred to as Paper I]{Men'shchikov2021} presented a multicomponent, multiwavelength
benchmark resembling the images observed by \emph{Herschel} in star-forming regions. The benchmark images contain a realistic
filamentary cloud and hundreds of starless and protostellar cores computed by radiative transfer modeling. Fully known properties
of all components allow conclusive comparisons of different methods by evaluating their extraction completeness, reliability, and
goodness, along with the detection and measurement accuracies. The benchmark images, together with the truth catalogs, are made
publicly available and proposed as the standard benchmark for existing and future extraction methods.

Besides the benchmark, Paper I presented \textsl{getsf}, the multiscale, multiwavelength source- and filament-extraction
method\footnote{\url{http://irfu.cea.fr/Pisp/alexander.menshchikov/}}, replacing the older \textsl{getsources},
\textsl{getfilaments}, and \textsl{getimages} algorithms \citep{Men'shchikov_etal2012,Men'shchikov2013,Men'shchikov2017};
throughout this paper, the three predecessors of \textsl{getsf} are collectively named \textsl{getold}. The new method handles both
sources and filaments consistently, separating the structural components from each other and from their backgrounds, thereby
facilitating their extraction problem. The method produces flattened detection images with uniform levels of the residual
background and noise fluctuations, which allows the use of global thresholds for detecting the structures. Independent information
contained in the multiwaveband images is combined in the detection images, preserving the higher angular resolutions. Properties of
the detected sources and filaments are measured in their background-subtracted images and cataloged.

This paper presents benchmarking results for source and filament extraction with \textsl{getsf} and for source extraction with
\textsl{getold}, using the new benchmark from Paper I and the old benchmark from \cite{Men'shchikov_etal2012}. Instead of
describing benchmarking results for an arbitrary selection of existing source-extraction tools, this paper provides researchers in
star formation with an extraction quality evaluation system and the source-extraction results obtained with \textsl{getsf} and
\textsl{getold} for five variants of the benchmarks with increasing complexity levels. Such an approach enables researchers to
benchmark any number of source-extraction tools of their choice and evaluate improved or newly developed methods in the future. It
is not unusual that researchers prefer to conduct their own benchmarking and analysis, which often is more convincing.

Extraction of filaments is more problematic than extraction of sources. Filaments are observed as the two-dimensional projections
that are really hard to decipher and relate to their complex three-dimensional structure. Their appearance, identification, and
measurements depend on the spatial scales of interest (cf. Sect. 3.4.5 in Paper I) and they usually contain sources that are either
formed within the filaments or appear on them in projection. They are often heavily curved and blended, but no filament deblending
algorithm is available, and their physically meaningful lengths and masses are hard to determine. Setting aside the difficult
problems to the further dedicated studies, this paper presents the benchmark filament extraction with \textsl{getsf}. No such
results are presented for \textsl{getold}, because this method was unable to reconstruct the filament with any acceptable level of
accuracy.

Section~\ref{skybench} summarizes all properties of the old and new multiwavelength benchmarks. Section~\ref{evalqual} introduces a
system of quantities for evaluating performances of source-extraction methods. Section~\ref{benches} presents the benchmarking
results for several variants of the benchmark. Section~\ref{conclusions} concludes this work.

Following Paper I, images are represented by the capital calligraphic characters (e.g., $\mathcal{A}, \mathcal{B}, \mathcal{C}$)
and software names and numerical methods are typeset slanted (e.g., \textsl{getsf}) to distinguish them from other emphasized
words. The curly brackets $\{\}$ are used to collectively refer to either of the characters, separated by vertical lines. For
example, $\{a|b\}$ refers to $a$ or $b$ and $\{A|B\}_{\rm \{a|b\}c}$ expands to $A_{\rm \{a|b\}c}$ or $B_{\rm \{a|b\}c}$, as well
as to $A_{\rm ac}$, $A_{\rm bc}$, $B_{\rm ac}$, or $B_{\rm bc}$.


\section{Benchmarks for extraction methods}
\label{skybench}

The benchmark from \cite{Men'shchikov_etal2012} (Benchmark A) includes a relatively simple background and many blended sources,
whereas the benchmark from Paper I (Benchmark B) features a more complex, strongly fluctuating, filamentary background, but it does
not have blended sources. All structural components were added to each other, without any attempt to account for the physical
picture that the star-forming cores are the integral parts of the filaments that, in turn, are the integral parts of the molecular
clouds. This is unnecessary for a benchmark, because the existing extraction tools do not discriminate between the embedded
structures and chance projections of the structural components along the line of sight.

\subsection{Benchmark A}
\label{skybenchA}

The multicomponent, multiwavelength benchmark, described by \cite{Men'shchikov_etal2012}, was constructed in 2009, before the
launch of \emph{Herschel}, at slightly nonstandard wavelengths ($\lambda$ of $75$, $110$, $170$, $250$, $350$, and
$500$\,${\mu}$m). The images on a $1800\times 1800$ grid of $2${\arcsec} pixels cover $1{\degr}\!\times 1{\degr}$ or $2.4$\,pc at a
distance $D = 140$\,pc. They include three independent structural components: the background $\mathcal{B}_{\lambda}$, sources
$\mathcal{S}_{\lambda}$, and small-scale instrumental noise $\mathcal{N}_{\lambda}$.

The backgrounds $\mathcal{B}_{\lambda}$ were computed from a synthetic scale-free image $\mathcal{D}_{\rm B}$. The image was scaled
at each wavelength to the typical intensities of molecular clouds in the nearby star-forming regions, adopting a planar image of
dust temperatures decreasing from $20$ to $15$\,K between the upper-left and lower-right corners, with a constant value of
$17.5$\,K along the other diagonal.

The component $\mathcal{S}_{\lambda}$ of sources was computed from the radiative transfer models of starless cores and protostellar
cores with a range of masses from $0.01$ to $6$\,$M_{\sun}$ and half-maximum sizes from ${\sim\,}0.001$ to $0.1$\,pc. The
individual model images of $360$ starless and $107$ protostellar cores were distributed quasi-randomly, preferentially in the
brighter areas of the background $\mathcal{B}_{\lambda}$, allowing them to overlap without any restrictions. A broken power-law
function with the slopes ${\rm d}N/{\rm d{\log_{10}}}M$ of $0.3$ for $M\le 0.08$\,{$M_{\sun}$}, $-0.3$ for $M \le
0.5$\,{$M_{\sun}$}, and $-1.3$ for $M > 0.5$\,{$M_{\sun}$} was used to determine the numbers of models per mass bin
$\delta{\log_{10}\!M}\approx 0.1$.

The final benchmark images $\mathcal{I}_{\!\lambda}$ were obtained by adding different realizations of the random Gaussian noise
$\mathcal{N}_{\lambda}$ at $75$, $110$, $170$, $250$, $350$, and $500$\,{${\mu}$m} and convolving them to the slightly nonstandard
\emph{Herschel} resolutions of $O_{\lambda}$ of $5$, $7$, $11$, $17$, $24$, and $35${\arcsec}. In this paper, the set of benchmark
images is extended with an additional image $\mathcal{I}_{{\!\lambdabar}}\equiv \mathcal{D}_{11{\arcsec}}$ of surface density at a
high angular resolution $O_{{\rm H}} = 11${\arcsec}, derived from $\mathcal{I}_{\!\lambda}$ at $170{-}500$\,{${\mu}$m} using the
algorithm \textsl{hires} described in Sect.~3.1.2 of Paper I.

\subsection{Benchmark B}
\label{skybenchB}

The multicomponent, multiwavelength benchmark from Paper~I is based on images of a simulated star-forming region at a distance $D =
140$\,pc. The images in all \emph{Herschel} wavebands ($\lambda$ of $70$, $100$, $160$, $250$, $350$, and $500$\,${\mu}$m) on a
$2690\times 2690$ grid of $2${\arcsec} pixels cover $1.5{\degr}\!\times 1.5{\degr}$ or $3.7$\,pc. They include emission of four
independent structural components: the background cloud $\mathcal{B}_{\lambda}$, long filament $\mathcal{F}_{\lambda}$, round
sources $\mathcal{S}_{\lambda}$, and small-scale instrumental noise $\mathcal{N}_{\lambda}$. A sum of the first two components
$\mathcal{C}_{\lambda}$ represents the emission of the filamentary background.

The benchmark images were computed from the adopted surface densities and dust temperatures of the structural components (Figs.
2\,--\,4 of Paper I). The background cloud $\mathcal{D}_{\rm B}$ from Benchmark A was scaled to produce the surface densities
$N_{{\rm H}_2}$ from $1.5\times 10^{21}$ to $4.8\times 10^{22}$\,cm$^{-2}$ and fluctuation levels differing by two orders of
magnitude in its diffuse and dense areas. The spiral filament $\mathcal{D}_{\rm F}$ has a crest density of $N_{0} =
10^{23}$\,cm$^{-2}$, a full width of $W = 0.1$\,pc ($150${\arcsec}) at half-maximum (FWHM), and a power-law profile ${N_{{\rm
H}_2}(\theta)\propto\theta^{\;\!-3}}$ at large distances $\theta$ from the crest. The filament is self-touching, because the two
sides of the tightly curved spiral touch each other (Fig. \ref{filamentsB4}), but the filament is not self-blending: there is no
additive mutual contribution of the two sides. This allows the benchmark filament to have unaltered radial profiles on both sides,
to test the extraction methods' ability to reproduce the profiles without any filament deblending algorithm. The filament mass
$M_{\rm F} = 3.04\times10^{3}$\,$M_{\sun}$ and length $L_{\rm F} = 10.5$\,pc correspond to the linear density $\Lambda_{\rm F}
= 290$\,$M_{\sun}$\,pc$^{-1}$.

The resulting surface densities $\mathcal{D}_{\rm C} = \mathcal{D}_{\rm B} + \mathcal{D}_{\rm F}$ of the filamentary cloud are in
the range of $1.7\times 10^{21}$ to $1.4\times 10^{23}$\,cm$^{-2}$. The dust temperatures $\mathcal{T}_{\!\rm C}$ have values from
$15$\,K in the densest central areas of the filamentary cloud to $20$\,K in its diffuse outer parts. The surface densities
$\mathcal{D}_{\rm C}$ and temperatures $\mathcal{T}_{\!\rm C}$ were used to compute the cloud images $\mathcal{C}_{\lambda}$ in all
\emph{Herschel} wavebands, assuming optically thin dust emission.

The component $\mathcal{D}_{\rm S}$ of sources was computed from radiative transfer models of starless cores and protostellar
cores, very similar to those in Benchmark A, in a wide range of masses (from $0.05$ to $2$\,$M_{\sun}$) and half-maximum sizes
(from ${\sim\,}0.001$ to $0.1$\,pc). Individual surface density images of the models of 828 starless and 91 protostellar cores were
distributed in the dense areas ($N_{{\rm H}_2}\ge 5\times 10^{21}$\,cm$^{-2}$) of the filamentary cloud $\mathcal{D}_{\rm C}$.
They were added quasi-randomly, without overlapping, at positions, where their peak density exceeded that of the cloud $N_{{\rm
H}_2}\!$ value. A power-law function with a slope ${\rm d}N/{\rm d{\log_{10}}}M$ of $-0.7$ was used to define the numbers of
models per mass bin $\delta{\log_{10}\!M}\approx 0.1$.

This resulted in the surface densities $\mathcal{D}_{\rm S}$, the intensities $\mathcal{S}_{\lambda}$ of sources, and the emission
$\mathcal{C}_{\lambda} + \mathcal{S}_{\lambda}$ of the simulated star-forming region. The complete benchmark images
$\mathcal{I}_{\!\lambda}$ were obtained by adding different realizations of the random Gaussian noise $\mathcal{N}_{\lambda}$ at
$70$, $100$, $160$, $250$, $350$, and $500$\,{${\mu}$m} and convolving the images to the \emph{Herschel} angular resolutions
$O_{\lambda}$ of $8.4$, $9.4$, $13.5$, $18.2$, $24.9$, and $36.3${\arcsec}, respectively. The set of benchmark images is extended
with an additional image $\mathcal{I}_{{\!\lambdabar}}\equiv \mathcal{D}_{13{\arcsec}}$ of surface density at a high angular
resolution $O_{{\rm H}} = 13.5${\arcsec} derived from $\mathcal{I}_{\!\lambda}$ at $160{-}500$\,{${\mu}$m} using the algorithm
\textsl{hires} described in Sect.~3.1.2 of Paper I.


\section{Quality evaluation system for source extractions}
\label{evalqual}

For comparisons of different source-extraction methods using benchmarks, it is necessary to define several quantities that would
evaluate an extraction quality by comparing the positions of detected sources and their measured properties with the true values.
Such a formalism was developed by the author in collaboration with Ph.\,Andr{\'e} a decade ago (2010, unpublished) and used to
compare \textsl{getsources} with seven other methods \citep[listed in Sect.~1.1 of][]{Men'shchikov_etal2012}. That quality
evaluation system has been slightly improved and is now described below and applied to assess performances of \textsl{getsf} and
\textsl{getold} in the benchmark extractions. Source extraction methods can be quantitatively compared with each other, using the
definitions below and the truth catalogs of the benchmarks.

It is convenient to denote $N_{\rm T}$ the true number of sources in a benchmark, $N_{{\rm D}{\lambda}}$ the number of detected
sources (acceptable at wavelength $\lambda$) whose peak coordinates match those of the model sources from the truth catalog,
$N_{{\rm G}{\lambda}}$ the number of sources among $N_{{\rm D}{\lambda}}$ that have good measurements, and $N_{{\rm S}{\lambda}}$
the number of spurious sources, that is the number of sources in $N_{{\rm D}{\lambda}}$ that do not have any positional match in
the truth catalog. A measurement is considered as good, if the measured quantities (fluxes, sizes) are within a factor of $2^{1/2}$
from its true model value; otherwise, the measurement is regarded as bad and the corresponding number of bad sources is $N_{{\rm
B}{\lambda}} = N_{{\rm D}{\lambda}} - N_{{\rm G}{\lambda}}$.

In the multiwavelength extraction catalogs, sources can be prominent in one waveband and completely undetectable or not measurable
in another one. In the above definitions, a source $n$ is deemed acceptable at wavelength $\lambda$, if
\begin{eqnarray} 
\left.\begin{aligned}
&{\Xi_{{\lambda}{n}} > 1} \,\land\, {\Gamma_{{\lambda}{n}} > 1} \,\land\, {\Omega_{{\lambda}{n}} > 2} \,\land\, 
{\Psi_{\!{\lambda}{n}} > 2} \,\land\, \\
&{A_{{\lambda}{n}} < 2 B_{{\lambda}{n}}} \,\land\, A_{{\rm F}{\lambda}{n}} > 1.15 A_{{\lambda}{n}},
\end{aligned}\right.
\label{acceptable} 
\end{eqnarray} 
where $\Xi_{{\lambda}{n}}$ is the source detection significance, $\Gamma_{{\lambda}{n}}$ is the source goodness,
$\Omega_{{\lambda}{n}}$ and $\Psi_{\!{\lambda}{n}}$ are the signal-to-noise ratios related to the peak intensity $F_{{\rm
P}{\lambda}{n}}$ and total flux $F_{{\rm T}{\lambda}{n}}$, respectively (cf. Eqs.~(41) and (42) of Paper I),
$\{A|B\}_{{\lambda}{n}}$ are the source FWHM sizes, and $A_{{\rm F}{\lambda}{n}}$ is the major diameter of the source footprint.
The last inequality discards the sources with unrealistically small ratios $A_{{\rm F}{\lambda}{n}}/A_{{\lambda}{n}}$ of their
footprint and half-maximum sizes. The empirical set of conditions in Eq.~(\ref{acceptable}) ensures that the selected subset of
sources is reliable (not contaminated by significant numbers of spurious sources) and that selected sources have acceptably
accurate measurements.

With the above definitions of $N_{\rm T}$ and $N_{{\rm \{D|G|S\}}{\lambda}}$, it makes sense to define the source extraction
completeness $C_{\lambda}$, reliability $R_{\lambda}$, and goodness $G_{\lambda}$ as
\begin{equation} 
C_{\lambda} = \frac{N_{{\rm D}{\lambda}}}{N_{\rm T}}, \,\,\, 
R_{\lambda} = \left(1 + 1200\left(\frac{N_{{\rm S}{\lambda}}}{N_{{\rm D}{\lambda}}}\right)^{2}\right)^{-1/2}, \,\,\,
G_{\lambda} = \frac{N_{{\rm G}{\lambda}}}{N_{\rm T}},
\label{qualities0}
\end{equation} 
where $R_{\lambda}$ has been updated with respect to the original version of the system, where it was defined as $1/N_{{\rm
S}{\lambda}}$. The newly defined reliability is the Moffat (Plummer) profile (cf. Eq.~(2) in Paper I), with $\Theta = 0.05 N_{{\rm
D}{\lambda}}$ and $\zeta = 1/2$. It has a Gaussian-like peak at $N_{{\rm S}{\lambda}} = 0$, slowly descends to $0.5$ when $5${\%}
of $N_{{\rm D}{\lambda}}$ are spurious sources, and decreases as $1/N_{{\rm S}{\lambda}}$ for $N_{{\rm S}{\lambda}} \gg 0.05
N_{{\rm D}{\lambda}}$.

It is useful to compute the ratios of the measured quantities to their true model values, for each acceptable source, and evaluate
their mean values among $N_{{\rm G}{\lambda}}$ sources with good measurements:
\begin{eqnarray} 
\left.\begin{aligned}
\varrho_{{\rm P}{\lambda}} &= \langle\frac{F_{{\rm P}{\lambda}{n}}}{F_{{\rm P}{\lambda}{n}{\rm T}}}\rangle, \,\,
\varrho_{{\rm T}{\lambda}} = \langle\frac{F_{{\rm T}{\lambda}{n}}}{F_{{\rm T}{\lambda}{n}{\rm T}}}\rangle, \\
\varrho_{{\rm A}{\lambda}} &= \langle\frac{A_{{\lambda}{n}}}{A_{{\lambda}{n}{\rm T}}}\rangle, \,\,
\varrho_{{\rm B}{\lambda}} = \langle\frac{B_{{\lambda}{n}}}{B_{{\lambda}{n}{\rm T}}}\rangle, \,\,
\varrho_{{\rm E}{\lambda}} = \langle\frac{A_{{\lambda}{n}}B_{{\lambda}{n}}}
{A_{{\lambda}{n}{\rm T}}B_{{\lambda}{n}{\rm T}}}\rangle,
\end{aligned}\right.
\label{devfactors}
\end{eqnarray} 
where $\varrho_{{\rm E}{\lambda}}$ evaluates the accuracy of the source area. The mean ratios with their standard deviations
$\sigma_{\{{\rm P|T|A|B}\}{\lambda}}$ can be used to define the qualities of the measured source parameters as
\begin{equation} 
Q_{\{{\rm P|T|E}\}{\lambda}} = \left(\frac{\max\,(1,\varrho_{\{{\rm P|T|E}\}{\lambda}})}
{\min\,(1,\varrho_{\{{\rm P|T|E}\}{\lambda}})} + \sigma_{\{{\rm P|T|E}\}{\lambda}}\right)^{-1}.
\label{qualities3}
\end{equation} 
Denoting $\delta_{{\rm D}{\lambda}}\equiv \langle D_{{\lambda}{n}}\rangle$ the mean distance of the well-measurable sources from
the true model peaks and $\sigma_{{\rm D}{\lambda}}$ the corresponding standard deviation, the positional quality is defined as
\begin{equation} 
Q_{{\rm D}{\lambda}} = \left(\max\,(1, \delta_{{\rm D}{\lambda}}) + \sigma_{{\rm D}{\lambda}}\right)^{-1}.
\label{qualities1}
\end{equation} 
It is convenient to define the detection quality $Q_{{\rm CR}{\lambda}}$ and the measurement quality $Q_{{\rm PTE}{\lambda}}$
combining the qualities related to the independent source detection and measurement steps, as well as the overall quality
$Q_{\lambda}$ of a source extraction, 
\begin{eqnarray} 
\left.\begin{aligned}
Q_{{\rm CR}{\lambda}} &= C_{\lambda}\, R_{\lambda}, \\
Q_{{\rm PTE}{\lambda}} &= Q_{{\rm P}{\lambda}}\, Q_{{\rm T}{\lambda}}\, Q_{{\rm E}{\lambda}}, \\
Q_{\lambda} &= G_{\lambda}\, Q_{{\rm D}{\lambda}}\, Q_{{\rm CR}{\lambda}}\, Q_{{\rm PTE}{\lambda}}.
\end{aligned}\right.
\label{finalqualities} 
\end{eqnarray} 
The quantities defined by Eqs.~(\ref{qualities0})\,--\,(\ref{finalqualities}) have values in the range $[0,1]$ that become unity
for an imaginary perfect extraction tool that would extract all simulated sources and measure their parameters with no deviations
from the true model values. The absolute values of the quantities are arbitrary and meaningless for a single extraction with a
single method. The values become quite useful, however, when comparing the relative extraction qualities of two or more methods or
of several extractions with a single method (with different parameters).

The quality evaluation system represented by Eqs.~(\ref{qualities0})\,--\,(\ref{finalqualities}) is not unique and other formalisms
might be devised and applied to the benchmark truth catalogs and the \textsl{getsf} and \textsl{getold} extraction catalogs found
on the benchmarking page of the \textsl{getsf} website\footnote{\url{http://irfu.cea.fr/Pisp/alexander.menshchikov/\#bench}}.


\section{Benchmarking}
\label{benches}

The benchmark names with subscripts are used to indicate the number of the structural components. For example, Benchmark A$_3$
contains three components (background, sources, and noise) and Benchmark B$_4$ has four components (background, filament, sources,
and noise). There are also three simpler variants of the benchmarks: B$_3$ has no filament, A$_2$ and B$_2$ have no background.
Below, the source extractions in $\{\mathrm{A},\mathrm{B}\}_2$, $\{\mathrm{A},\mathrm{B}\}_3$, and B$_4$ are presented in a
sequence of their increasing complexity and followed by the filament extraction in B$_4$.

\begin{figure*}
\centering
\centerline{
  \resizebox{0.328\hsize}{!}{\includegraphics{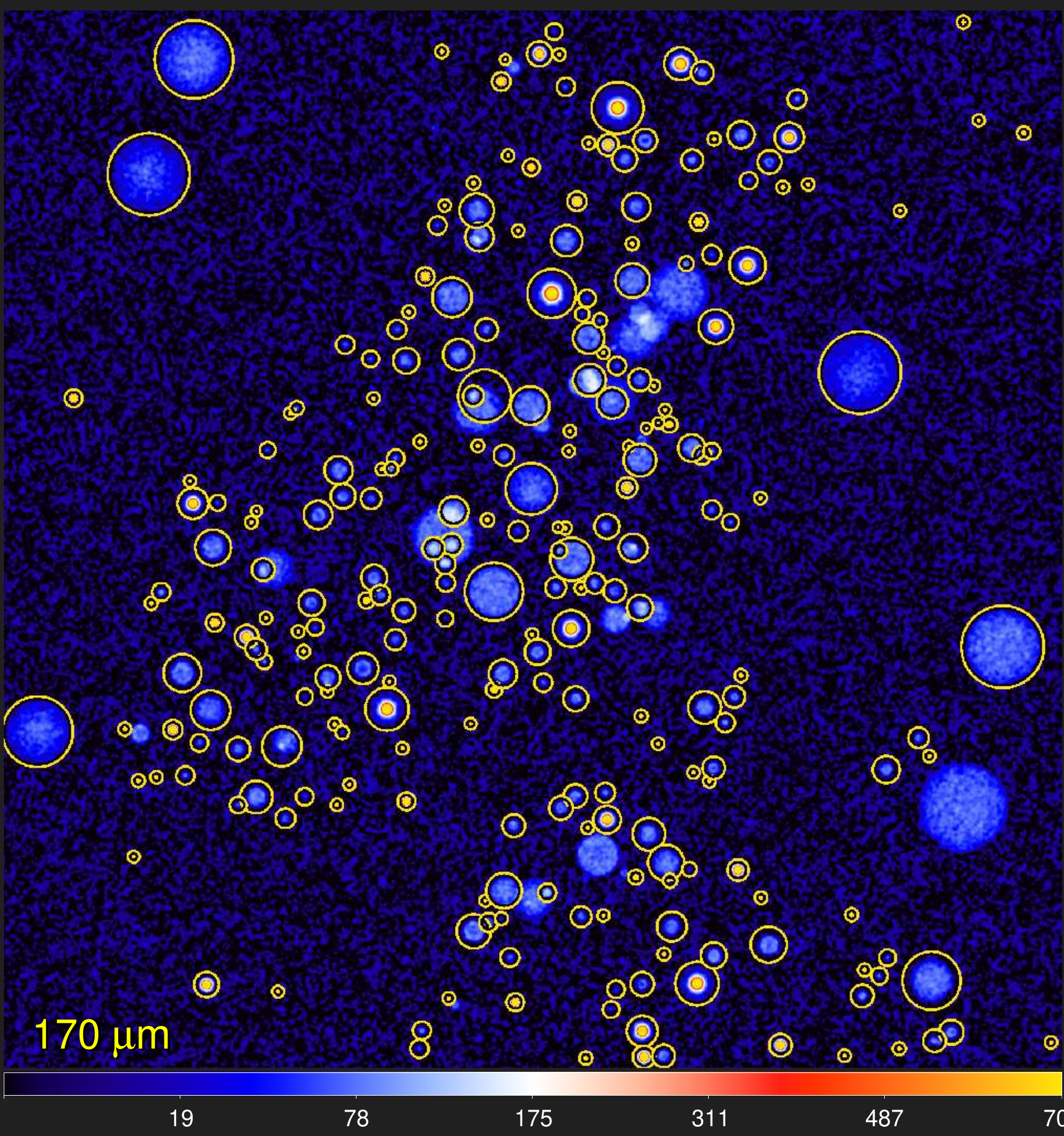}}
  \resizebox{0.328\hsize}{!}{\includegraphics{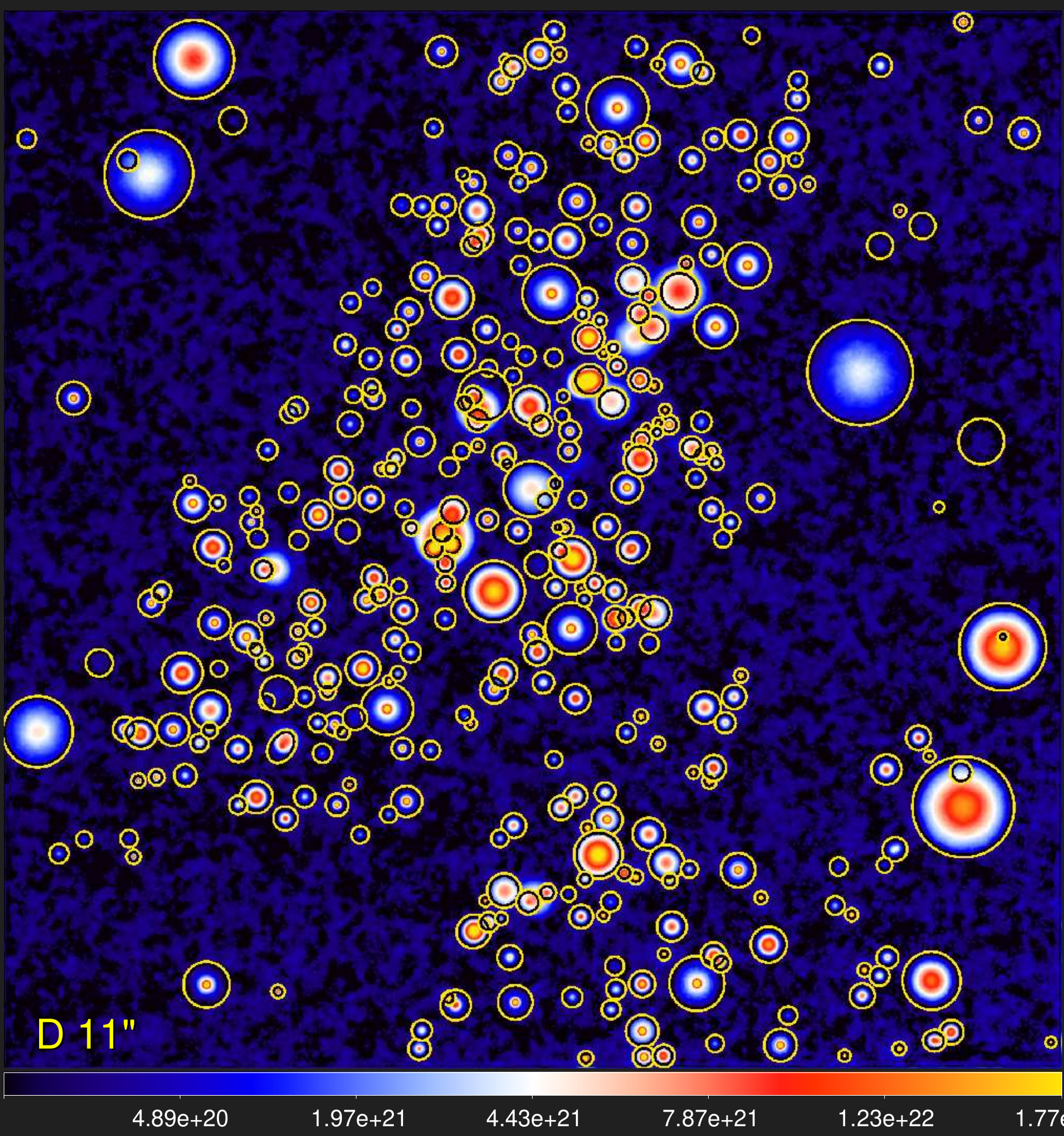}}
  \resizebox{0.328\hsize}{!}{\includegraphics{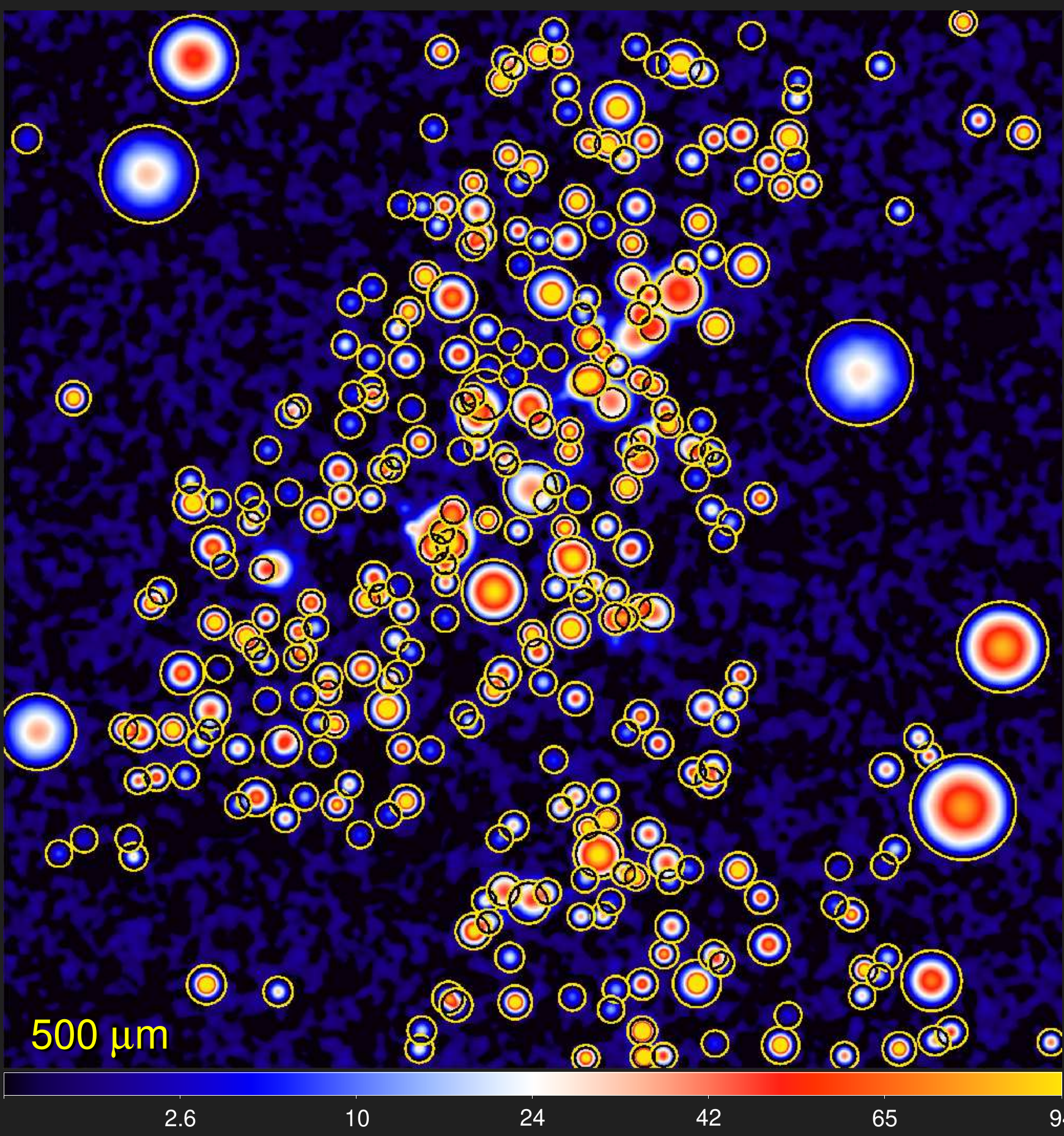}}}
\vspace{0.5mm}
\centerline{
  \resizebox{0.328\hsize}{!}{\includegraphics{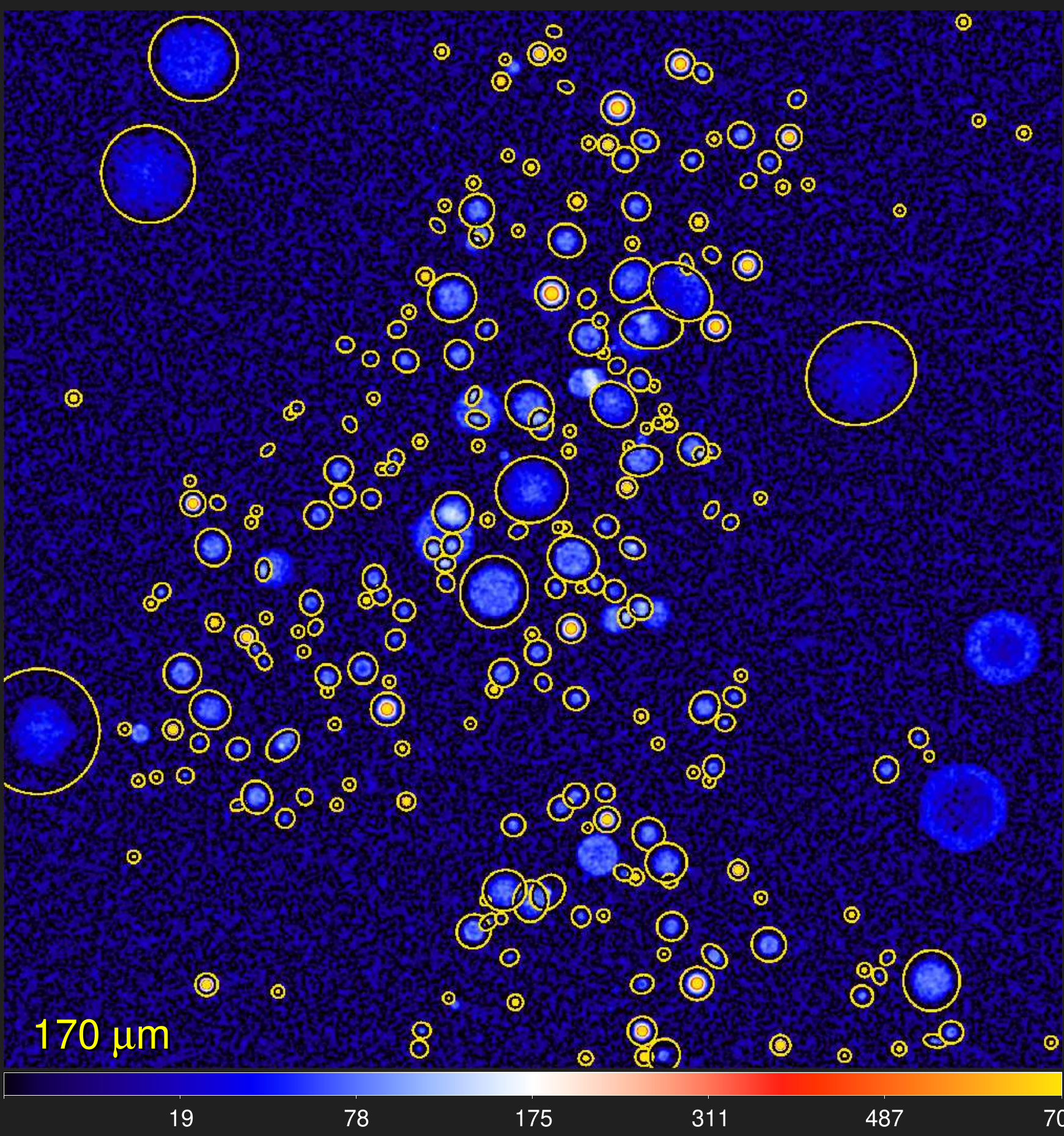}}
  \resizebox{0.328\hsize}{!}{\includegraphics{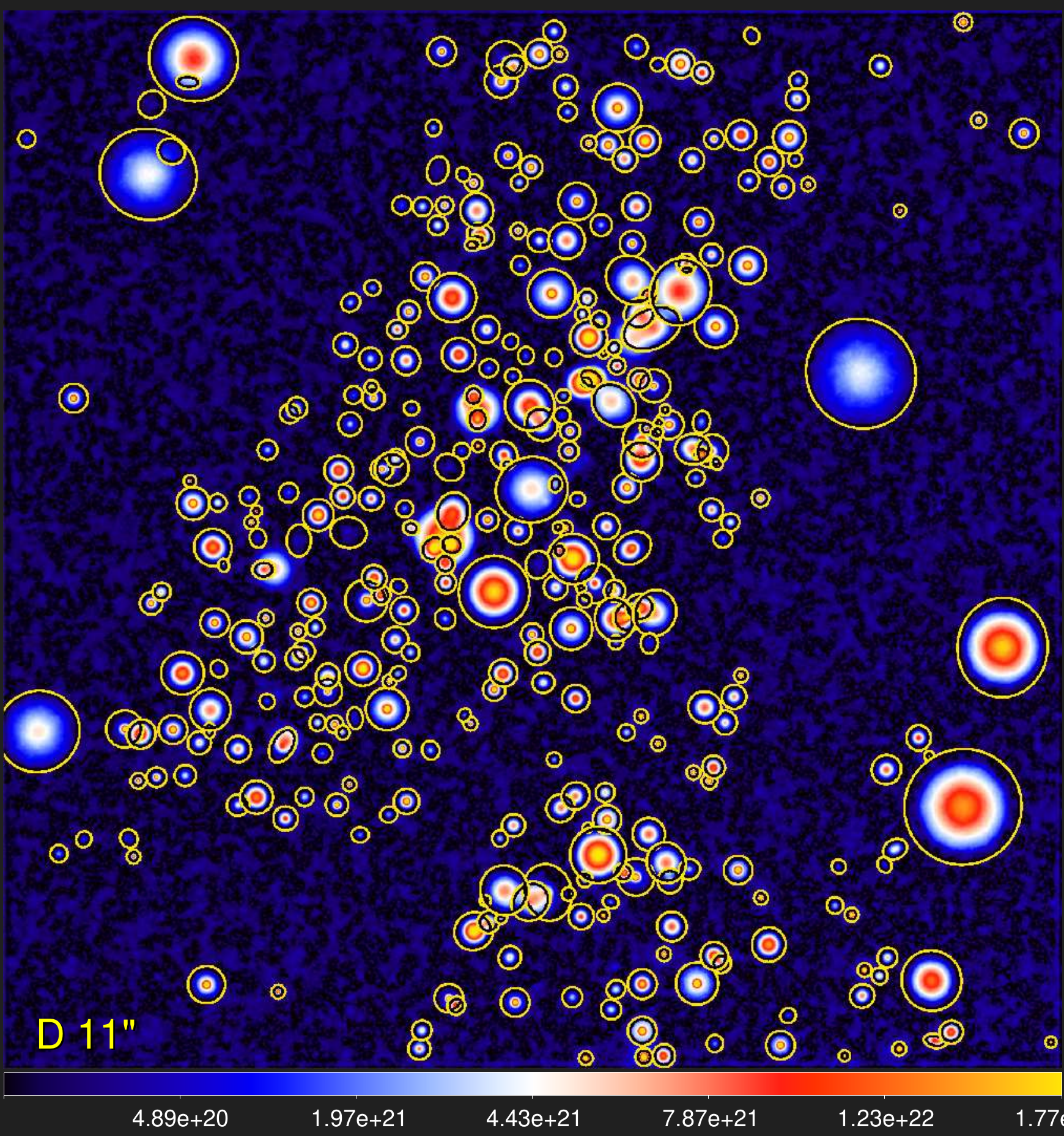}}
  \resizebox{0.328\hsize}{!}{\includegraphics{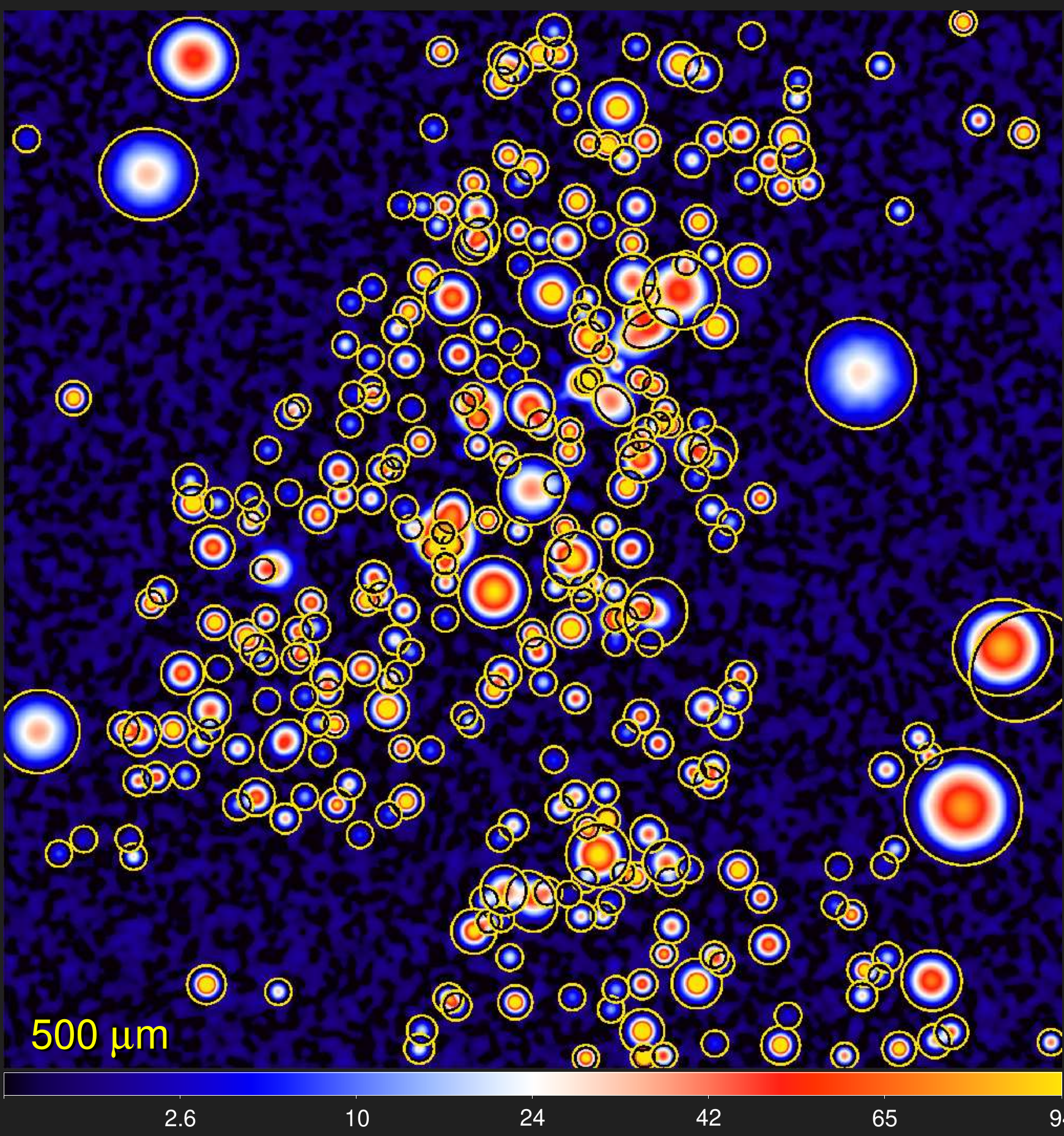}}}
\caption
{ 
Benchmark A$_2$ extraction of sources with \textsl{getsf} and \textsl{getold}. The original $\mathcal{I}_{\!\lambda}$ are overlaid
with the footprint ellipses from the measurement step. In the \textsl{getold} extraction (\emph{bottom}), the nonexistent
large-scale background was determined and subtracted in a preliminary run of \textsl{getimages}, in order to keep the general
extraction scheme unaltered for all benchmarks. The images are displayed with a square-root color mapping.
} 
\label{imagesA2}
\end{figure*}

\begin{figure*}                                                               
\centering
\centerline{
  \resizebox{0.328\hsize}{!}{\includegraphics{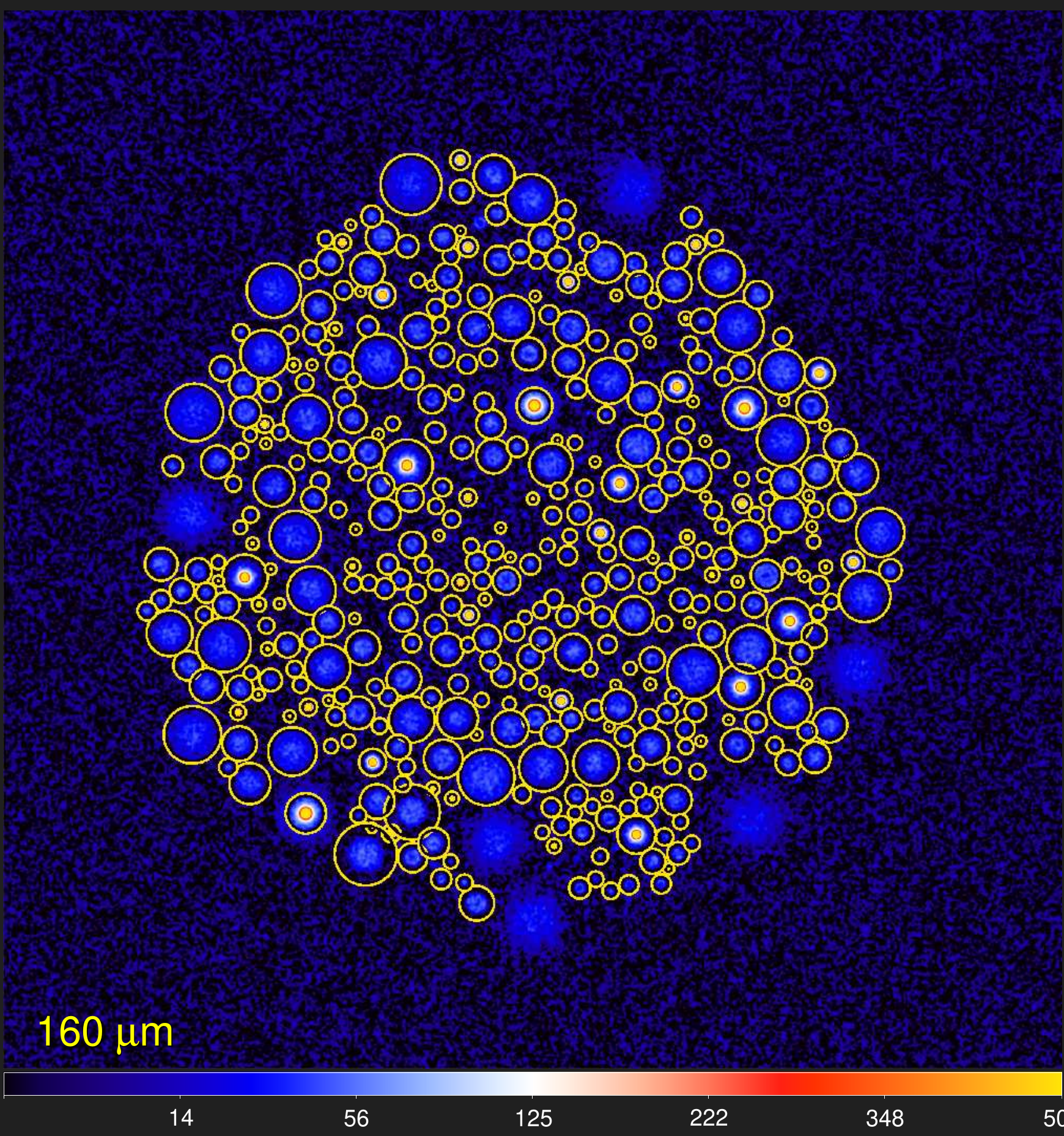}}
  \resizebox{0.328\hsize}{!}{\includegraphics{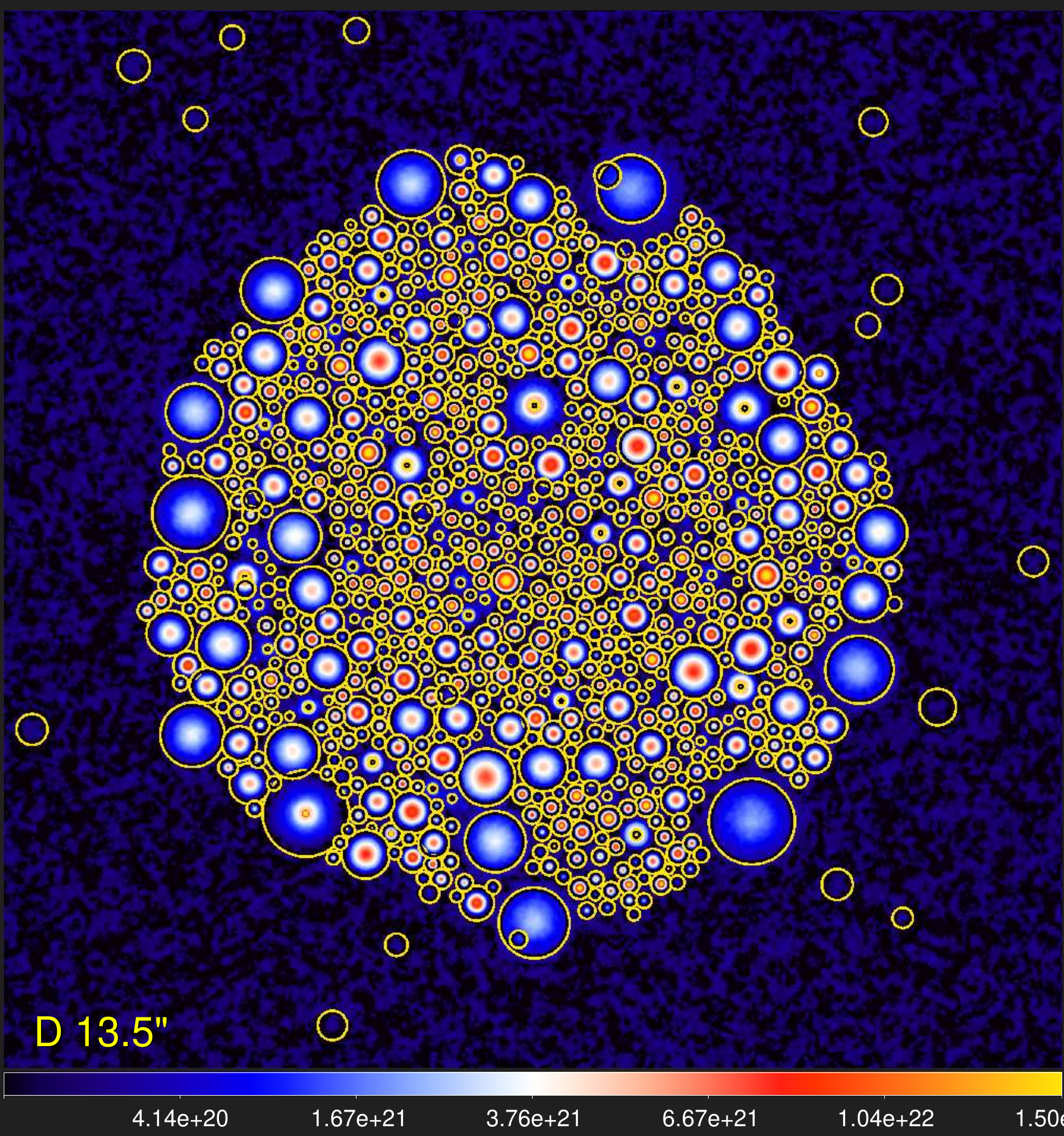}}
  \resizebox{0.328\hsize}{!}{\includegraphics{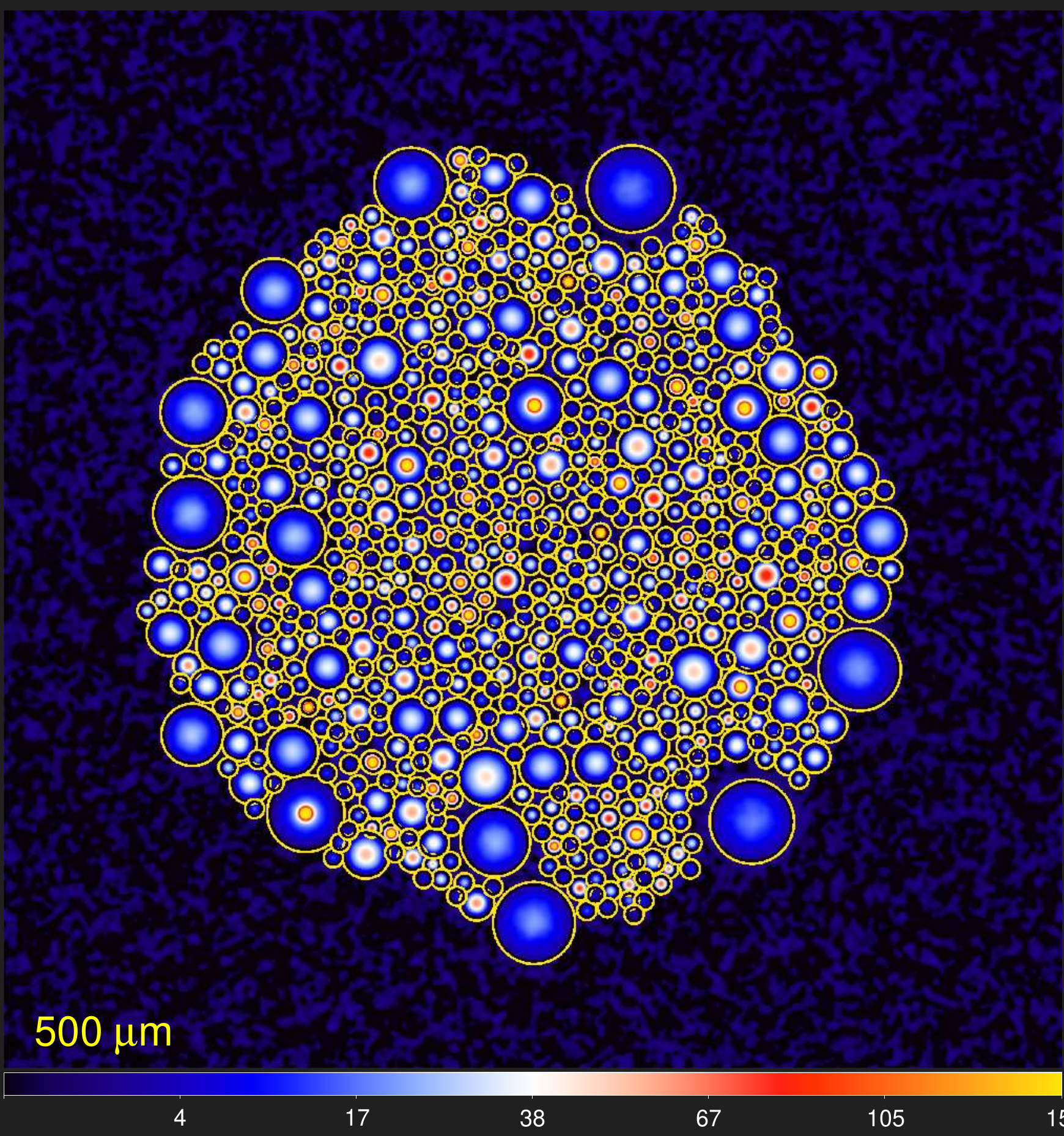}}}
\vspace{0.5mm}
\centerline{
  \resizebox{0.328\hsize}{!}{\includegraphics{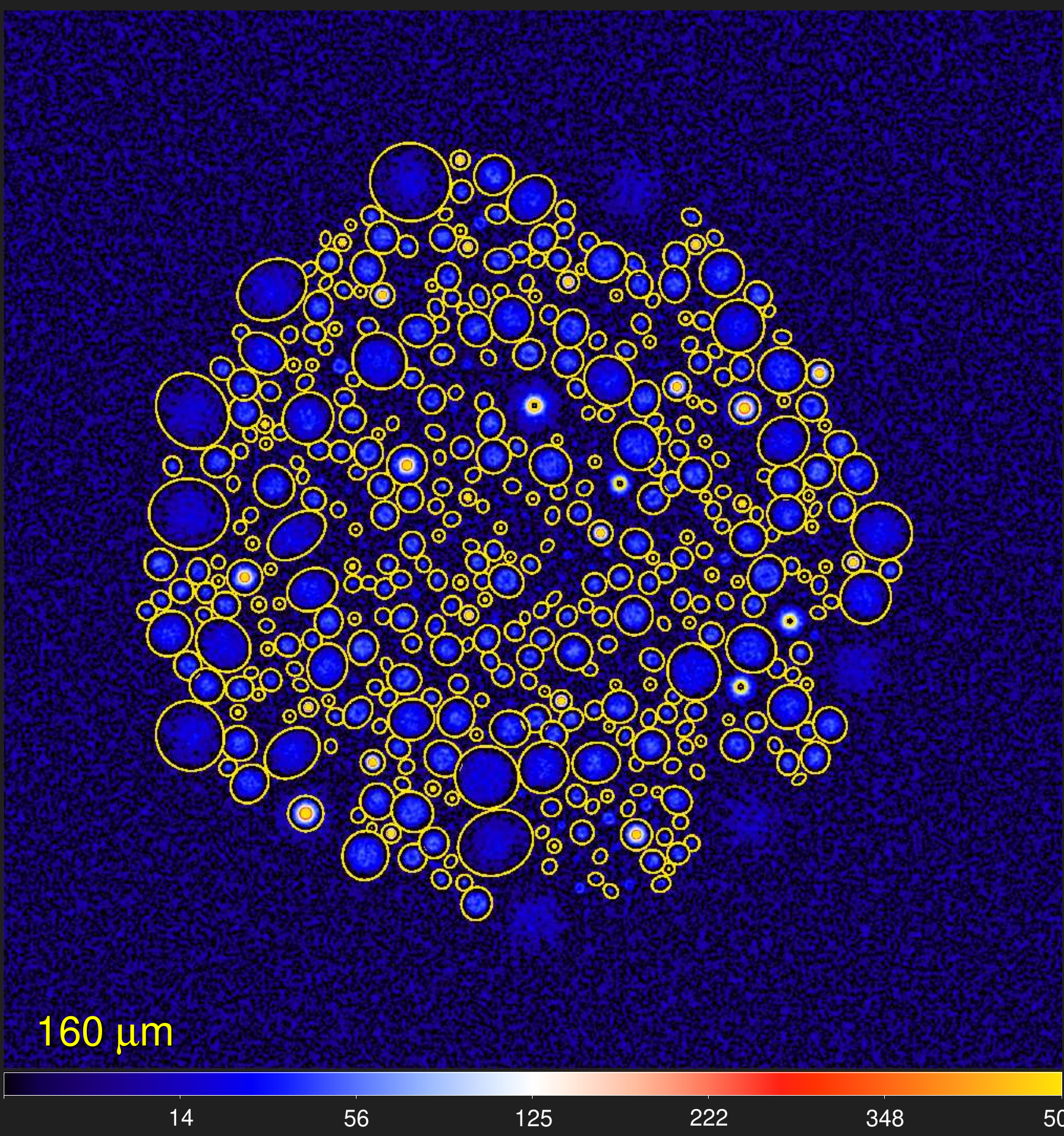}}
  \resizebox{0.328\hsize}{!}{\includegraphics{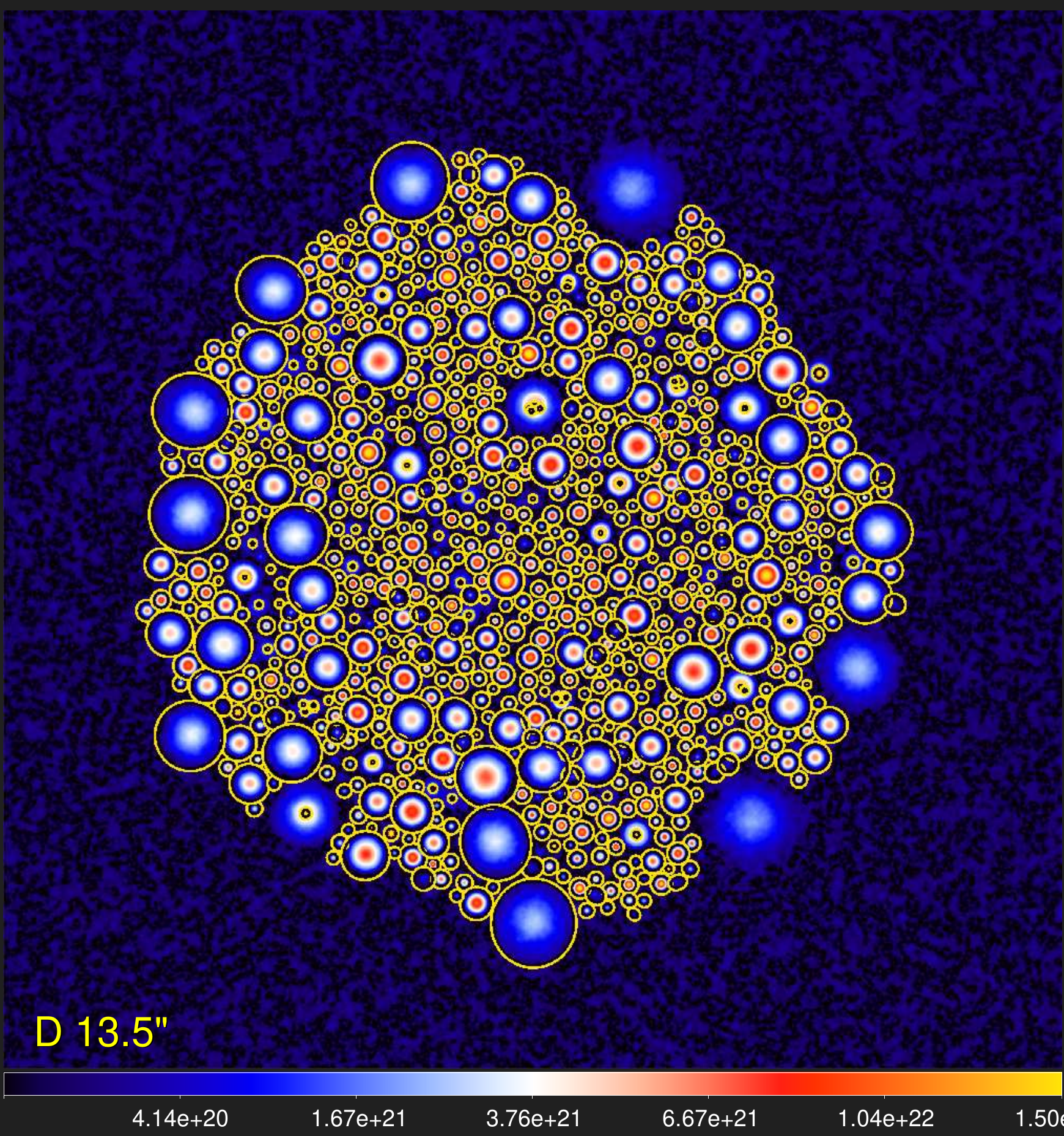}}
  \resizebox{0.328\hsize}{!}{\includegraphics{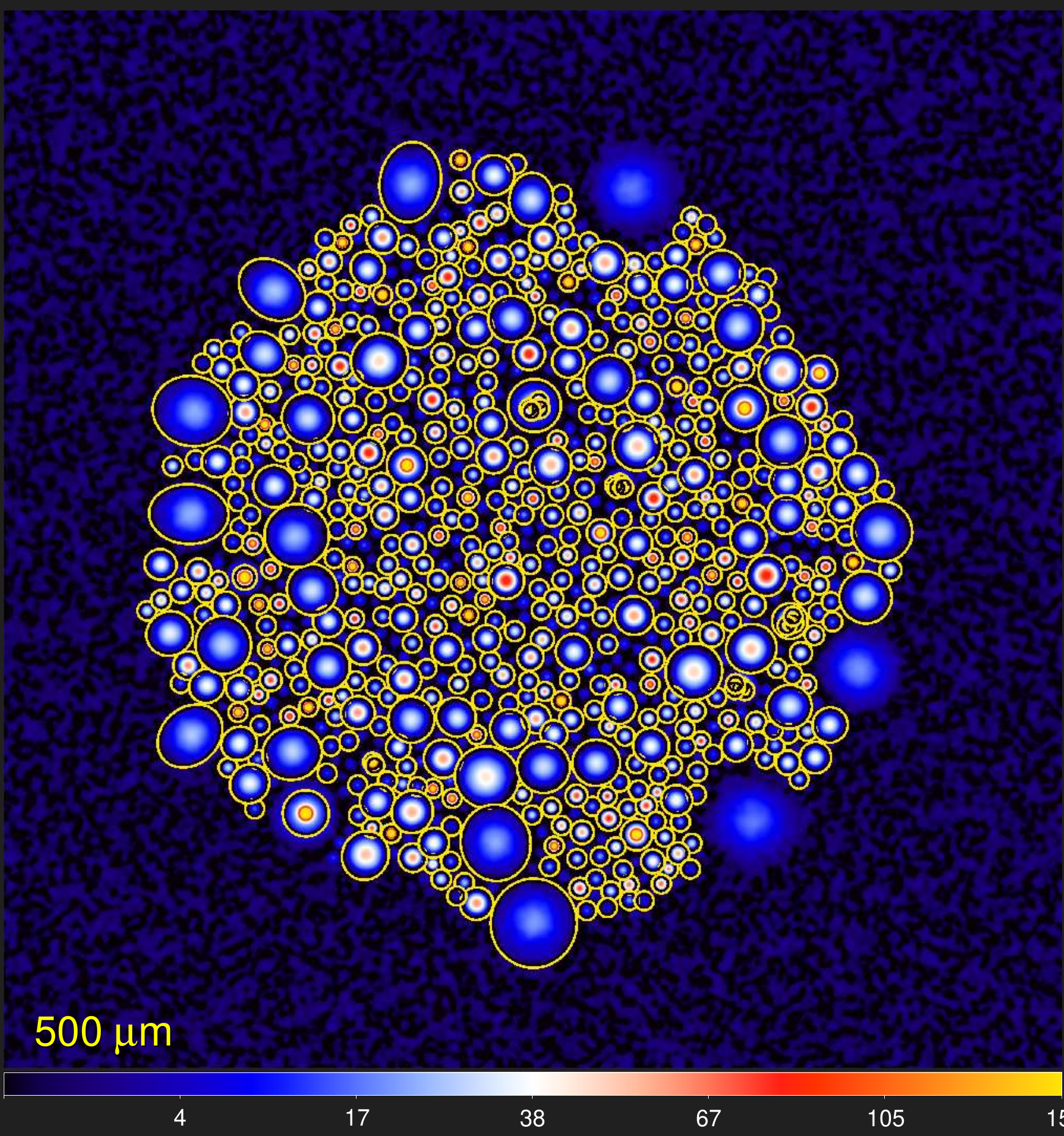}}}
\caption
{ 
Benchmark B$_2$ extraction of sources with \textsl{getsf} and \textsl{getold}. The original $\mathcal{I}_{\!\lambda}$ are overlaid
with the footprint ellipses from the measurement step. In the \textsl{getold} extraction (\emph{bottom}), the nonexistent
large-scale background was determined and subtracted in a preliminary run of \textsl{getimages}, in order to keep the general
extraction scheme unaltered for all benchmarks. The images are displayed with a square-root color mapping.
} 
\label{imagesB2}
\end{figure*}

The simplest benchmarks $\{\mathrm{A},\mathrm{B}\}_2$ (Figs. \ref{imagesA2} and \ref{imagesB2}) contain only two components, the
model cores and noise. Most sources are clearly visible against the noise, and therefore they must be uncomplicated to detect for a
variety of extraction methods. The model sources have a wide range of the FWHM sizes from the angular resolution $O_{\lambda}$ up
to $A_{{\lambda}{n}}\approx 200${\arcsec}; therefore the methods that limit the largest sizes of extractable sources by only a few
beams are expected to miss many larger sources. The resolved models and real objects have non-Gaussian intensity distributions;
therefore, the methods, assuming that all sources have Gaussian shapes, are expected to produce less accurate measurements.

\begin{figure*}
\centering
\centerline{
  \resizebox{0.328\hsize}{!}{\includegraphics{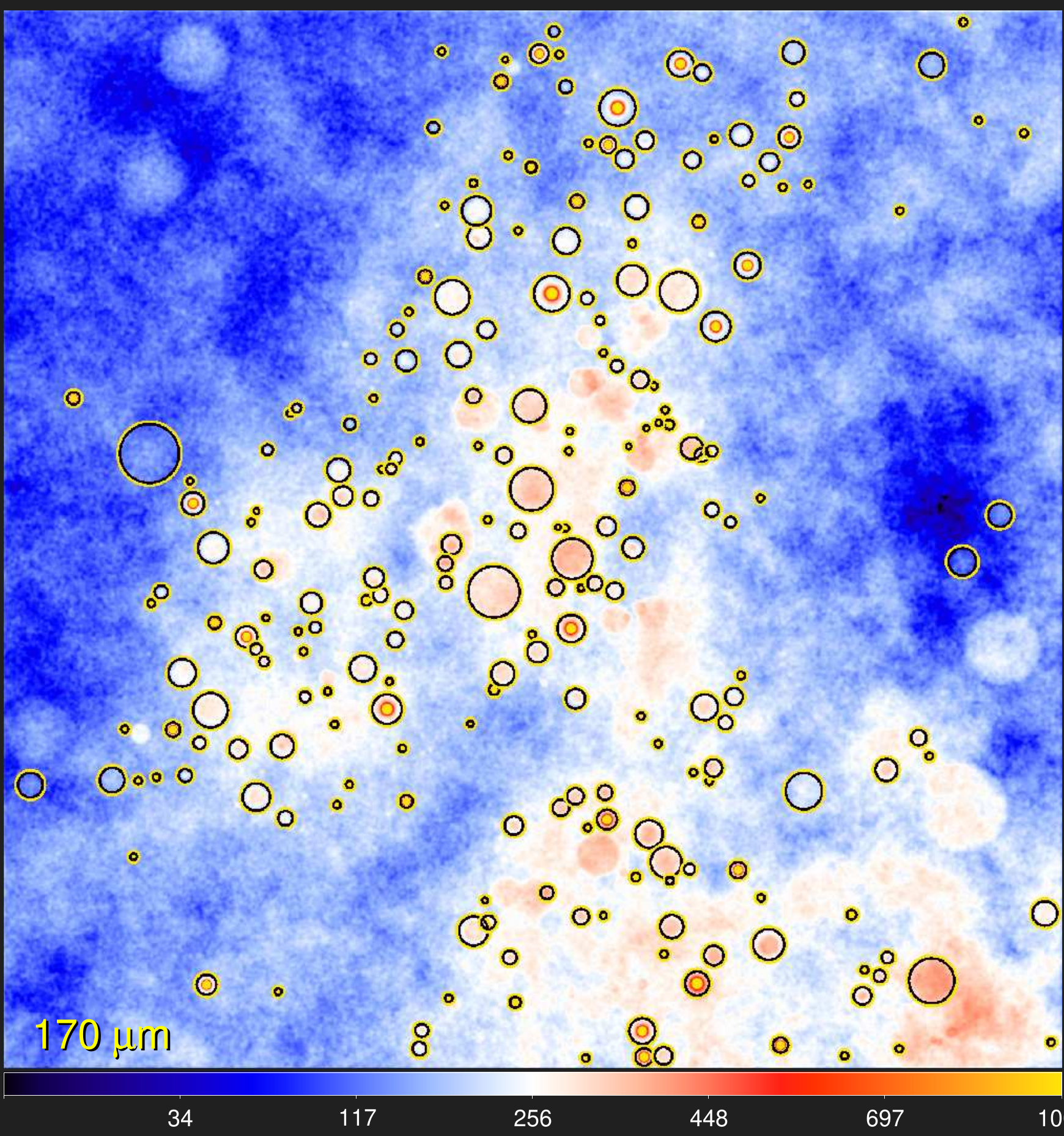}}
  \resizebox{0.328\hsize}{!}{\includegraphics{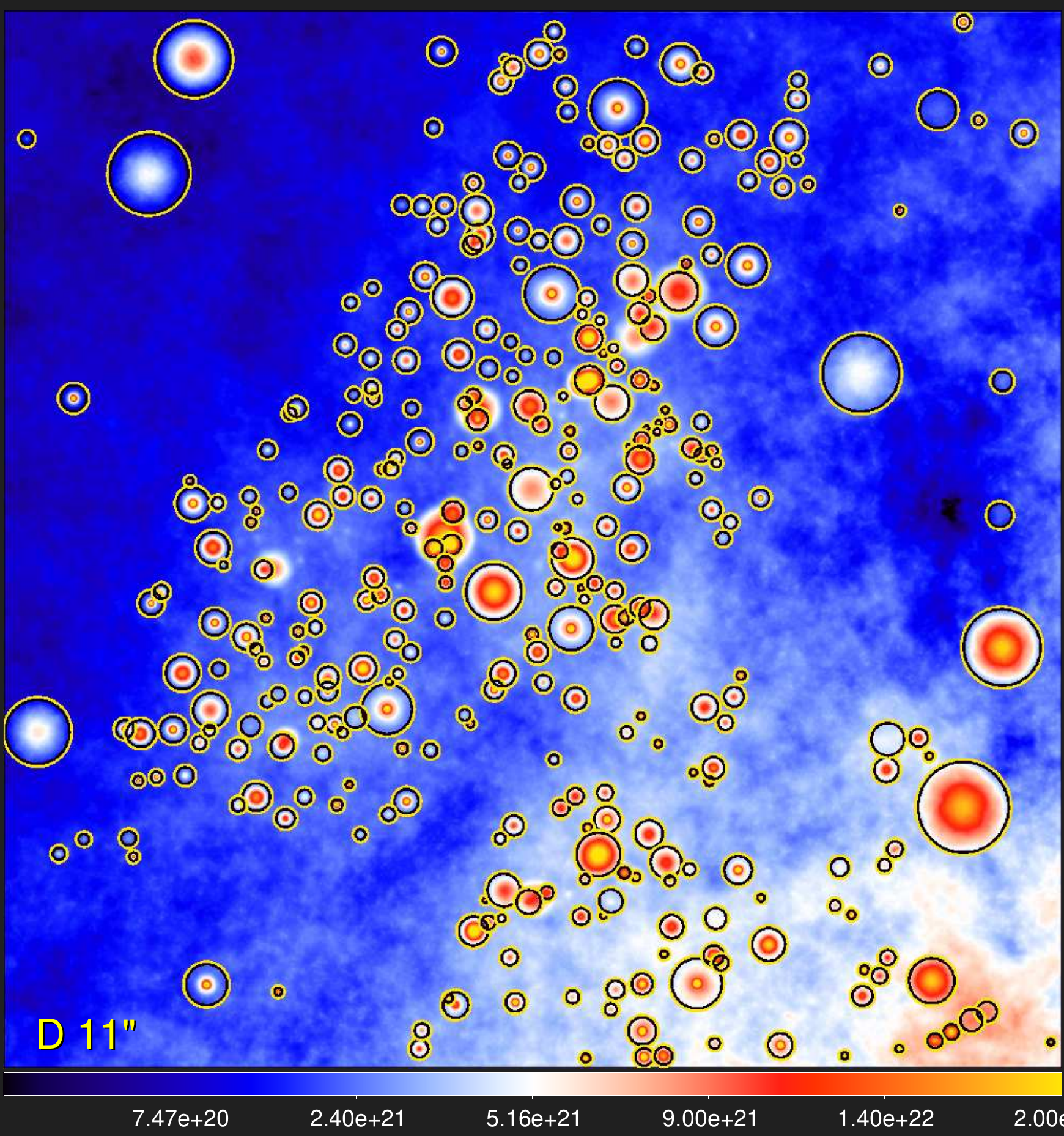}}
  \resizebox{0.328\hsize}{!}{\includegraphics{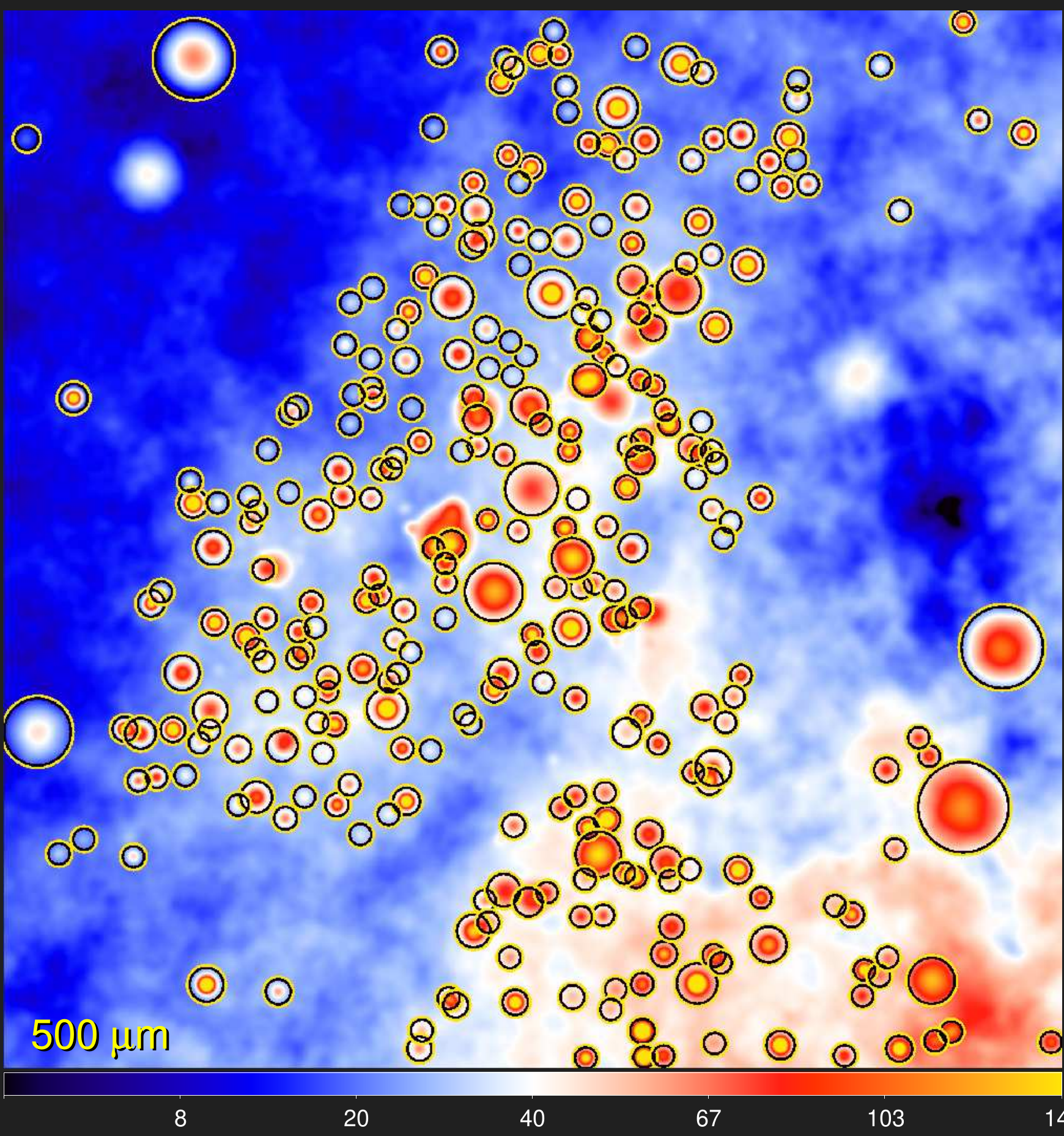}}}
\vspace{0.5mm}
\centerline{
  \resizebox{0.328\hsize}{!}{\includegraphics{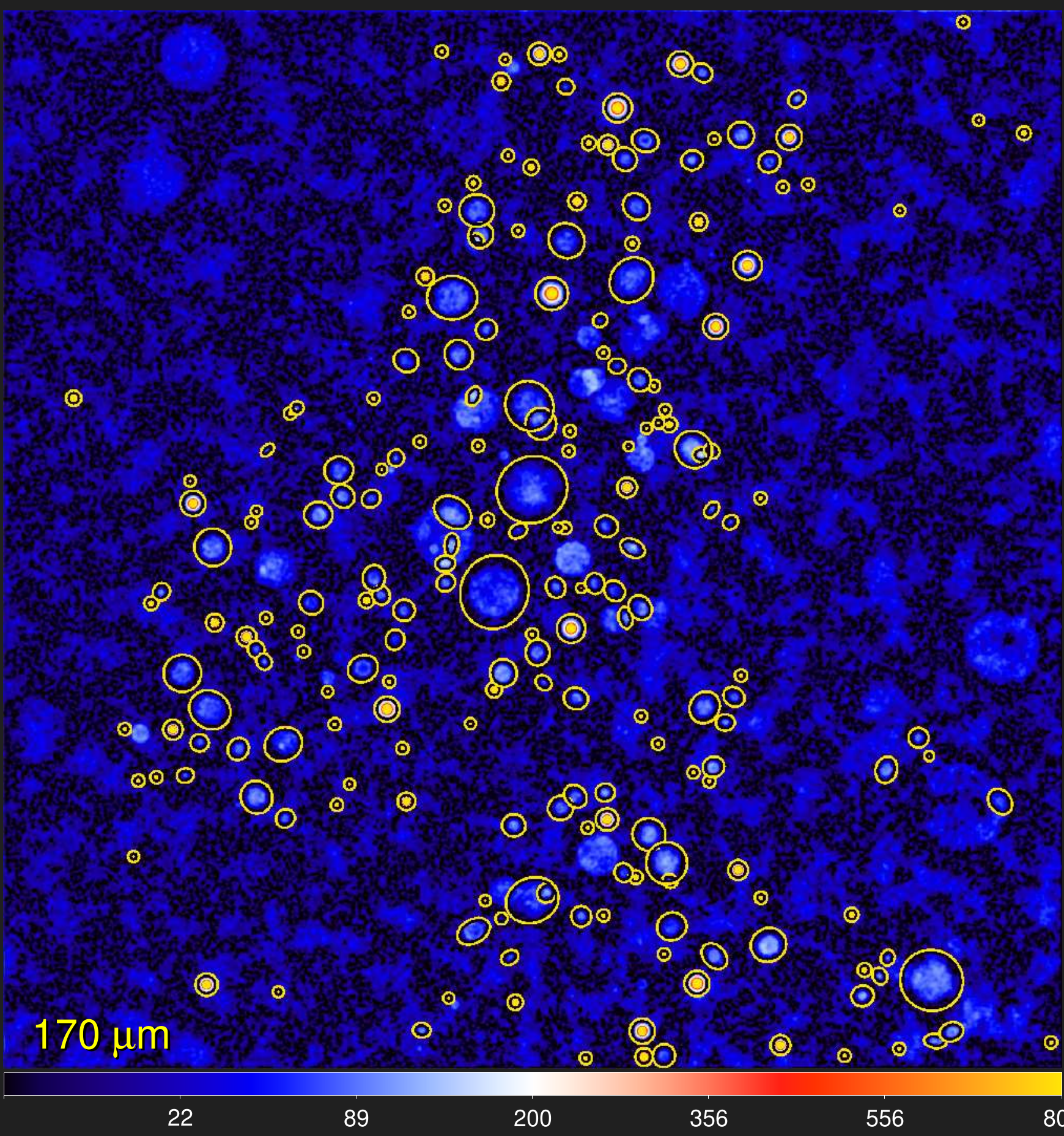}}
  \resizebox{0.328\hsize}{!}{\includegraphics{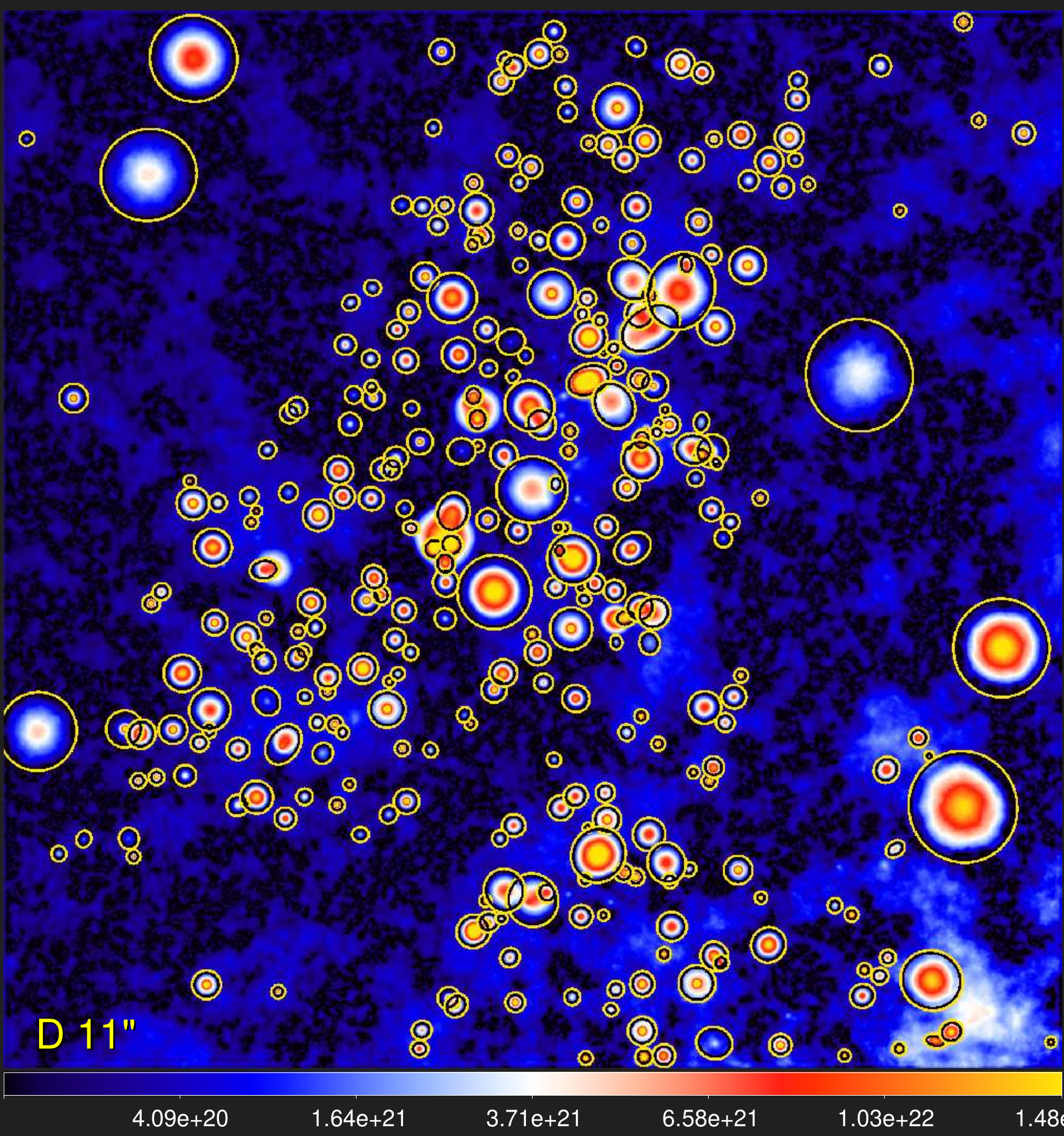}}
  \resizebox{0.328\hsize}{!}{\includegraphics{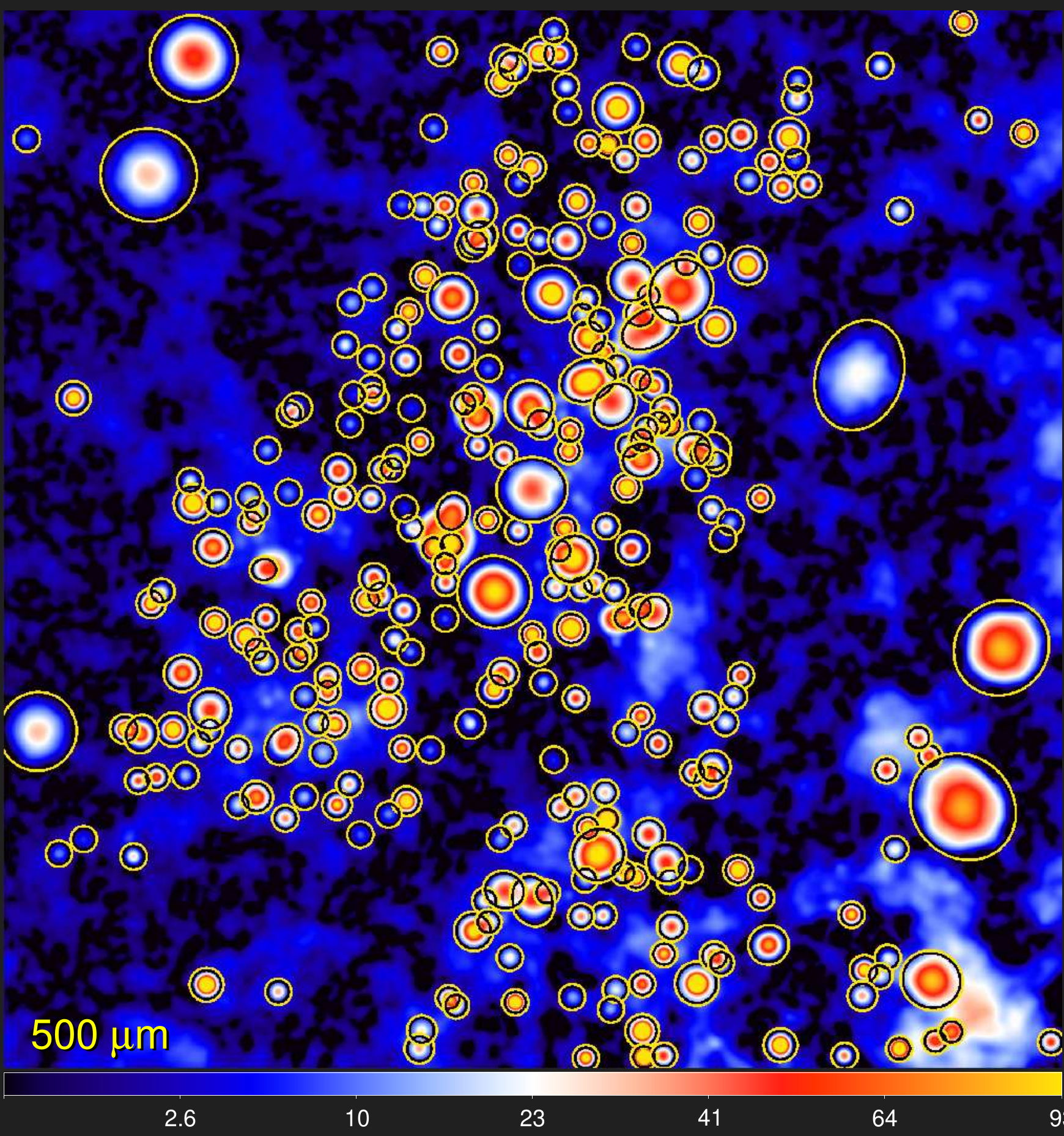}}}
\caption
{ 
Benchmark A$_3$ extraction of sources with \textsl{getsf} and \textsl{getold}. The original $\mathcal{I}_{\!\lambda}$ are overlaid
with the footprint ellipses from the measurement step. In the \textsl{getold} extraction (\emph{bottom}), the large-scale
background was determined and subtracted by \textsl{getimages}. The images are displayed with a square-root color mapping.
} 
\label{imagesA3}
\end{figure*}

\begin{figure*}                                                               
\centering
\centerline{
  \resizebox{0.328\hsize}{!}{\includegraphics{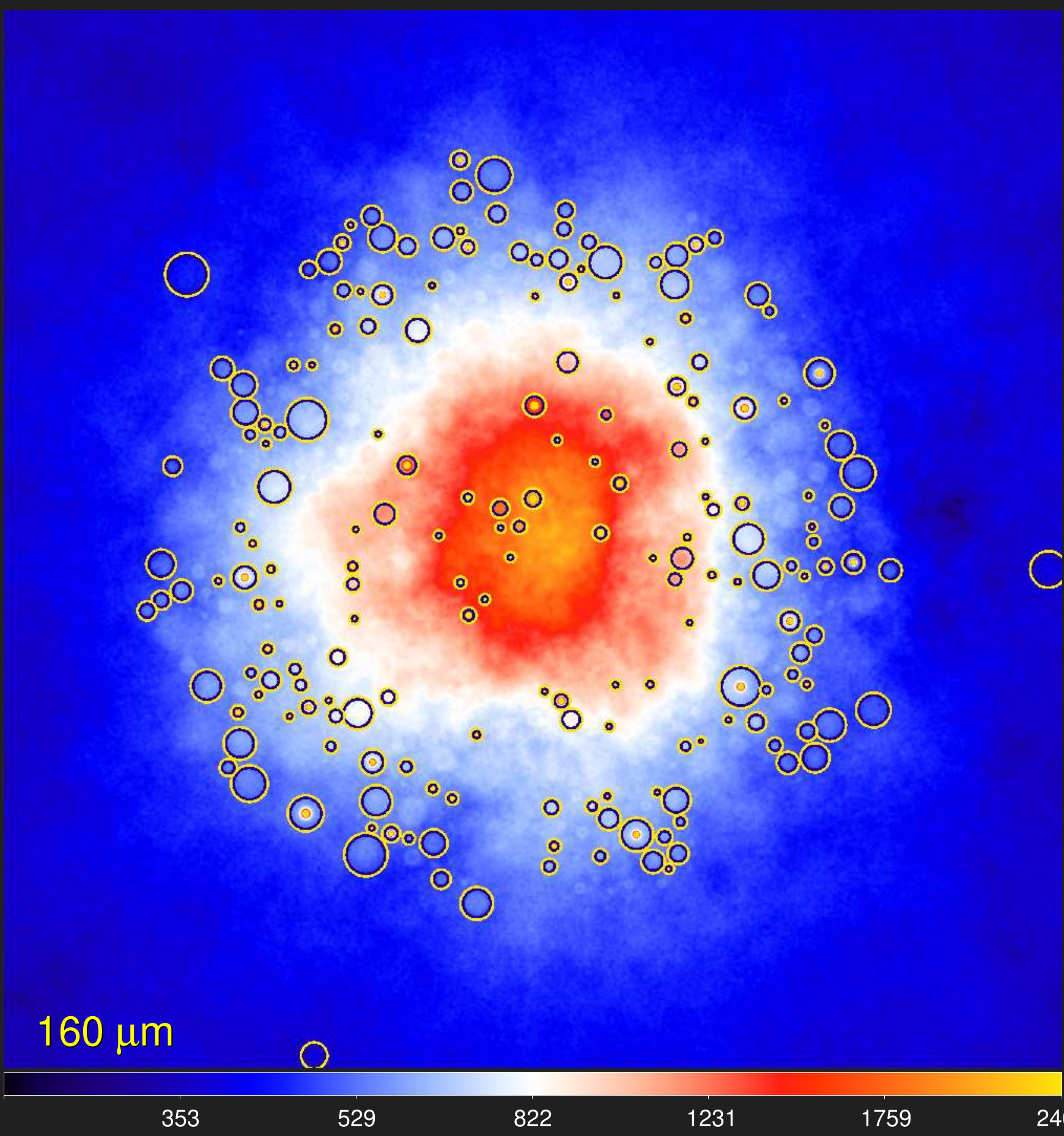}}
  \resizebox{0.328\hsize}{!}{\includegraphics{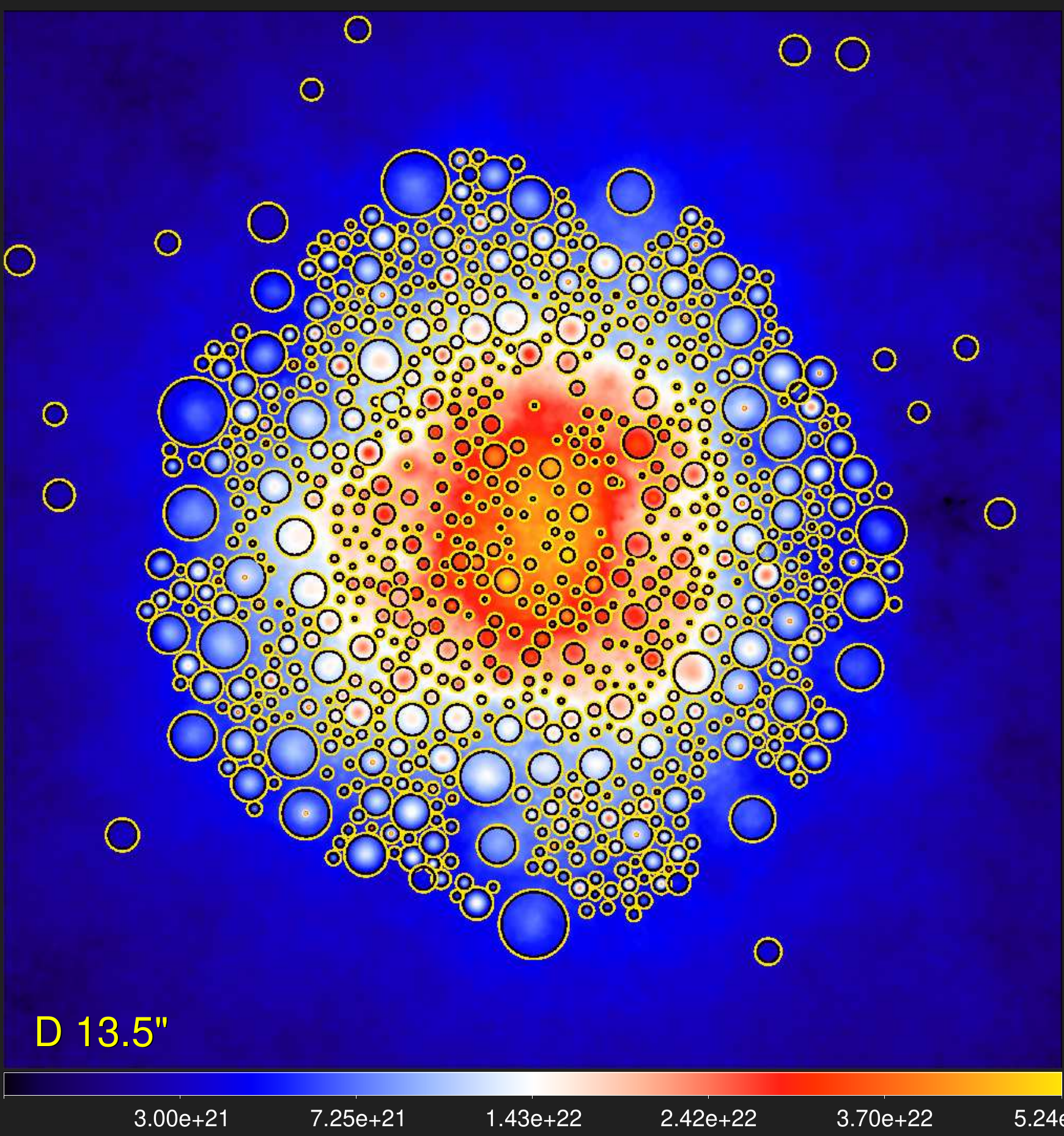}}
  \resizebox{0.328\hsize}{!}{\includegraphics{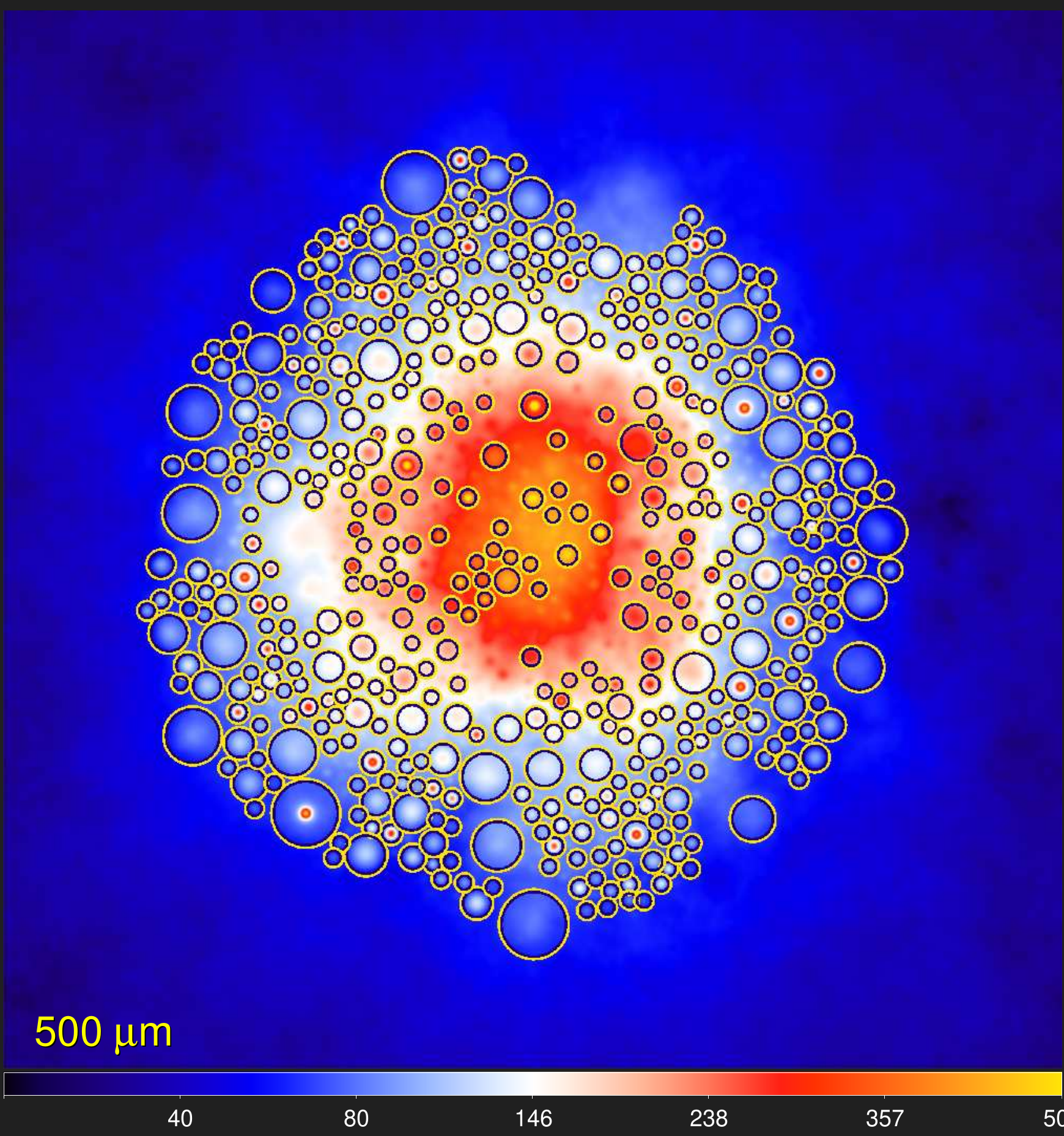}}}
\vspace{0.5mm}
\centerline{
  \resizebox{0.328\hsize}{!}{\includegraphics{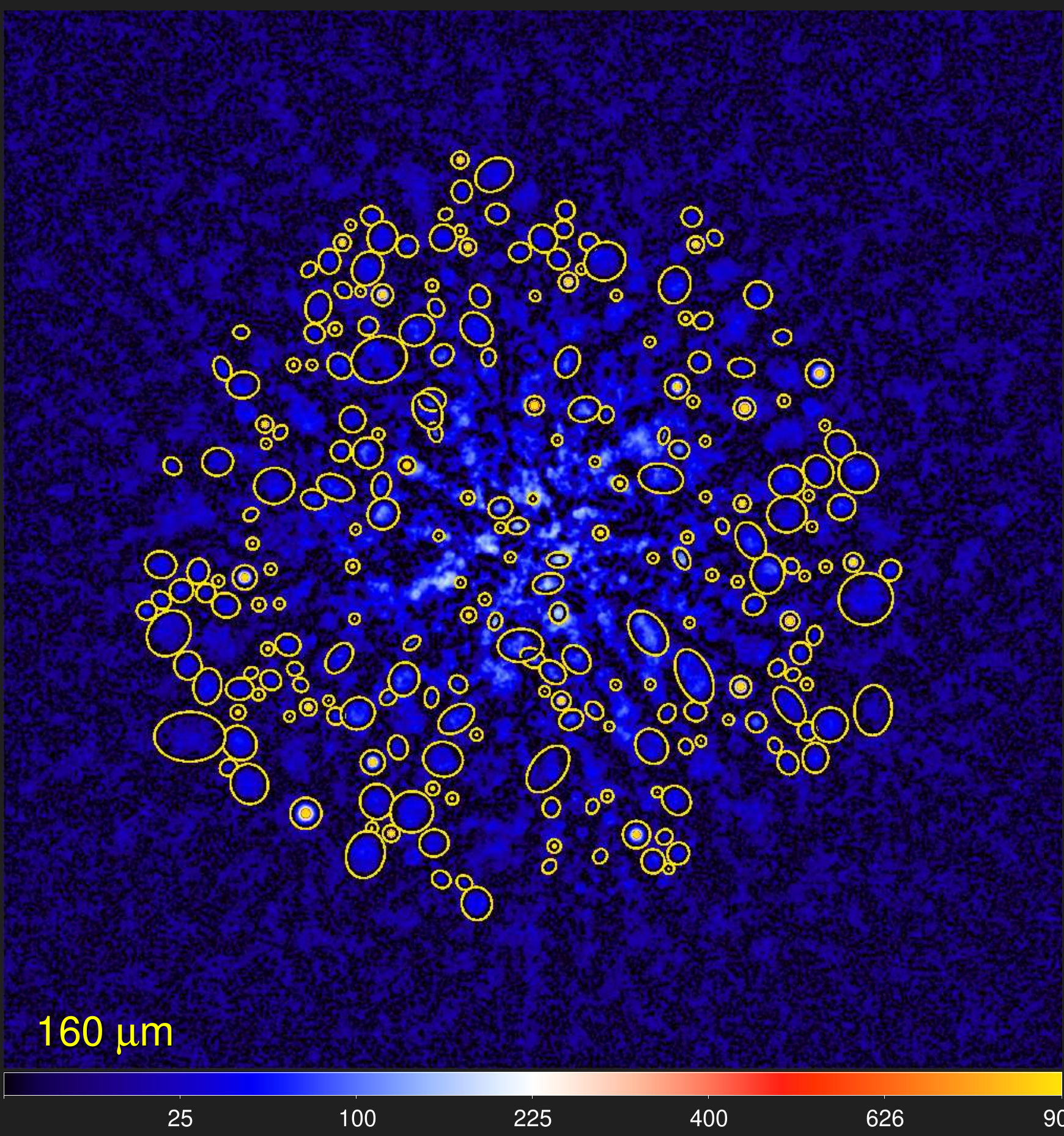}}
  \resizebox{0.328\hsize}{!}{\includegraphics{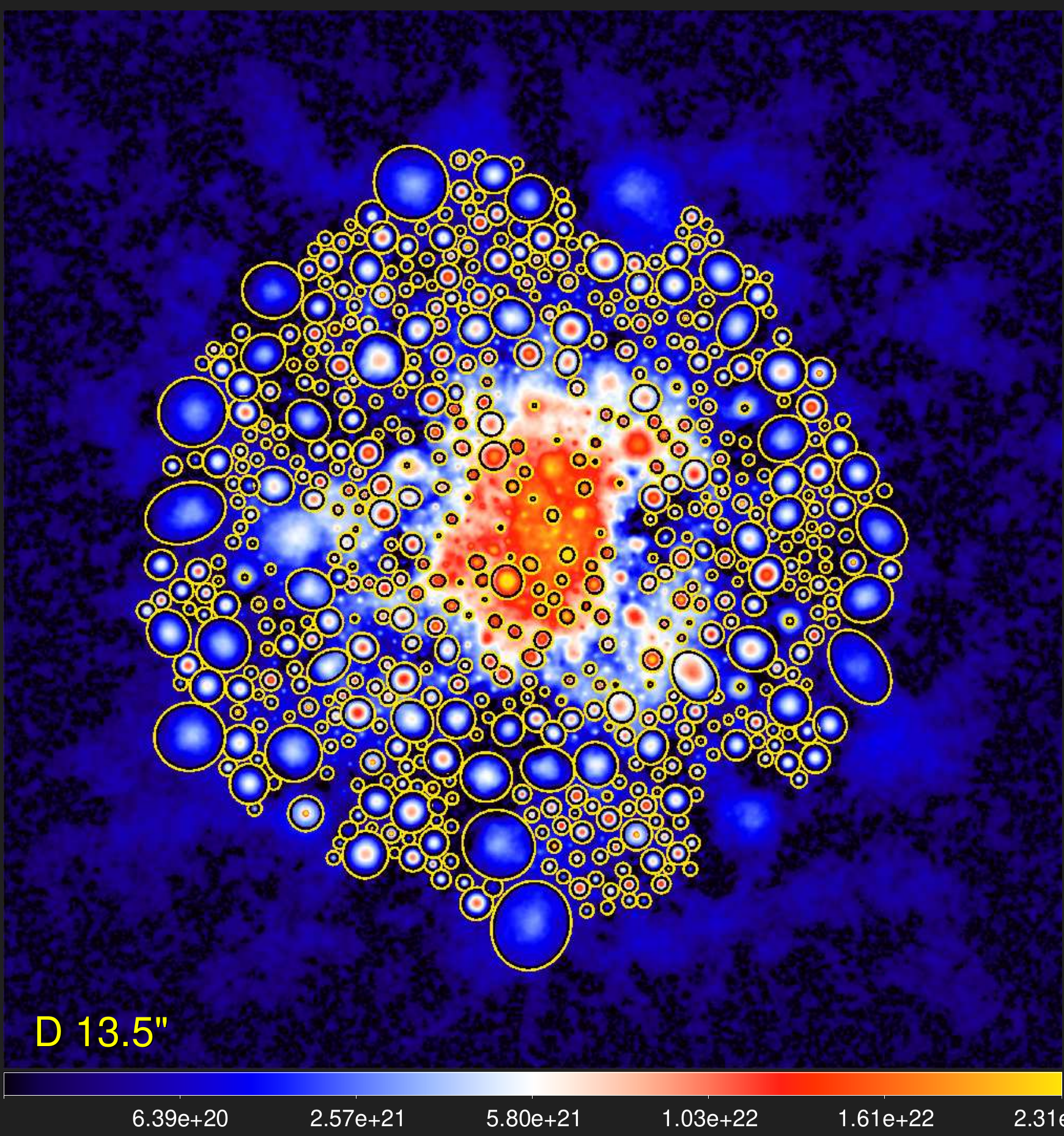}}
  \resizebox{0.328\hsize}{!}{\includegraphics{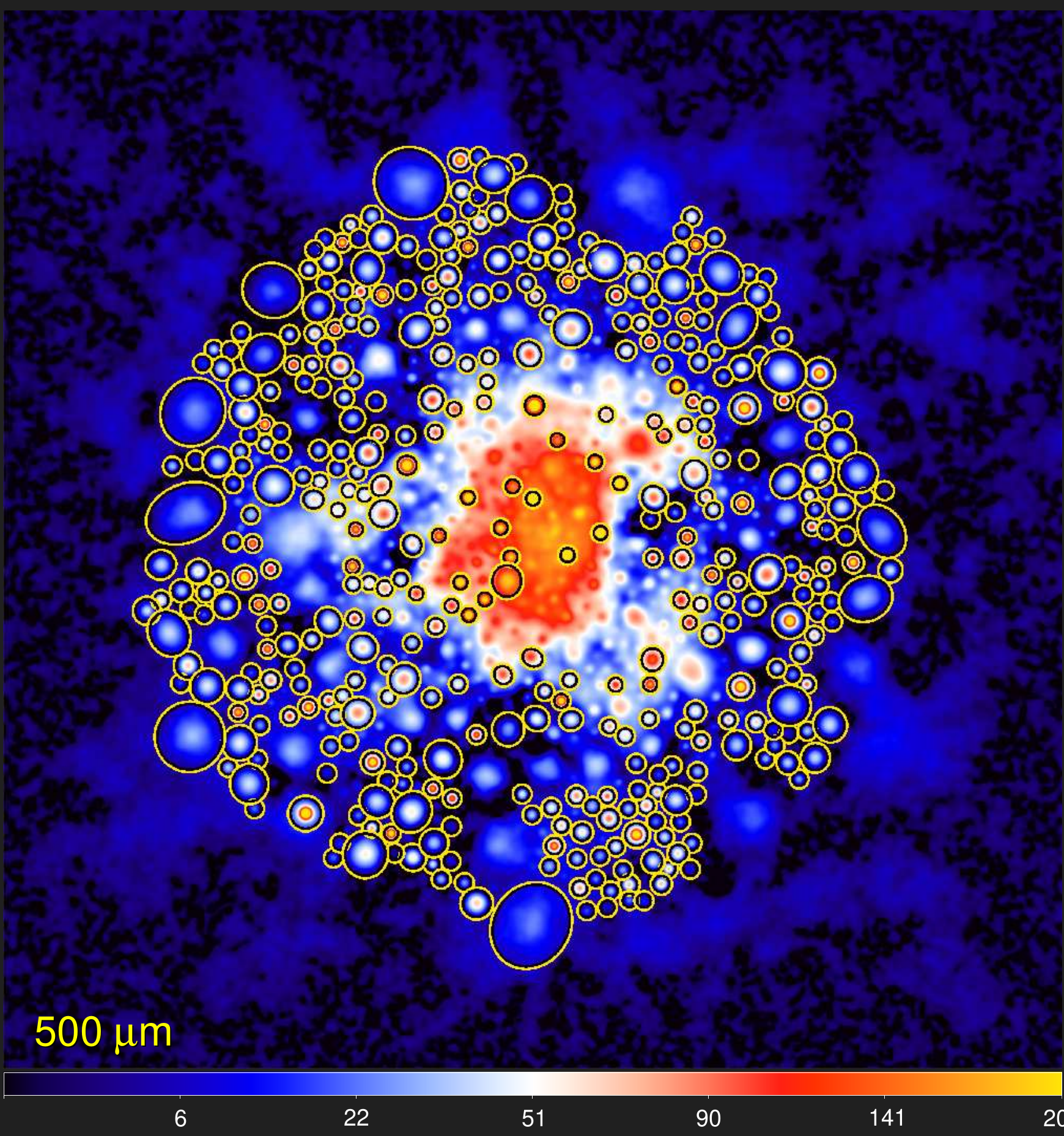}}}
\caption
{ 
Benchmark B$_3$ extraction of sources with \textsl{getsf} and \textsl{getold}. The original $\mathcal{I}_{\!\lambda}$ are overlaid
with the footprint ellipses from the measurement step. In the \textsl{getold} extraction (\emph{bottom}), the large-scale
background was determined and subtracted by \textsl{getimages}. The images are displayed with a square-root color mapping.
} 
\label{imagesB3}
\end{figure*}

The benchmark variants $\{\mathrm{A},\mathrm{B}\}_3$ (Figs. \ref{imagesA3} and \ref{imagesB3}) contain three components
(background, sources, and noise), adding fluctuating backgrounds to the sources and uniform noise of $\{\mathrm{A},\mathrm{B}\}_2$.
The background fluctuations in A$_3$ are similar in both diffuse and dense areas, whereas in B$_3$ they they progressively increase
in the denser areas. In the presence of the background clouds, more of the sources are expected to remain undetected and possibly
more spurious sources to become cataloged. Extraction methods may perform well in A$_3$ with its relatively simple background, but
some of them would experience greater problems in B$_3$. The benchmarks could present serious problems to those extraction tools
that are not designed to handle complex backgrounds.

\begin{figure*}                                                               
\centering
\centerline{
  \resizebox{0.328\hsize}{!}{\includegraphics{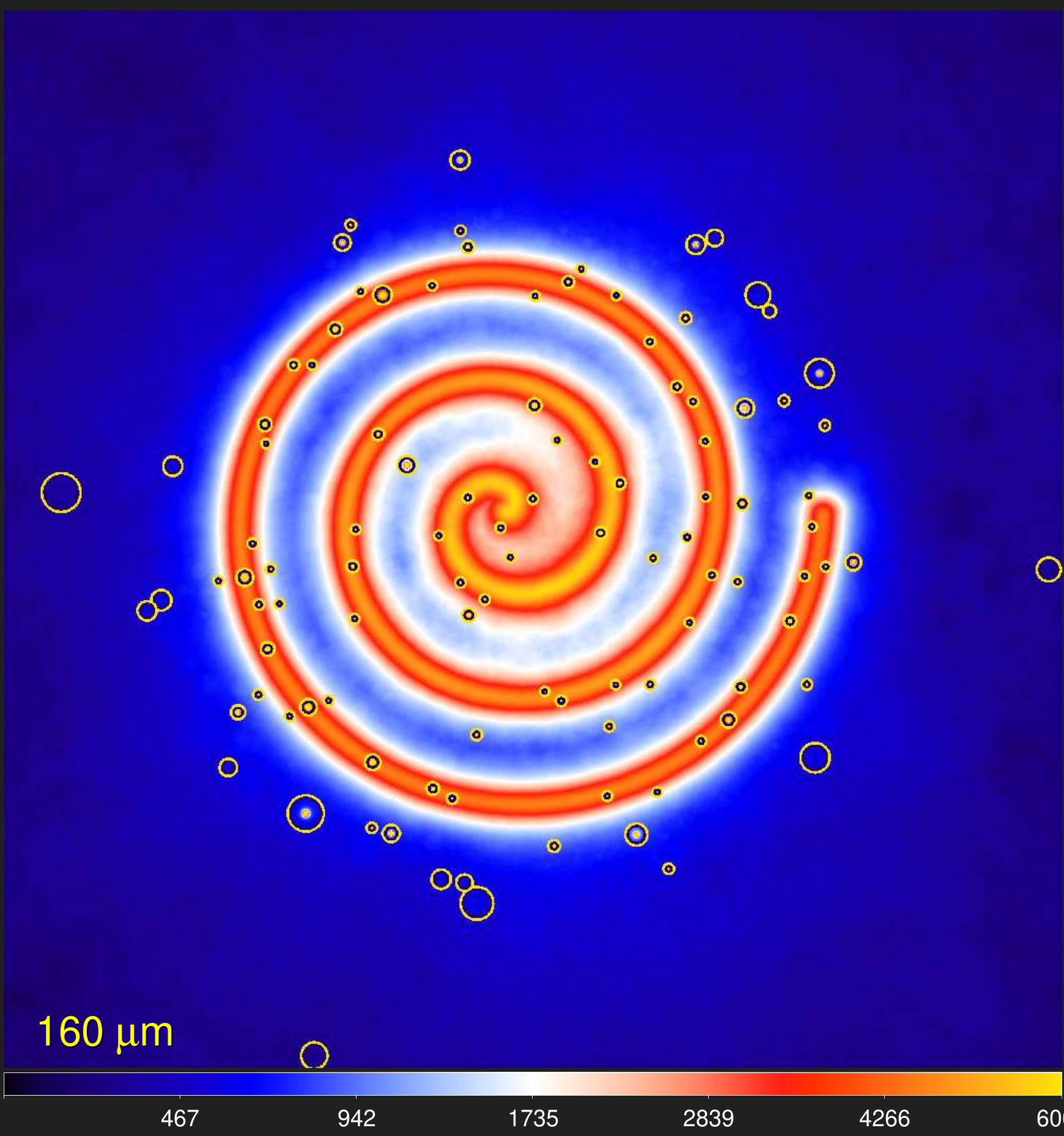}}
  \resizebox{0.328\hsize}{!}{\includegraphics{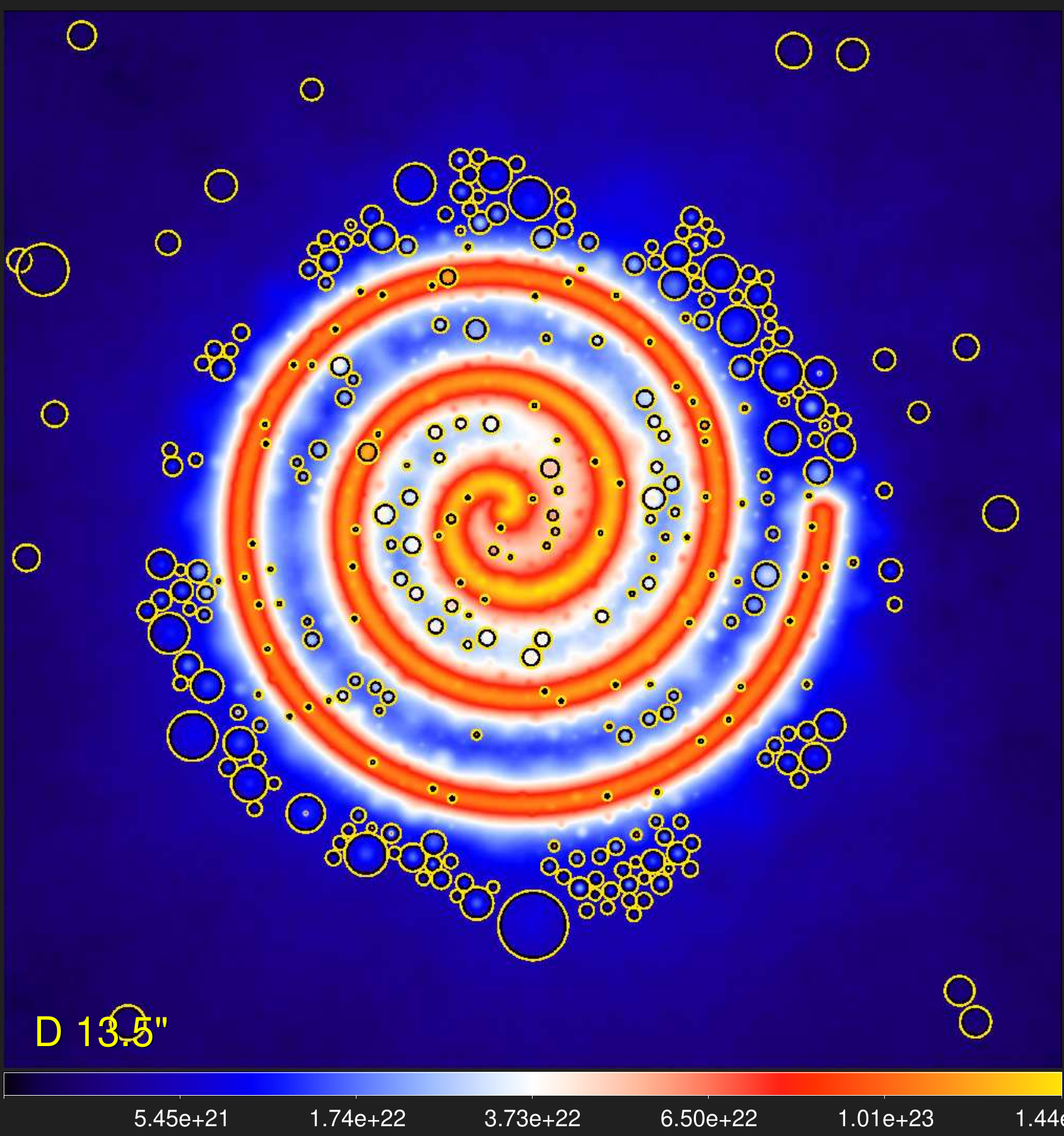}}
  \resizebox{0.328\hsize}{!}{\includegraphics{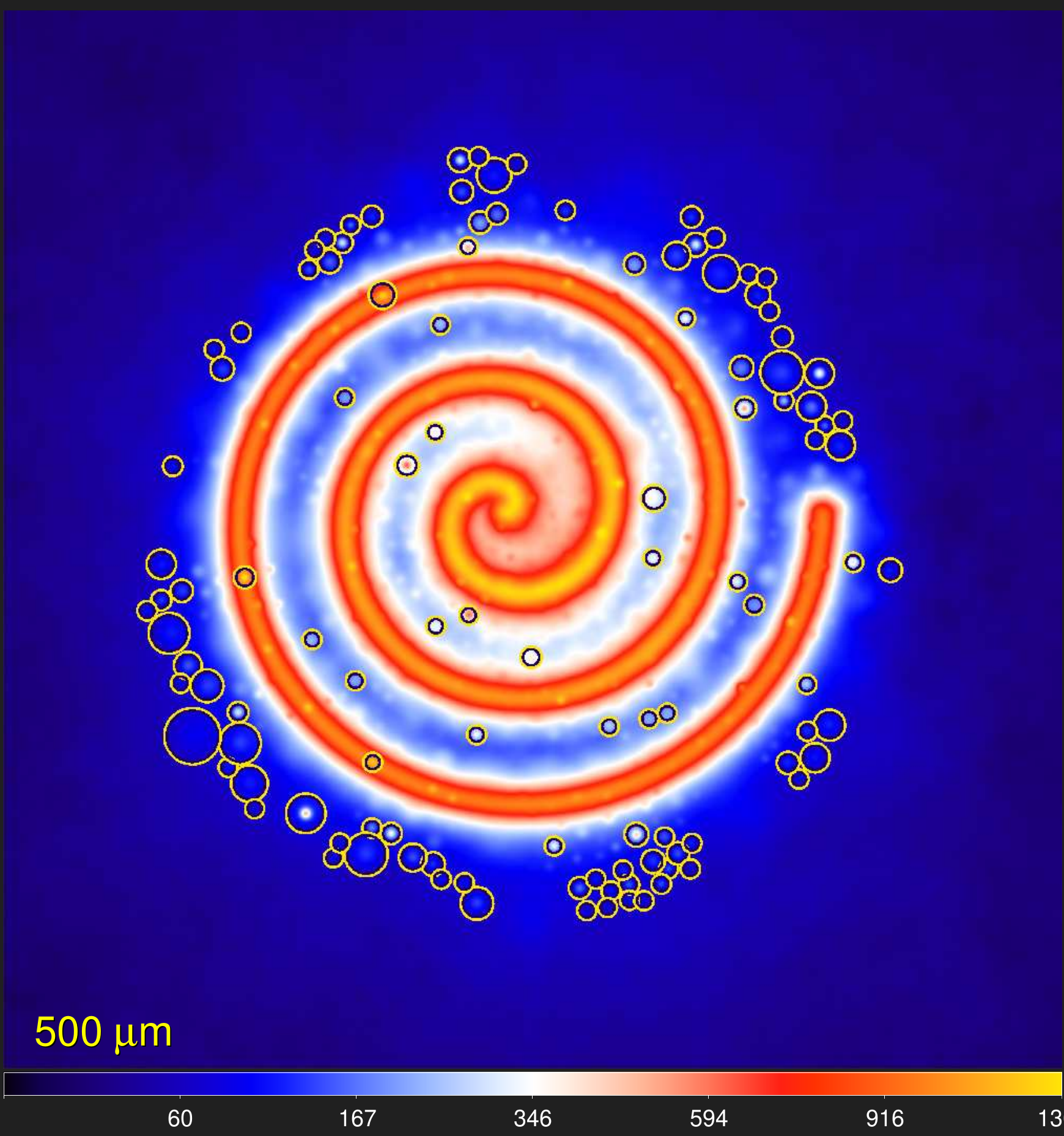}}}
\vspace{0.5mm}
\centerline{
  \resizebox{0.328\hsize}{!}{\includegraphics{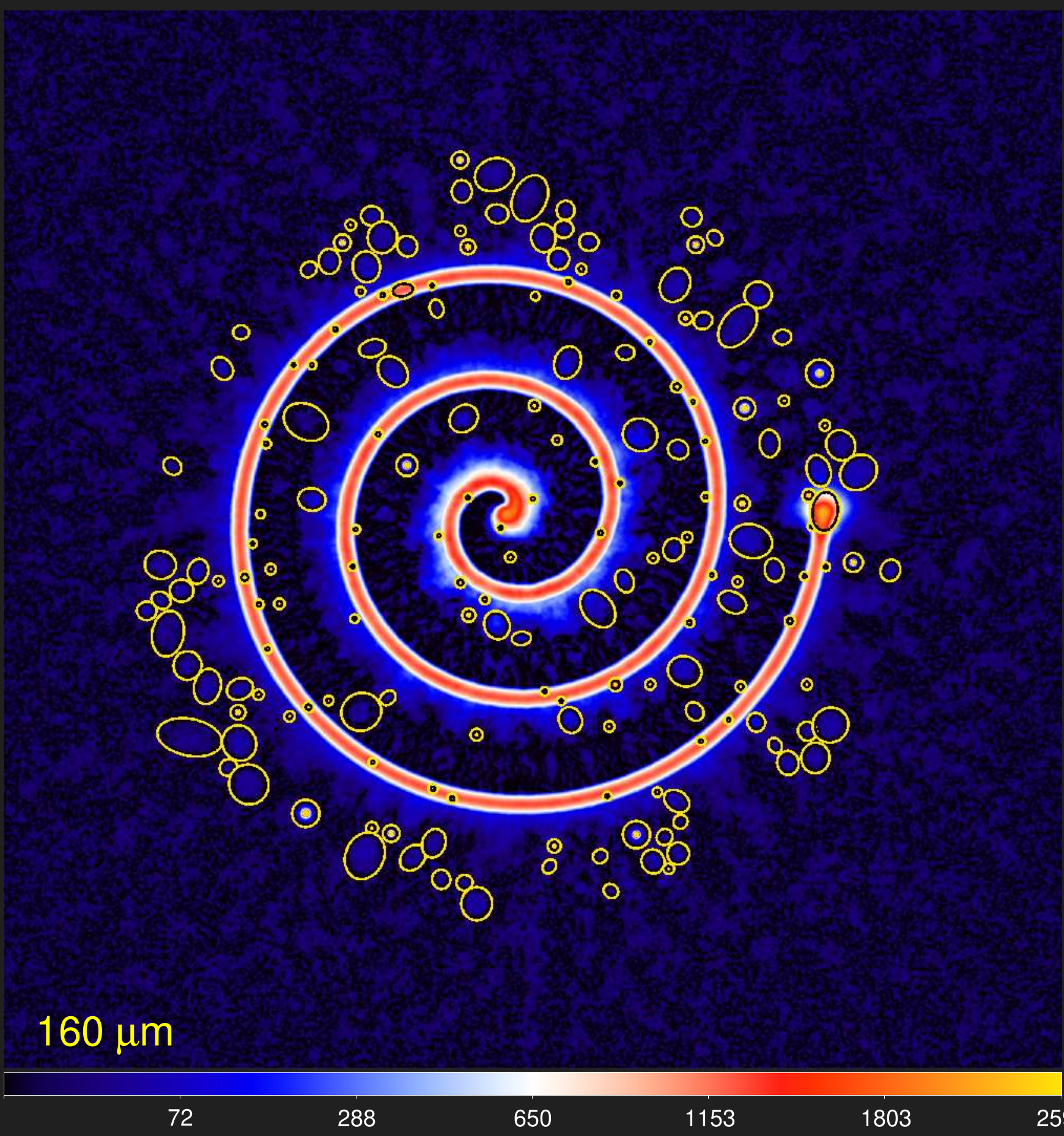}}
  \resizebox{0.328\hsize}{!}{\includegraphics{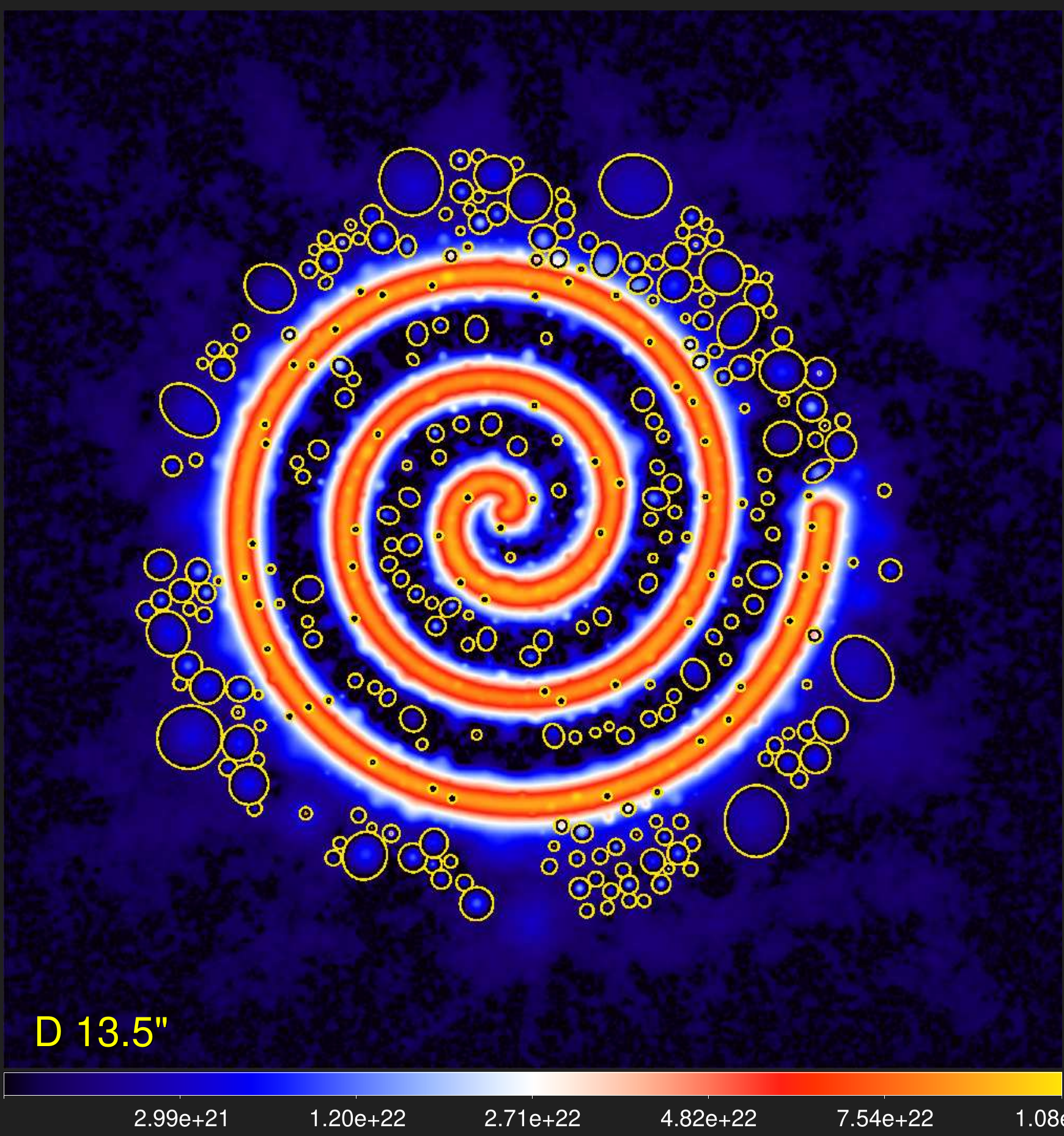}}
  \resizebox{0.328\hsize}{!}{\includegraphics{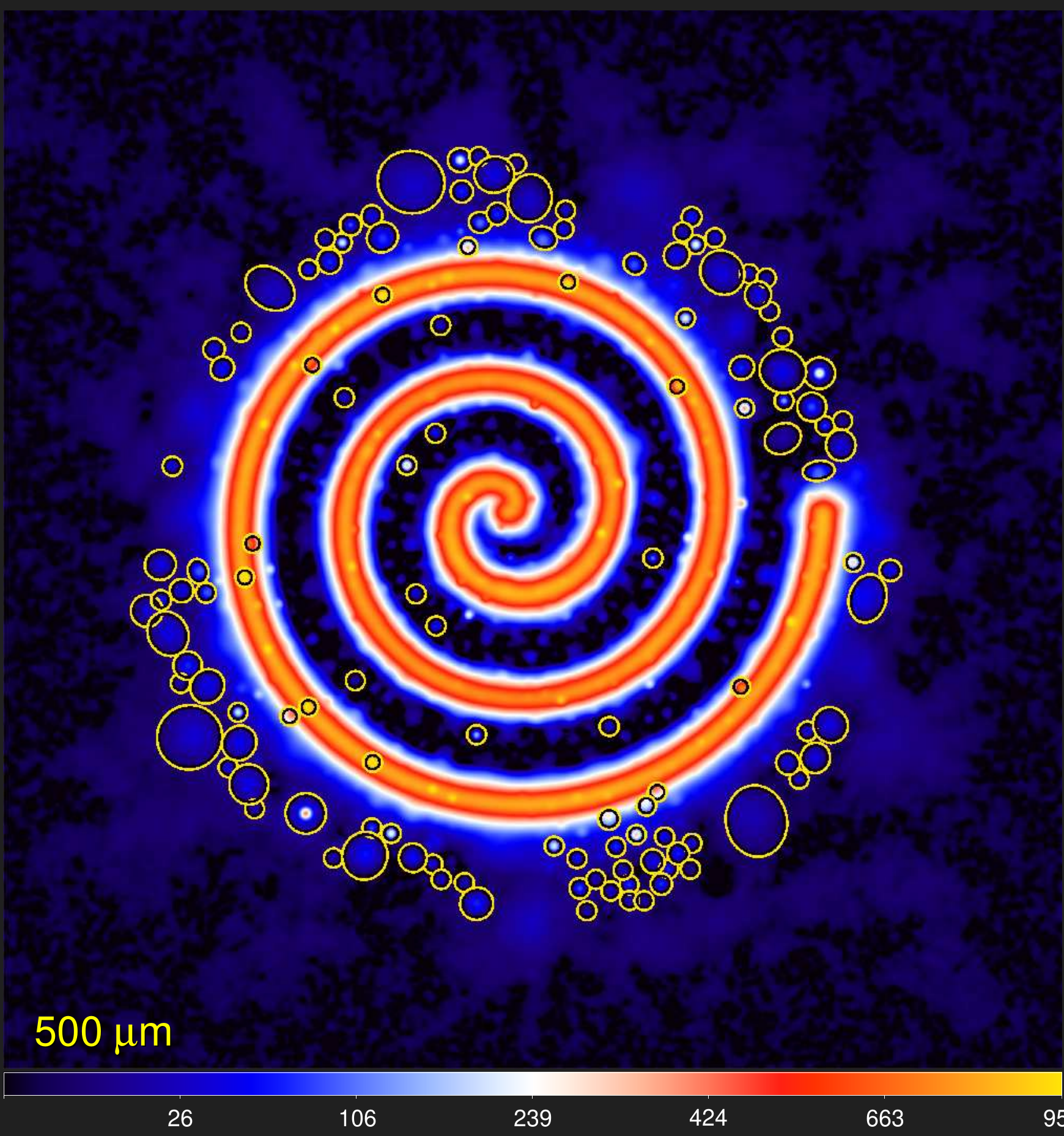}}}
\caption
{ 
Benchmark B$_4$ extraction of sources with \textsl{getsf} and \textsl{getold}. The original $\mathcal{I}_{\!\lambda}$ are overlaid
with the footprint ellipses from the measurement step. In the \textsl{getold} extraction (\emph{bottom}), the large-scale
background was determined and subtracted by \textsl{getimages}. The images are displayed with a square-root color mapping.
} 
\label{imagesB4}
\end{figure*}

The most complex variant B$_4$ (Fig. \ref{imagesB4}) contains 4 components (background, filament, sources, and noise), adding the
dense spiral filament to the structural components of B$_3$. The filamentary background of the sources becomes much denser and it
acquires markedly different anisotropic properties (e.g., along the filament crest and in the orthogonal directions), in addition
to the strong and nonuniform background fluctuations of B$_3$. Better resembling the complexity of the interstellar clouds revealed
by the \emph{Herschel} observations, it further complicates the source extraction problem. Among all benchmark variants, the
largest numbers of model sources are expected to vanish in the filamentary background cloud of B$_4$.


\begin{figure*}
\centering
\centerline{\resizebox{0.2695\hsize}{!}{\includegraphics{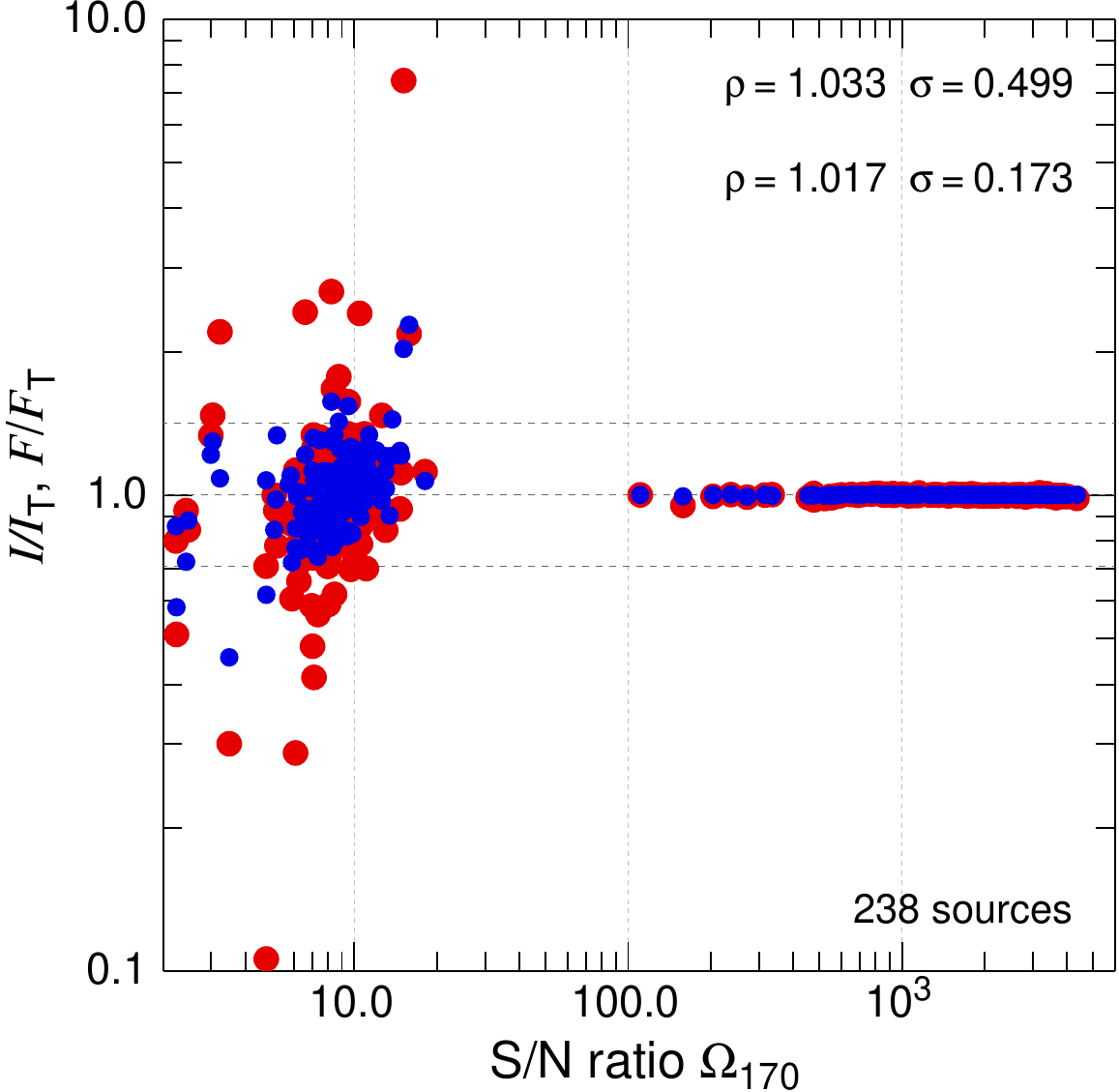}}
            \resizebox{0.2315\hsize}{!}{\includegraphics{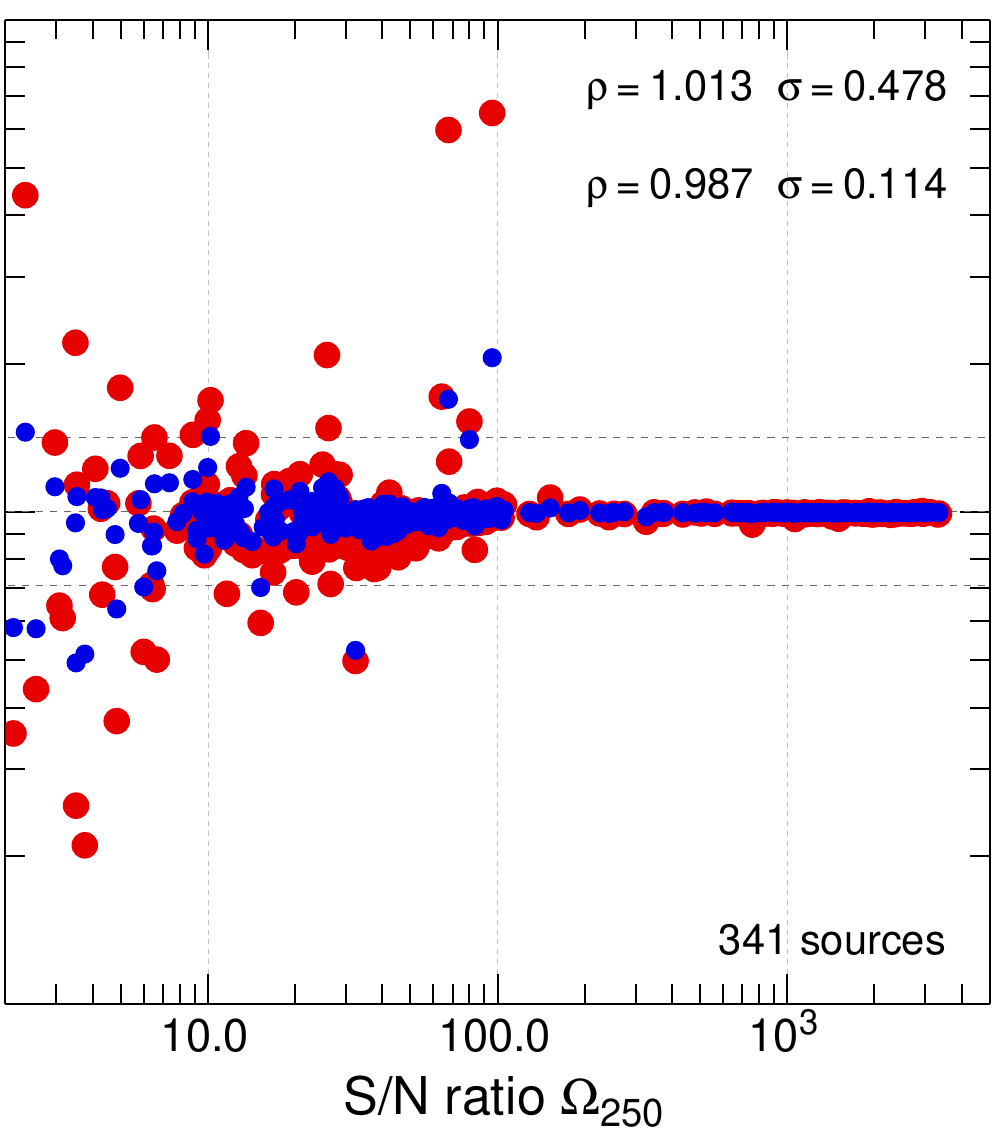}}
            \resizebox{0.2315\hsize}{!}{\includegraphics{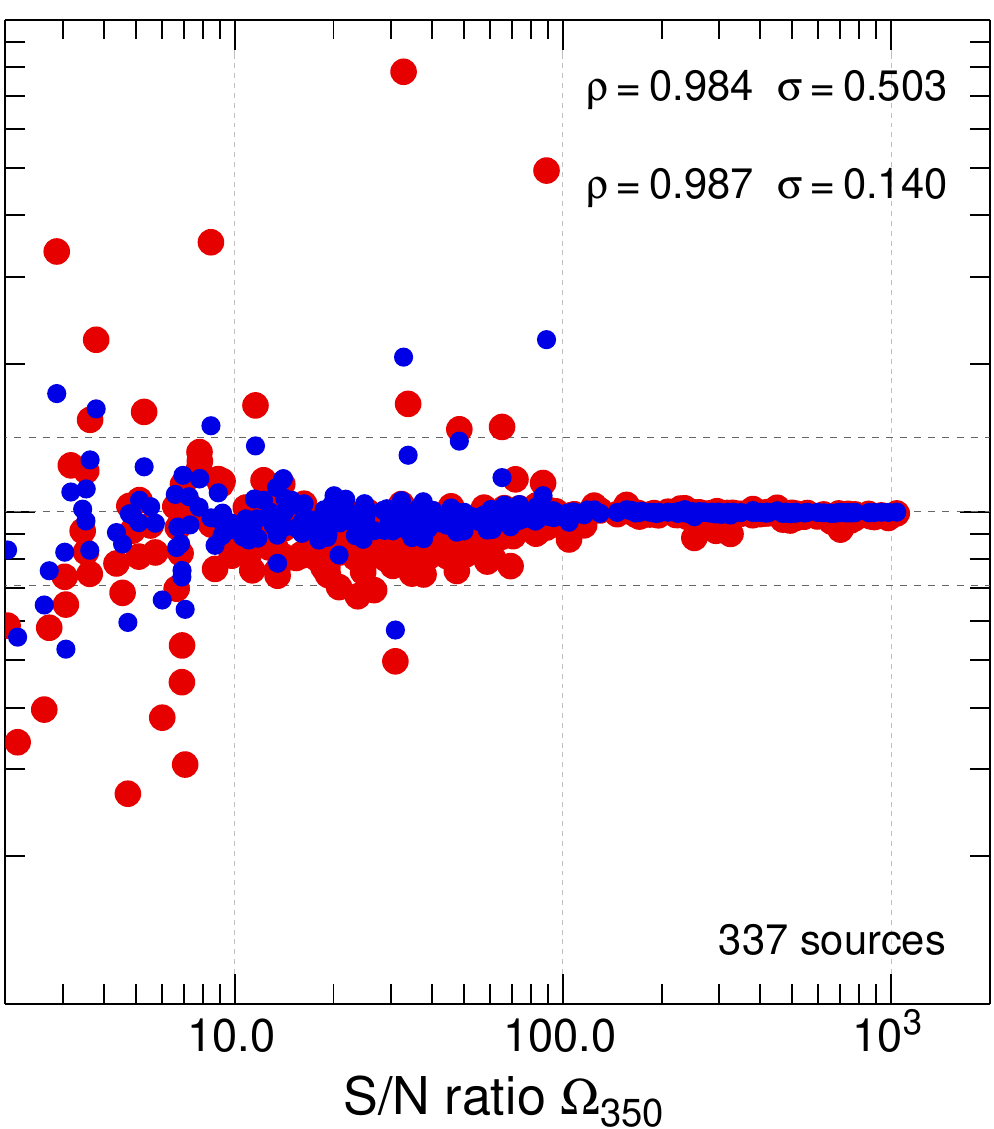}}
            \resizebox{0.2435\hsize}{!}{\includegraphics{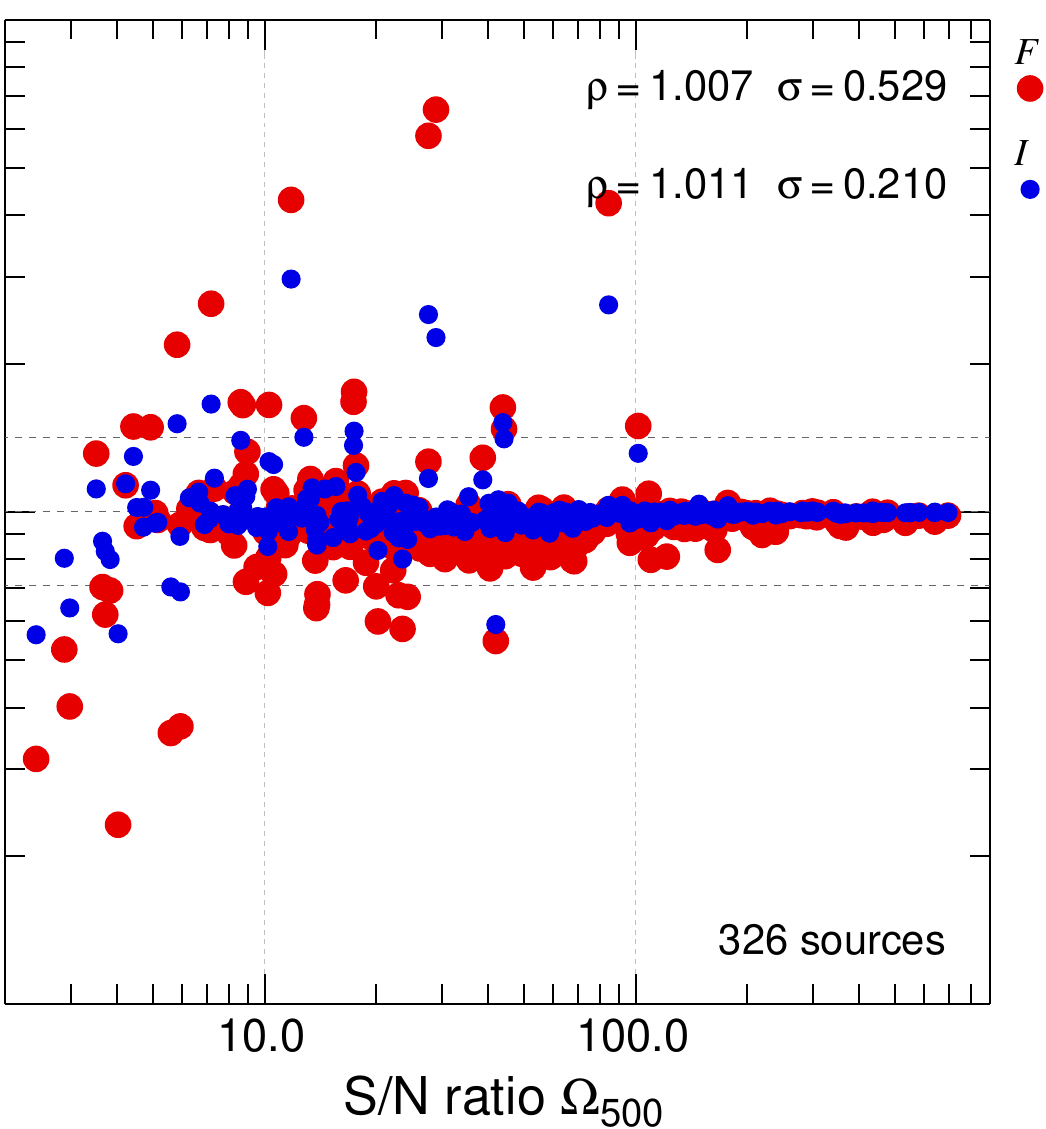}}}
\vspace{1.0mm}
\centerline{\resizebox{0.2695\hsize}{!}{\includegraphics{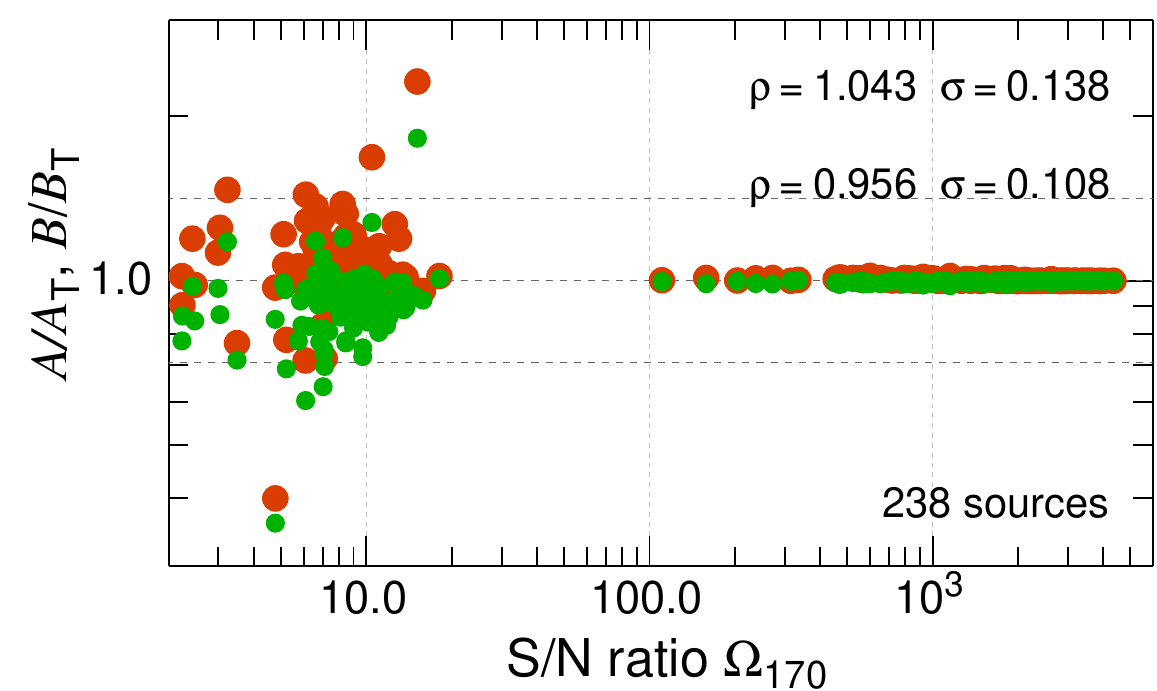}}
            \resizebox{0.2315\hsize}{!}{\includegraphics{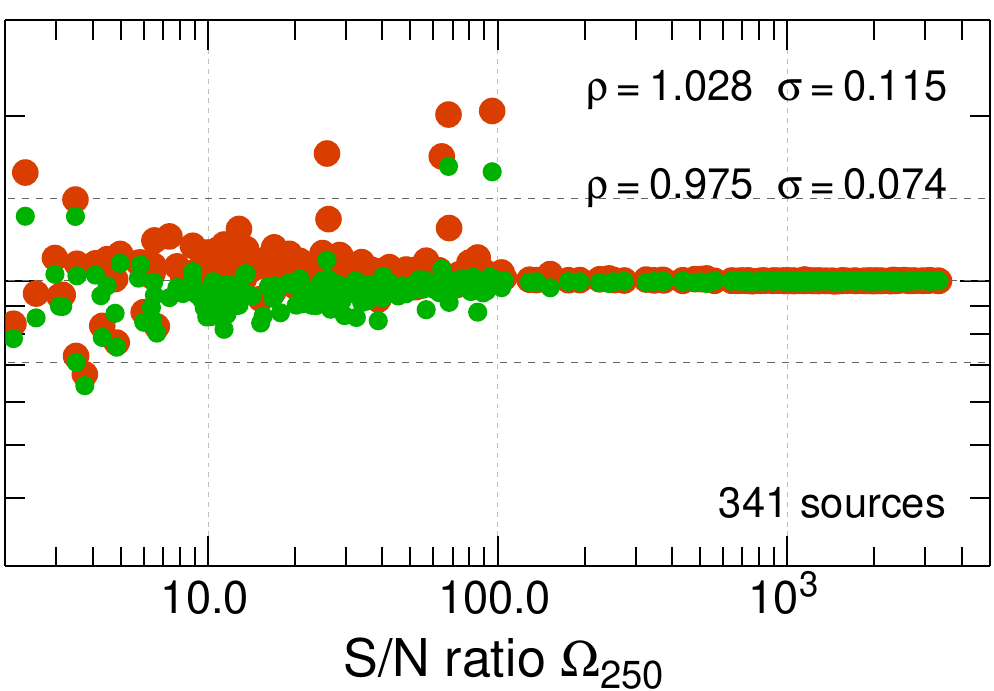}}
            \resizebox{0.2315\hsize}{!}{\includegraphics{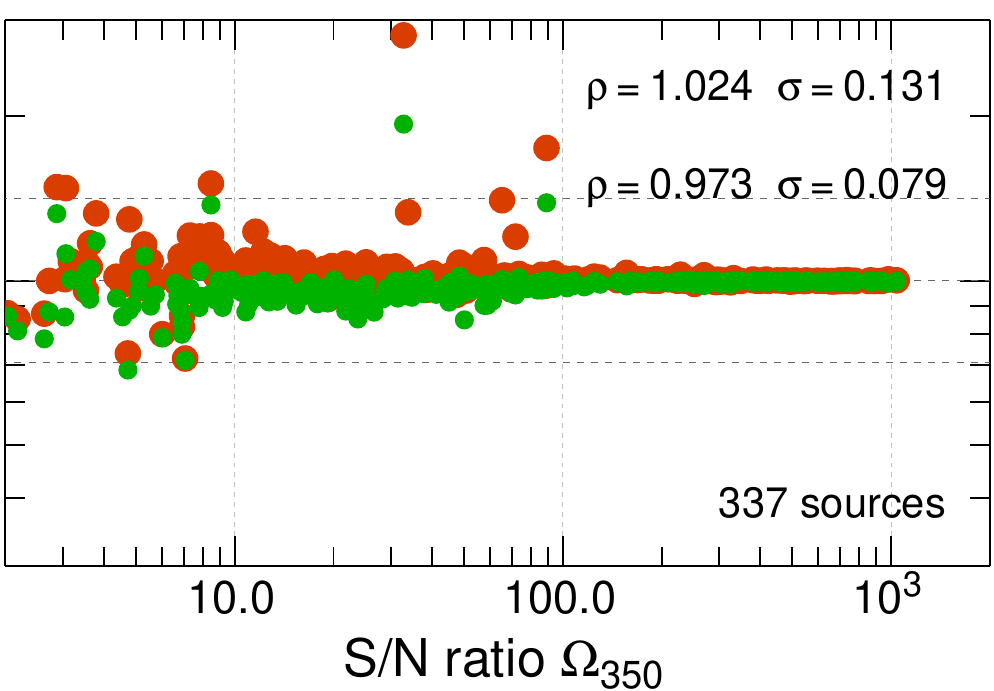}}
            \resizebox{0.2435\hsize}{!}{\includegraphics{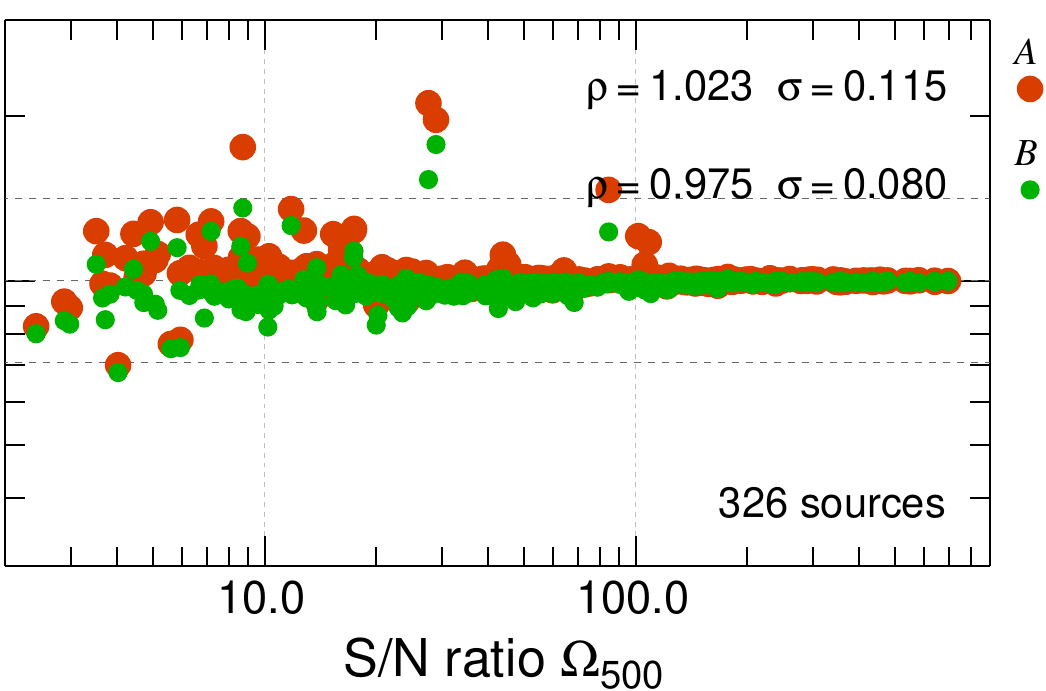}}}
\vspace{1.0mm}
\centerline{\resizebox{0.2695\hsize}{!}{\includegraphics{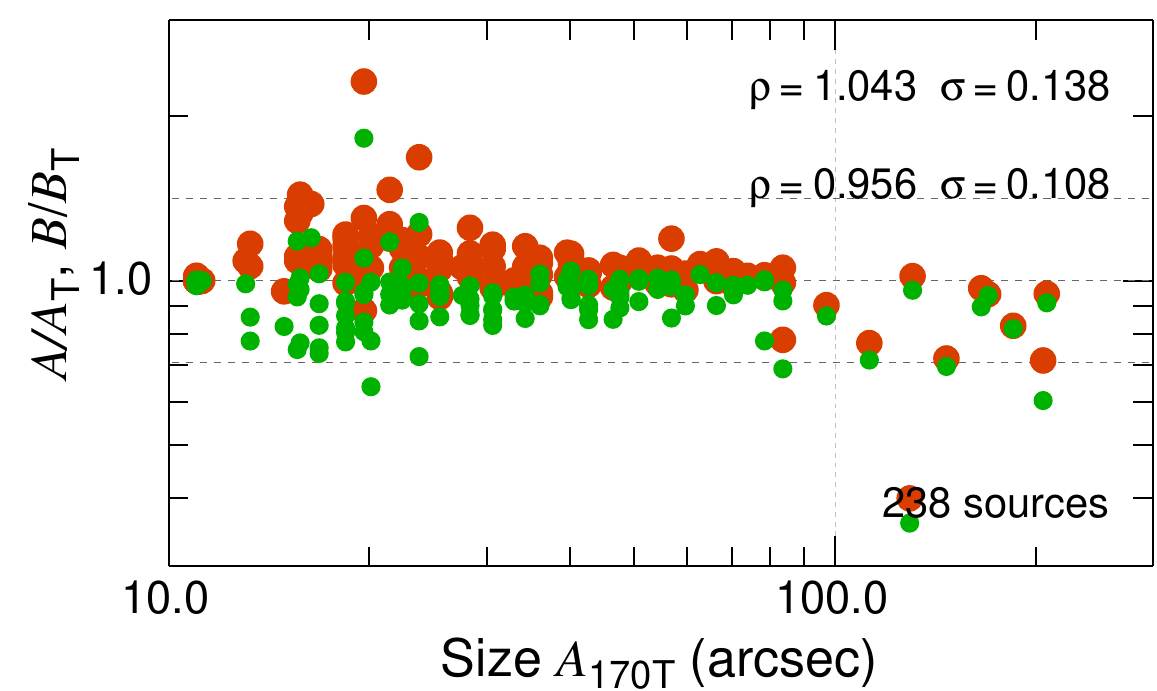}}
            \resizebox{0.2315\hsize}{!}{\includegraphics{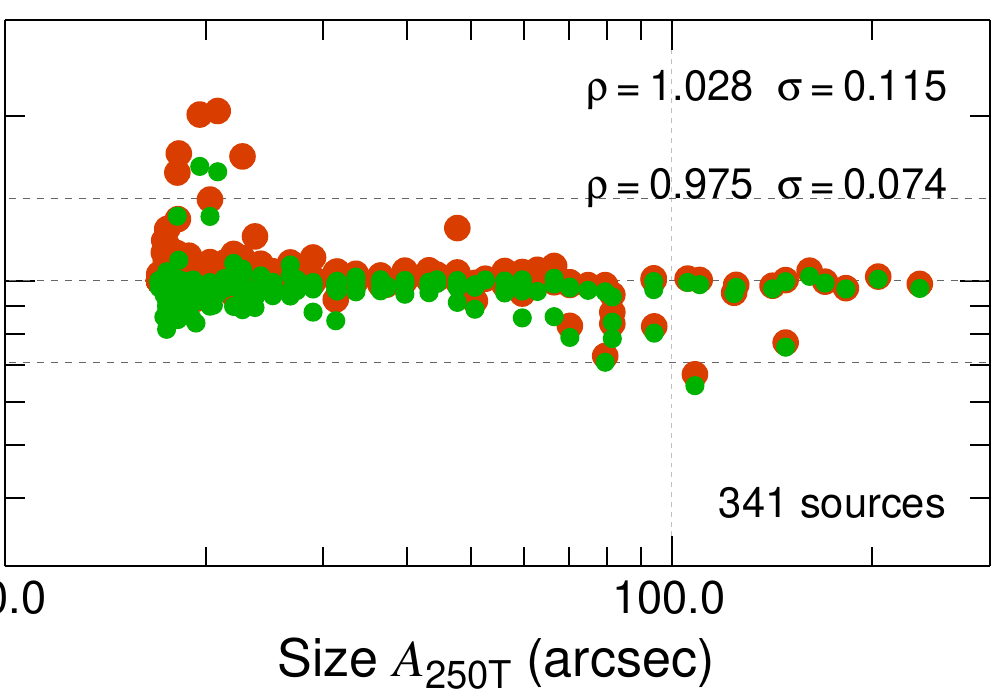}}
            \resizebox{0.2315\hsize}{!}{\includegraphics{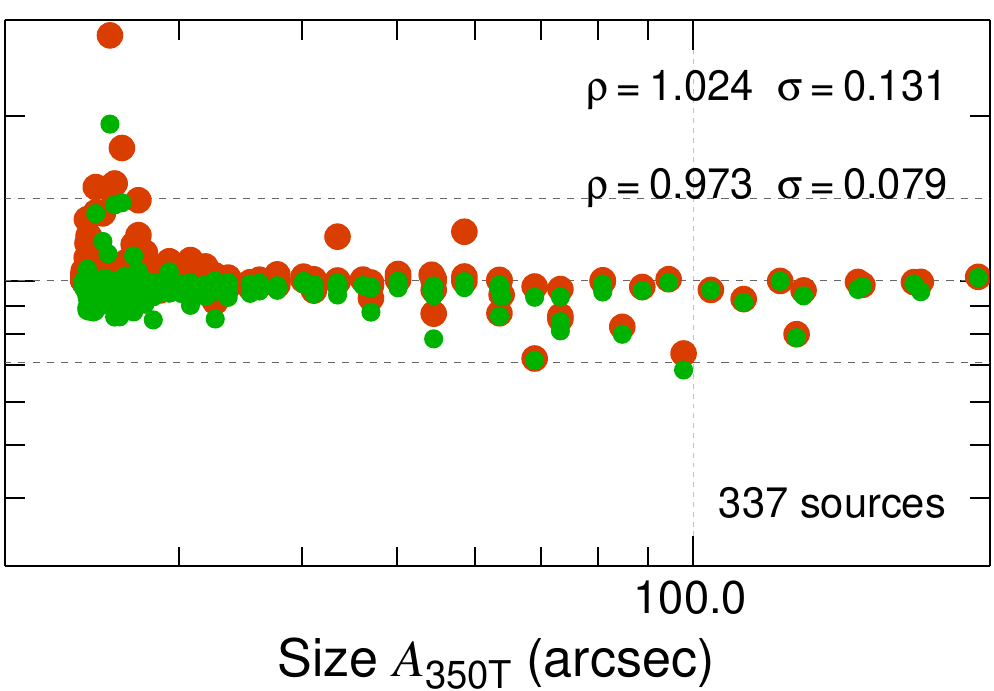}}
            \resizebox{0.2435\hsize}{!}{\includegraphics{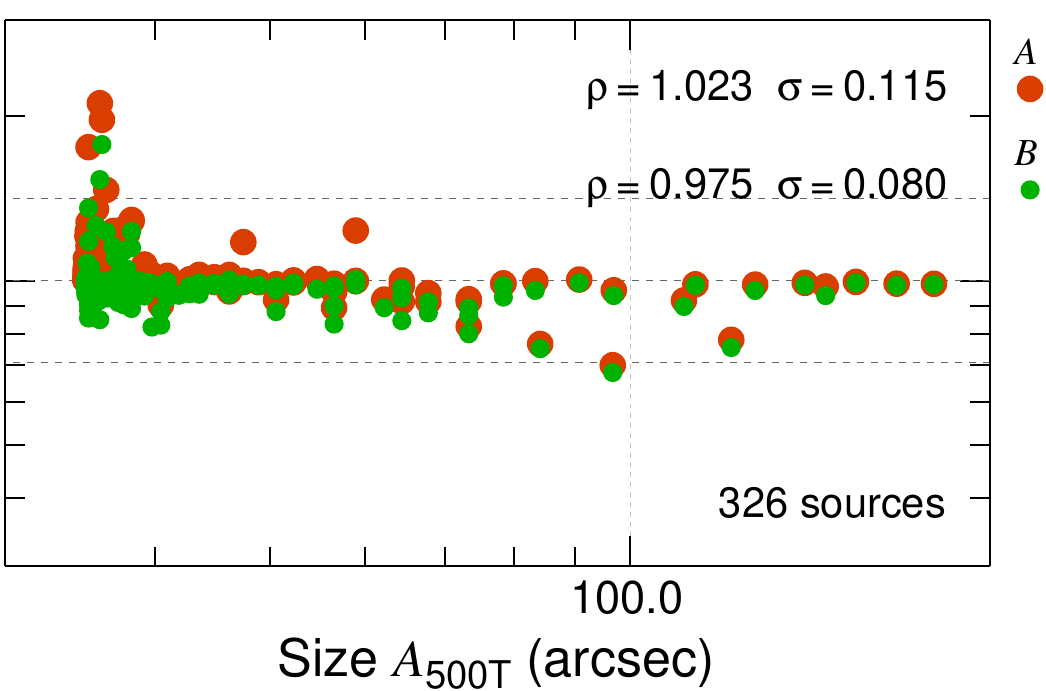}}}
\vspace{2.3mm}
\centerline{\resizebox{0.2695\hsize}{!}{\includegraphics{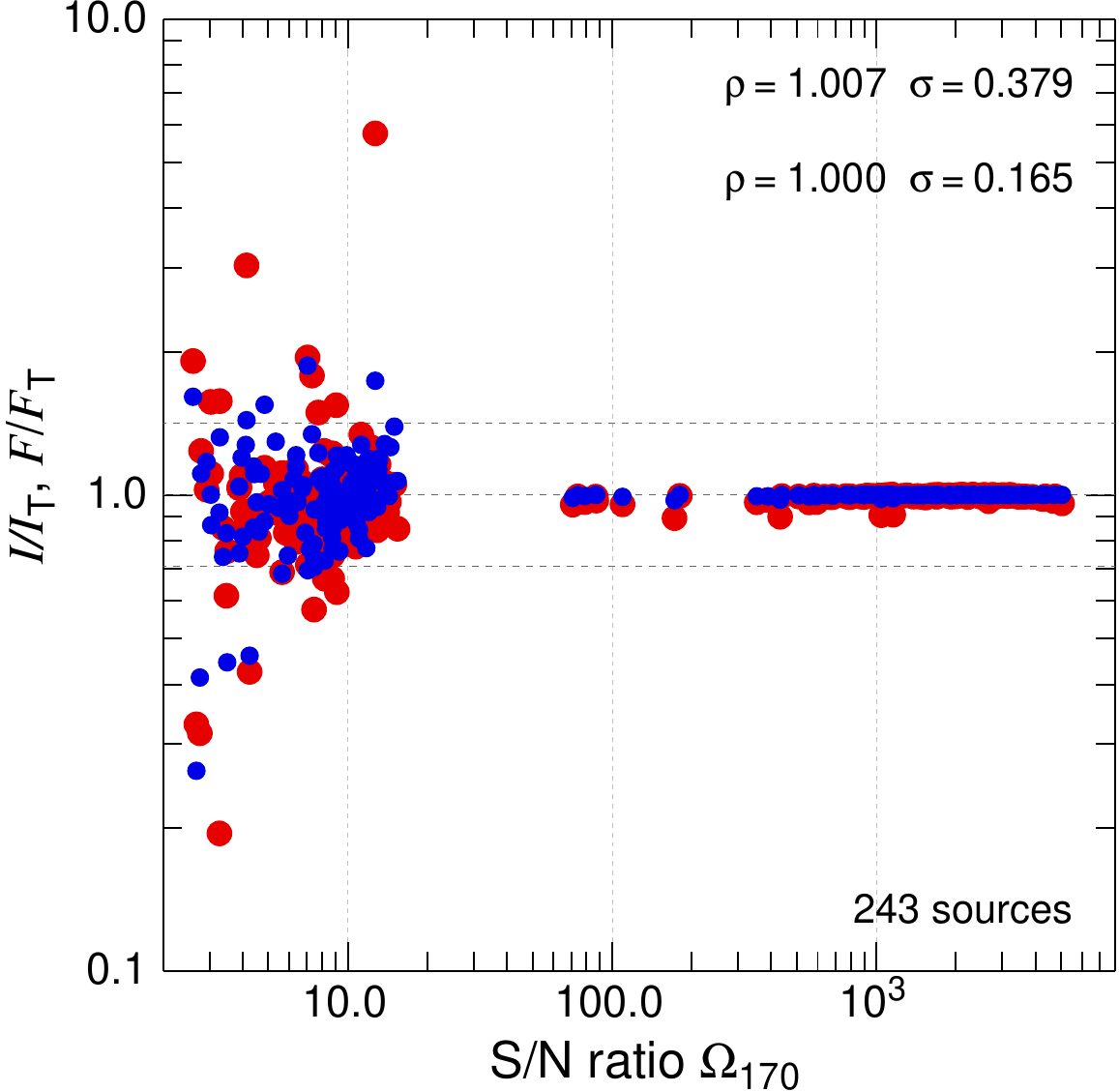}}
            \resizebox{0.2315\hsize}{!}{\includegraphics{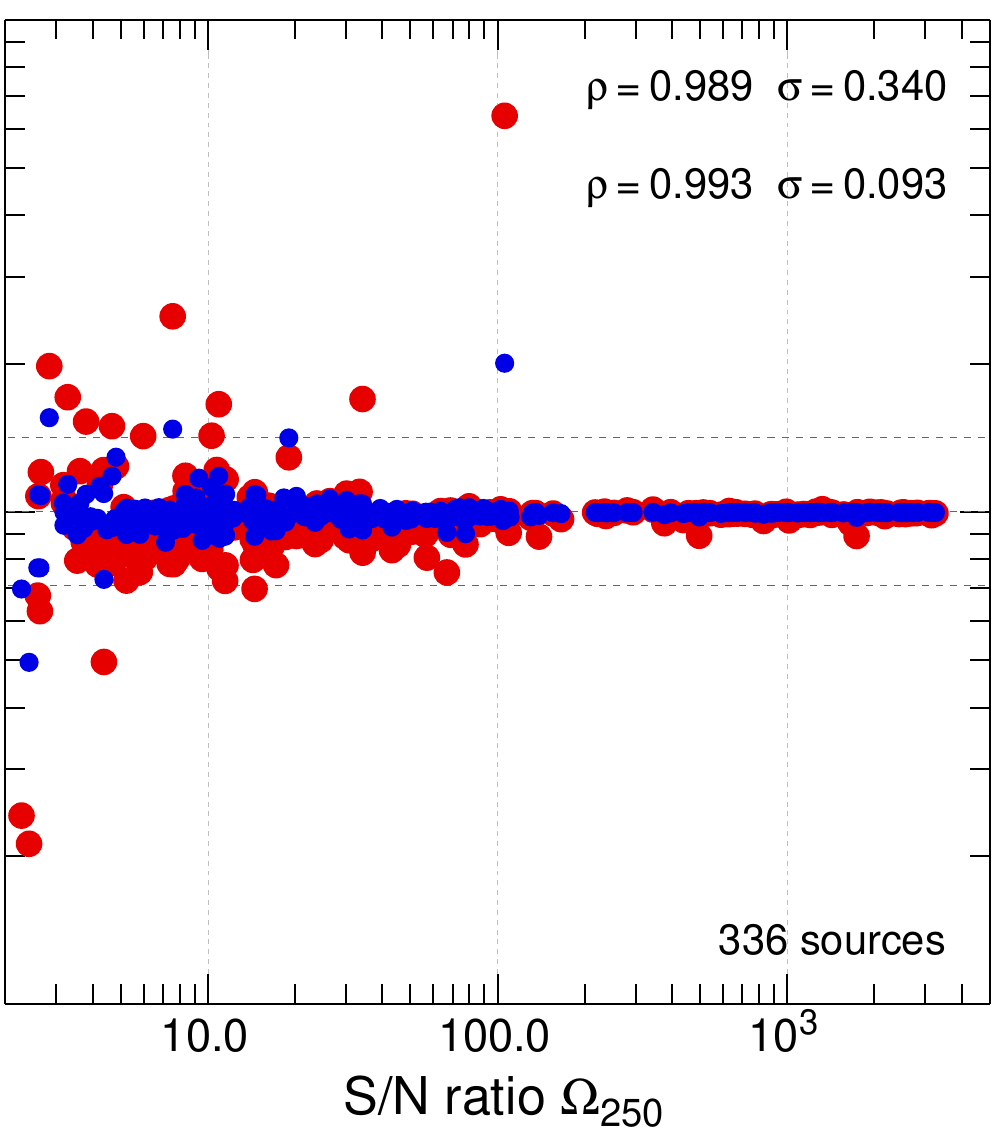}}
            \resizebox{0.2315\hsize}{!}{\includegraphics{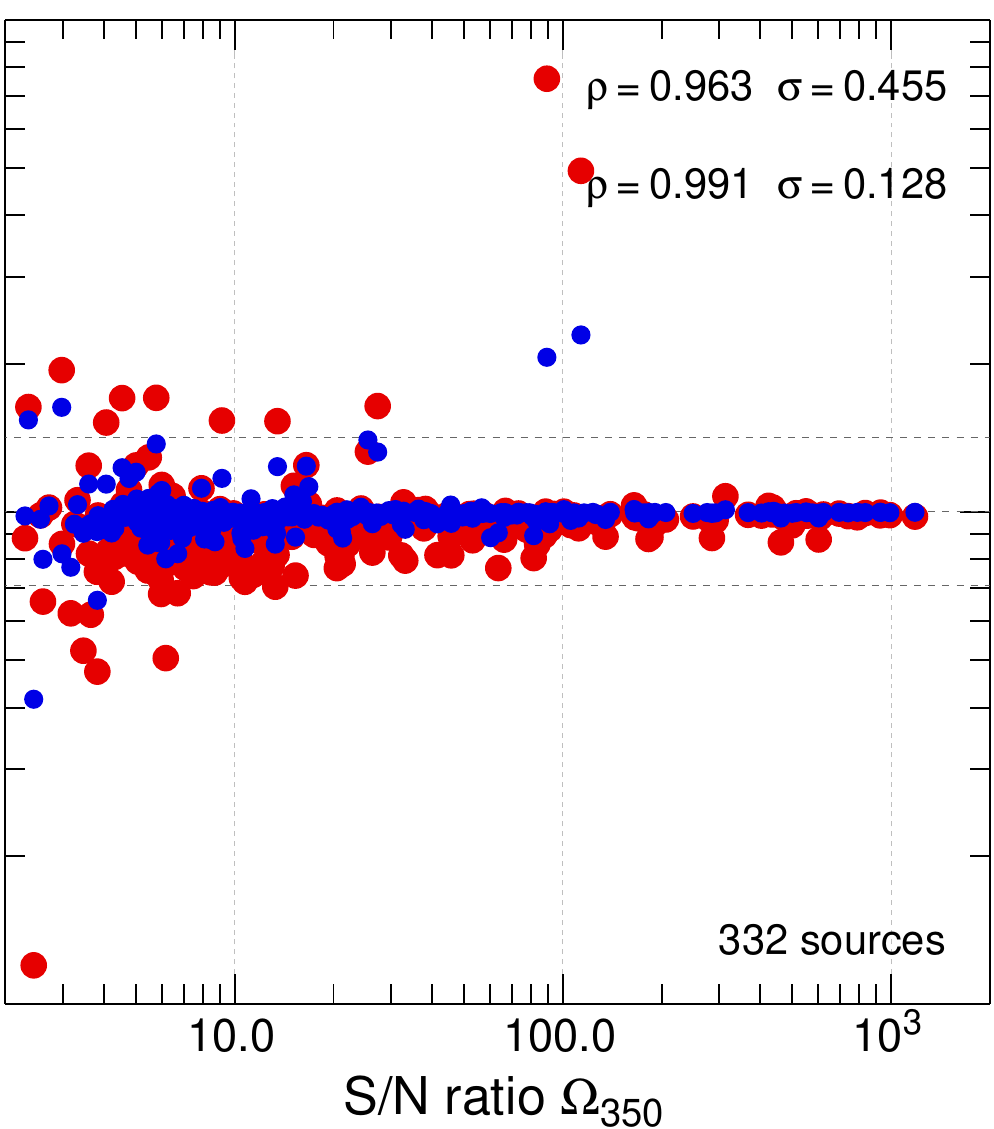}}
            \resizebox{0.2435\hsize}{!}{\includegraphics{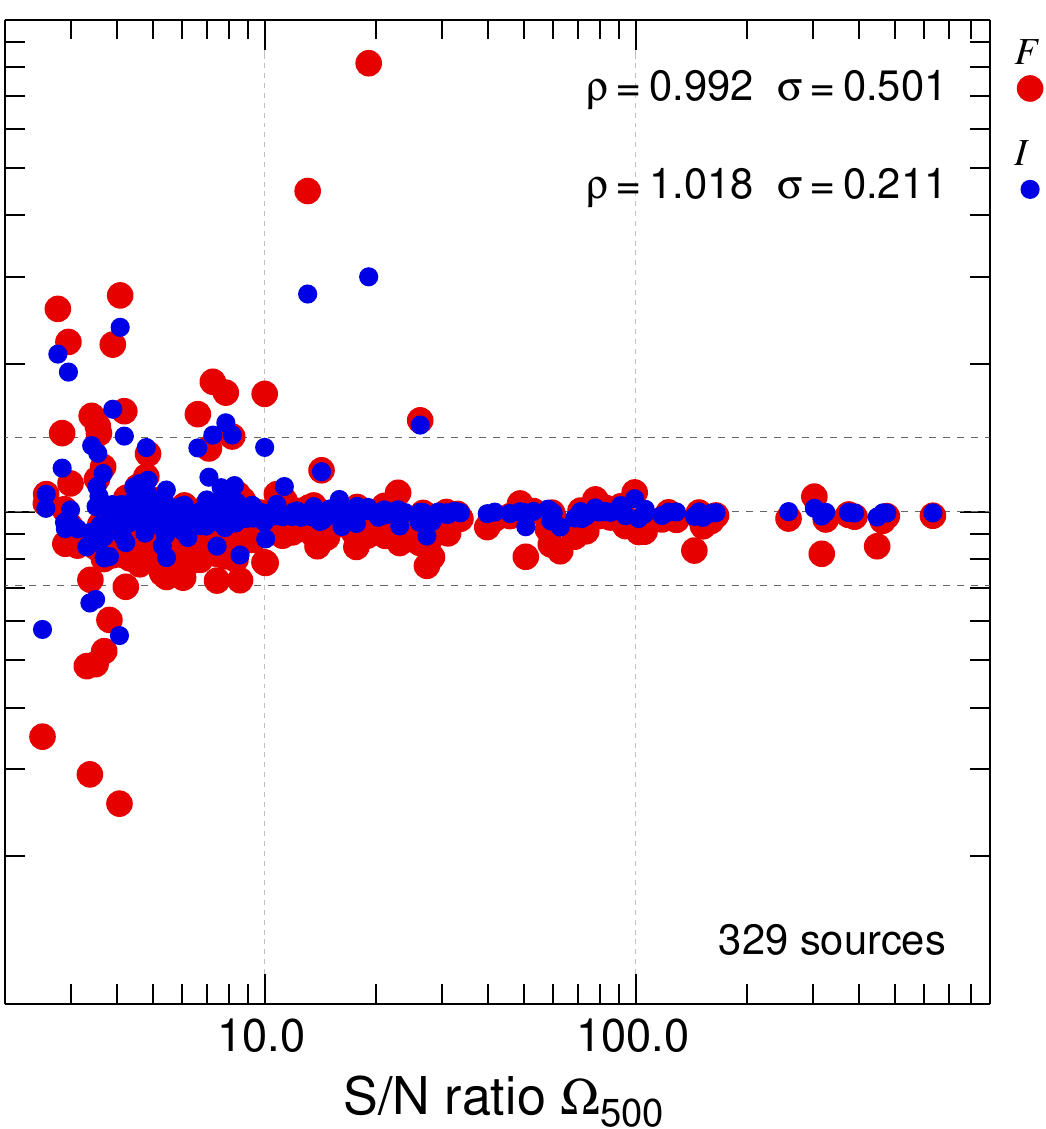}}}
\vspace{1.0mm}
\centerline{\resizebox{0.2695\hsize}{!}{\includegraphics{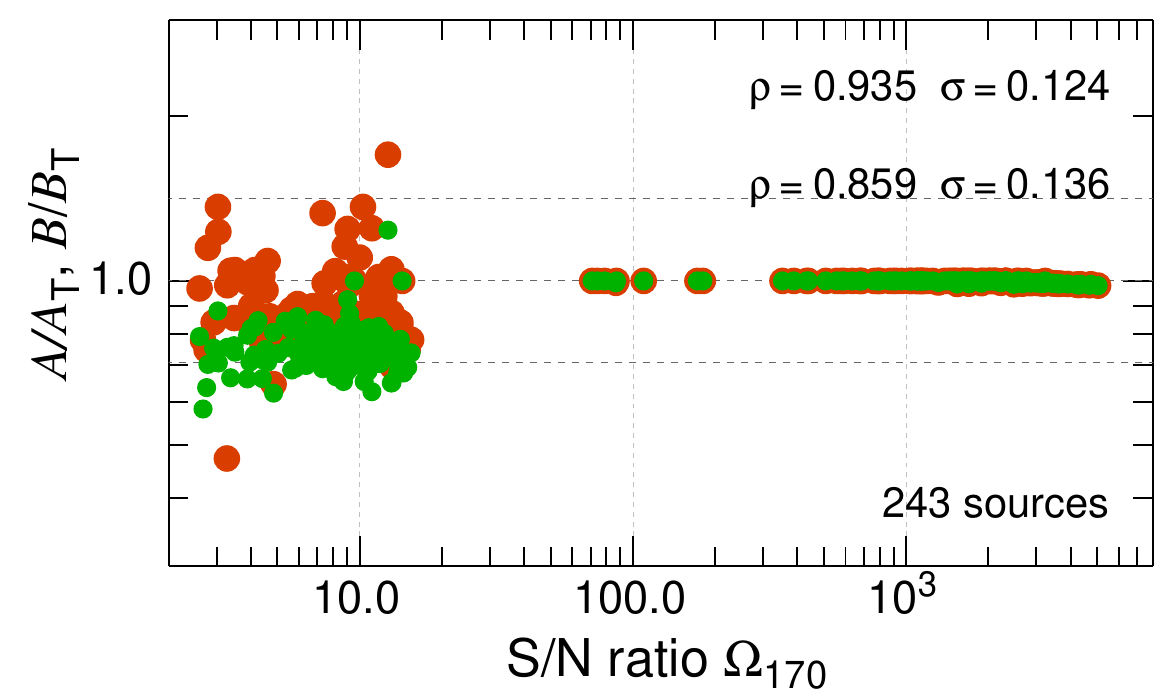}}
            \resizebox{0.2315\hsize}{!}{\includegraphics{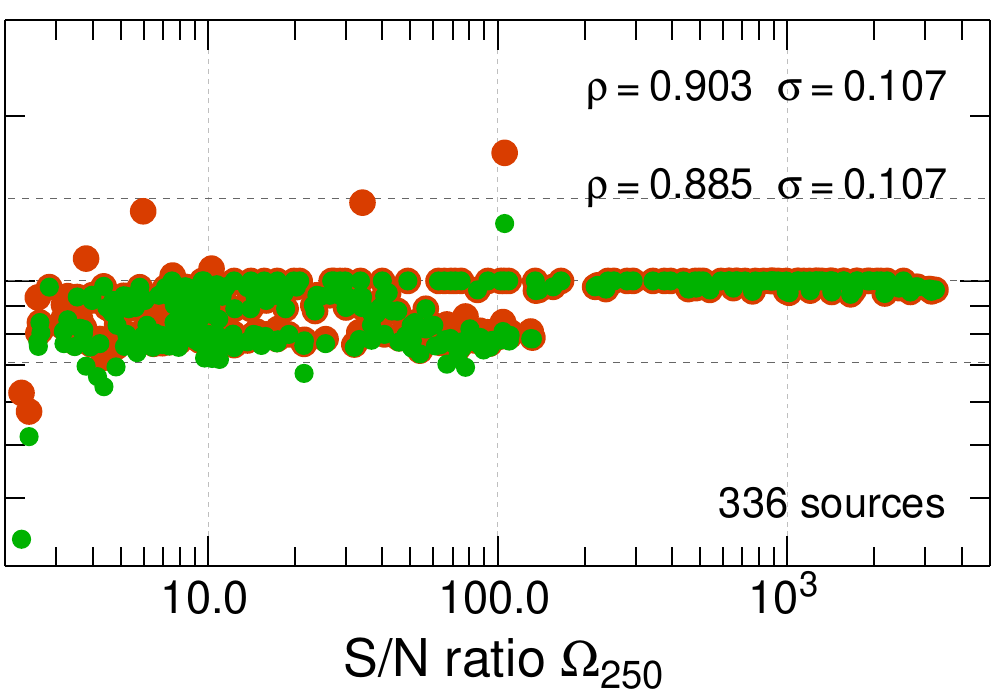}}
            \resizebox{0.2315\hsize}{!}{\includegraphics{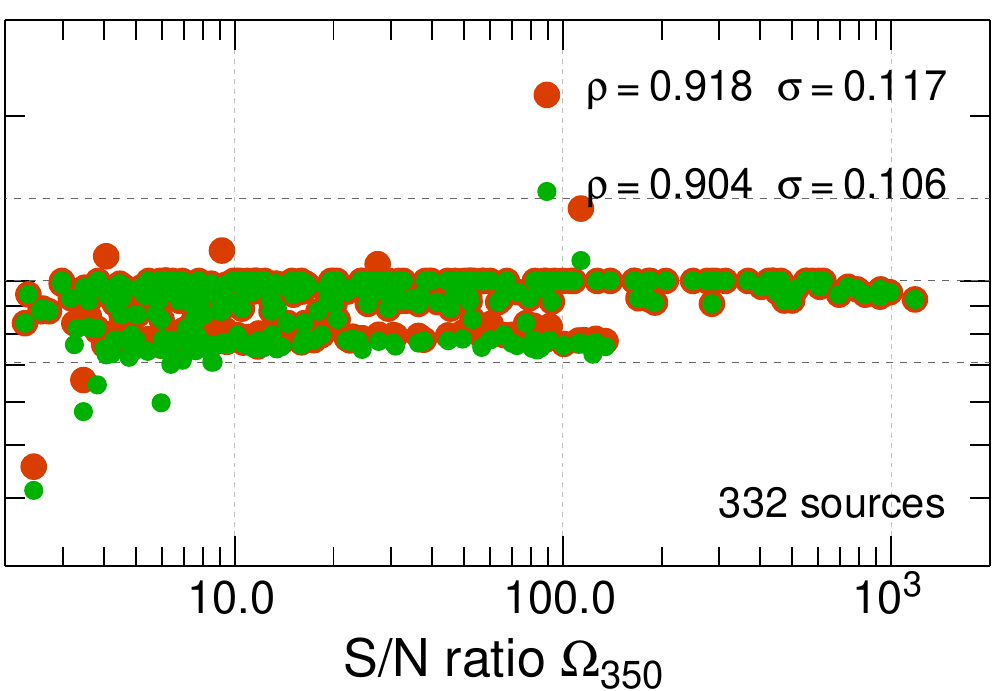}}
            \resizebox{0.2435\hsize}{!}{\includegraphics{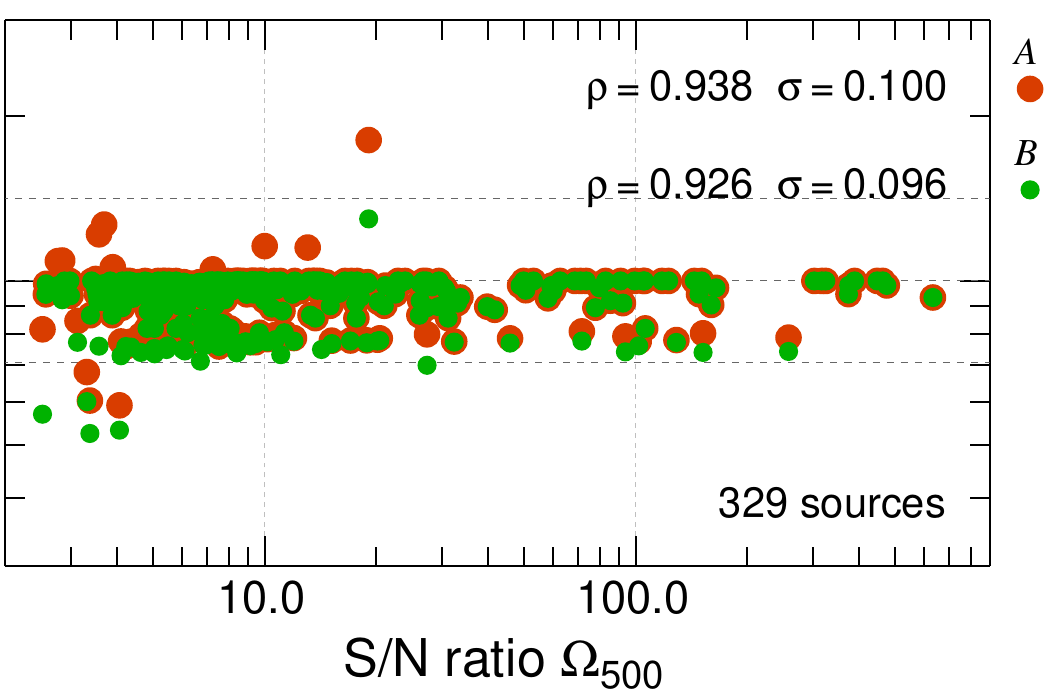}}}
\vspace{1.0mm}
\centerline{\resizebox{0.2695\hsize}{!}{\includegraphics{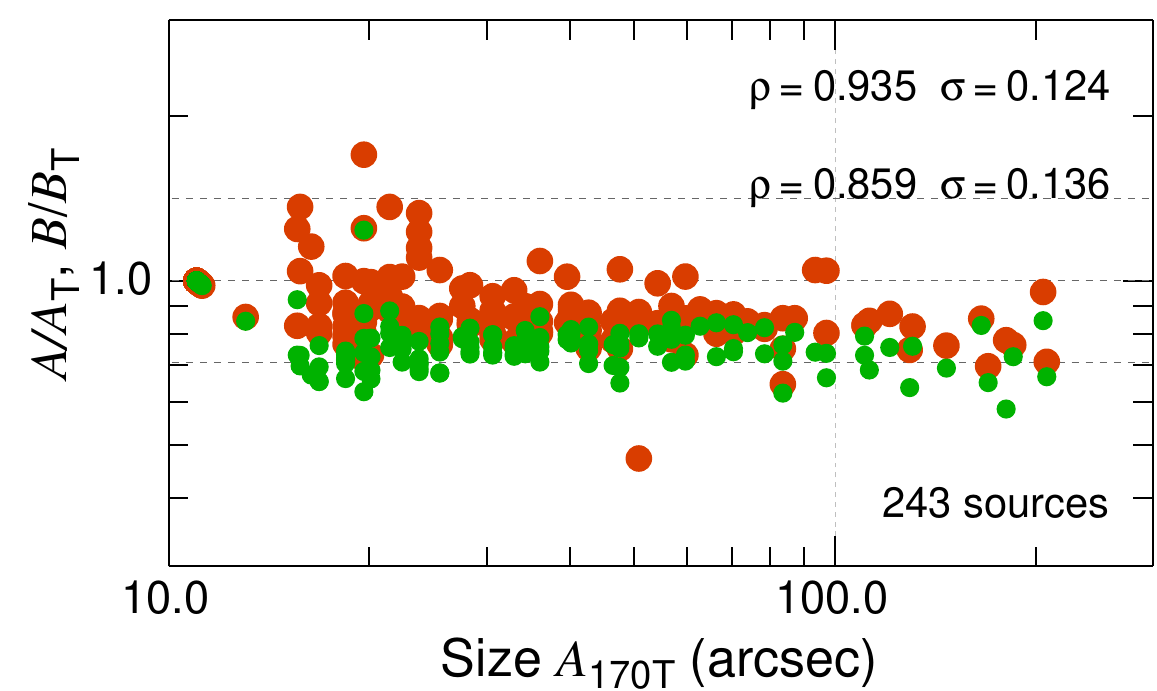}}
            \resizebox{0.2315\hsize}{!}{\includegraphics{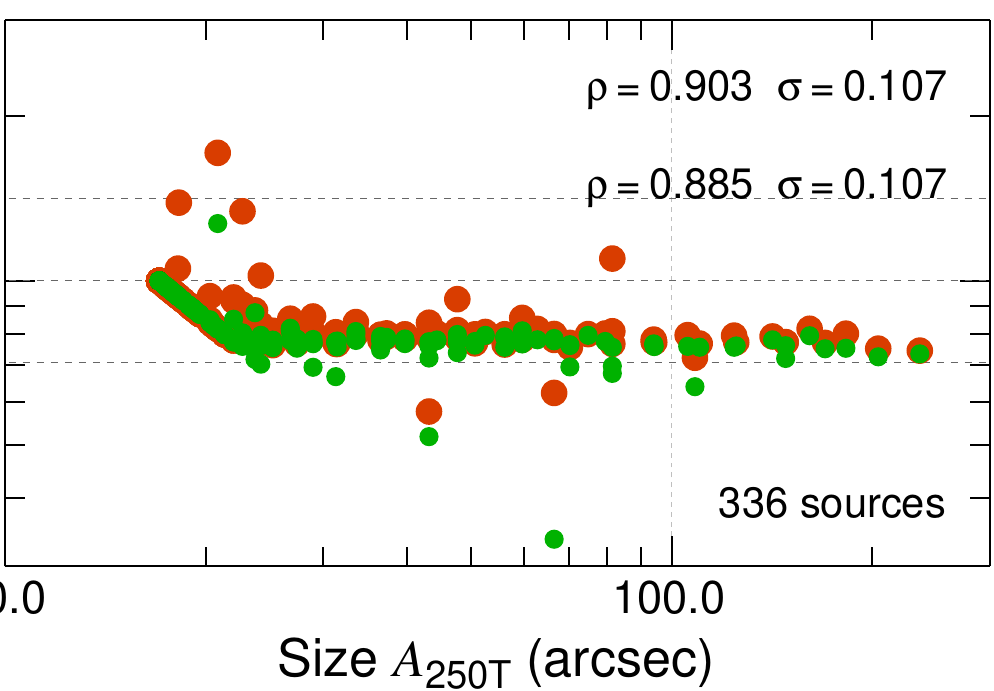}}
            \resizebox{0.2315\hsize}{!}{\includegraphics{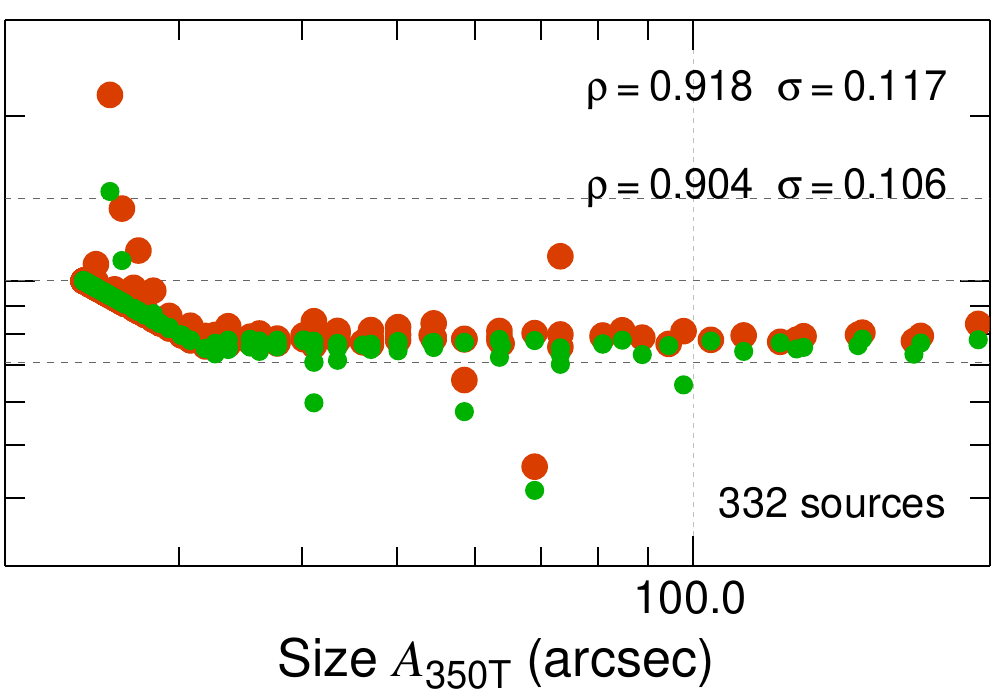}}
            \resizebox{0.2435\hsize}{!}{\includegraphics{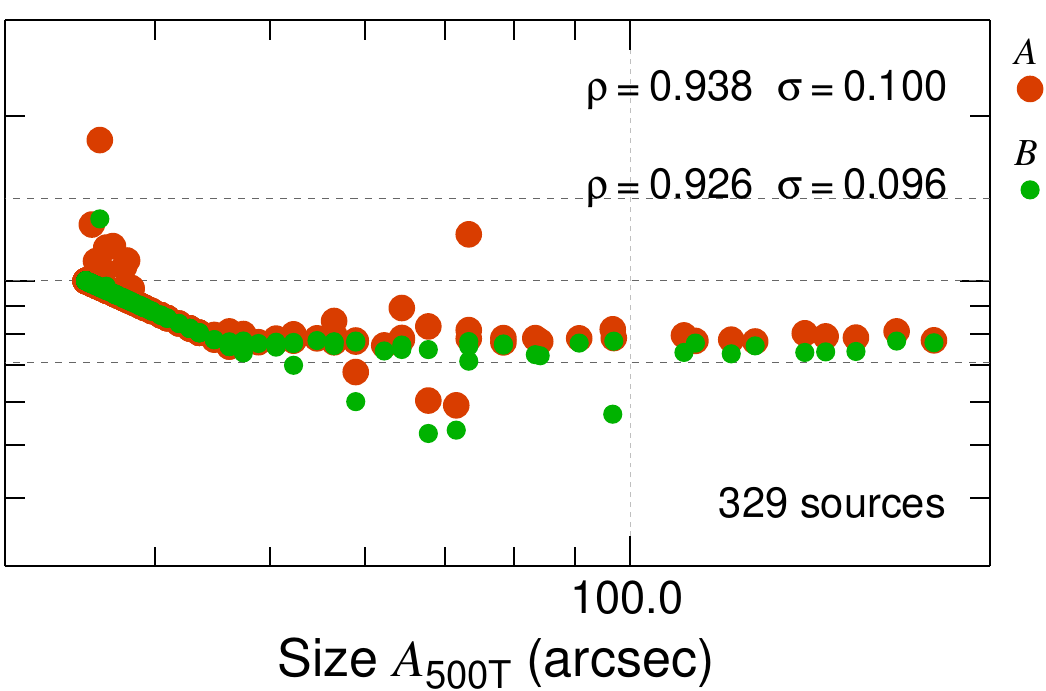}}}
\caption
{ 
Benchmark A$_2$ extraction with \textsl{getsf} (three \emph{top} rows) and \textsl{getold} (three \emph{bottom} rows). Ratios of
the measured fluxes $F_{{\rm T}{\lambda}{n}}$, peak intensities $F_{{\rm P}{\lambda}{n}}$, and sizes $\{A,B\}_{{\lambda}{n}}$ to
their true values ($F/F_{\rm T}$, $I/I_{\rm T}$, $A/A_{\rm T}$, and $B/B_{\rm T}$) are shown as a function of the S/N ratio
$\Omega_{{\lambda}{n}}$. The size ratios $A/A_{\rm T}$ and $B/B_{\rm T}$ are also shown as a function of the true sizes
$\{A,B\}_{{\lambda}{n}{\rm T}}$. The mean $\varrho_{{\rm \{P|T|A|B\}}{\lambda}}$ and standard deviation $\sigma_{{\rm
\{P|T|A|B\}}{\lambda}}$ of the ratios are displayed in the panels. Similar plots for $\lambda\le 110$\,$\mu$m with only bright
protostellar cores are not presented, because their measurements are quite accurate, with $\varrho_{\{{\rm
P|T|A|B}\}{\lambda}}\approx \{0.999|0.998|0.999|0.999\}$ and $\sigma_{\{{\rm P|T|A|B}\}{\lambda}}\approx
\{0.002|0.006|0.00004|0.00004\}$.
} 
\label{accuracyA2}
\end{figure*}


\begin{figure*}
\centering
\centerline{\resizebox{0.2695\hsize}{!}{\includegraphics{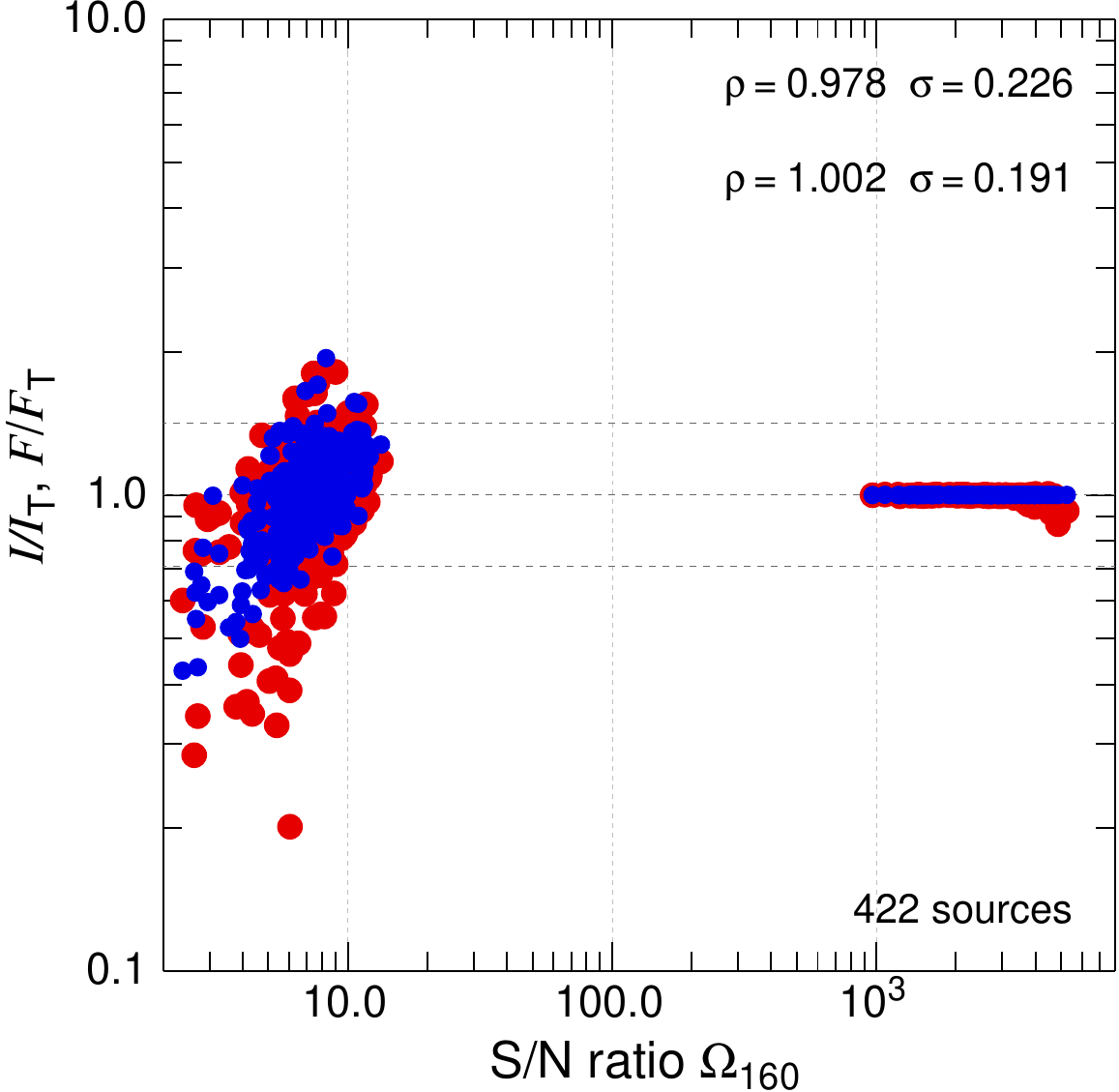}}
            \resizebox{0.2315\hsize}{!}{\includegraphics{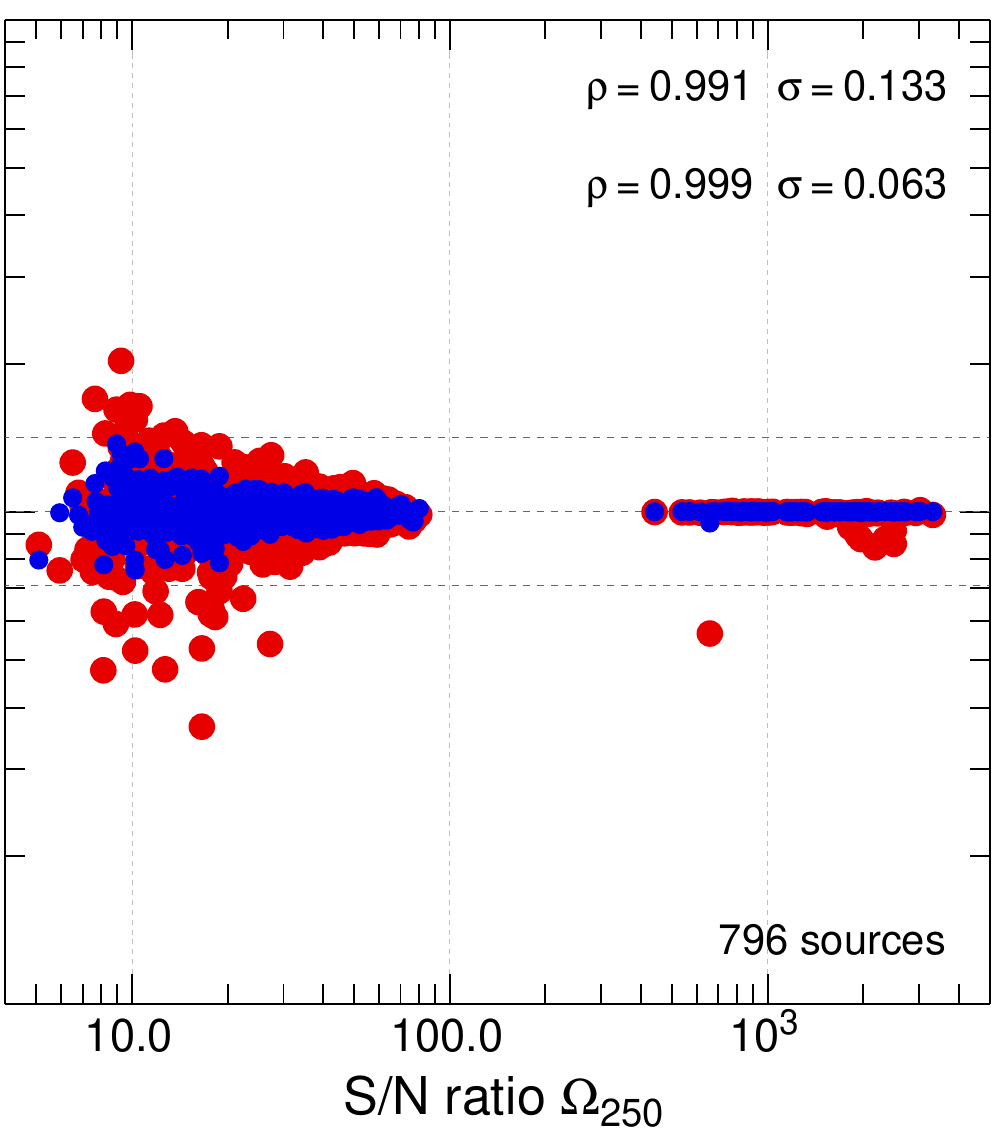}}
            \resizebox{0.2315\hsize}{!}{\includegraphics{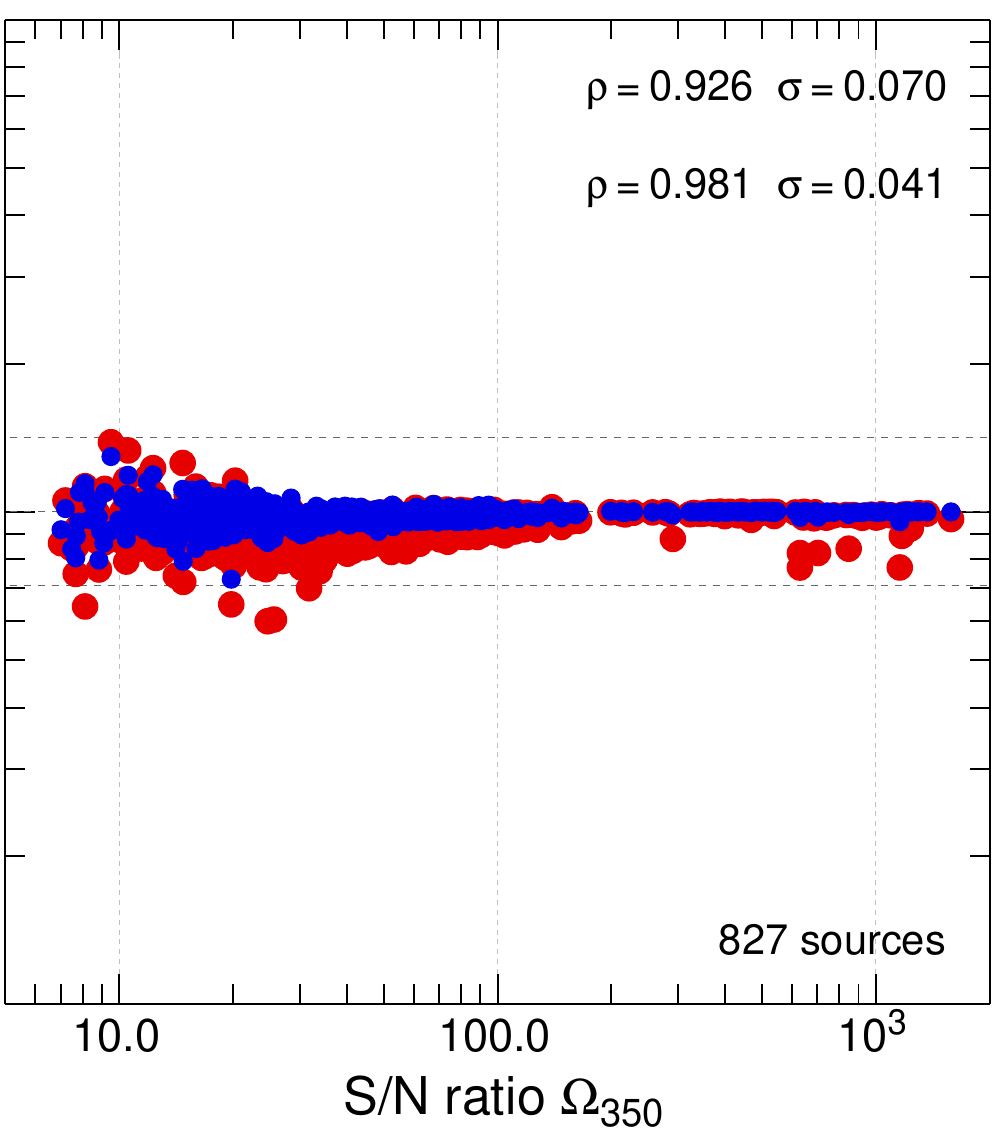}}
            \resizebox{0.2435\hsize}{!}{\includegraphics{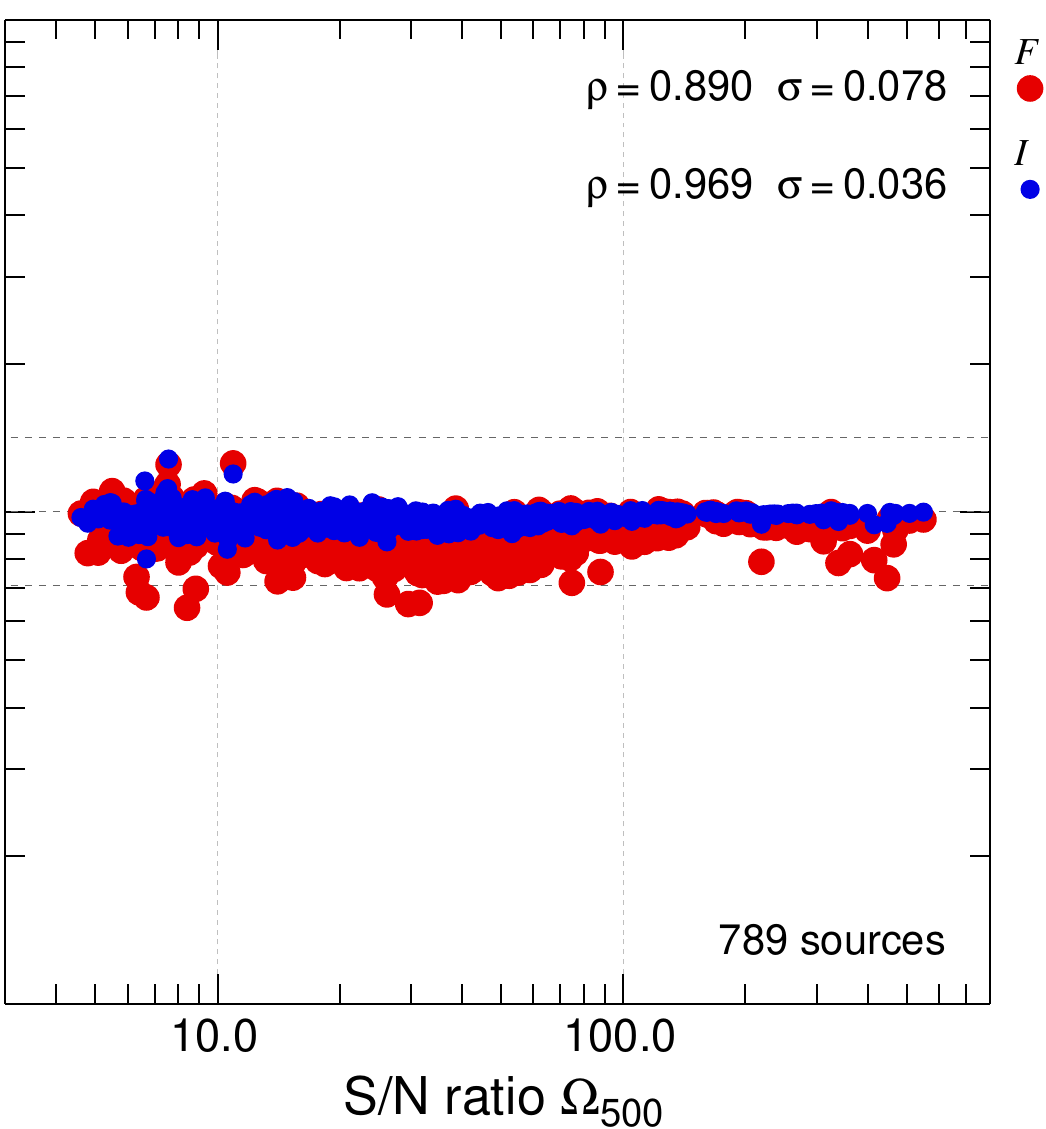}}}
\vspace{1.0mm}
\centerline{\resizebox{0.2695\hsize}{!}{\includegraphics{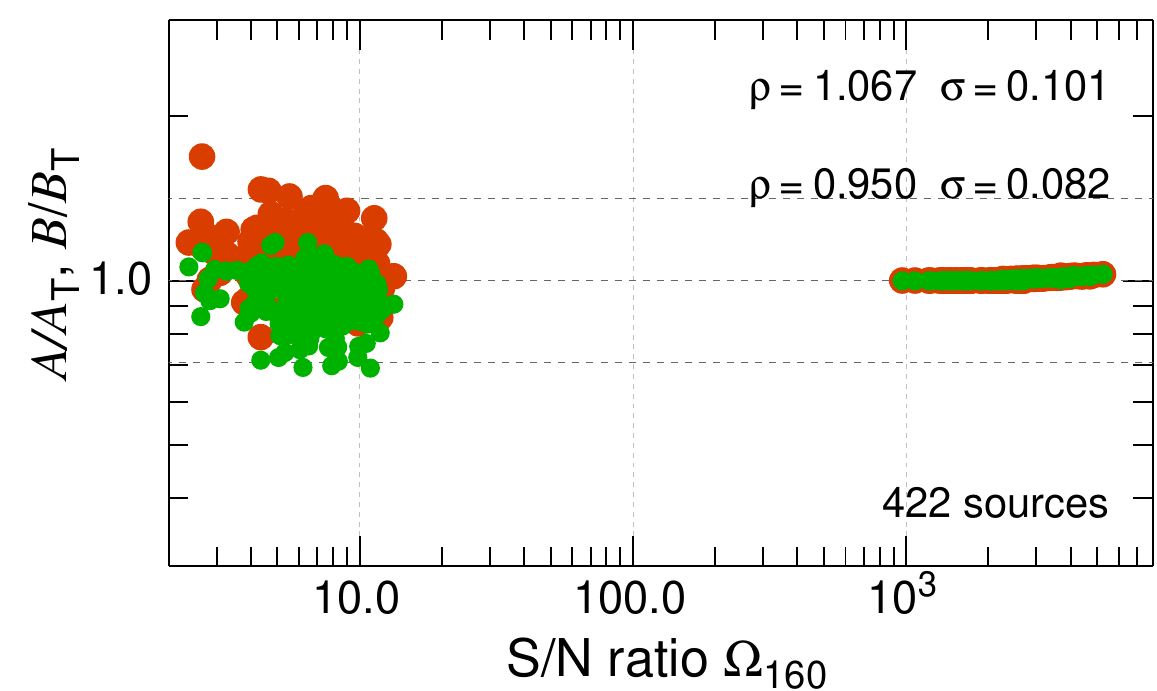}}
            \resizebox{0.2315\hsize}{!}{\includegraphics{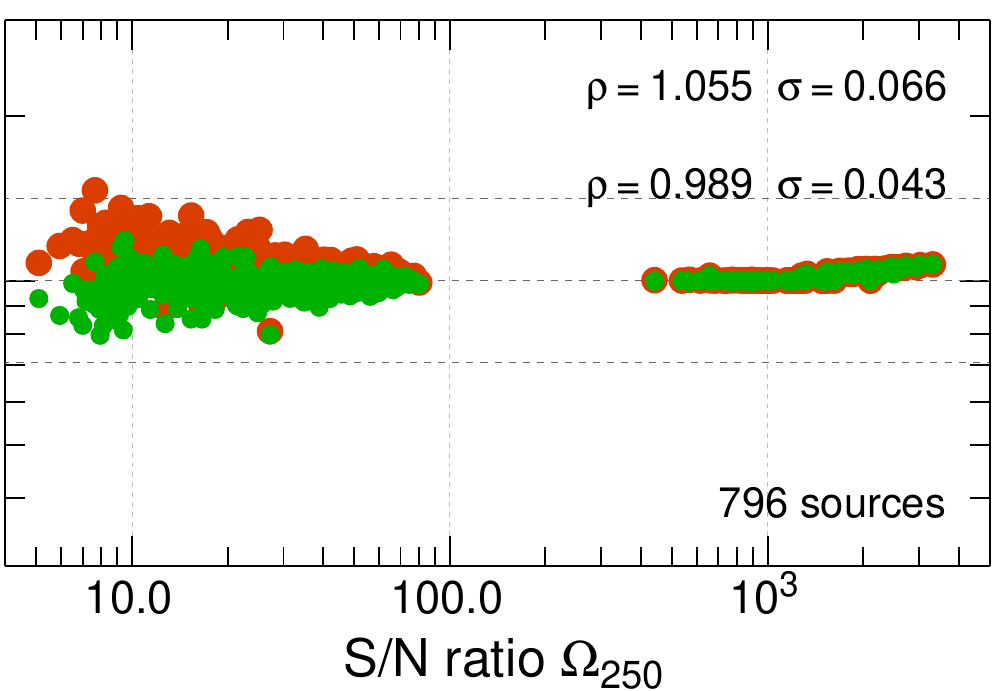}}
            \resizebox{0.2315\hsize}{!}{\includegraphics{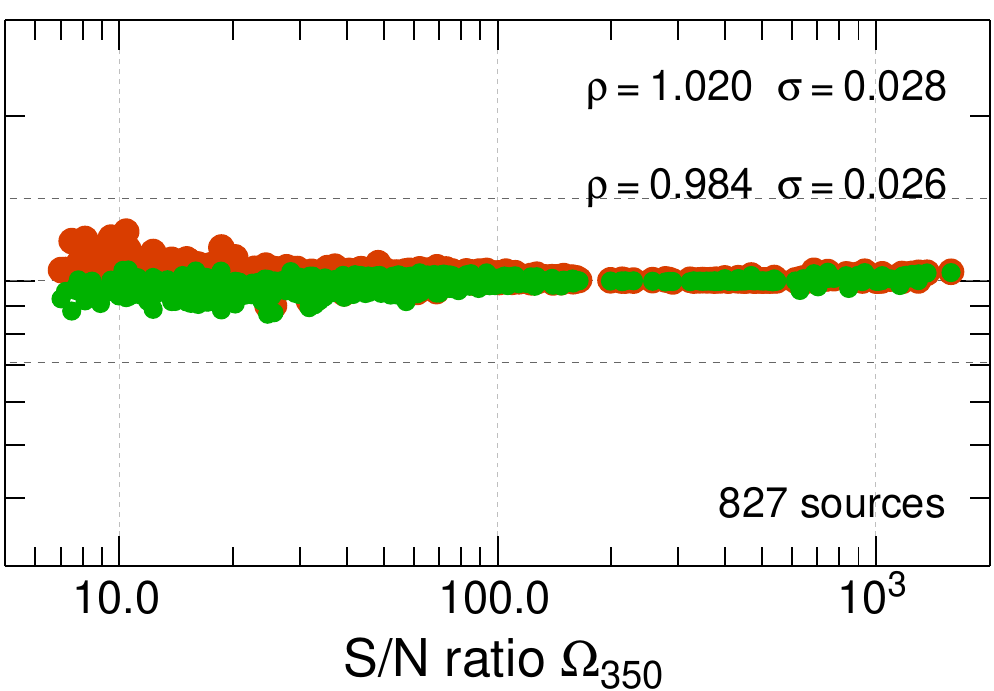}}
            \resizebox{0.2435\hsize}{!}{\includegraphics{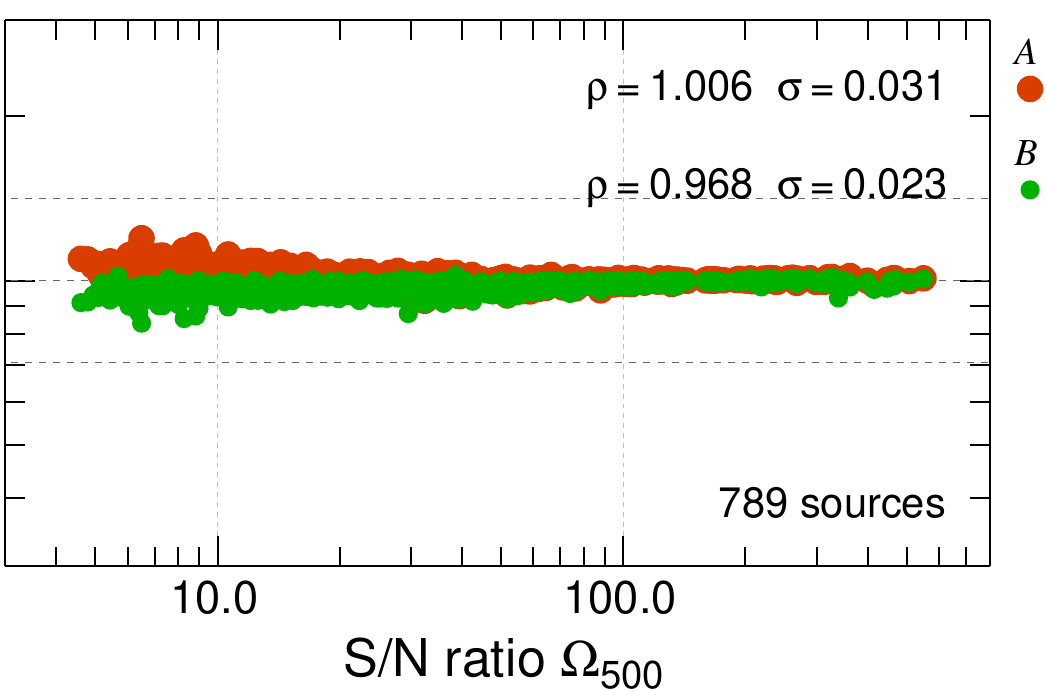}}}
\vspace{1.0mm}
\centerline{\resizebox{0.2695\hsize}{!}{\includegraphics{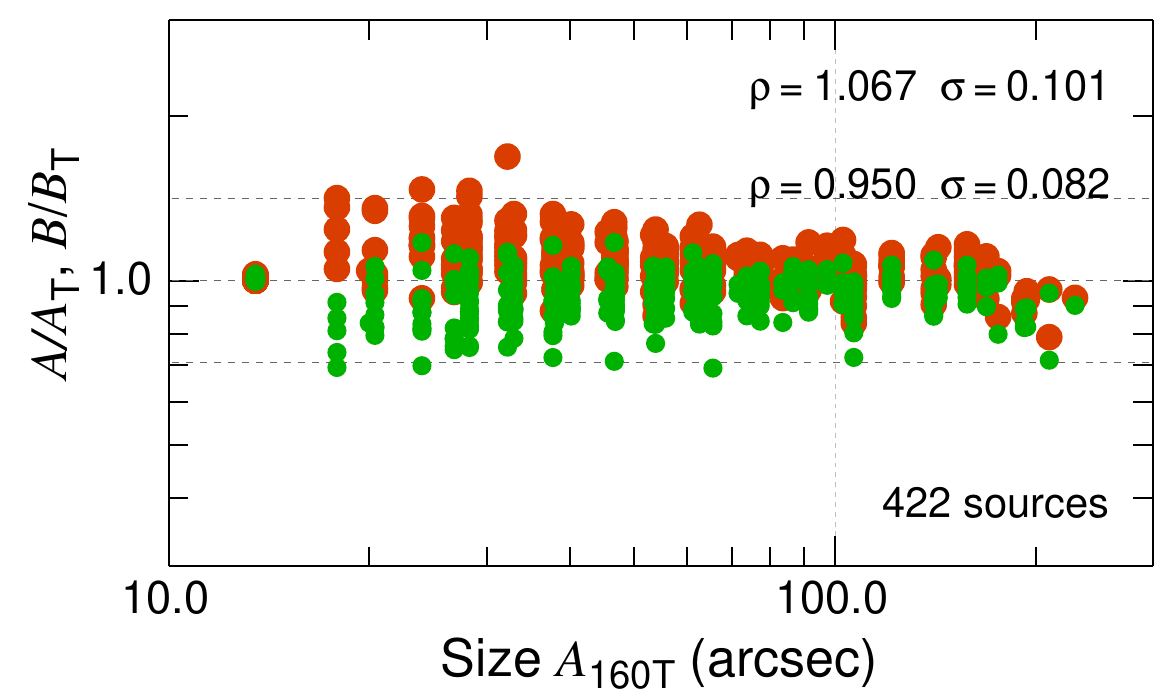}}
            \resizebox{0.2315\hsize}{!}{\includegraphics{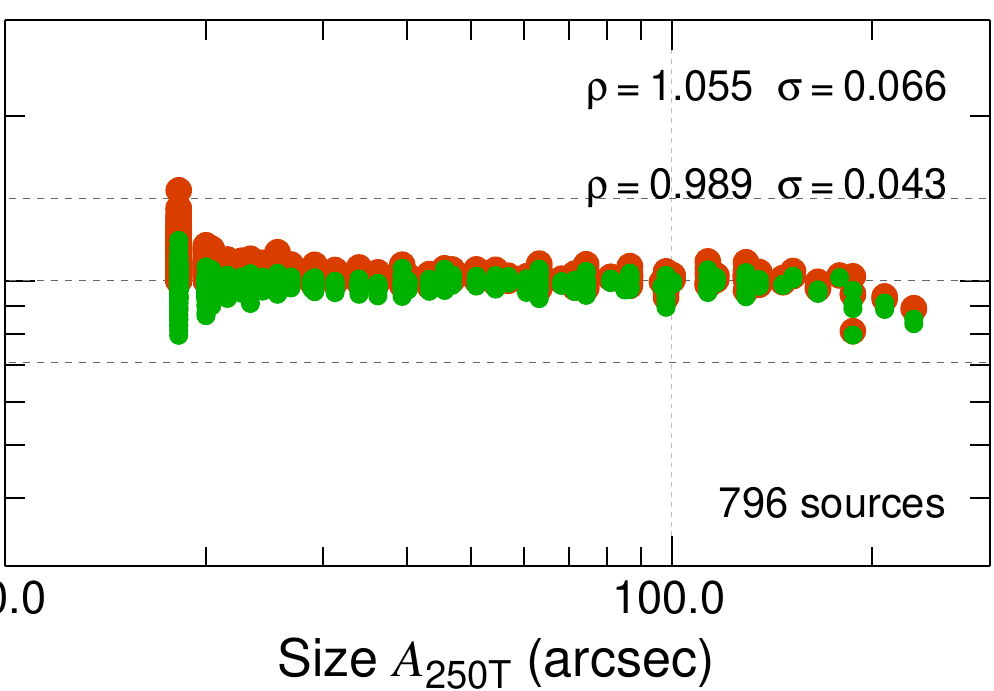}}
            \resizebox{0.2315\hsize}{!}{\includegraphics{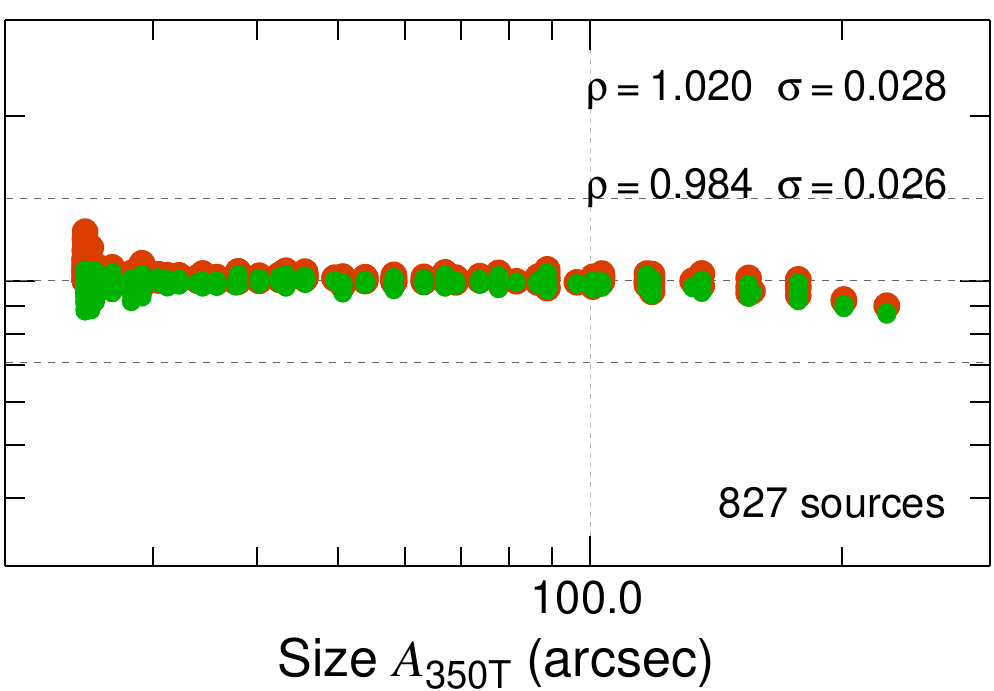}}
            \resizebox{0.2435\hsize}{!}{\includegraphics{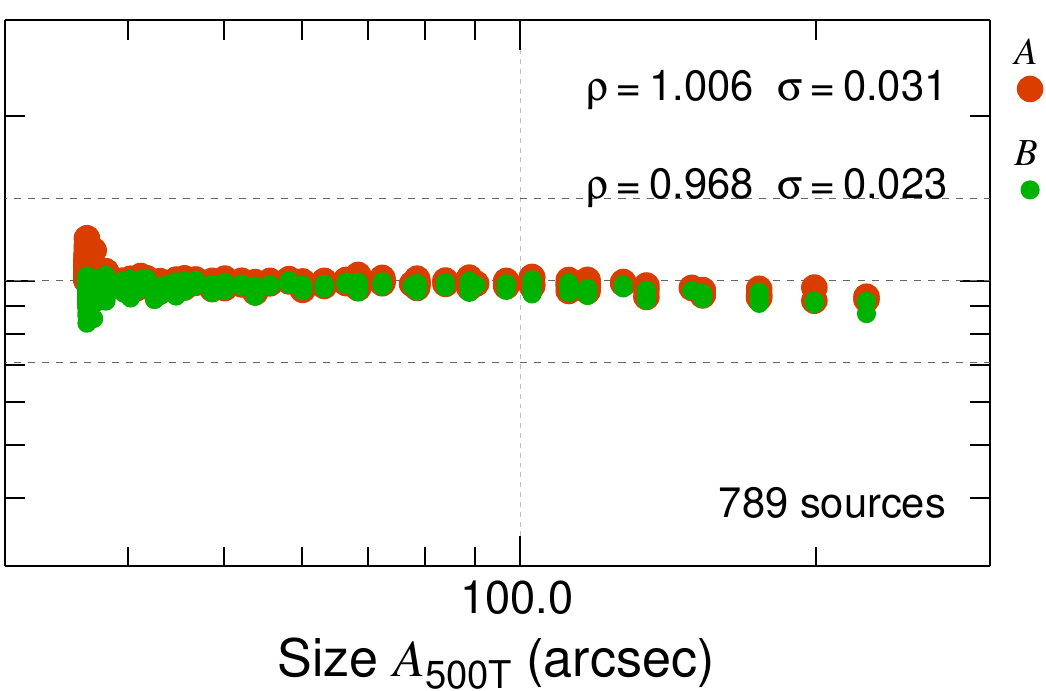}}}
\vspace{2.2mm}
\centerline{\resizebox{0.2695\hsize}{!}{\includegraphics{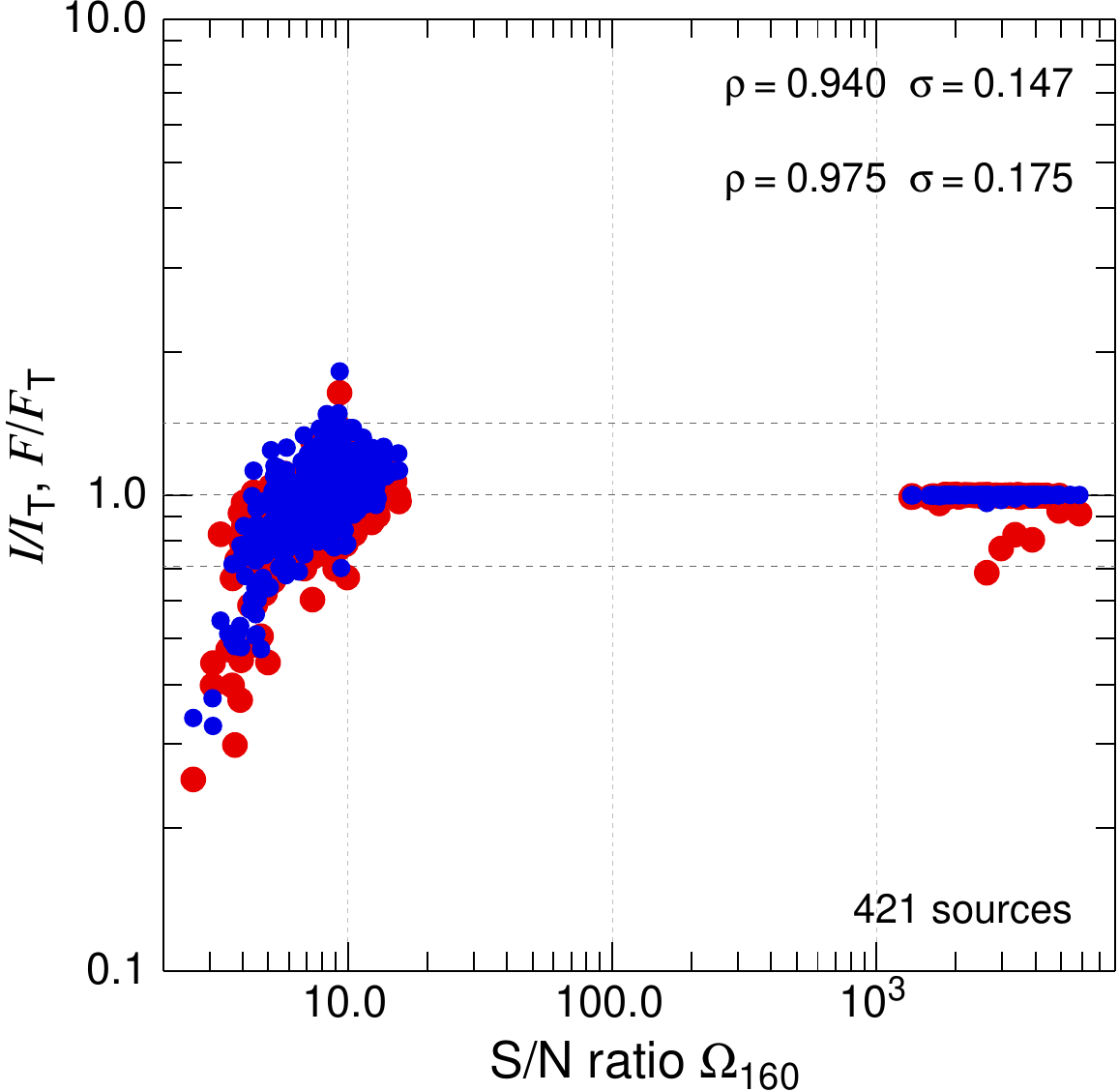}}
            \resizebox{0.2315\hsize}{!}{\includegraphics{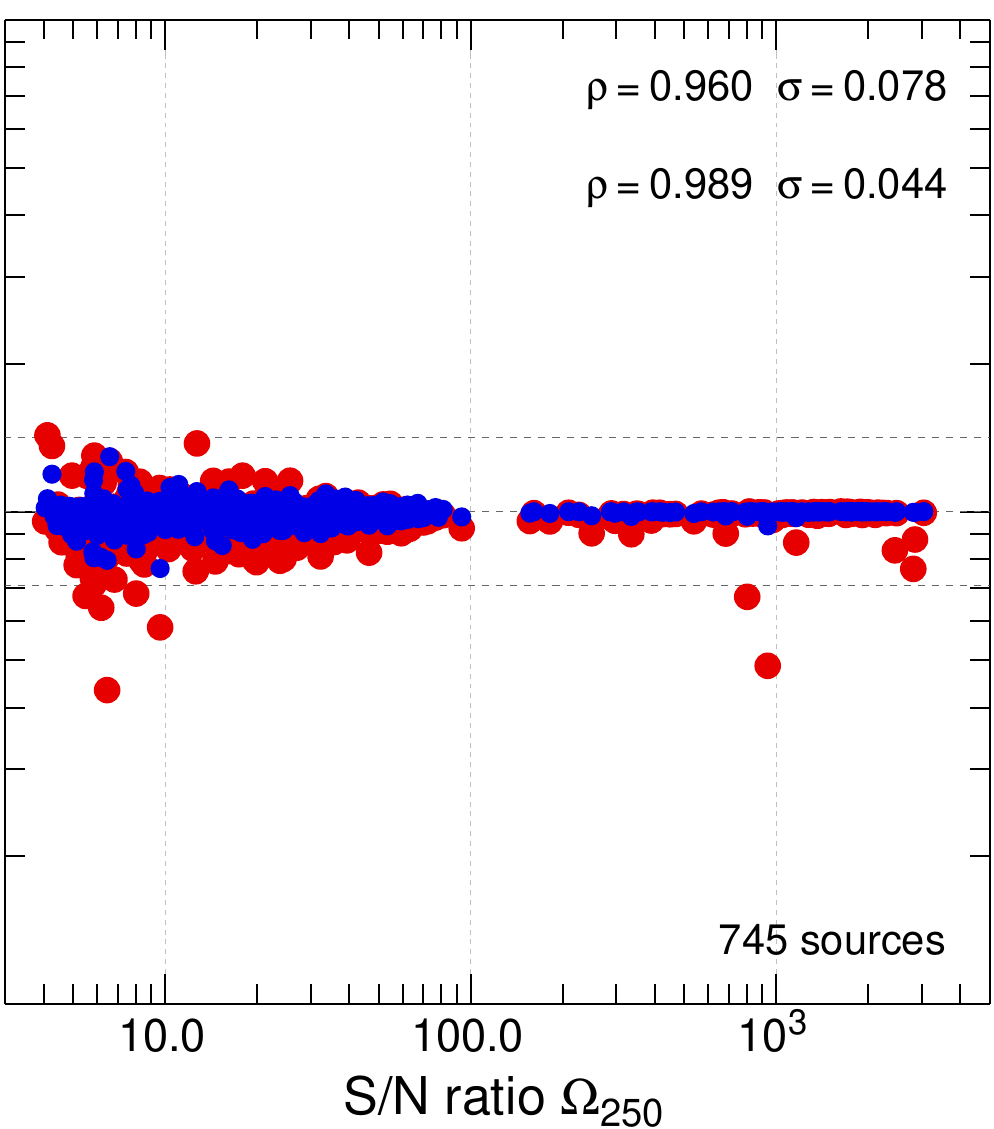}}
            \resizebox{0.2315\hsize}{!}{\includegraphics{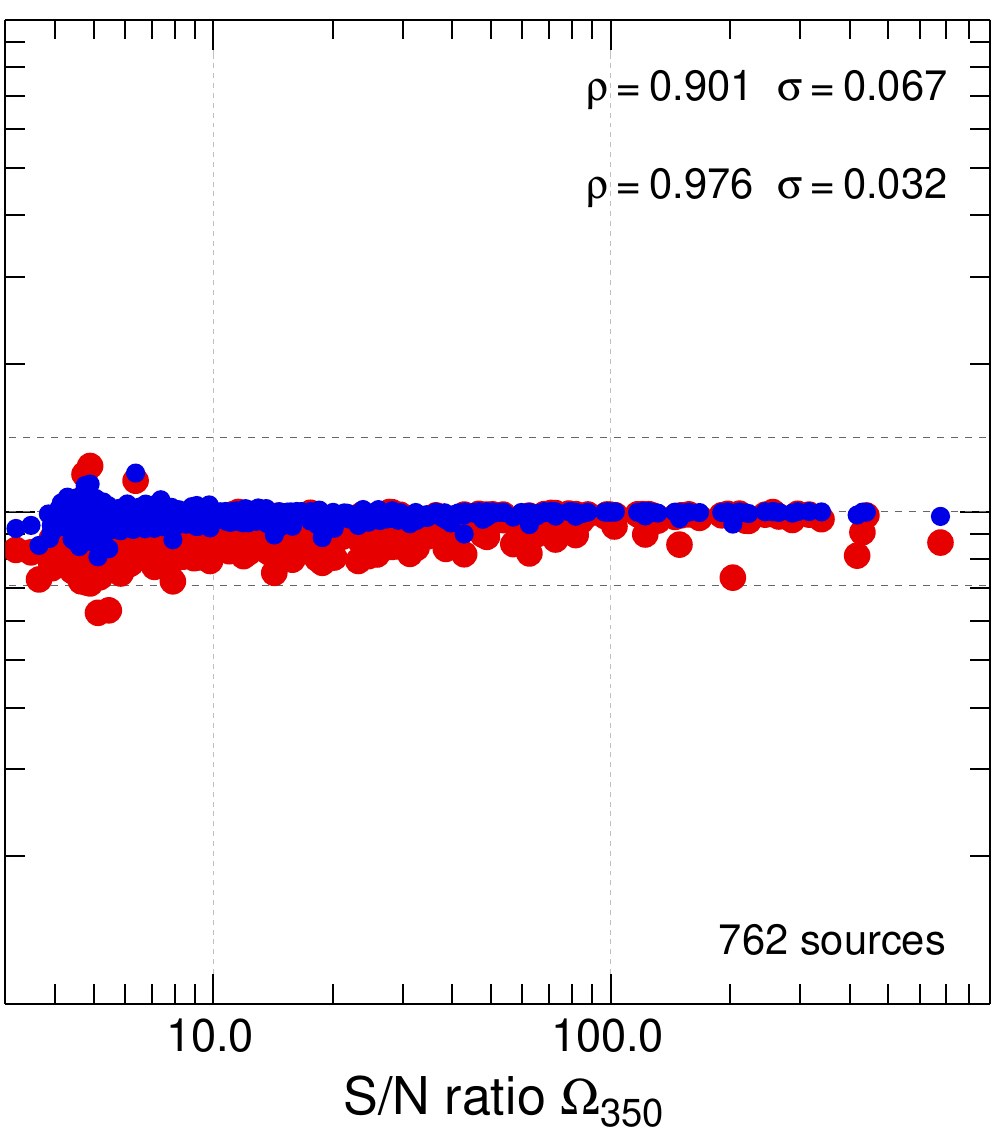}}
            \resizebox{0.2435\hsize}{!}{\includegraphics{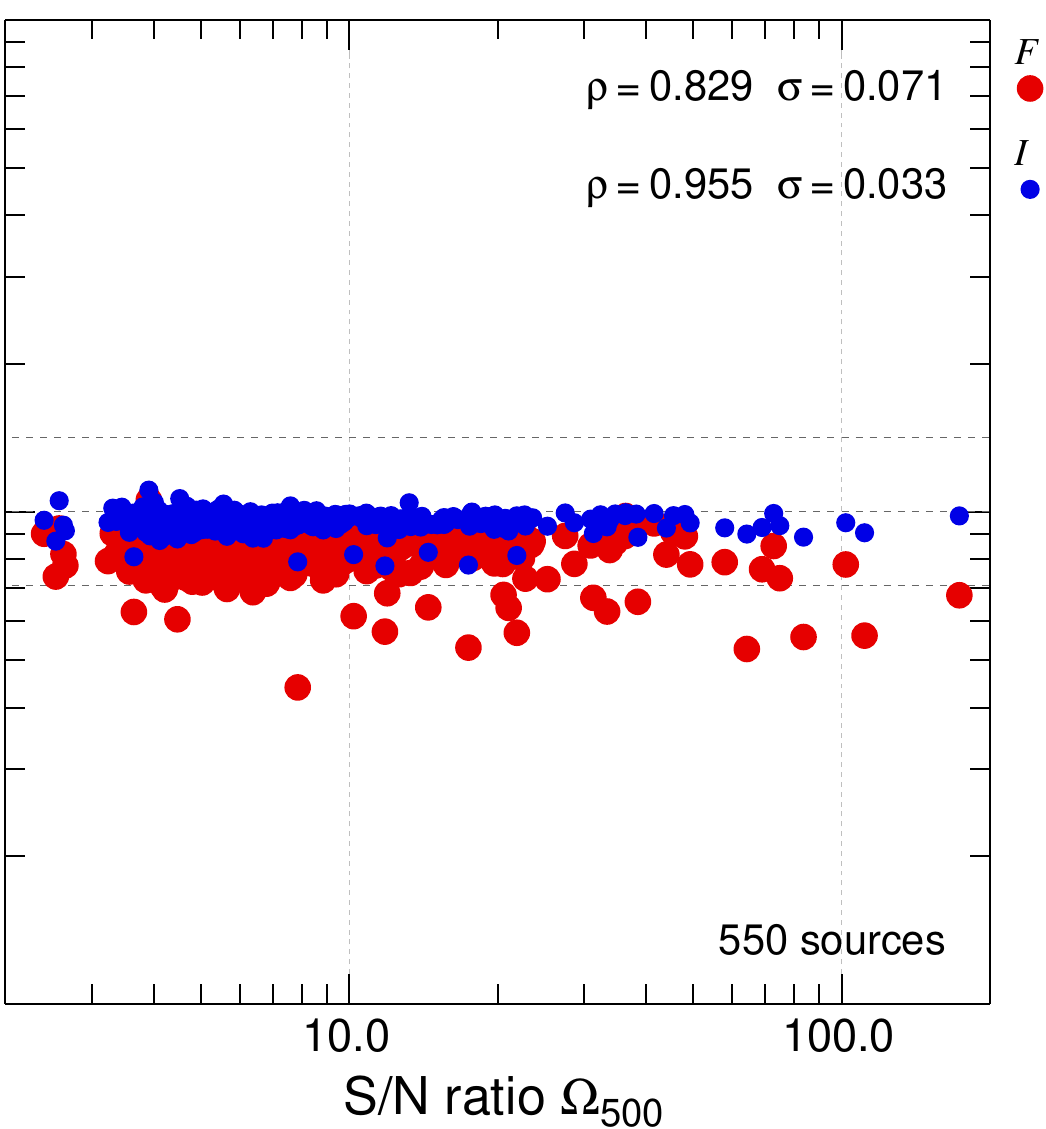}}}
\vspace{1.0mm}
\centerline{\resizebox{0.2695\hsize}{!}{\includegraphics{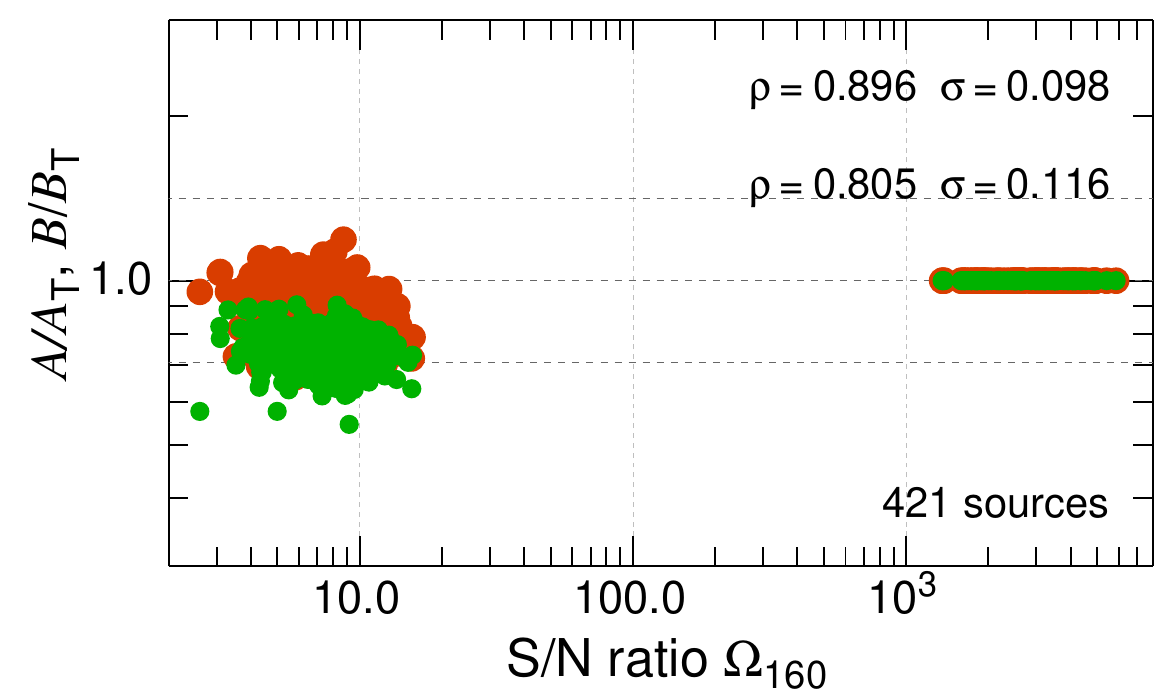}}
            \resizebox{0.2315\hsize}{!}{\includegraphics{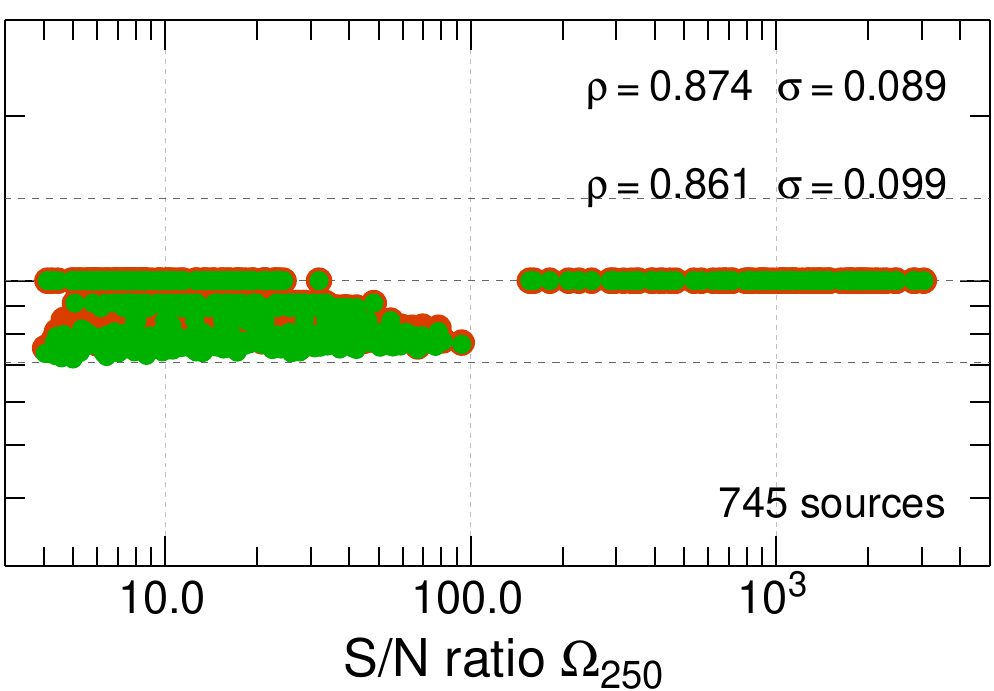}}
            \resizebox{0.2315\hsize}{!}{\includegraphics{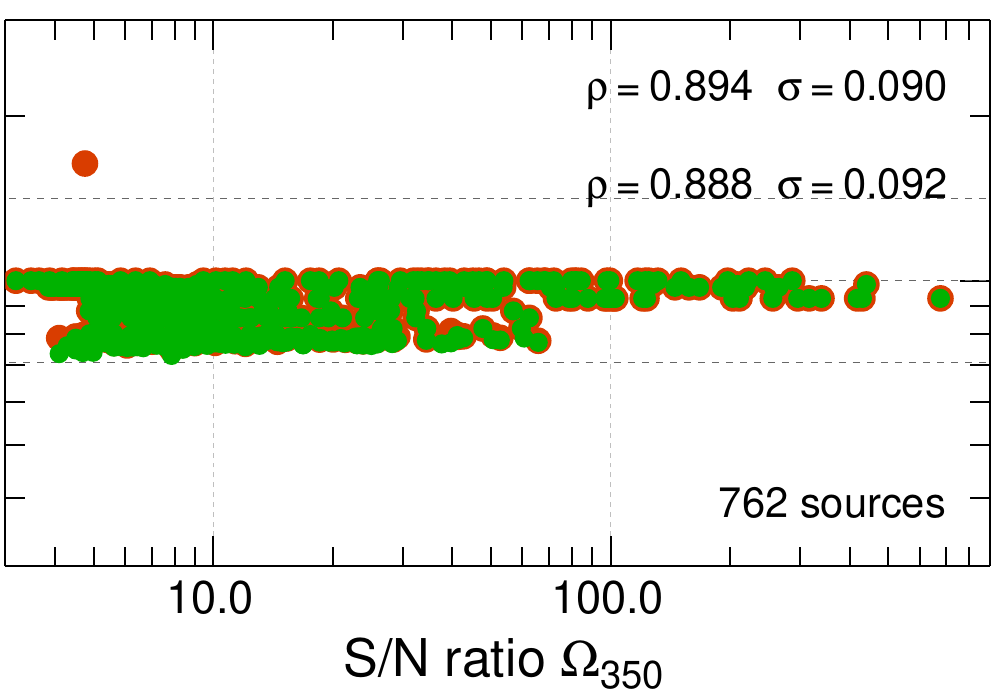}}
            \resizebox{0.2435\hsize}{!}{\includegraphics{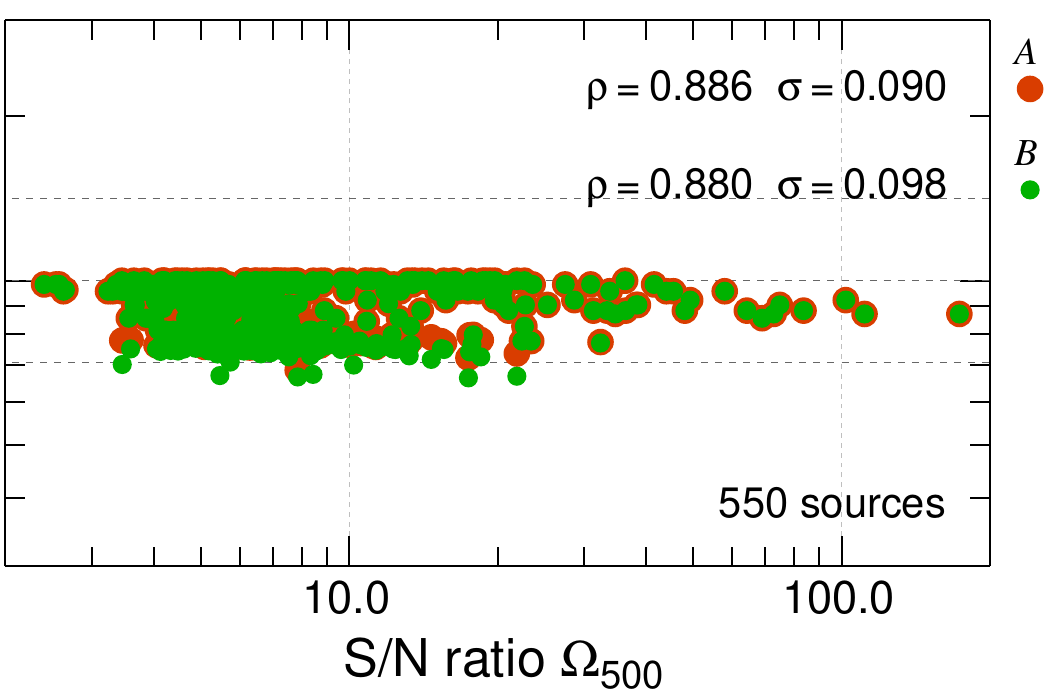}}}
\vspace{1.0mm}
\centerline{\resizebox{0.2695\hsize}{!}{\includegraphics{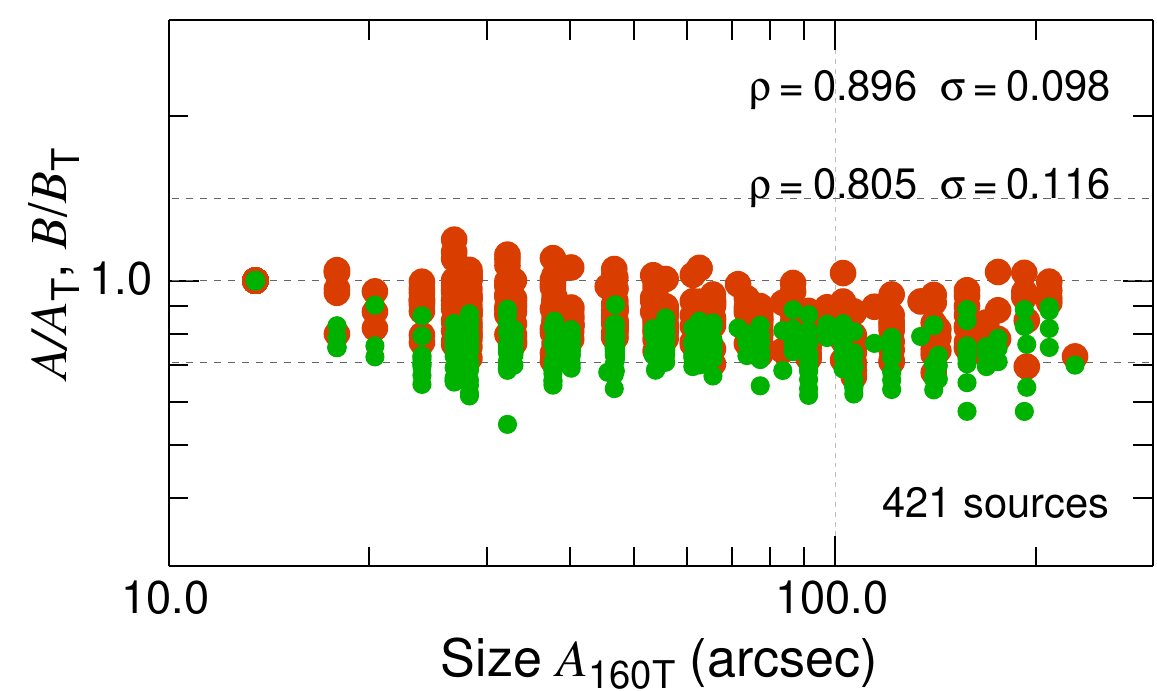}}
            \resizebox{0.2315\hsize}{!}{\includegraphics{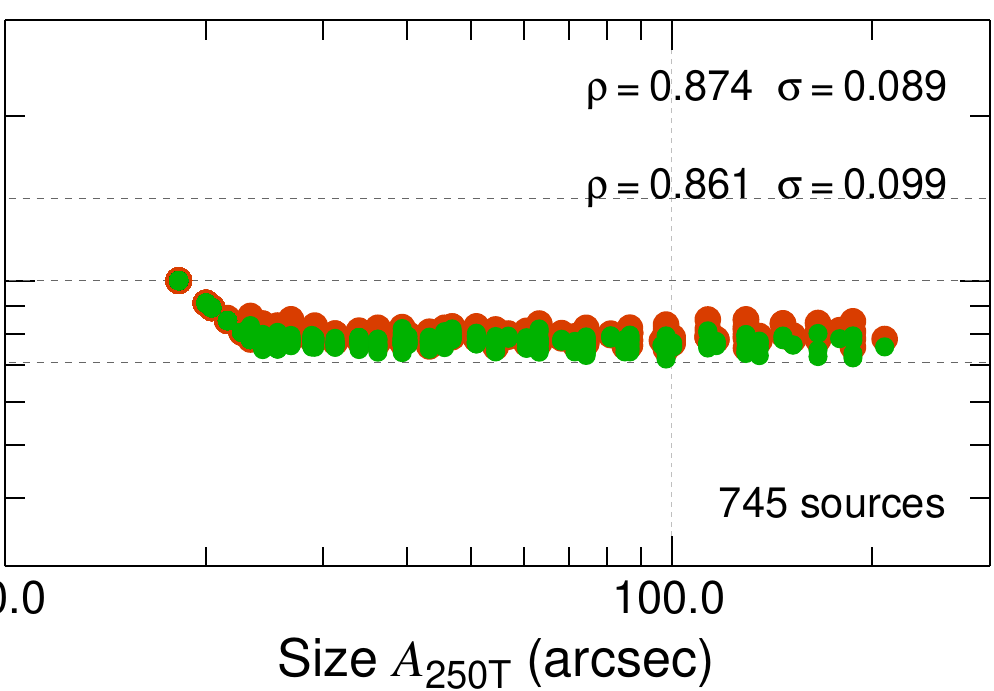}}
            \resizebox{0.2315\hsize}{!}{\includegraphics{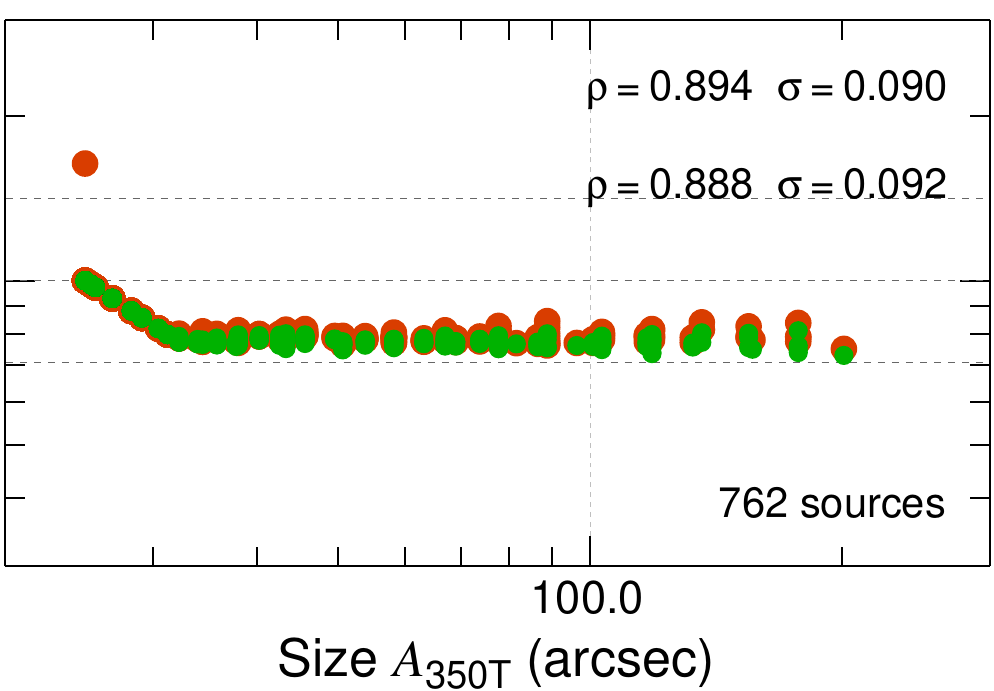}}
            \resizebox{0.2435\hsize}{!}{\includegraphics{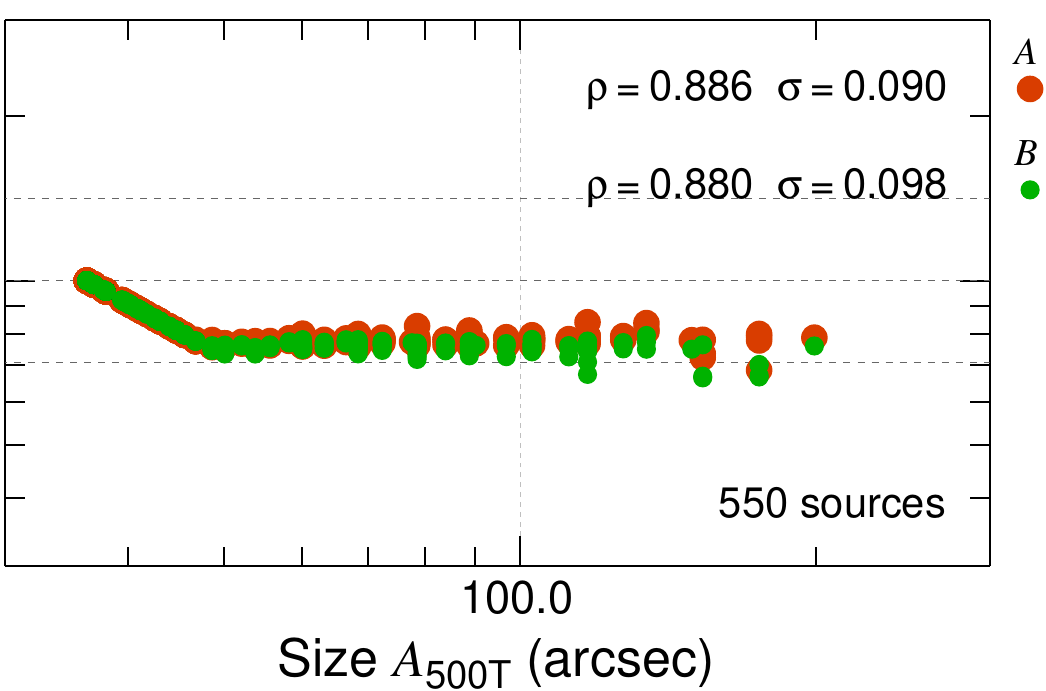}}}
\caption
{ 
Benchmark B$_2$ extraction with \textsl{getsf} (three \emph{top} rows) and \textsl{getold} (three \emph{bottom} rows). Ratios of
the measured fluxes $F_{{\rm T}{\lambda}{n}}$, peak intensities $F_{{\rm P}{\lambda}{n}}$, and sizes $\{A,B\}_{{\lambda}{n}}$ to
their true values ($F/F_{\rm T}$, $I/I_{\rm T}$, $A/A_{\rm T}$, and $B/B_{\rm T}$) are shown as a function of the S/N ratio
$\Omega_{{\lambda}{n}}$. The size ratios $A/A_{\rm T}$ and $B/B_{\rm T}$ are also shown as a function of the true sizes
$\{A,B\}_{{\lambda}{n}{\rm T}}$. The mean $\varrho_{{\rm \{P|T|A|B\}}{\lambda}}$ and standard deviation $\sigma_{{\rm
\{P|T|A|B\}}{\lambda}}$ of the ratios are displayed in the panels. Similar plots for $\lambda\le 100$\,$\mu$m with only bright
protostellar cores are not presented, because their measurements are quite accurate, with $\varrho_{\{{\rm
P|T|A|B}\}{\lambda}}\approx \{0.999|0.998|1.005|0.996\}$ and $\sigma_{\{{\rm P|T|A|B}\}{\lambda}}\approx
\{0.001|0.005|0.01|0.007\}$.
} 
\label{accuracyB2}
\end{figure*}


In extractions with \textsl{getsf} and \textsl{getold}, it is necessary to determine the structures of interest to extract and
specify their maximum sizes for each waveband \citep[$\{X|Y\}_{\lambda}$ in Paper I and $A^{\rm max}_{\lambda}$
in][]{Men'shchikov_etal2012}. In all benchmark extractions, this user-definable parameter was assigned the same values for both
methods. The maximum sizes $X_{\lambda}$ and $A^{\rm max}_{\lambda}$ for sources were $16$, $25$, $30$, $150$, $150$, and
$150${\arcsec} for the \emph{Herschel} wavebands and $150${\arcsec} for the surface density image. In Benchmark B, the maximum size
$Y_{\lambda}$ for filaments was $350${\arcsec} for all images. The \textsl{getold} extractions followed an improved scheme
\citep{Men'shchikov2017}: all benchmark images were first processed by \textsl{getimages} (using the above maximum sizes) that
subtracted their large-scale backgrounds and flattened residual background and noise fluctuations. The background-subtracted and
flattened images were then used in the \textsl{getsources} extractions. 


\subsection{Source extractions in Benchmarks A and B}
\label{benchAB}

In the standard approach to the multiwavelength benchmarking adopted in this paper, sources are detected in the
wavelength-independent images combined from all (seven) wavelengths. Effects of different combinations of images for source
detection on extraction qualities are discussed in Sect.~\ref{dependonimg}.

In the analysis of the extractions, all acceptable sources from the catalogs were positionally matched with the truth catalogs
using \textsl{stilts} \citep{Taylor2006}. The matching radius was essentially a quadratic mean of the angular resolution
$O_{\lambda}$ and the true FWHM size of the model core,
\begin{equation} 
R_{{\lambda}{n}} = 0.5 \left( O_{\lambda}^2 + A_{{\lambda}{n}{\rm T}} B_{{\lambda}{n}{\rm T}} \right)^{1/2}.
\label{quadmean}
\end{equation} 
The extracted sources with positions within the circles were considered the matches to the true model cores. Only those of them
with errors in measurements within a factor of $2^{1/2}$ were evaluated in Tables \ref{qualAB23}\,--\,\ref{qualB4} according to the
system outlined in Sect. \ref{evalqual}. For plotting the ratios of the measured and true parameters (cf. Figs.
\ref{accuracyA2}\,--\,\ref{accuracyB4}) the sources with measurement errors within a factor of $10$ were used.

\subsubsection{Qualitative comparisons}
\label{visuals}

Figures \ref{imagesA2}\,--\,\ref{imagesB4} visualize the source extraction results by means of three benchmark images overlaid with
the footprints ellipses $\{A,B,\omega\}_{{\rm F}{\lambda}{n}}$ of all acceptably good sources $n$, selected by
Eq.~(\ref{acceptable}). The two short-wavelength images at $\lambda < 160$\,$\mu$m are not shown, because only the strong
unresolved peaks of protostellar cores appear there and their extraction would be uncomplicated for most methods. The $250$ and
$350$\,$\mu$m images are not presented either, because of their similarity to the three images displayed. The starless cores are
best visible against their background in the surface densities $\mathcal{D}_{\{11|13\}{\arcsec}}$ that expose the dense sources
more clearly than the intensities do. In $\{\mathrm{A},\mathrm{B}\}_3$ and B$_4$, the images are dissimilar for \textsl{getsf} and
\textsl{getold}, because \textsl{getold} subtracts their large-scale backgrounds in a preliminary run of \textsl{getimages}.

Figures \ref{imagesA2}\,--\,\ref{imagesB4} illustrate the difficulties created by the starless cores. Such cores are totally
invisible or very faint at the higher resolutions of $\lambda < 160$\,$\mu$m and blended with the other sources (A$_2$ and A$_3$)
or backgrounds ($\{\mathrm{A},\mathrm{B}\}_3$ and B$_4$) at the lower resolutions of the longer wavelengths. The starless cores
with temperatures $T_{\rm D}\la 10$\,K produce little emission at $\lambda\la 100$\,$\mu$m; therefore, only the protostellar cores
are extractable at the short wavelengths. Although the starless sources appear stronger at $\lambda > 170$\,$\mu$m, progressively
lower resolutions spread their emission over larger footprints, which makes interpolation of the fluctuating background less
accurate.

The strongly overlapping sources in the crowded areas of Benchmark A become more heavily blended with each other and with their
background, which makes their deblending less accurate. The backgrounds and true extents of the footprints of such blended sources
are difficult, if not impossible, to determine reliably. Their footprints often become overestimated and the backgrounds
underestimated, which leads to excessively large measured fluxes. For other sources that are largely isolated, the measurements of
fluxes, sizes, and positions are usually more accurate. Increasing numbers of overlapping sources at the lower resolutions of
longer wavelengths degrade the quality of their backgrounds further, because much more distant source-free pixels are to be used in
the background interpolation.

Figures \ref{imagesA2} and \ref{imagesA3} demonstrate several difficult cases, when one or more narrow sources appear on top of the
much wider, well-resolved starless source, referred to as a sub-structured source. If the narrow source is located close to the
peak of the wide source, it is practically impossible for an automated extraction method to distinguish these two sources.
Depending on the intensity distributions, the wide source may be regarded as the background of the narrow source, remaining not
extracted, or it may be considered as belonging to the power-law outskirts of the narrow source. More often, such narrow sources
are located off-peak of the wide source, hence they can be detected as separate sources. In both cases, however, the benchmarks
reveal that their measurements are inaccurate, because of the incorrectly determined individual backgrounds of each source and an
approximate nature of their deblending. Backgrounds of sources are highly uncertain (cf. Appendix \ref{backgrounds}) and it is not
surprising that they are even less accurate for the blended sources.

An inspection of Figs. \ref{imagesA2}\,--\,\ref{imagesB4} reveals several spurious sources, those that do not exist in the
benchmarks. The spurious detections are partially or completely discarded from the final catalogs during measurements by the
acceptability criteria in Eq.~(\ref{acceptable}). Some spurious sources are found on the well-resolved starless sources, whose
large-scale intensity peak enhances the small-scale background and noise fluctuations, making them appear as real sources. When a
source extraction aims at the highest possible completeness, at finding the faintest sources, it is normal that some peaks,
produced by the background and noise fluctuations, are mistakenly identified as genuine sources. A good source extraction method
must, however, guarantee that the number $N_{{\rm S}{\lambda}}$ of spurious sources in the final catalog remains below a few
percent of the number of real sources $N_{{\rm D}{\lambda}}$. For some studies, it may be beneficial to require that a valid source
must be detected and acceptable in at least two wavebands. This strategy potentially removes most of the spurious sources, together
with some real sources, unfortunately. It is better not to apply such a condition when benchmarking source extraction methods,
because practical applications often require extractions in a single image.

In the surface density images $\mathcal{D}_{\{11|13\}{\arcsec}}$ in Figs. \ref{imagesA2} and \ref{imagesB2}, the footprints of
several unresolved peaks of quite extended protostellar cores appear too small. They correspond to just the unresolved central
peaks and not to the entire large cores with their power-law profiles. The same sources have large sizes of their extended
footprints at $\{160|170\}$ and $500$\,$\mu$m and in the benchmarks with background (Figs. \ref{imagesA3}\,--\,\ref{imagesB4}).
This abnormality is caused by the derivation algorithm of the images $\mathcal{D}_{\{11|13\}{\arcsec}}$, which employs fitting of
the spectral shapes $\Pi_{\lambda}$ of the pixels. The surface densities are known to be quite inaccurate in the pixels with strong
temperature gradients along the lines of sight (e.g., Appendix~A of Paper I). Such fitting problems lead to the overestimated
temperatures and underestimated surface densities around the unresolved peaks. The resulting strong depressions (local minima)
around the peaks of several protostellar cores in $\mathcal{D}_{\{11|13\}{\arcsec}}$ prevent the extraction methods from finding
the correct footprint sizes. This happens only in the simplest benchmarks $\{\mathrm{A},\mathrm{B}\}_2$ with just two components
(sources and noise), because the bright emission of the background and filament dilutes the temperature effect along the lines of
sight within the cores.


\begin{figure*}
\centering
\centerline{\resizebox{0.2695\hsize}{!}{\includegraphics{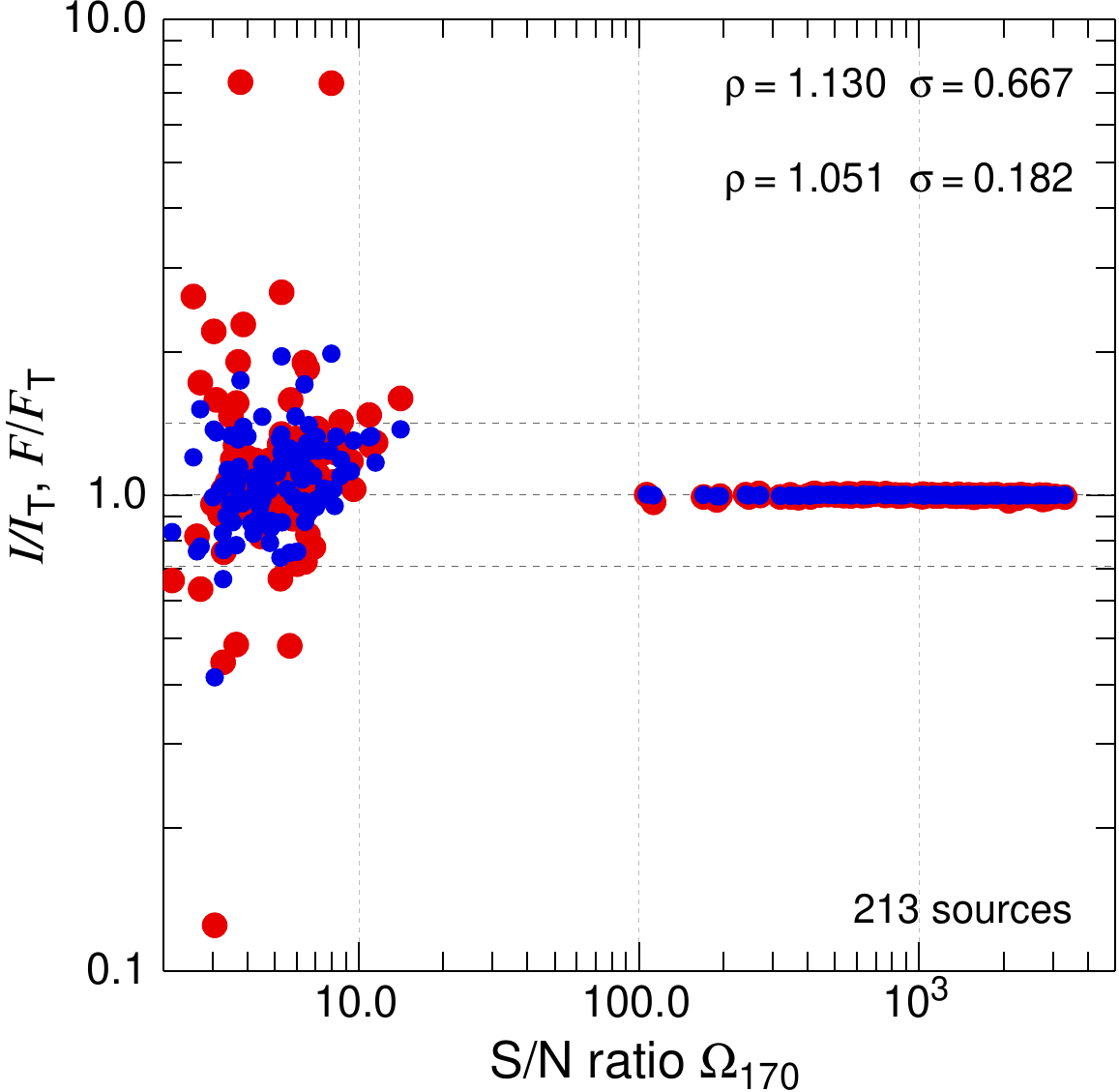}}
            \resizebox{0.2315\hsize}{!}{\includegraphics{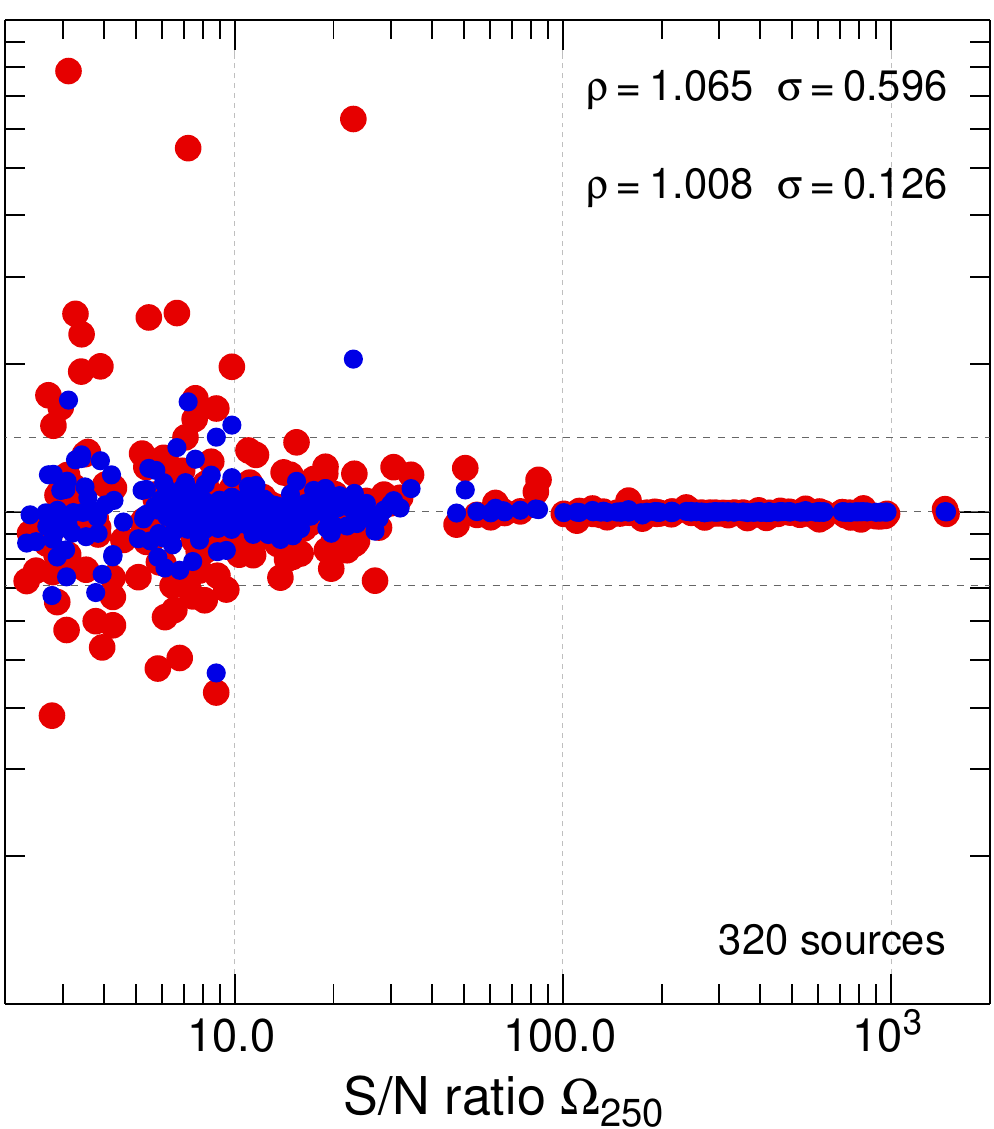}}
            \resizebox{0.2315\hsize}{!}{\includegraphics{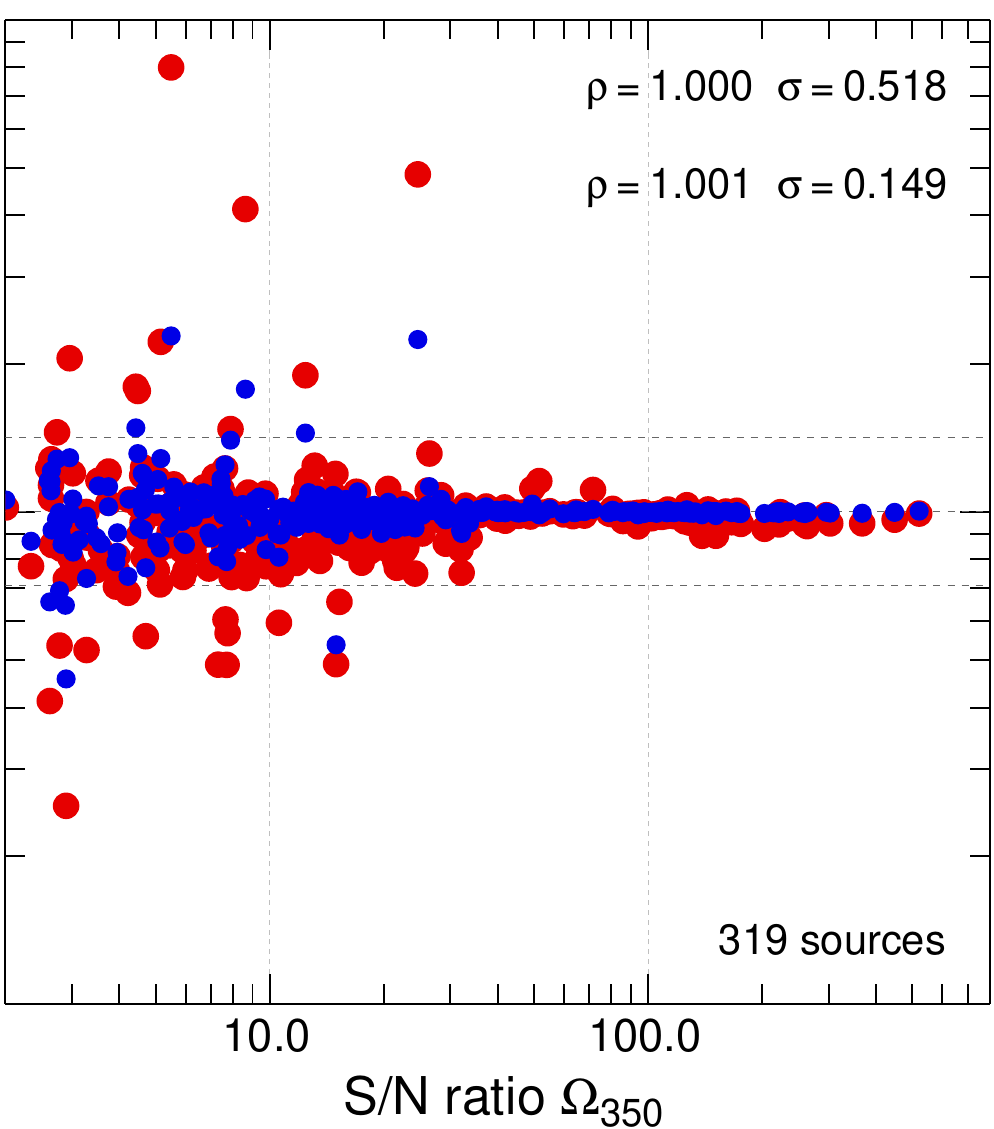}}
            \resizebox{0.2435\hsize}{!}{\includegraphics{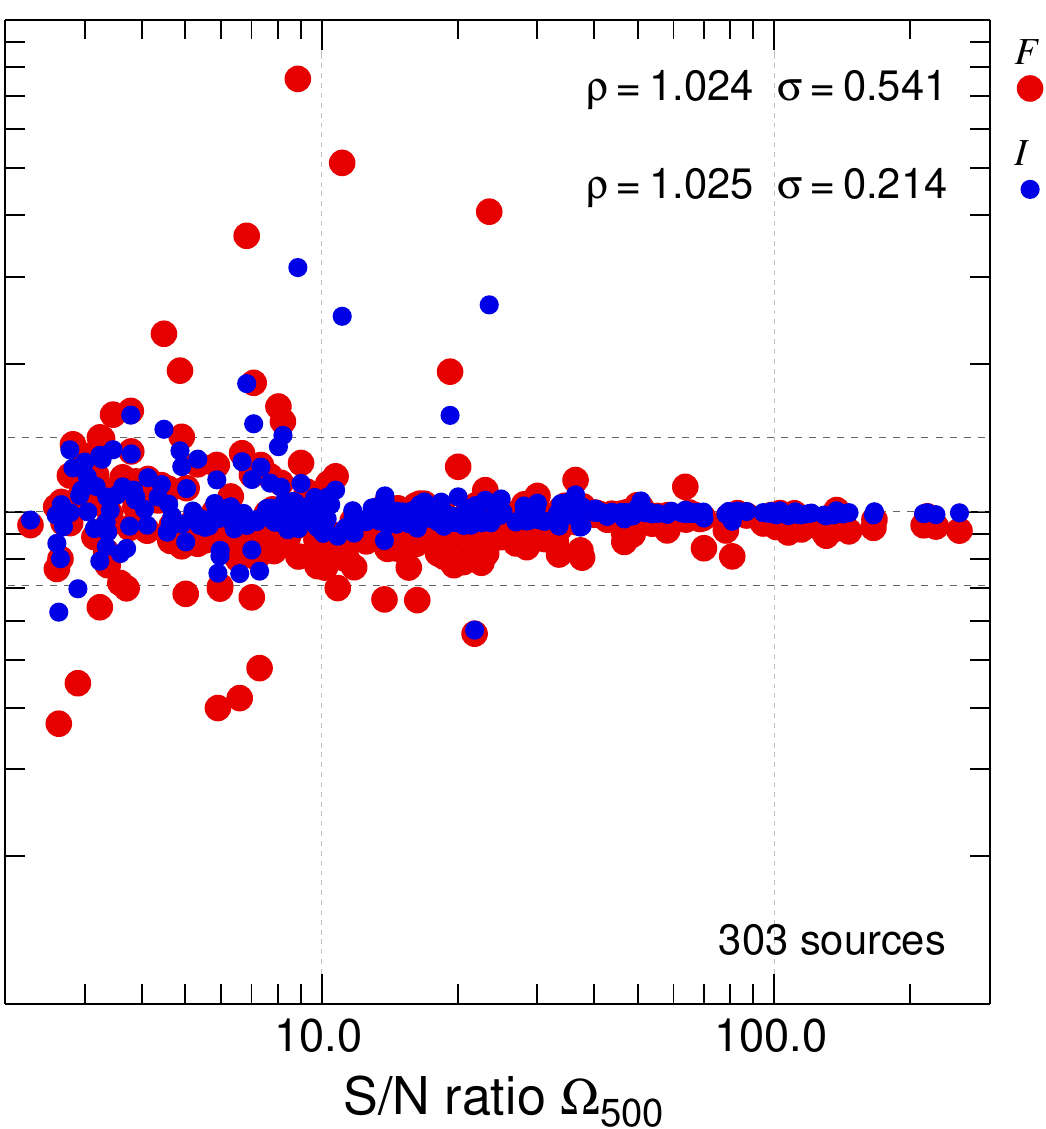}}}
\vspace{1.0mm}
\centerline{\resizebox{0.2695\hsize}{!}{\includegraphics{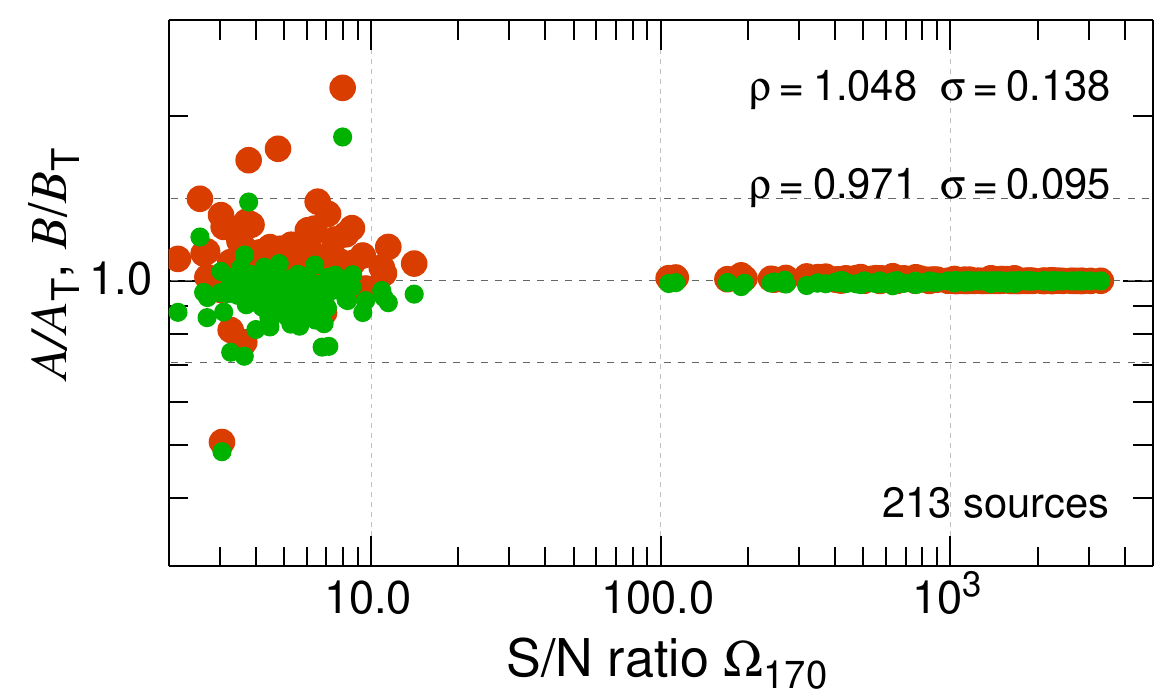}}
            \resizebox{0.2315\hsize}{!}{\includegraphics{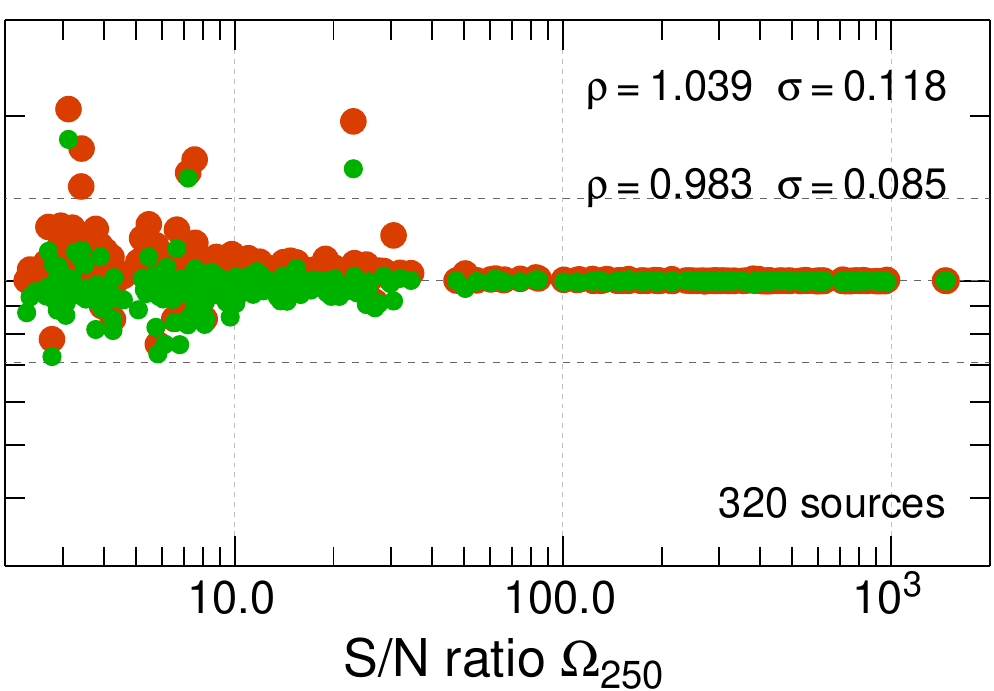}}
            \resizebox{0.2315\hsize}{!}{\includegraphics{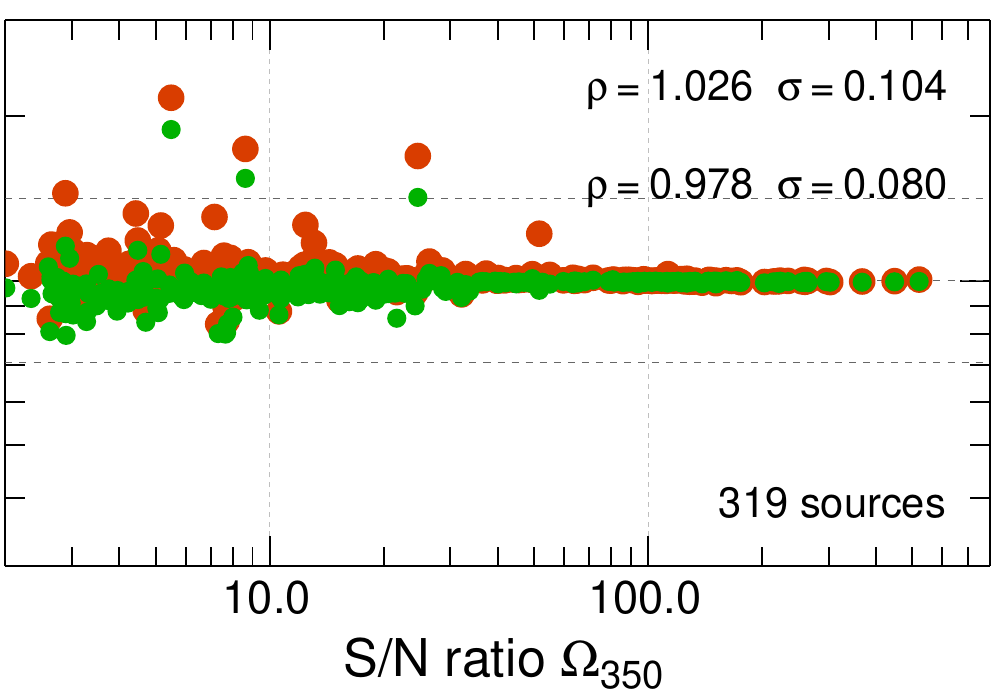}}
            \resizebox{0.2435\hsize}{!}{\includegraphics{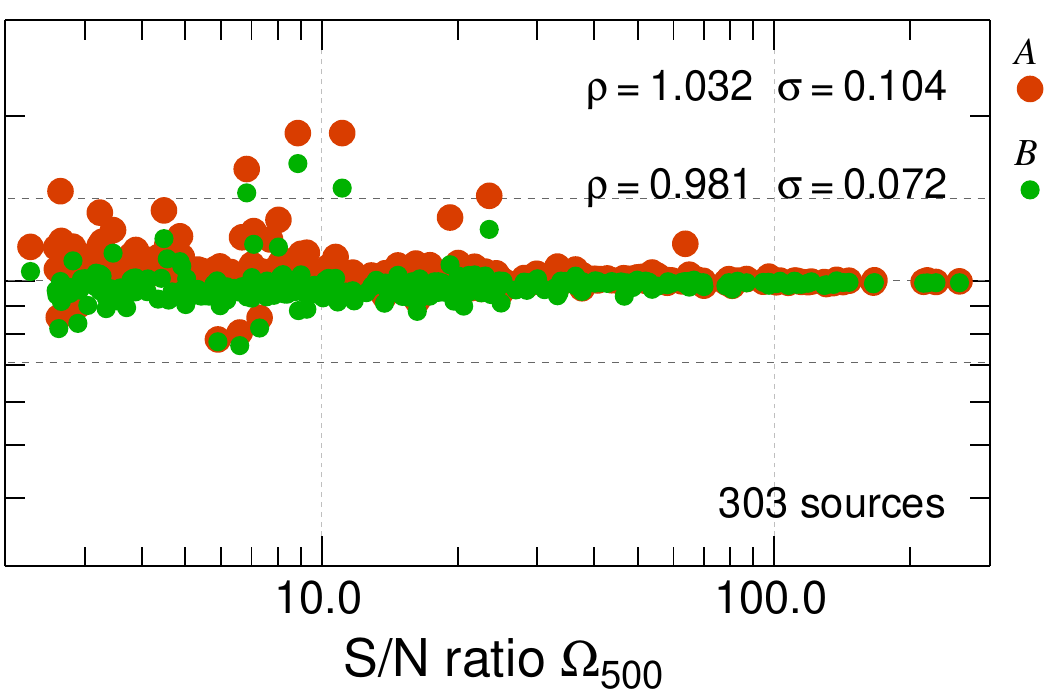}}}
\vspace{1.0mm}
\centerline{\resizebox{0.2695\hsize}{!}{\includegraphics{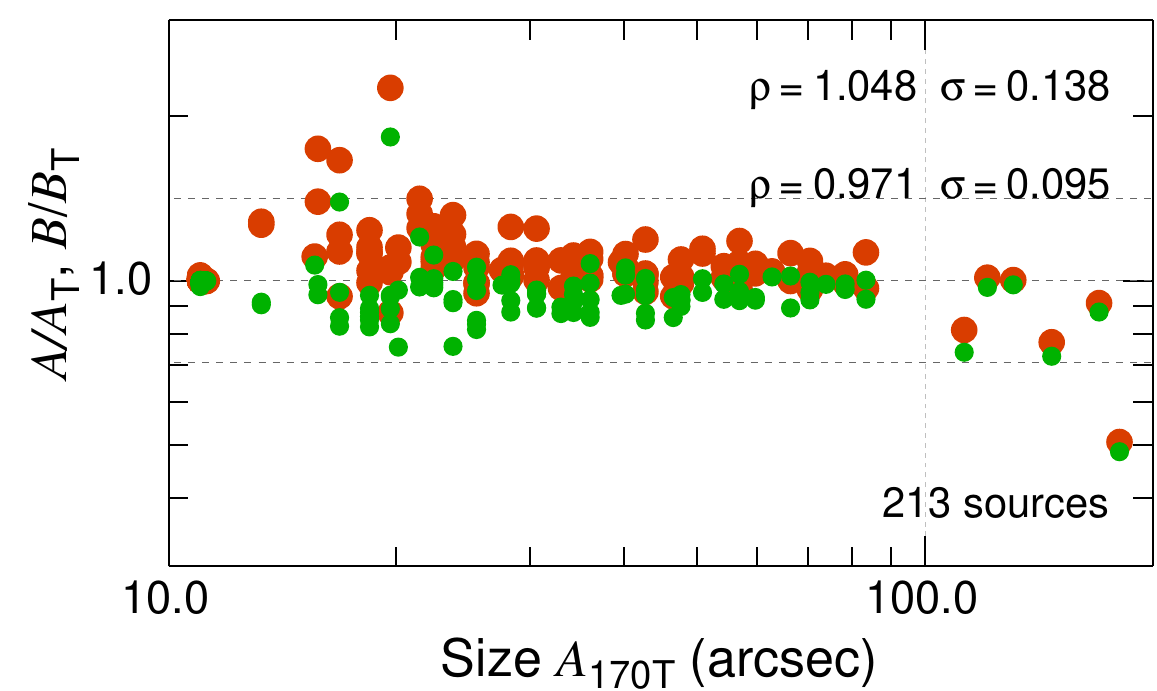}}
            \resizebox{0.2315\hsize}{!}{\includegraphics{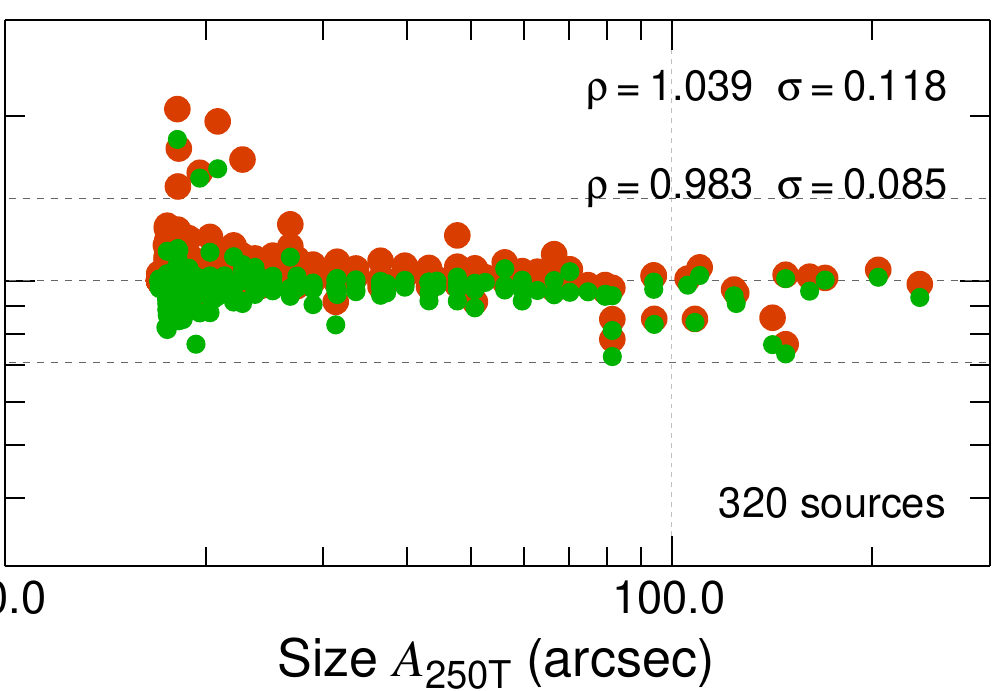}}
            \resizebox{0.2315\hsize}{!}{\includegraphics{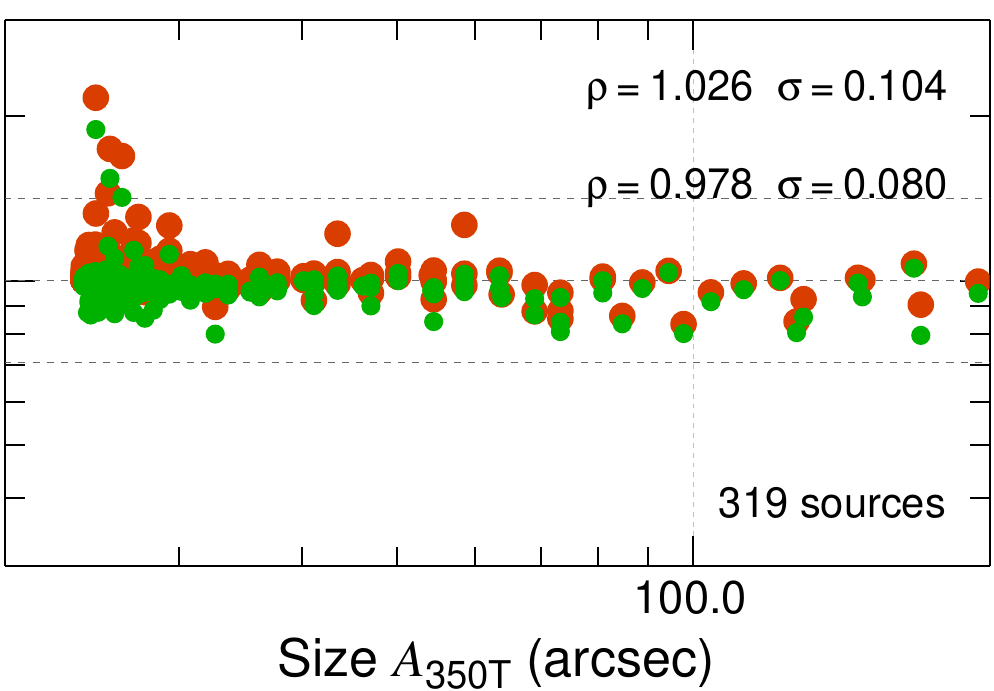}}
            \resizebox{0.2435\hsize}{!}{\includegraphics{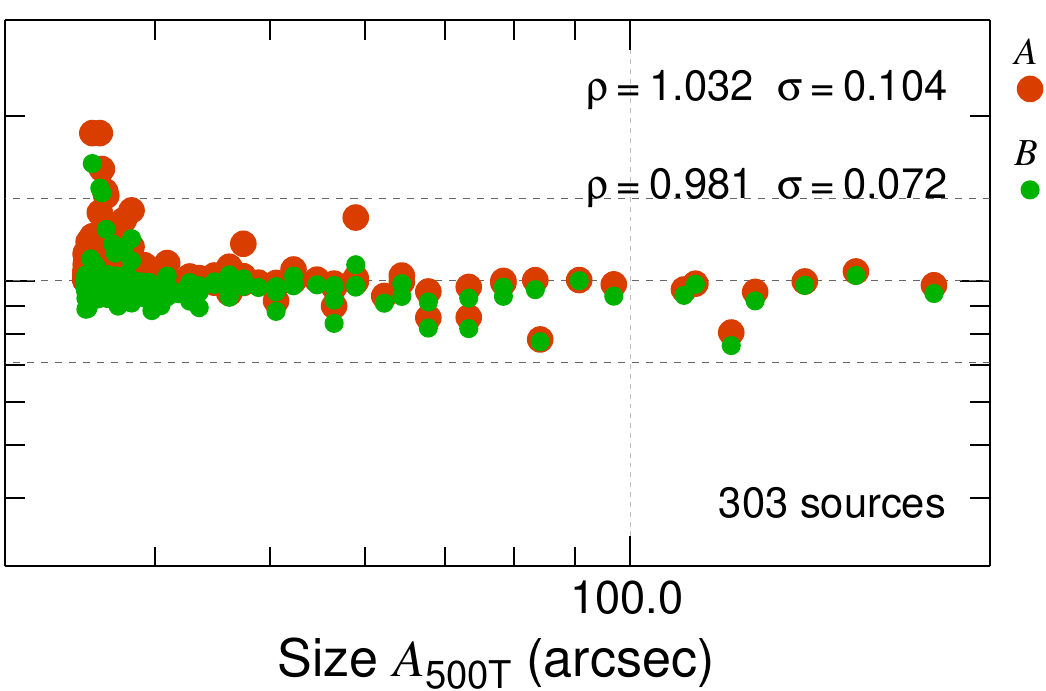}}}
\vspace{2.3mm}
\centerline{\resizebox{0.2695\hsize}{!}{\includegraphics{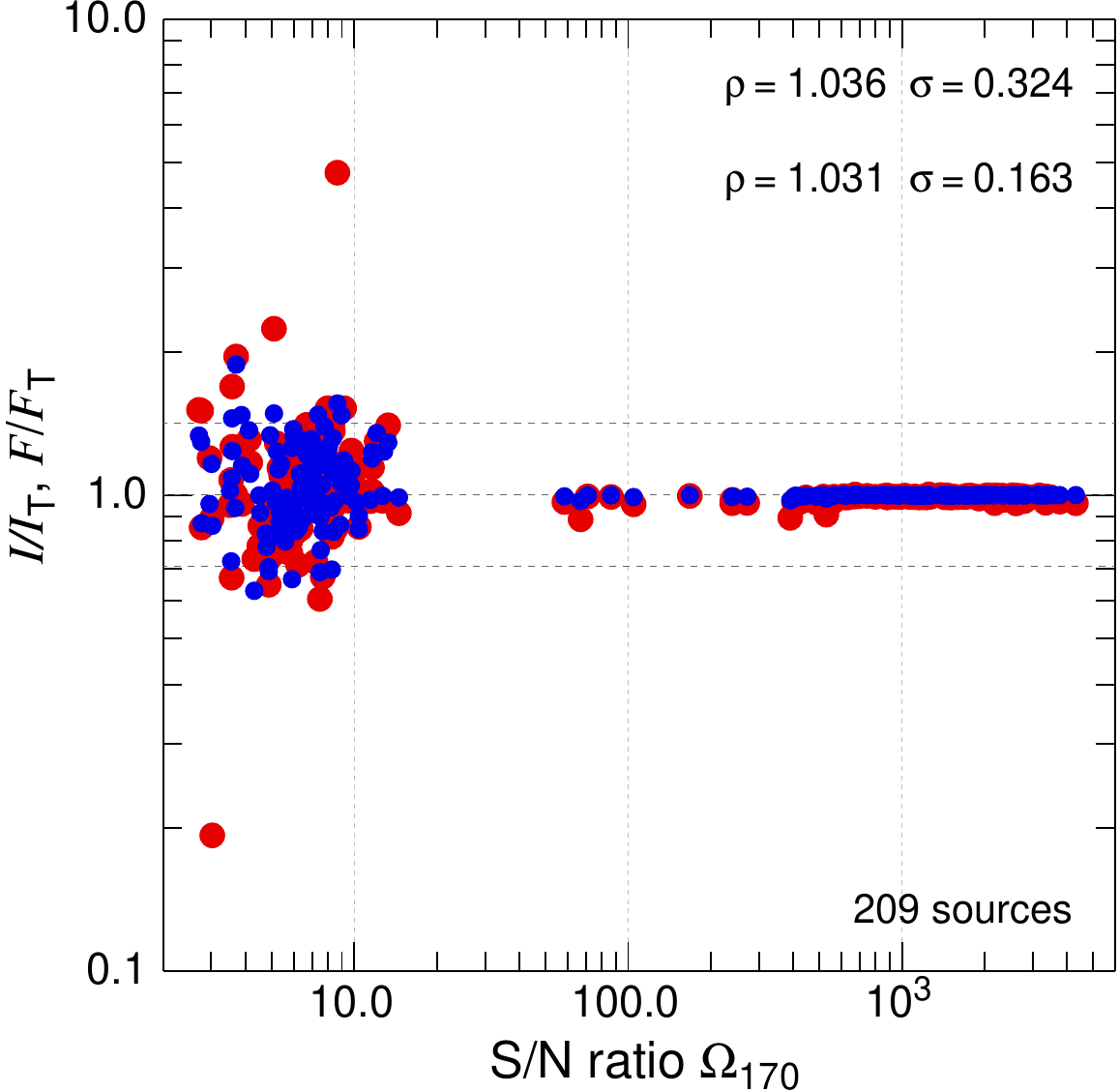}}
            \resizebox{0.2315\hsize}{!}{\includegraphics{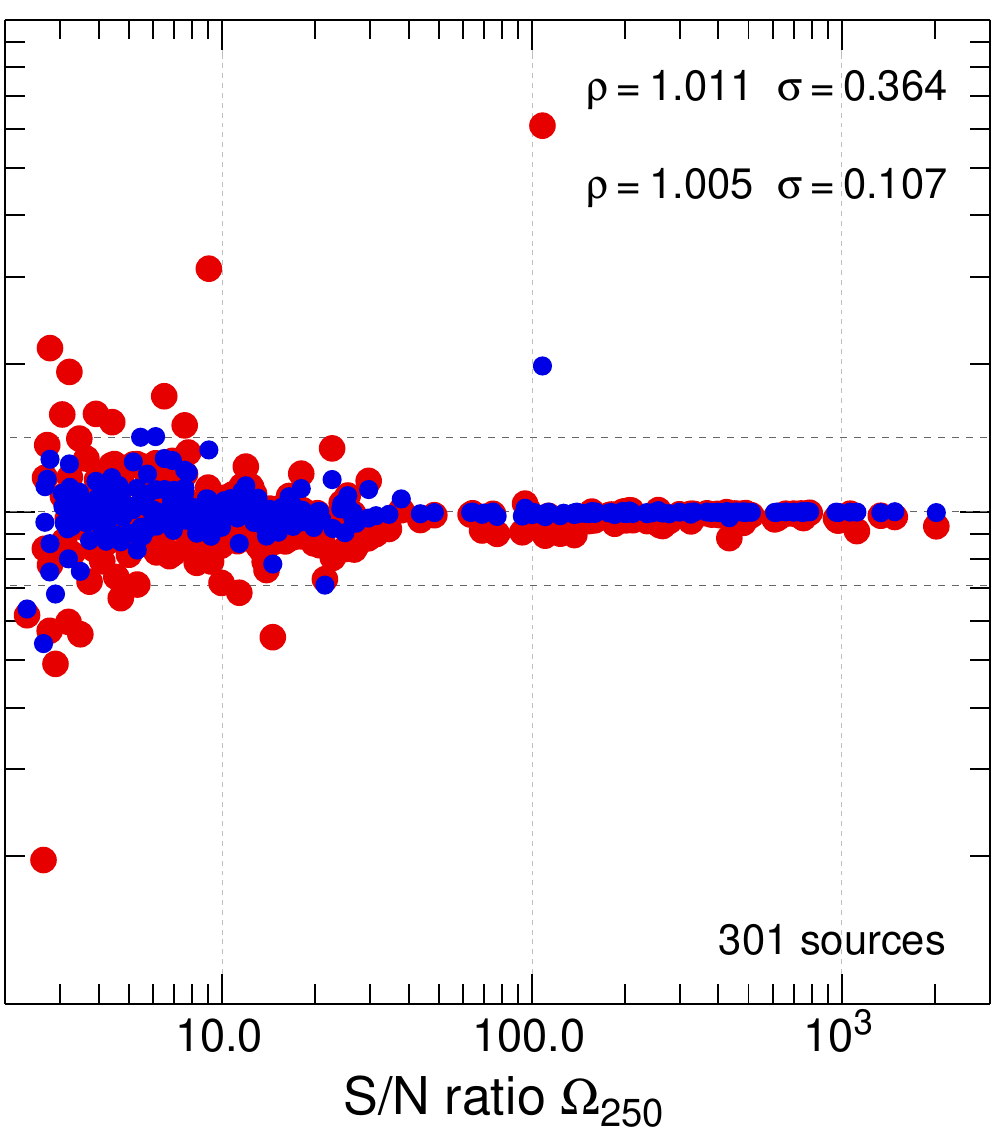}}
            \resizebox{0.2315\hsize}{!}{\includegraphics{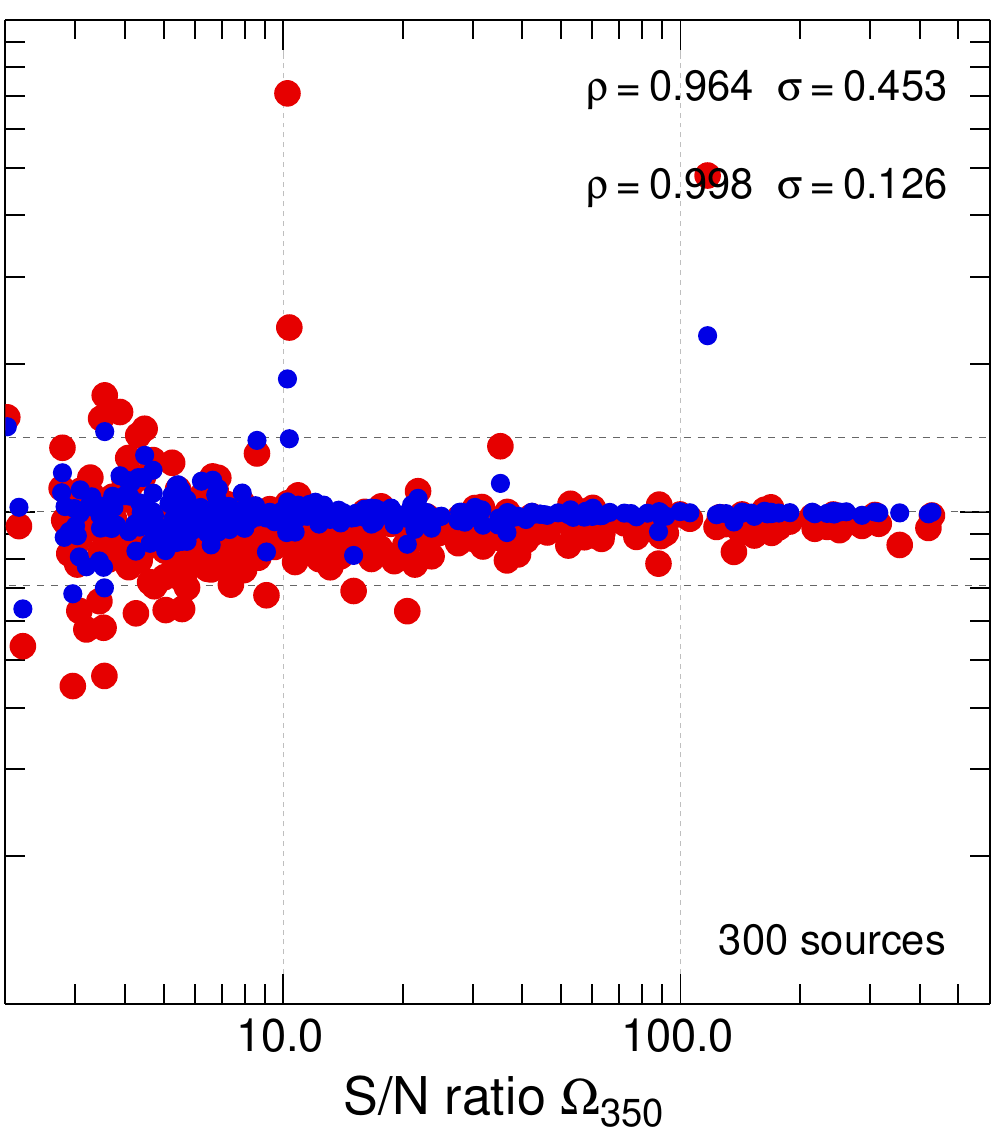}}
            \resizebox{0.2435\hsize}{!}{\includegraphics{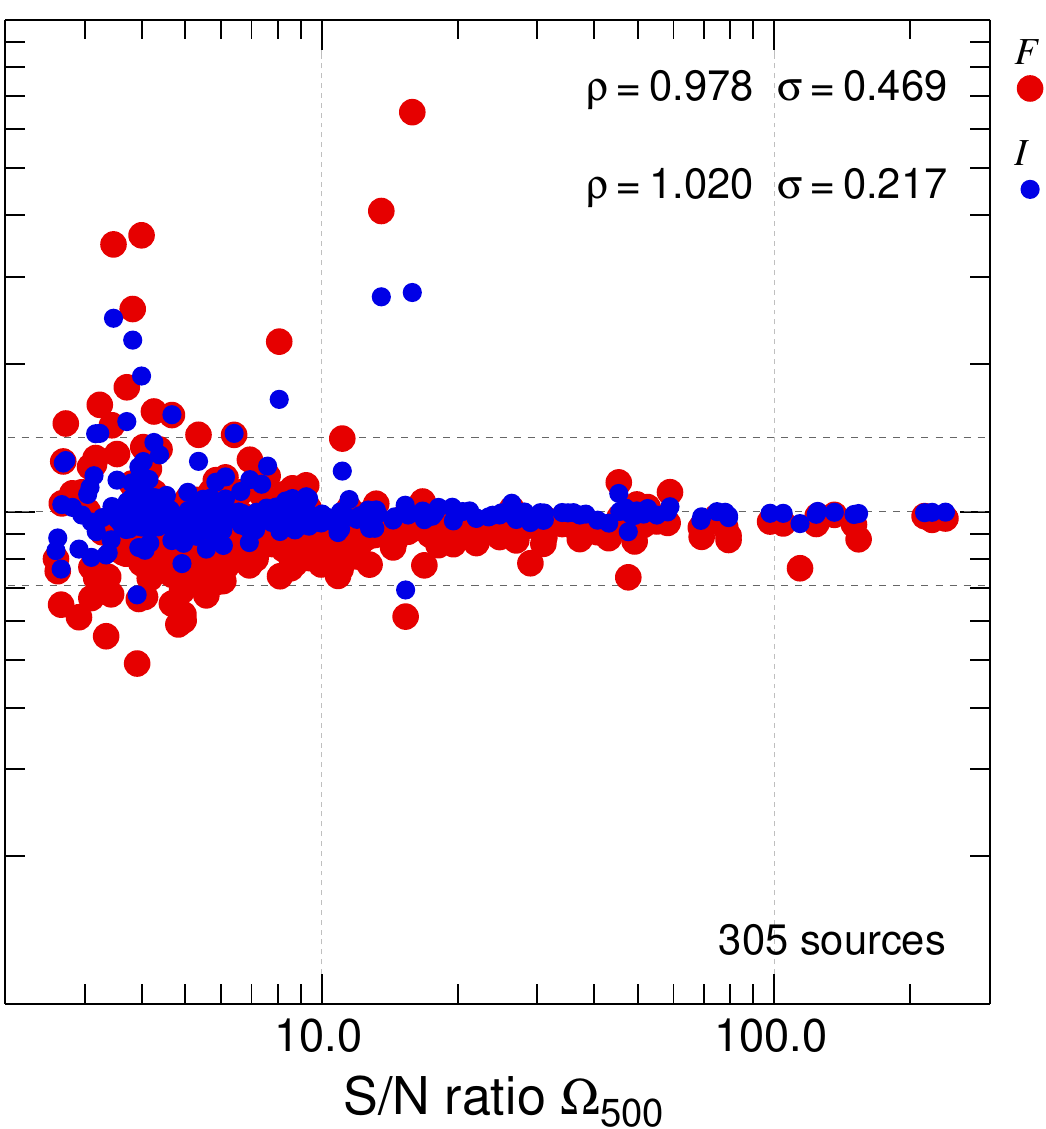}}}
\vspace{1.0mm}
\centerline{\resizebox{0.2695\hsize}{!}{\includegraphics{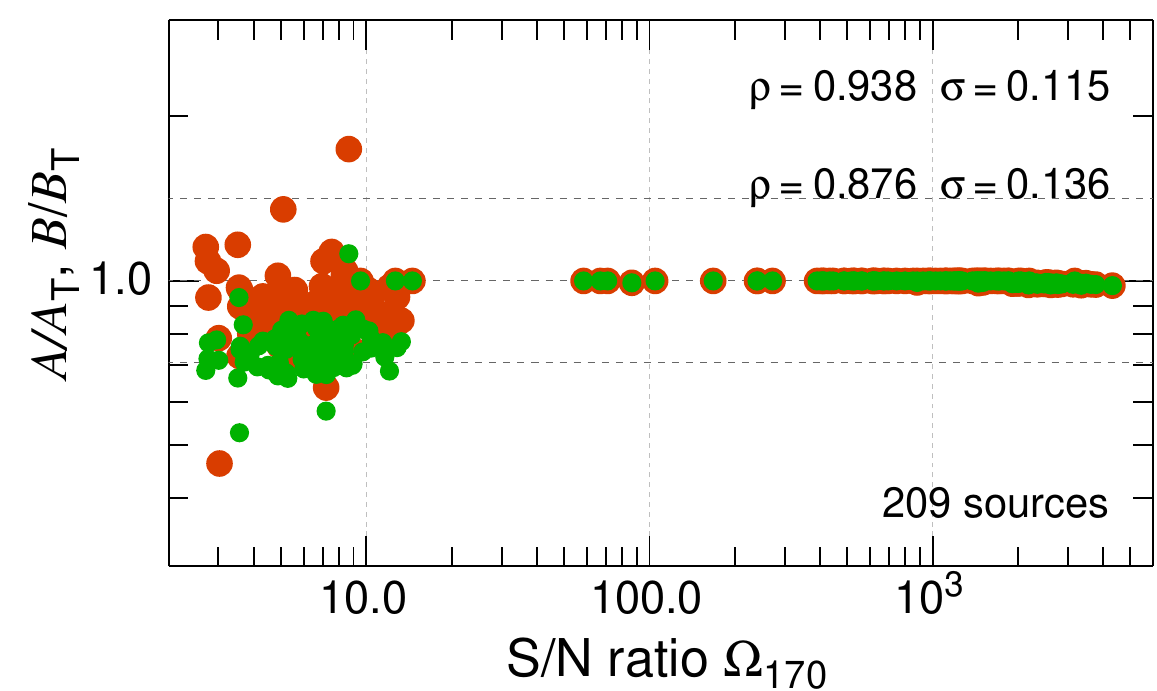}}
            \resizebox{0.2315\hsize}{!}{\includegraphics{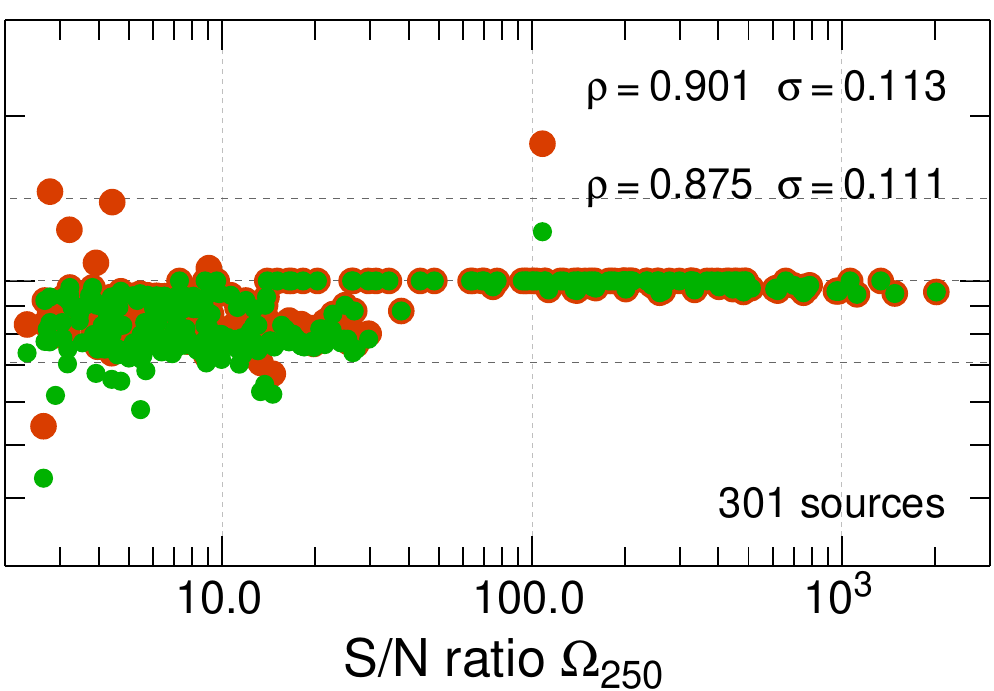}}
            \resizebox{0.2315\hsize}{!}{\includegraphics{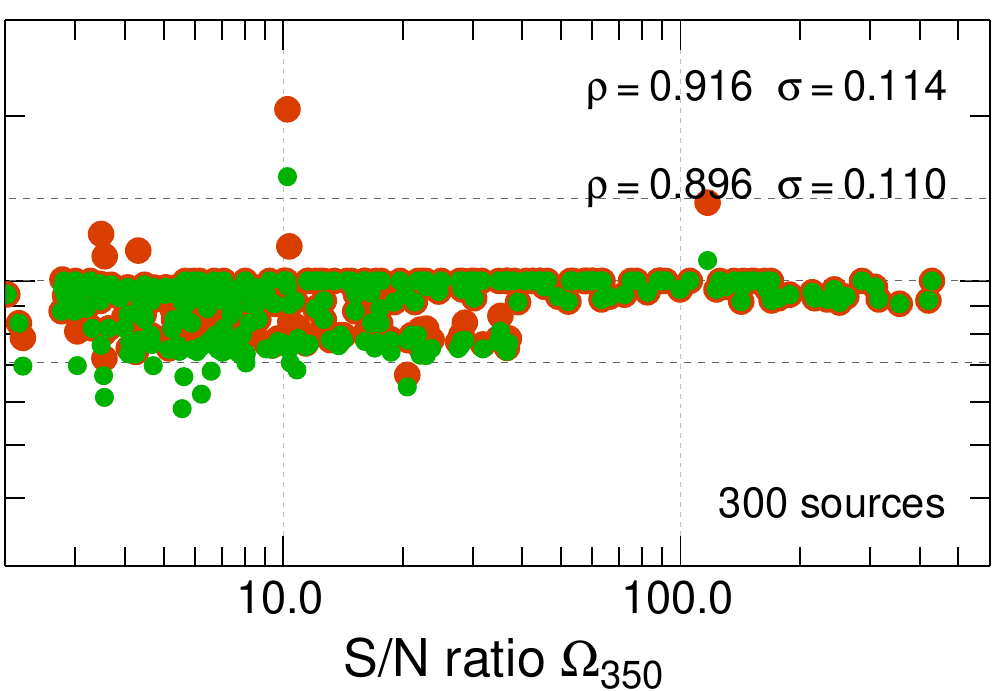}}
            \resizebox{0.2435\hsize}{!}{\includegraphics{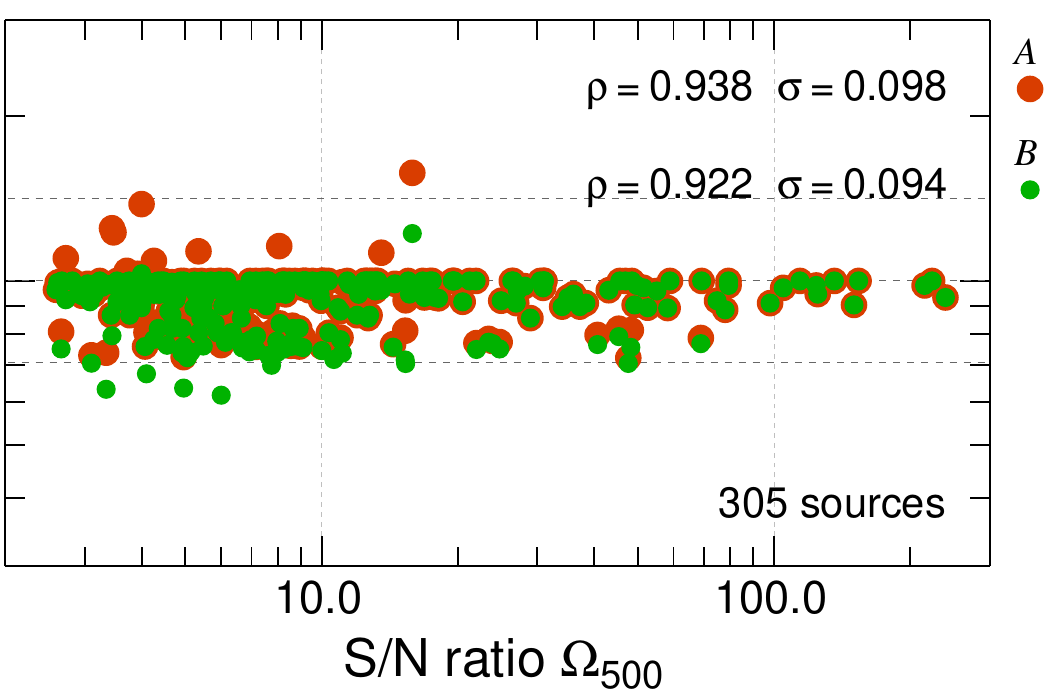}}}
\vspace{1.0mm}
\centerline{\resizebox{0.2695\hsize}{!}{\includegraphics{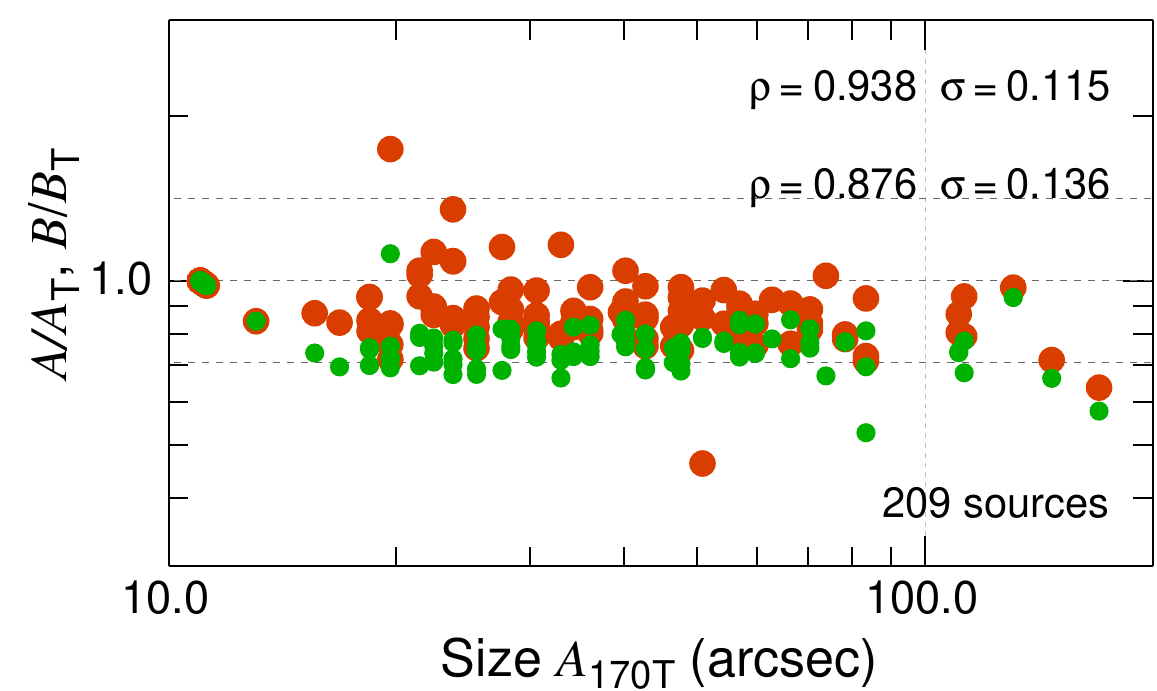}}
            \resizebox{0.2315\hsize}{!}{\includegraphics{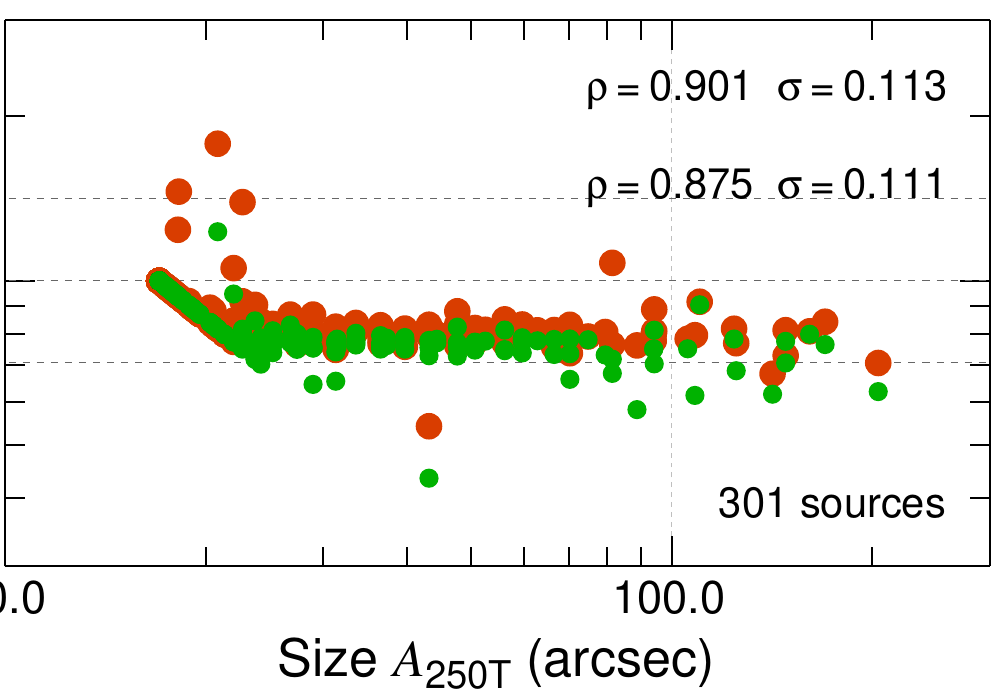}}
            \resizebox{0.2315\hsize}{!}{\includegraphics{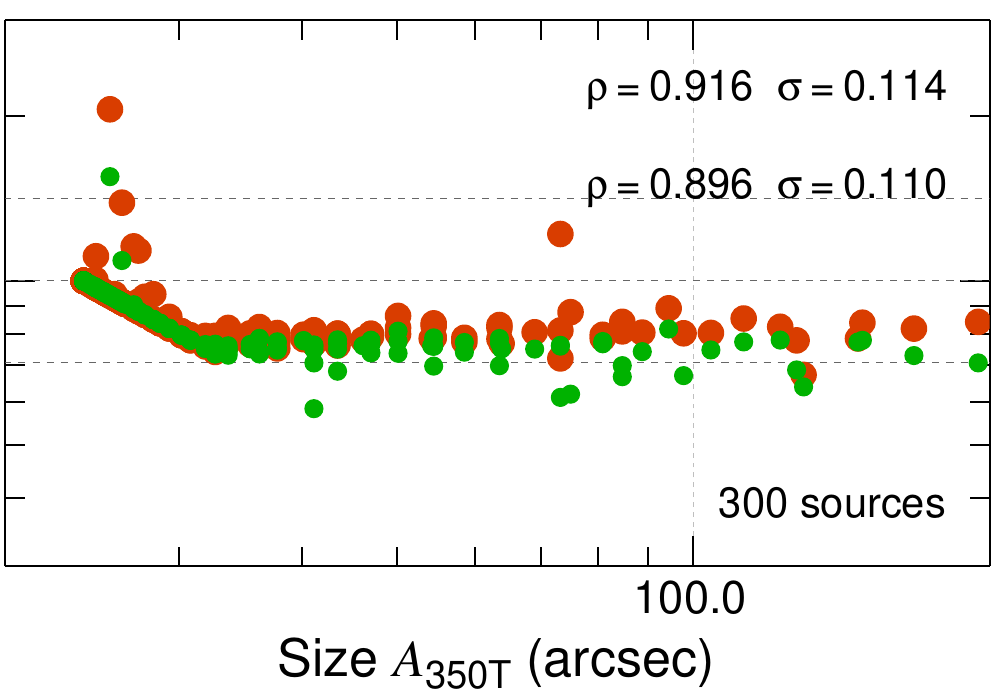}}
            \resizebox{0.2435\hsize}{!}{\includegraphics{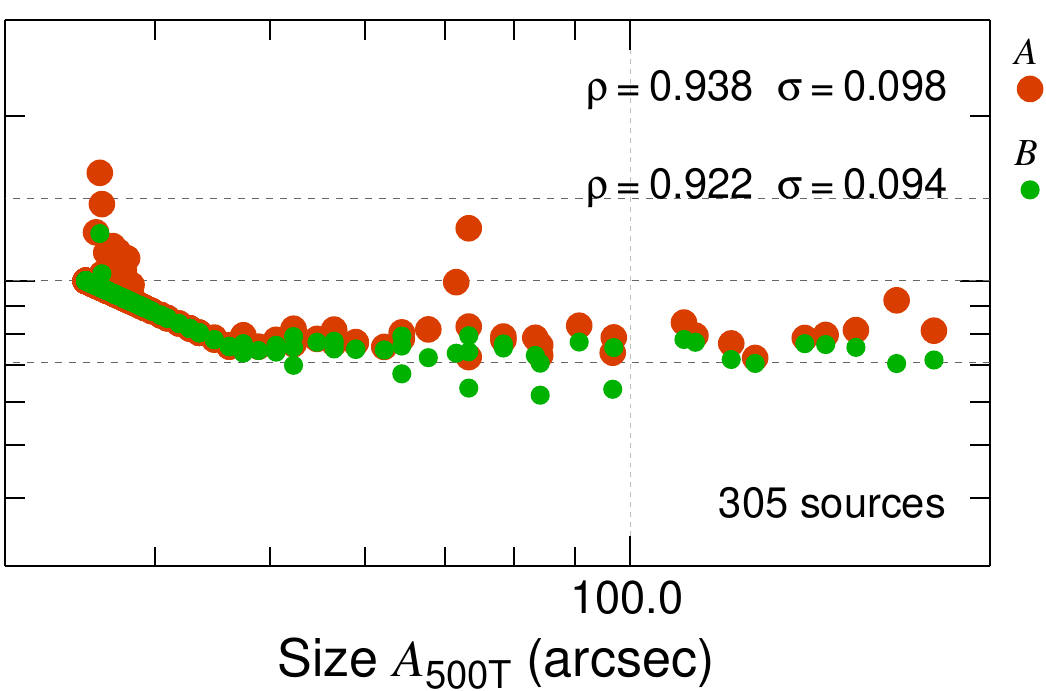}}}
\caption
{ 
Benchmark A$_3$ extraction with \textsl{getsf} (three \emph{top} rows) and \textsl{getold} (three \emph{bottom} rows). Ratios of
the measured fluxes $F_{{\rm T}{\lambda}{n}}$, peak intensities $F_{{\rm P}{\lambda}{n}}$, and sizes $\{A,B\}_{{\lambda}{n}}$ to
their true values ($F/F_{\rm T}$, $I/I_{\rm T}$, $A/A_{\rm T}$, and $B/B_{\rm T}$) are shown as a function of the S/N ratio
$\Omega_{{\lambda}{n}}$. The size ratios $A/A_{\rm T}$ and $B/B_{\rm T}$ are also shown as a function of the true sizes
$\{A,B\}_{{\lambda}{n}{\rm T}}$. The mean $\varrho_{{\rm \{P|T|A|B\}}{\lambda}}$ and standard deviation $\sigma_{{\rm
\{P|T|A|B\}}{\lambda}}$ of the ratios are displayed in the panels. Similar plots for $\lambda\le 110$\,$\mu$m with only bright
protostellar cores are not presented, because their measurements are quite accurate, with $\varrho_{\{{\rm
P|T|A|B}\}{\lambda}}\approx \{0.999|0.998|1.001|0.999\}$ and $\sigma_{\{{\rm P|T|A|B}\}{\lambda}}\approx
\{0.002|0.006|0.00004|0.00004\}$.
} 
\label{accuracyA3}
\end{figure*}


\begin{figure*}
\centering
\centerline{\resizebox{0.2695\hsize}{!}{\includegraphics{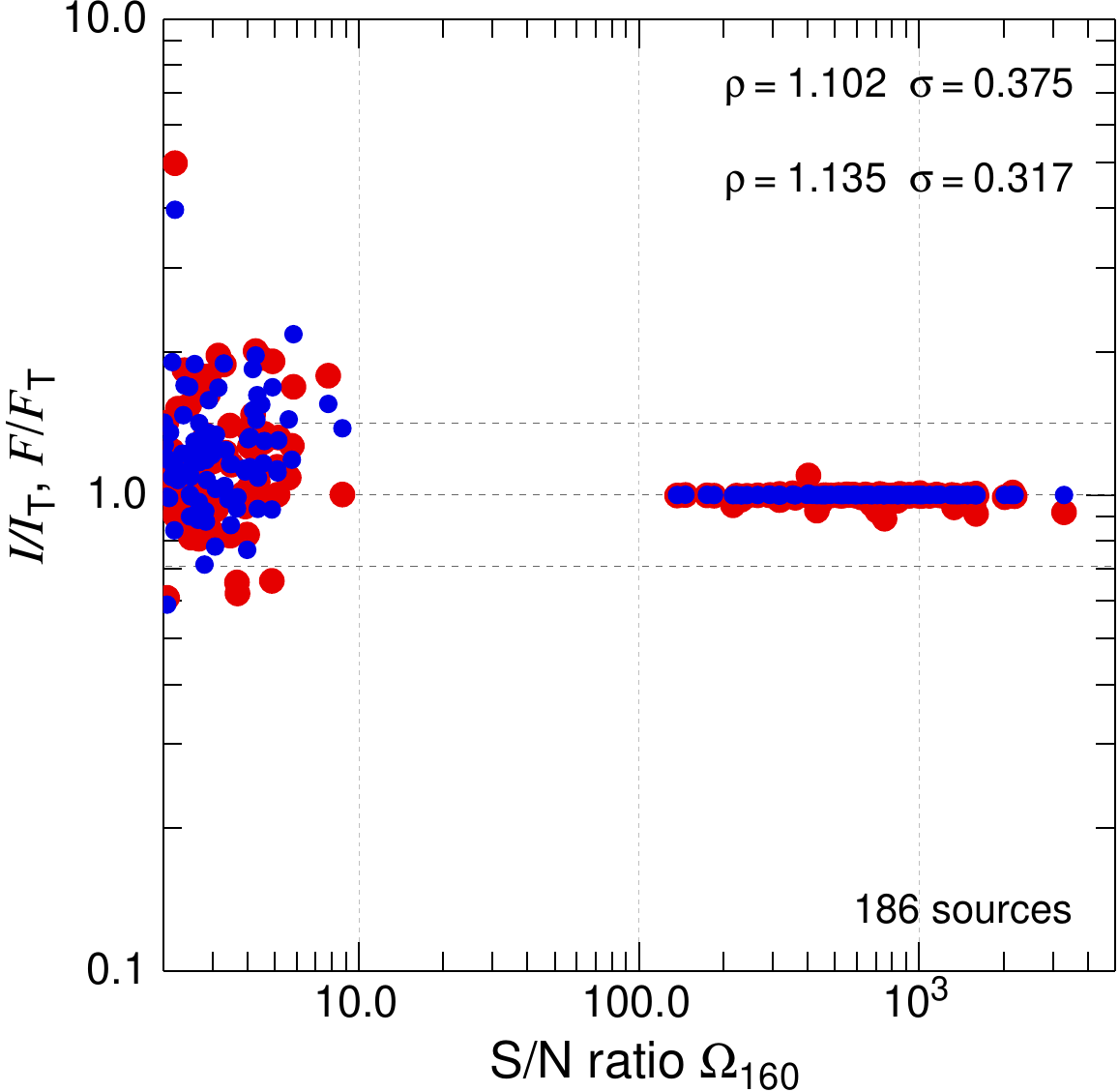}}
            \resizebox{0.2315\hsize}{!}{\includegraphics{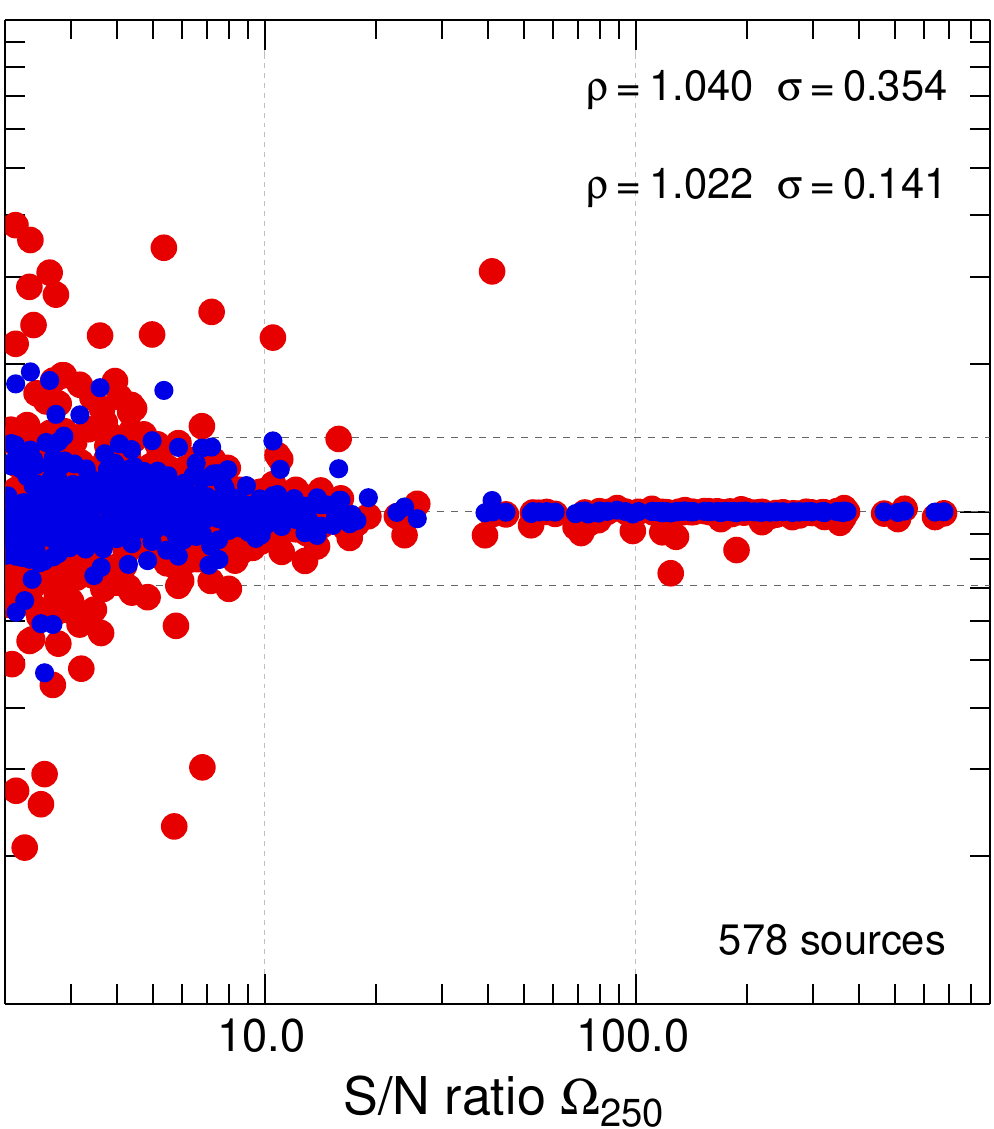}}
            \resizebox{0.2315\hsize}{!}{\includegraphics{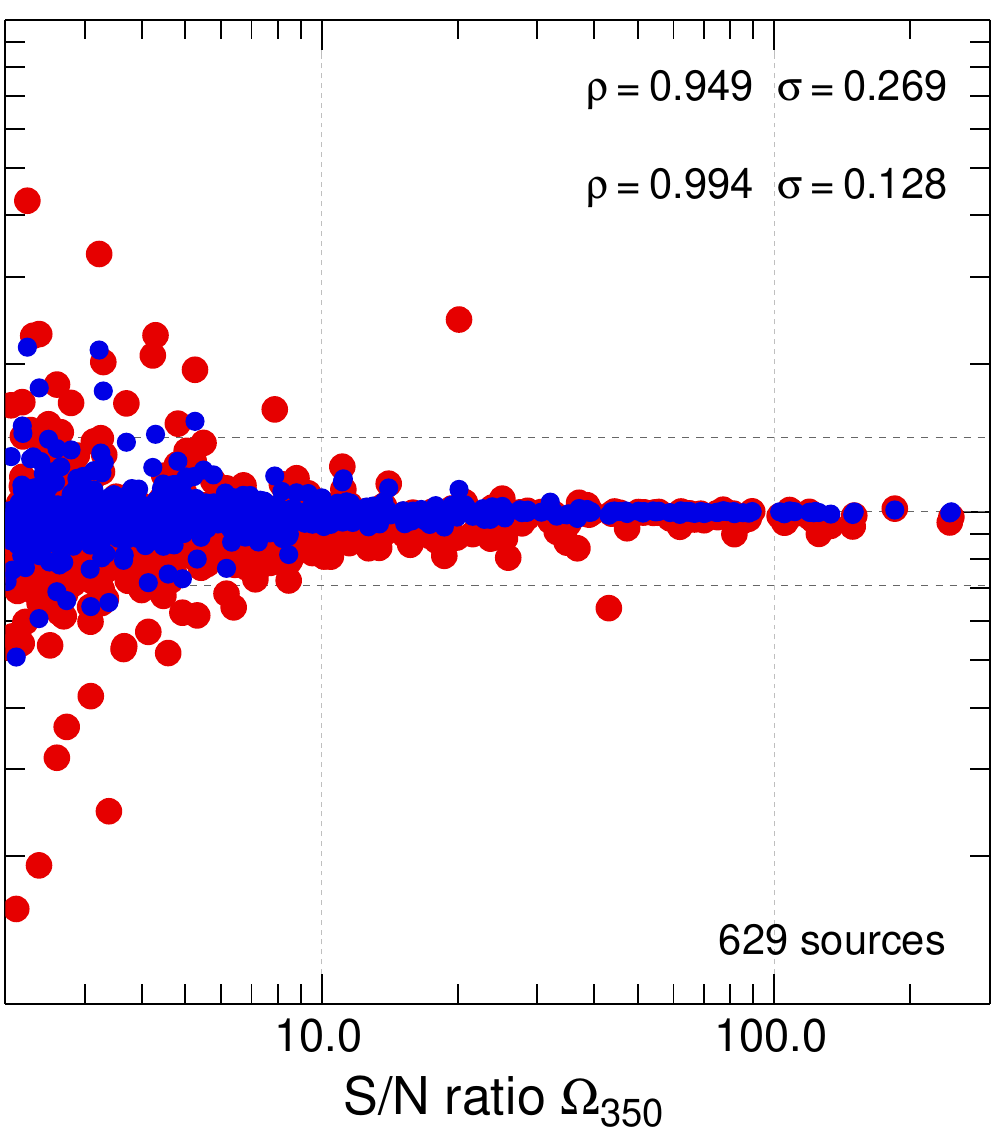}}
            \resizebox{0.2435\hsize}{!}{\includegraphics{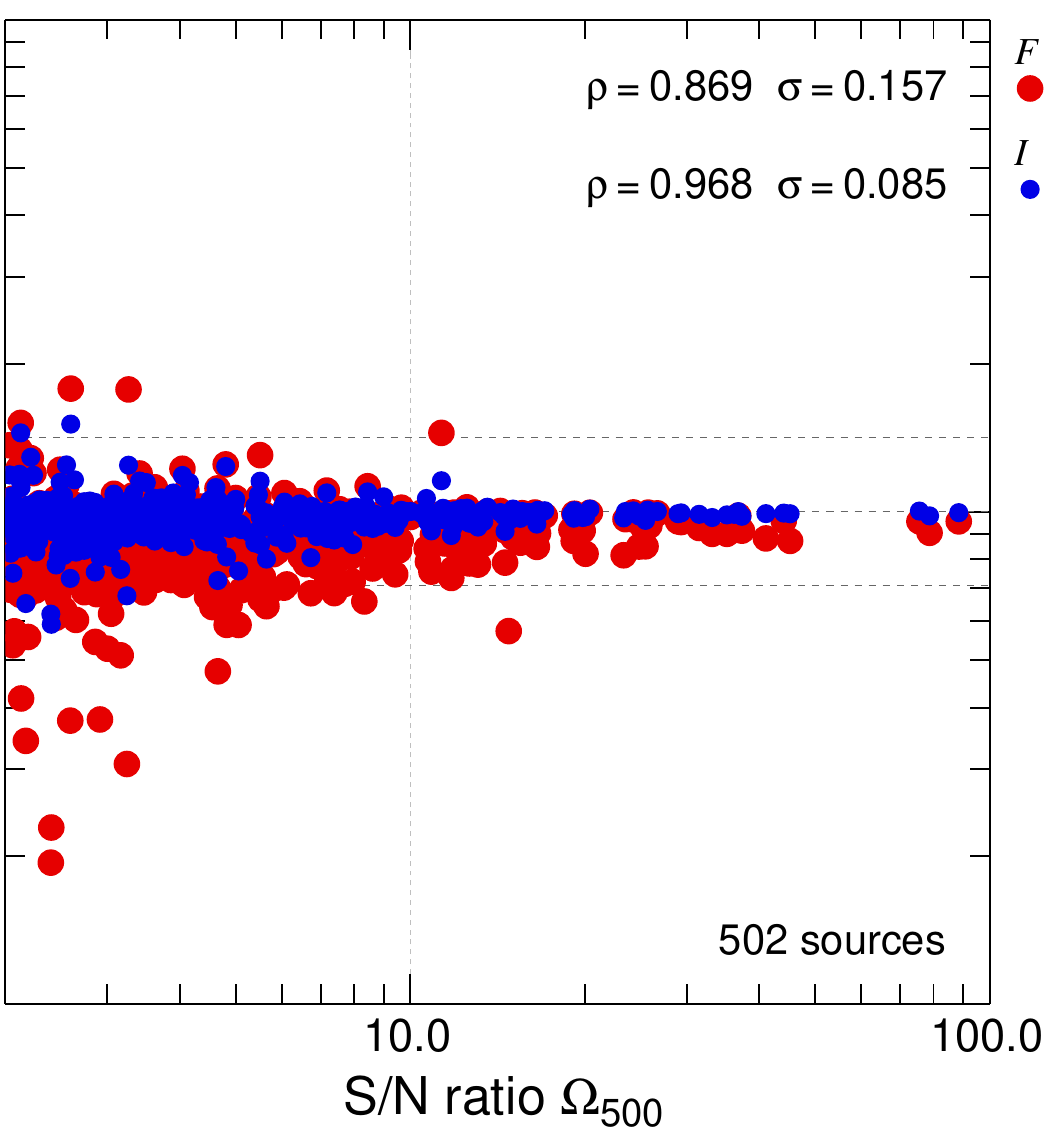}}}
\vspace{1.0mm}
\centerline{\resizebox{0.2695\hsize}{!}{\includegraphics{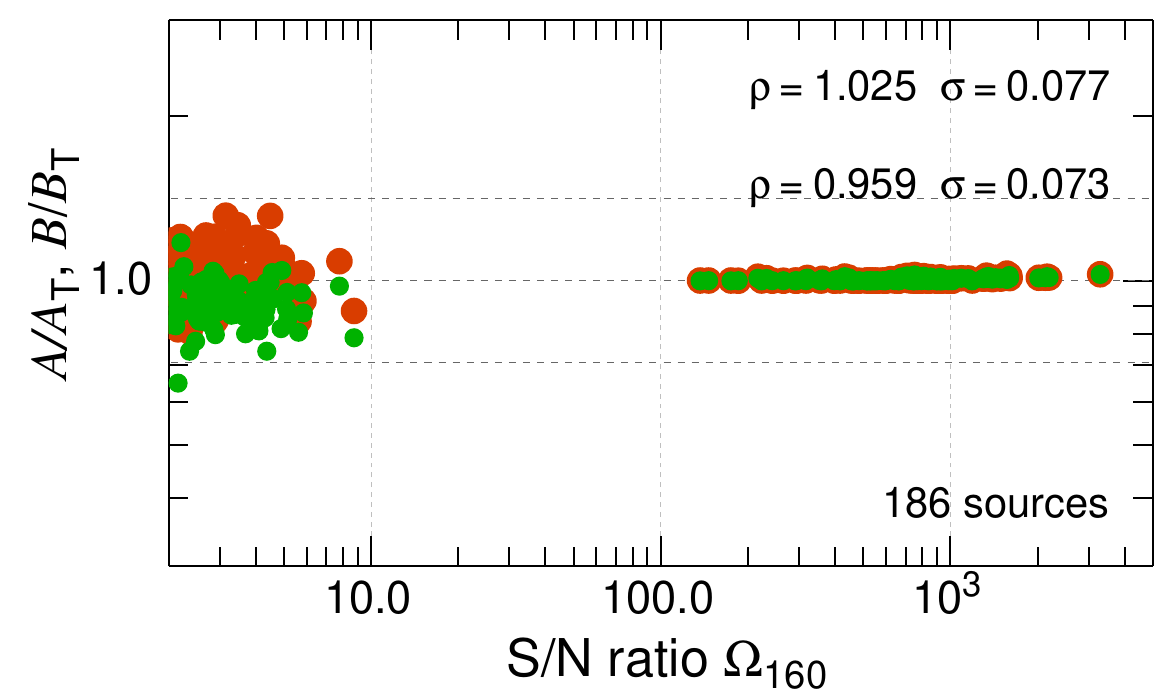}}
            \resizebox{0.2315\hsize}{!}{\includegraphics{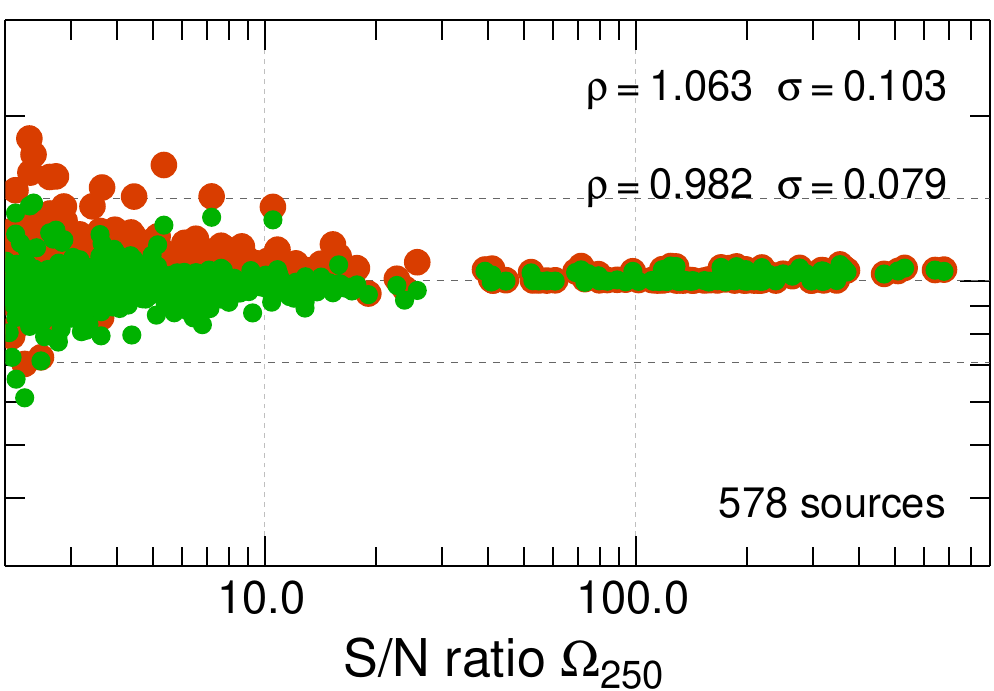}}
            \resizebox{0.2315\hsize}{!}{\includegraphics{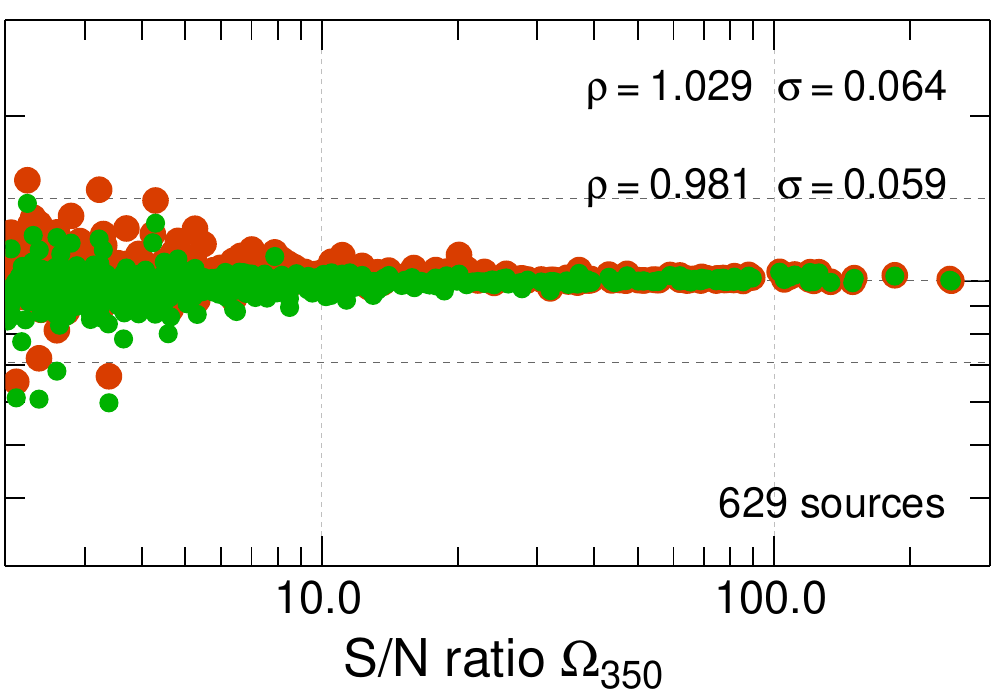}}
            \resizebox{0.2435\hsize}{!}{\includegraphics{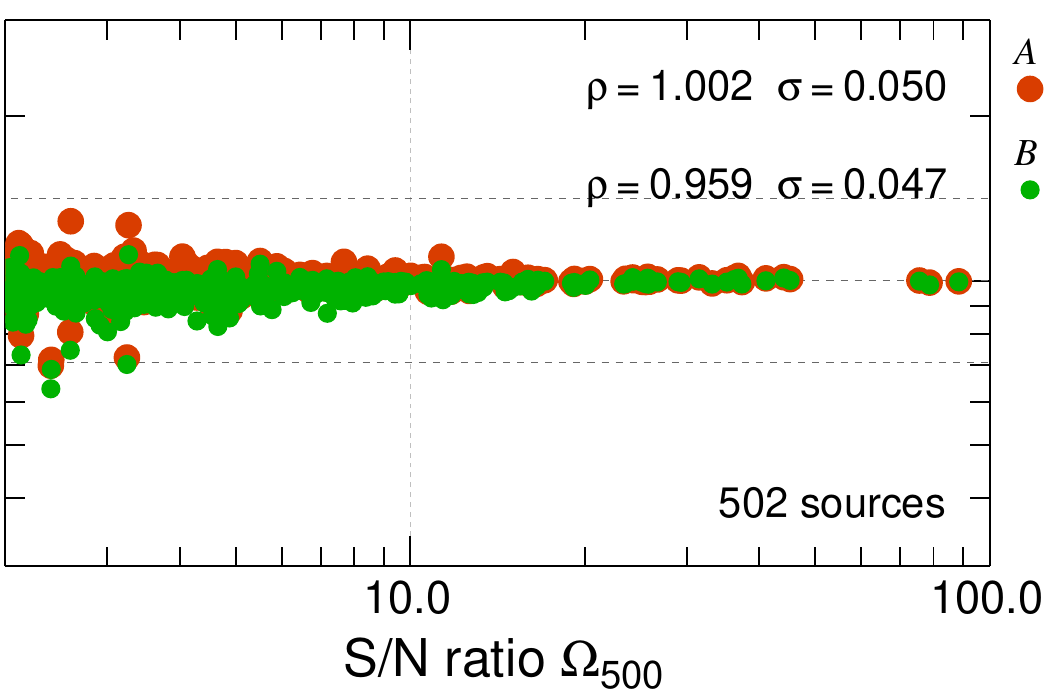}}}
\vspace{1.0mm}
\centerline{\resizebox{0.2695\hsize}{!}{\includegraphics{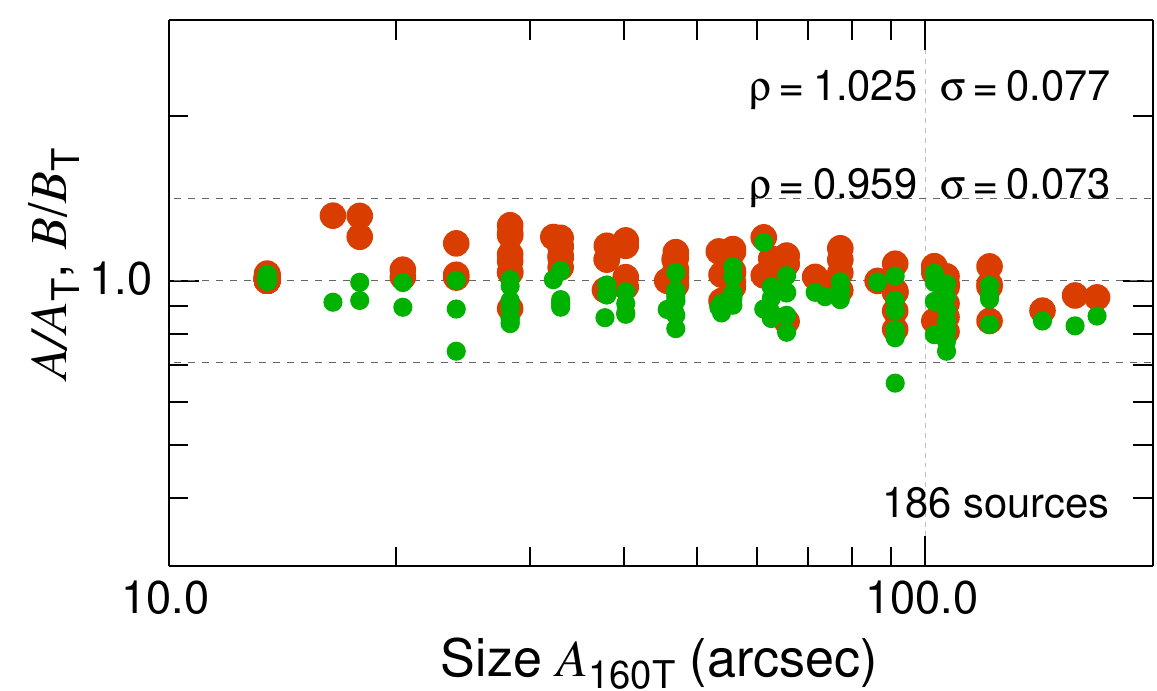}}
            \resizebox{0.2315\hsize}{!}{\includegraphics{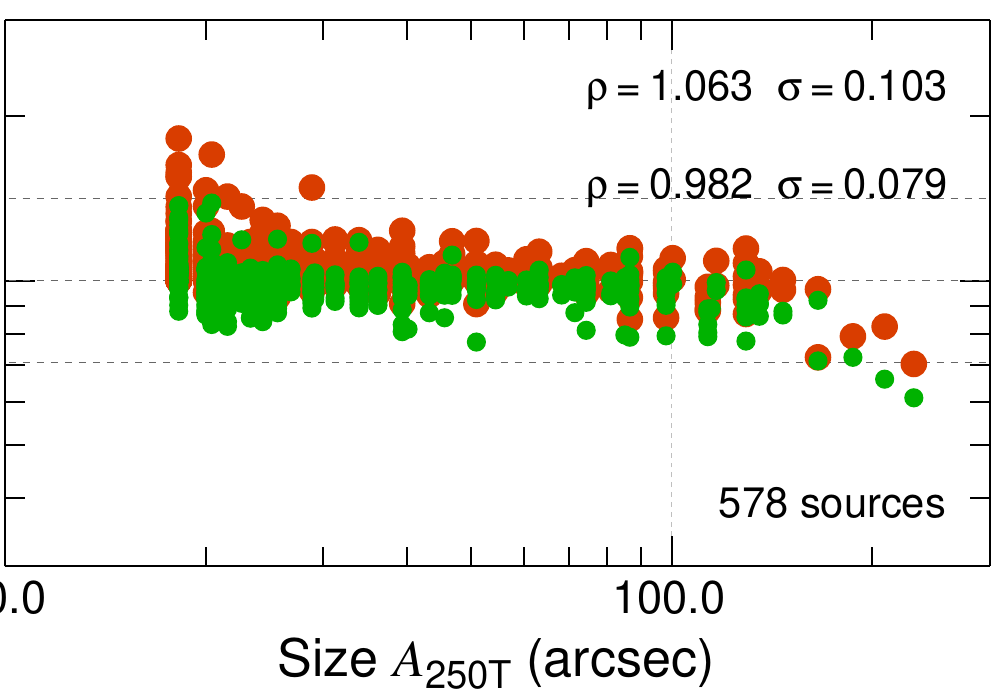}}
            \resizebox{0.2315\hsize}{!}{\includegraphics{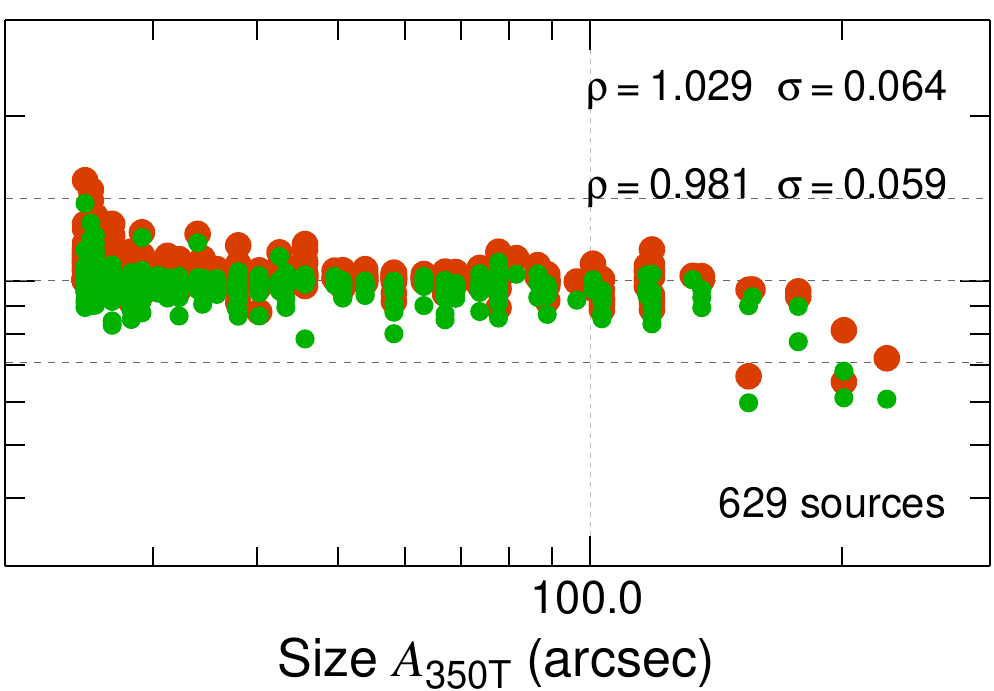}}
            \resizebox{0.2435\hsize}{!}{\includegraphics{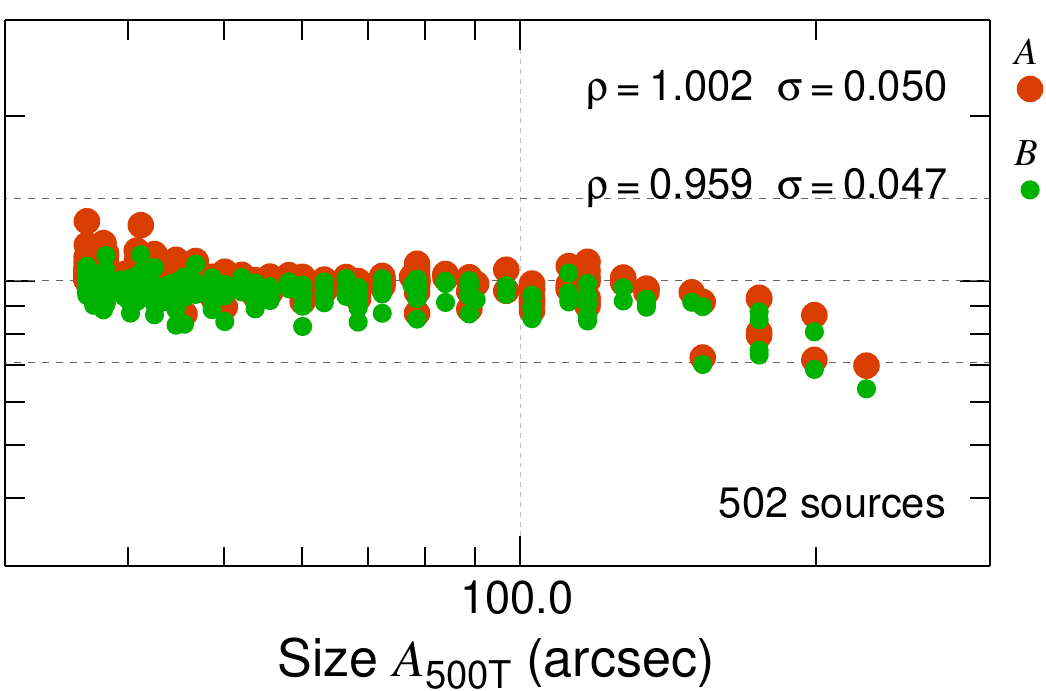}}}
\vspace{2.3mm}
\centerline{\resizebox{0.2695\hsize}{!}{\includegraphics{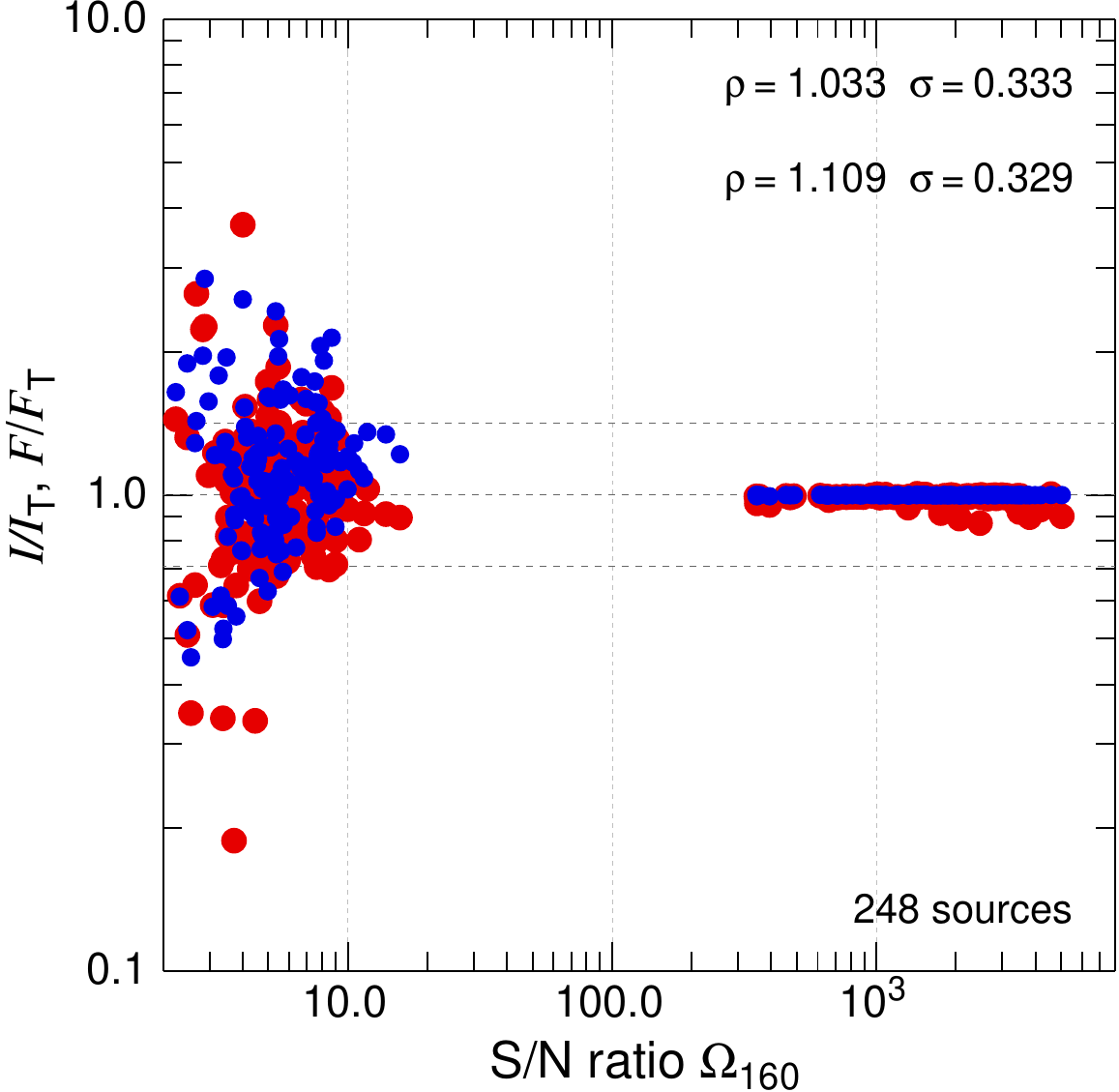}}
            \resizebox{0.2315\hsize}{!}{\includegraphics{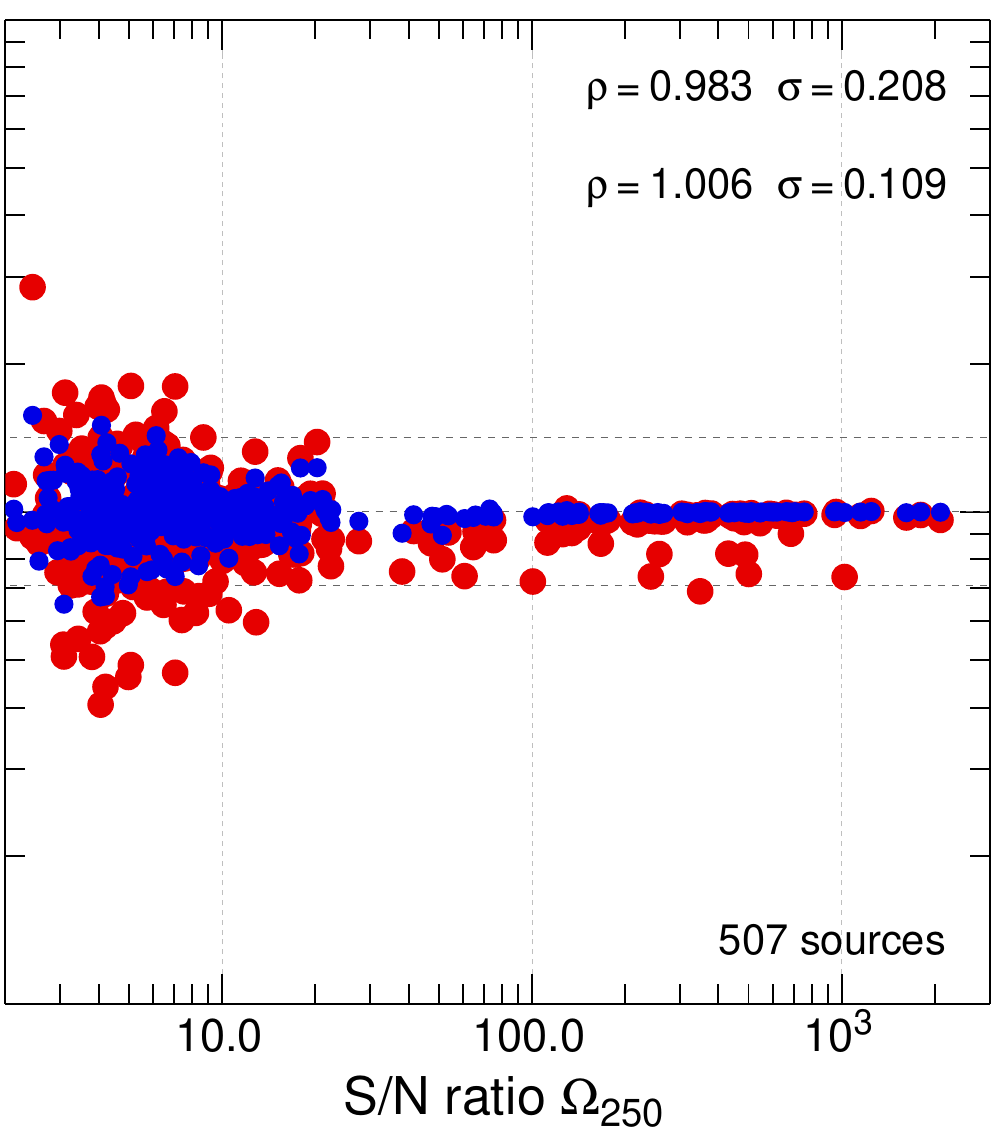}}
            \resizebox{0.2315\hsize}{!}{\includegraphics{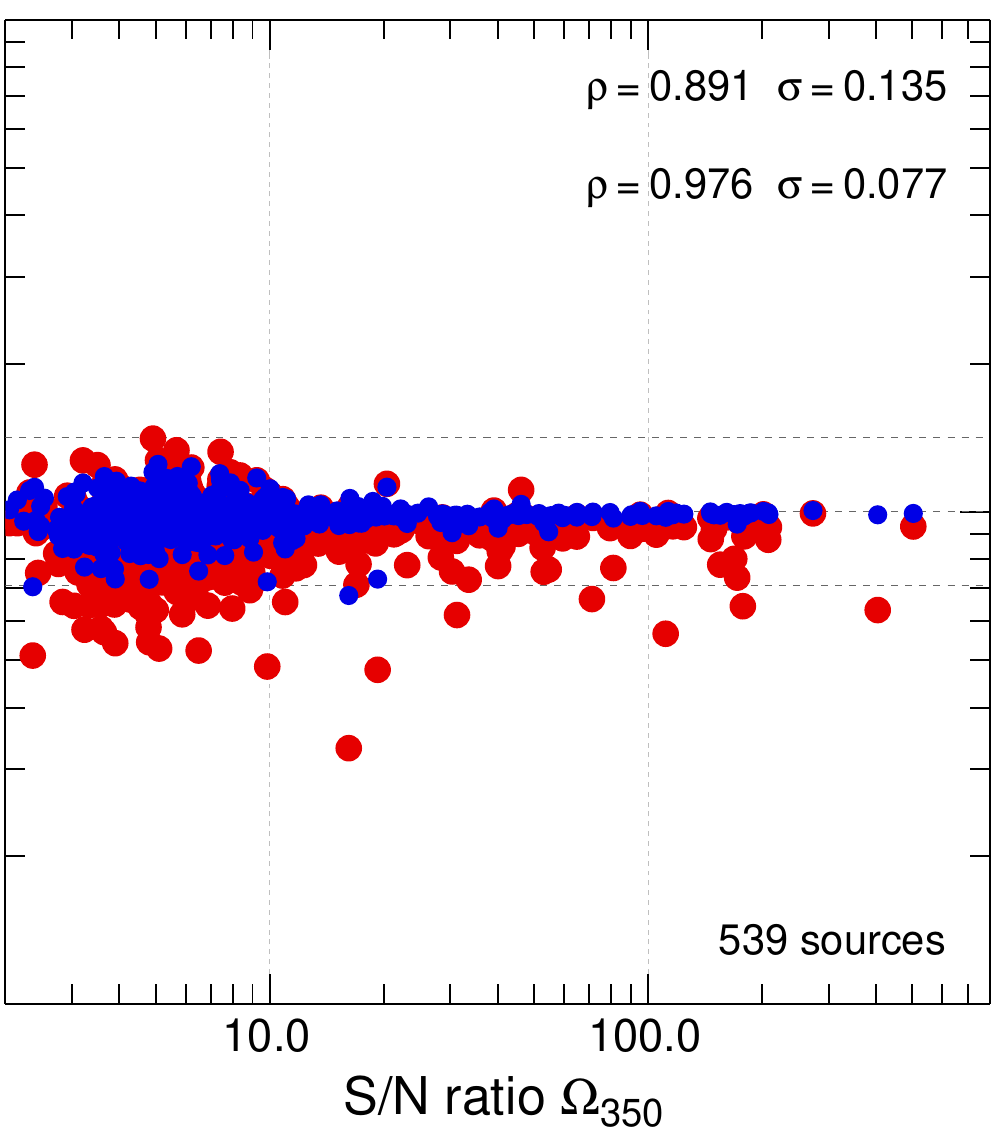}}
            \resizebox{0.2435\hsize}{!}{\includegraphics{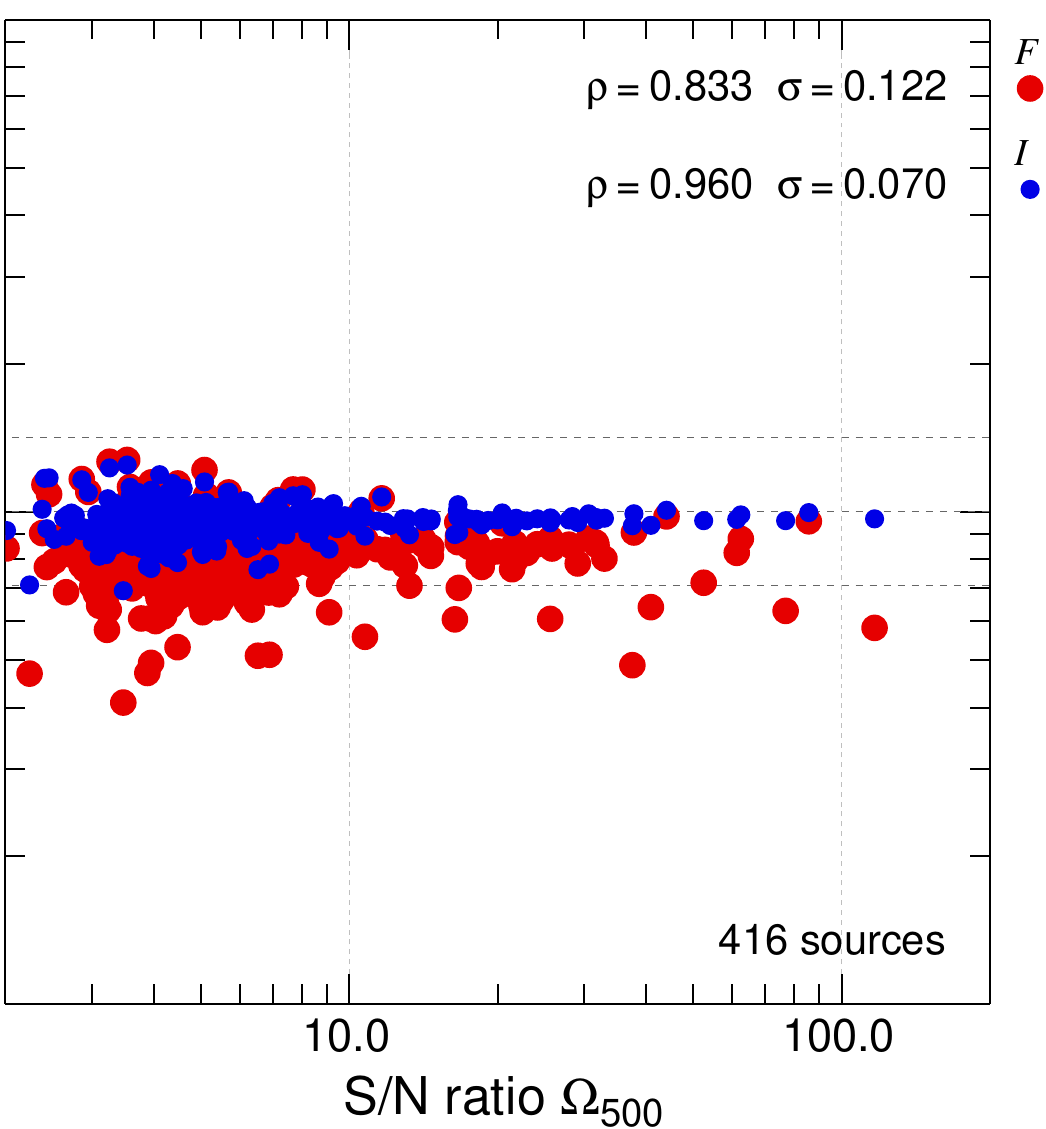}}}
\vspace{1.0mm}
\centerline{\resizebox{0.2695\hsize}{!}{\includegraphics{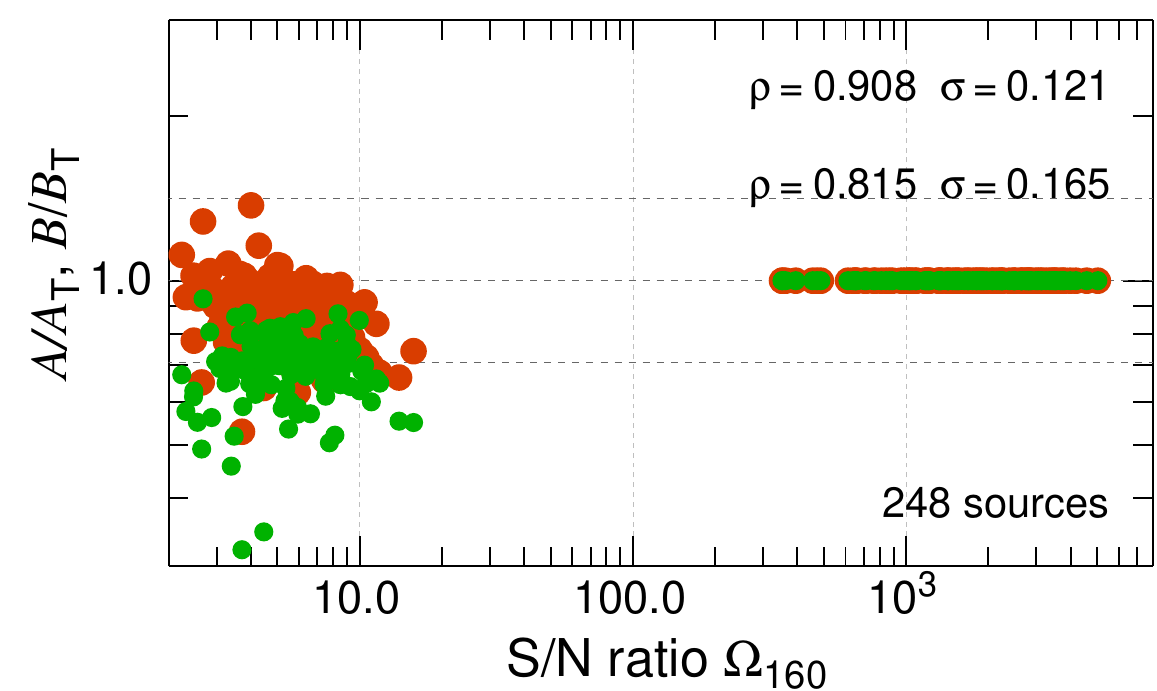}}
            \resizebox{0.2315\hsize}{!}{\includegraphics{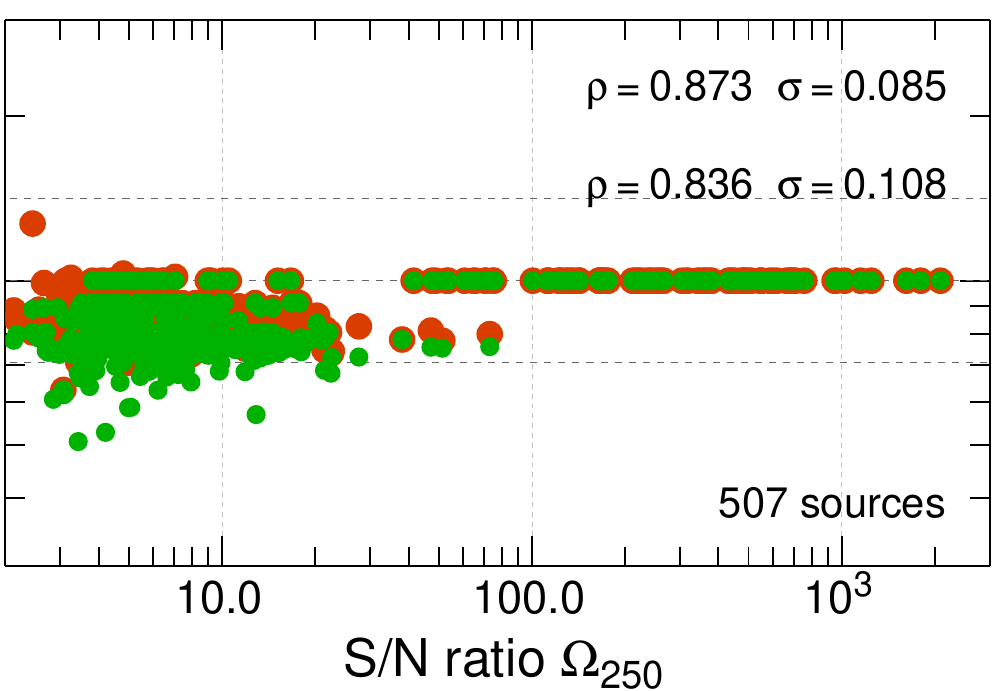}}
            \resizebox{0.2315\hsize}{!}{\includegraphics{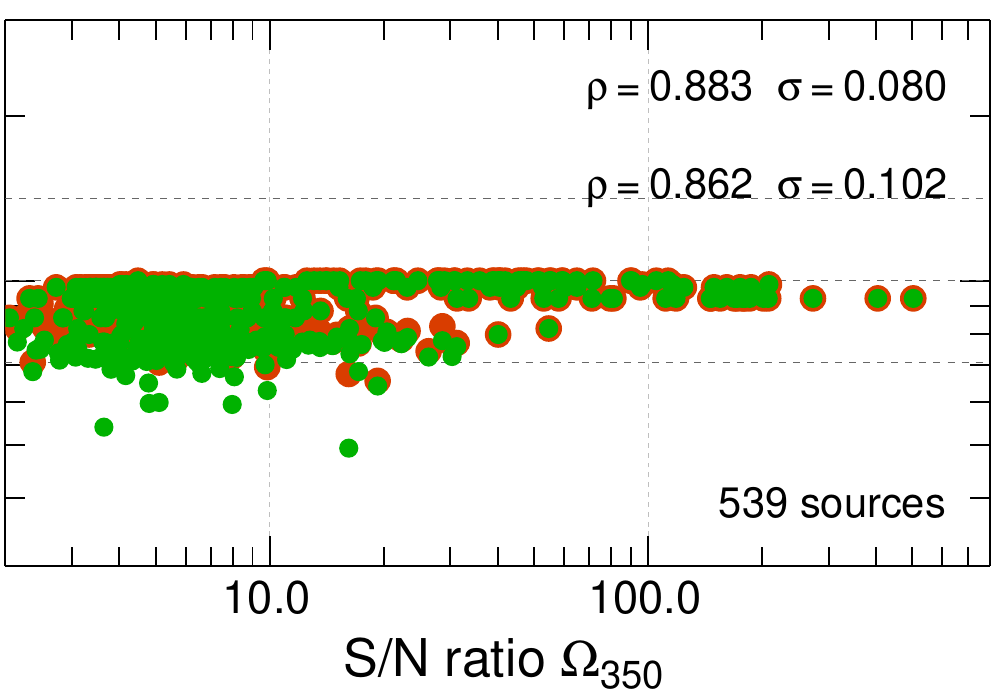}}
            \resizebox{0.2435\hsize}{!}{\includegraphics{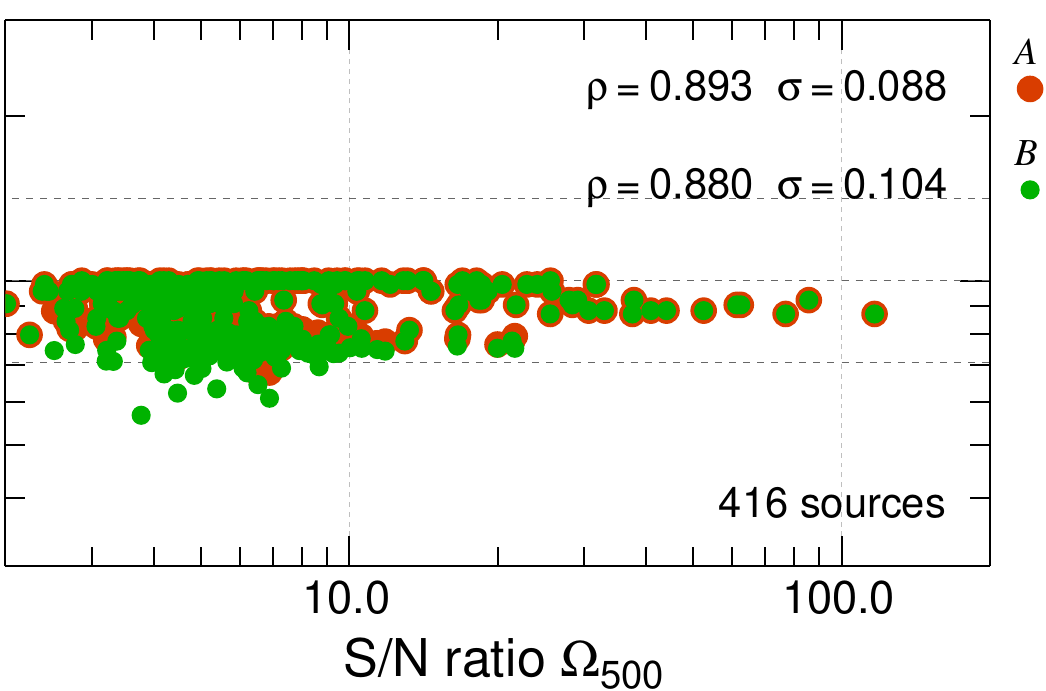}}}
\vspace{1.0mm}
\centerline{\resizebox{0.2695\hsize}{!}{\includegraphics{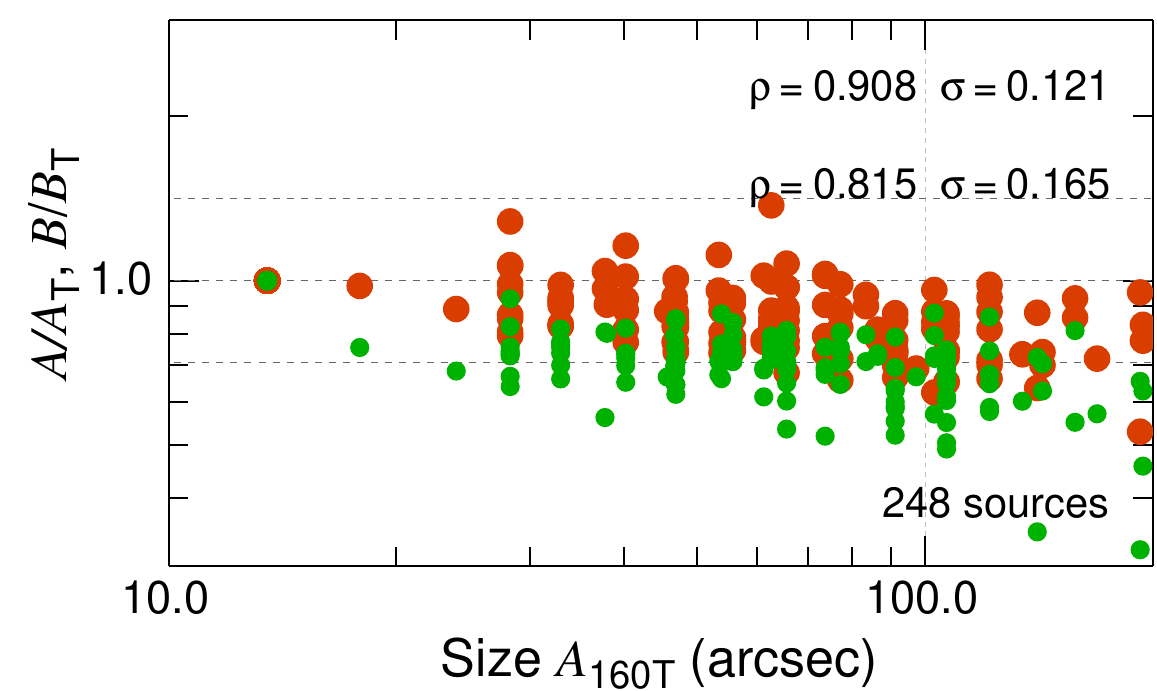}}
            \resizebox{0.2315\hsize}{!}{\includegraphics{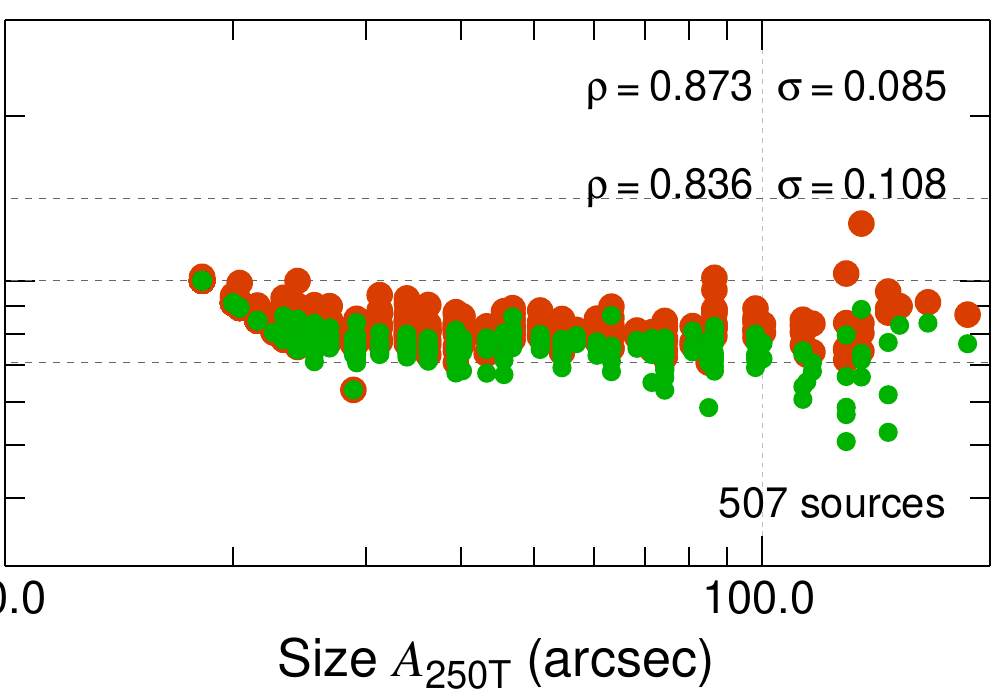}}
            \resizebox{0.2315\hsize}{!}{\includegraphics{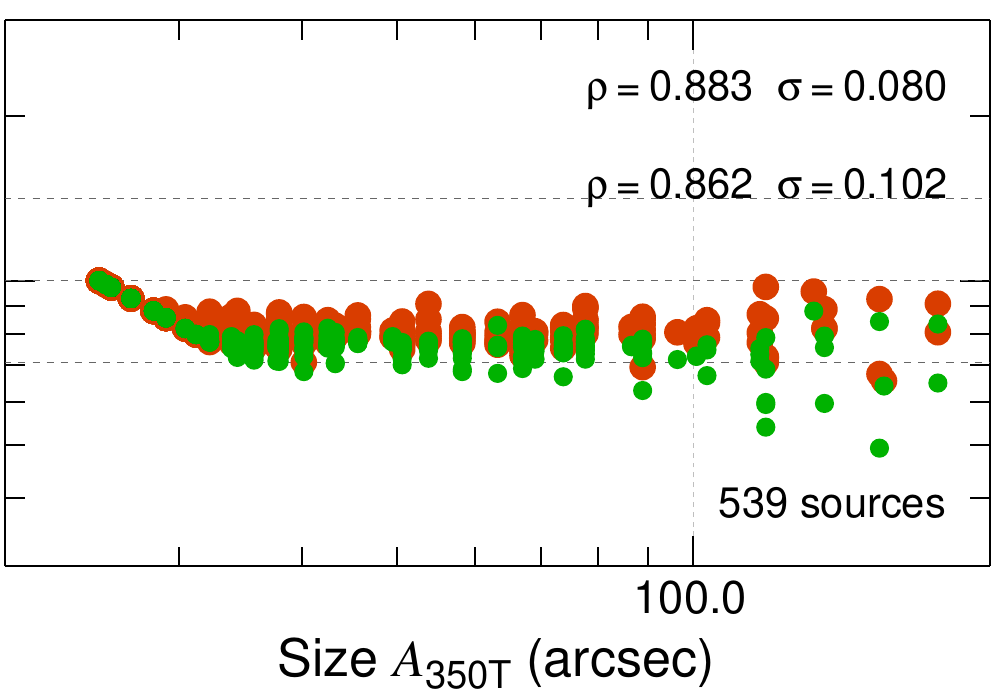}}
            \resizebox{0.2435\hsize}{!}{\includegraphics{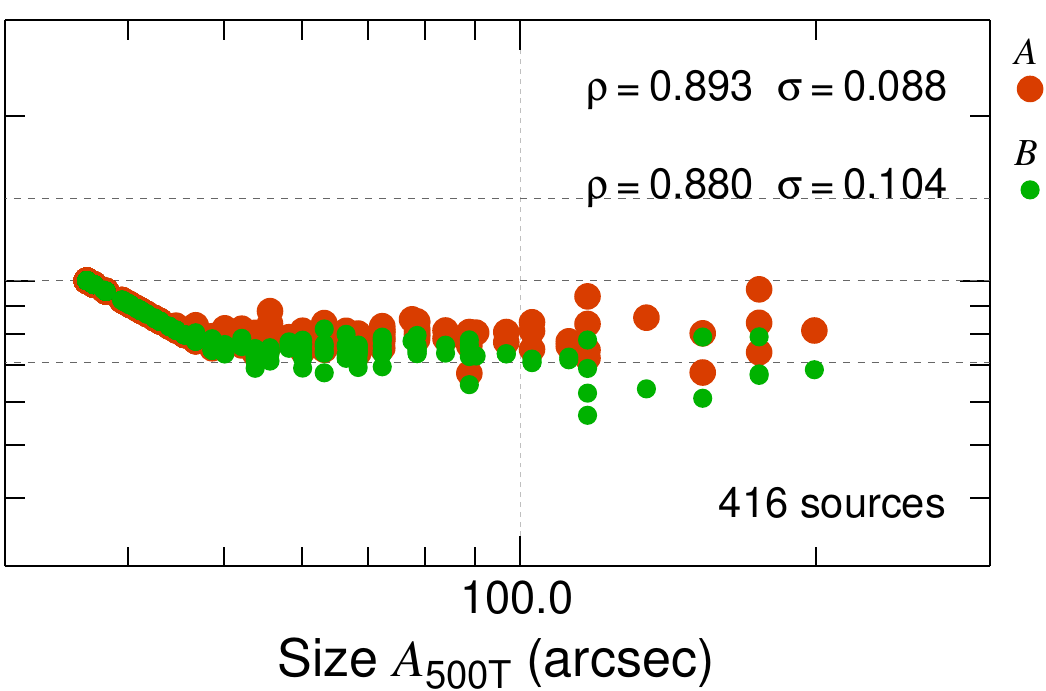}}}
\caption
{ 
Benchmark B$_3$ extraction with \textsl{getsf} (three \emph{top} rows) and \textsl{getold} (three \emph{bottom} rows). Ratios of
the measured fluxes $F_{{\rm T}{\lambda}{n}}$, peak intensities $F_{{\rm P}{\lambda}{n}}$, and sizes $\{A,B\}_{{\lambda}{n}}$ to
their true values ($F/F_{\rm T}$, $I/I_{\rm T}$, $A/A_{\rm T}$, and $B/B_{\rm T}$) are shown as a function of the S/N ratio
$\Omega_{{\lambda}{n}}$. The size ratios $A/A_{\rm T}$ and $B/B_{\rm T}$ are also shown as a function of the true sizes
$\{A,B\}_{{\lambda}{n}{\rm T}}$. The mean $\varrho_{{\rm \{P|T|A|B\}}{\lambda}}$ and standard deviation $\sigma_{{\rm
\{P|T|A|B\}}{\lambda}}$ of the ratios are displayed in the panels. Similar plots for $\lambda\le 100$\,$\mu$m with only bright
protostellar cores are not presented, because their measurements are quite accurate, with $\varrho_{\{{\rm
P|T|A|B}\}{\lambda}}\approx \{1.000|0.999|1.000|1.000\}$ and $\sigma_{\{{\rm P|T|A|B}\}{\lambda}}\approx
\{0.0004|0.001|0.0002|0.0002\}$.
} 
\label{accuracyB3}
\end{figure*}



\subsubsection{Measurement accuracies}
\label{accuracies}

\begin{figure*}
\centering
\centerline{\resizebox{0.2695\hsize}{!}{\includegraphics{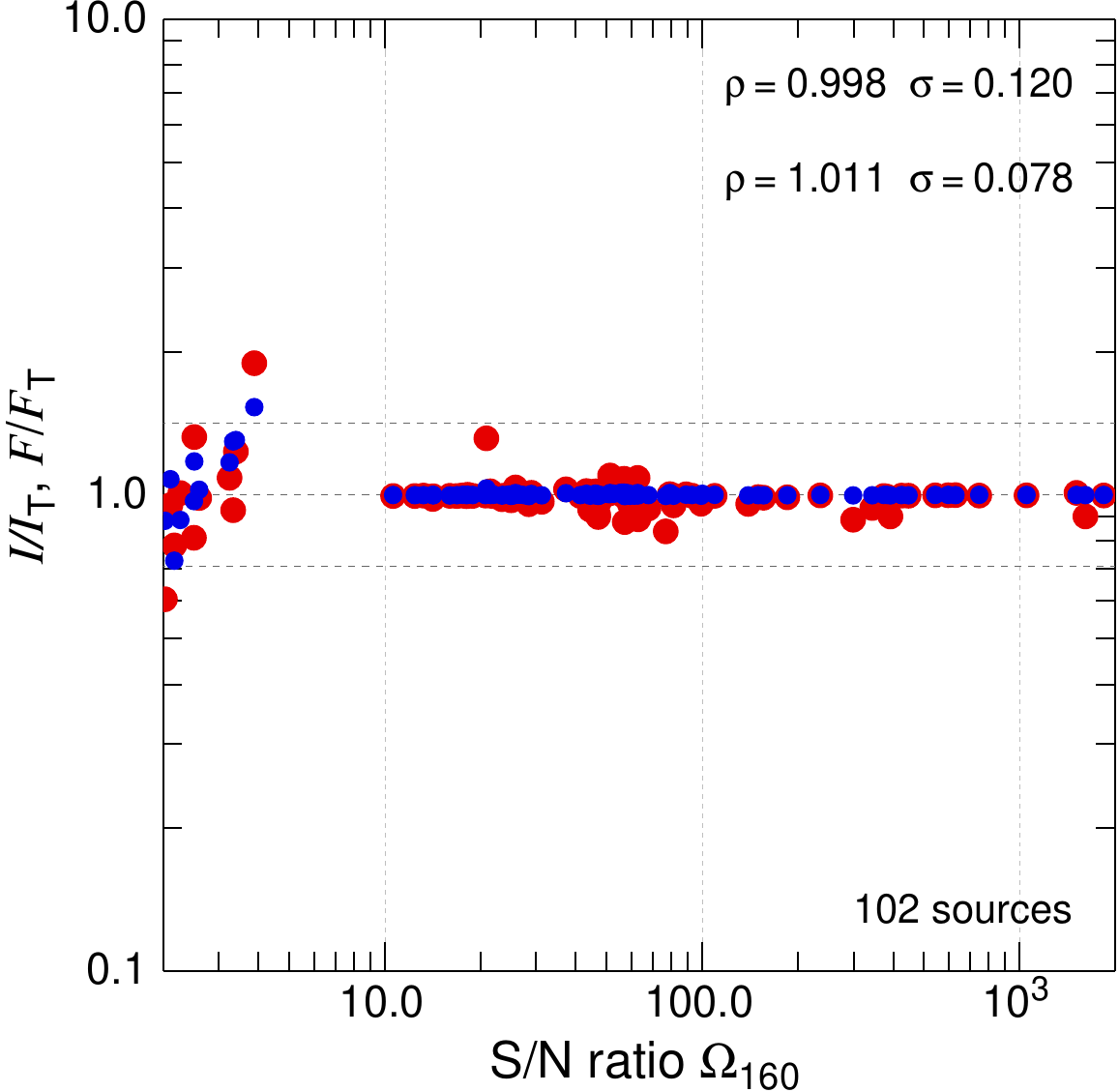}}
            \resizebox{0.2315\hsize}{!}{\includegraphics{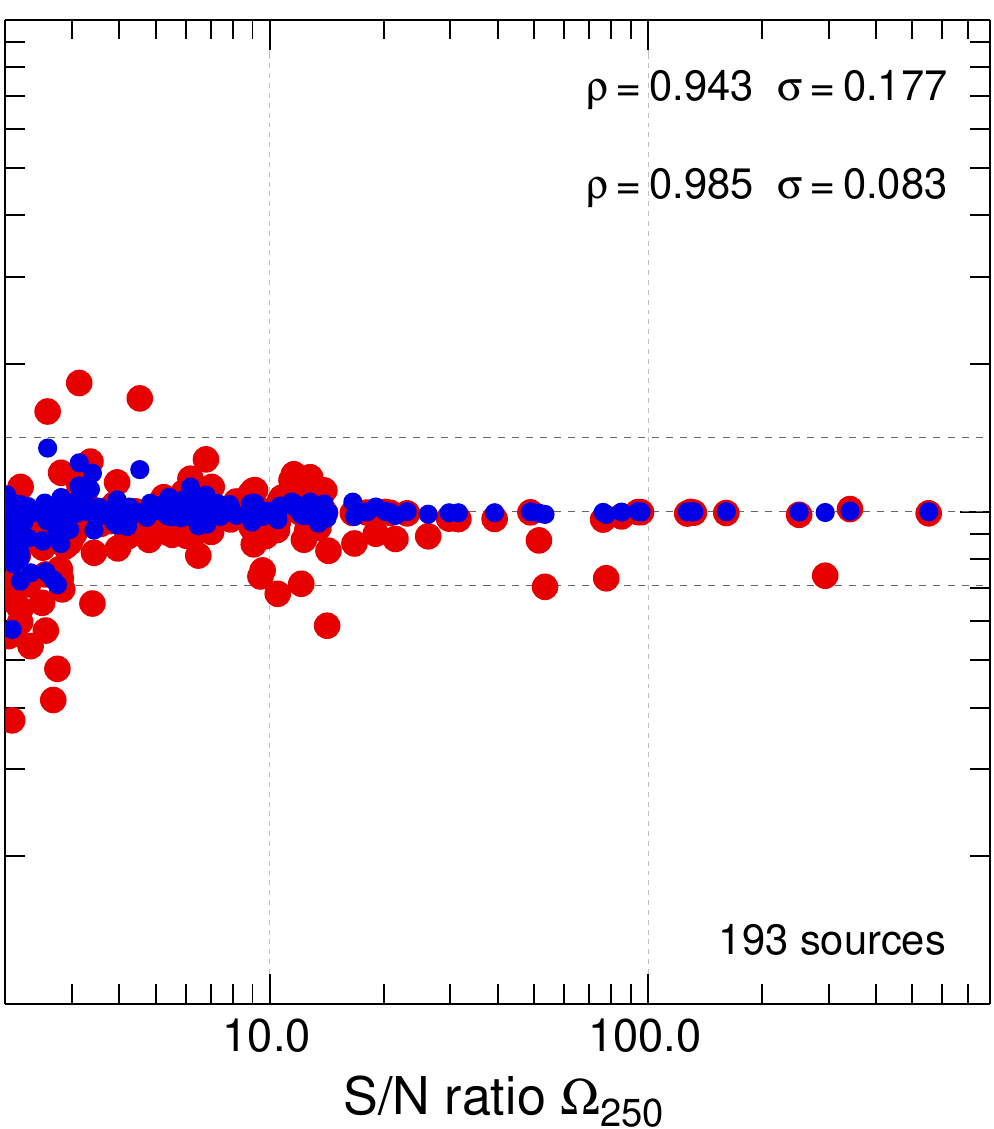}}
            \resizebox{0.2315\hsize}{!}{\includegraphics{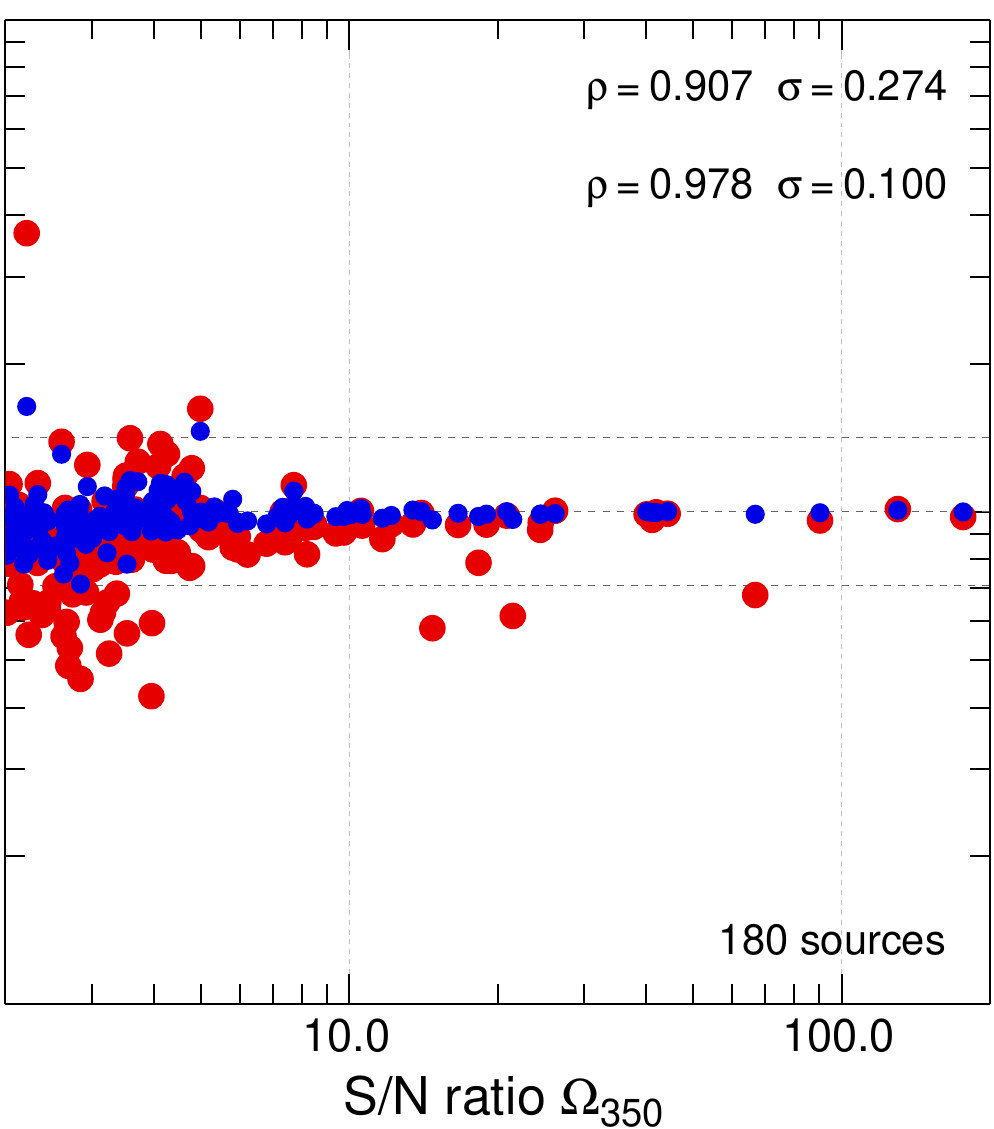}}
            \resizebox{0.2435\hsize}{!}{\includegraphics{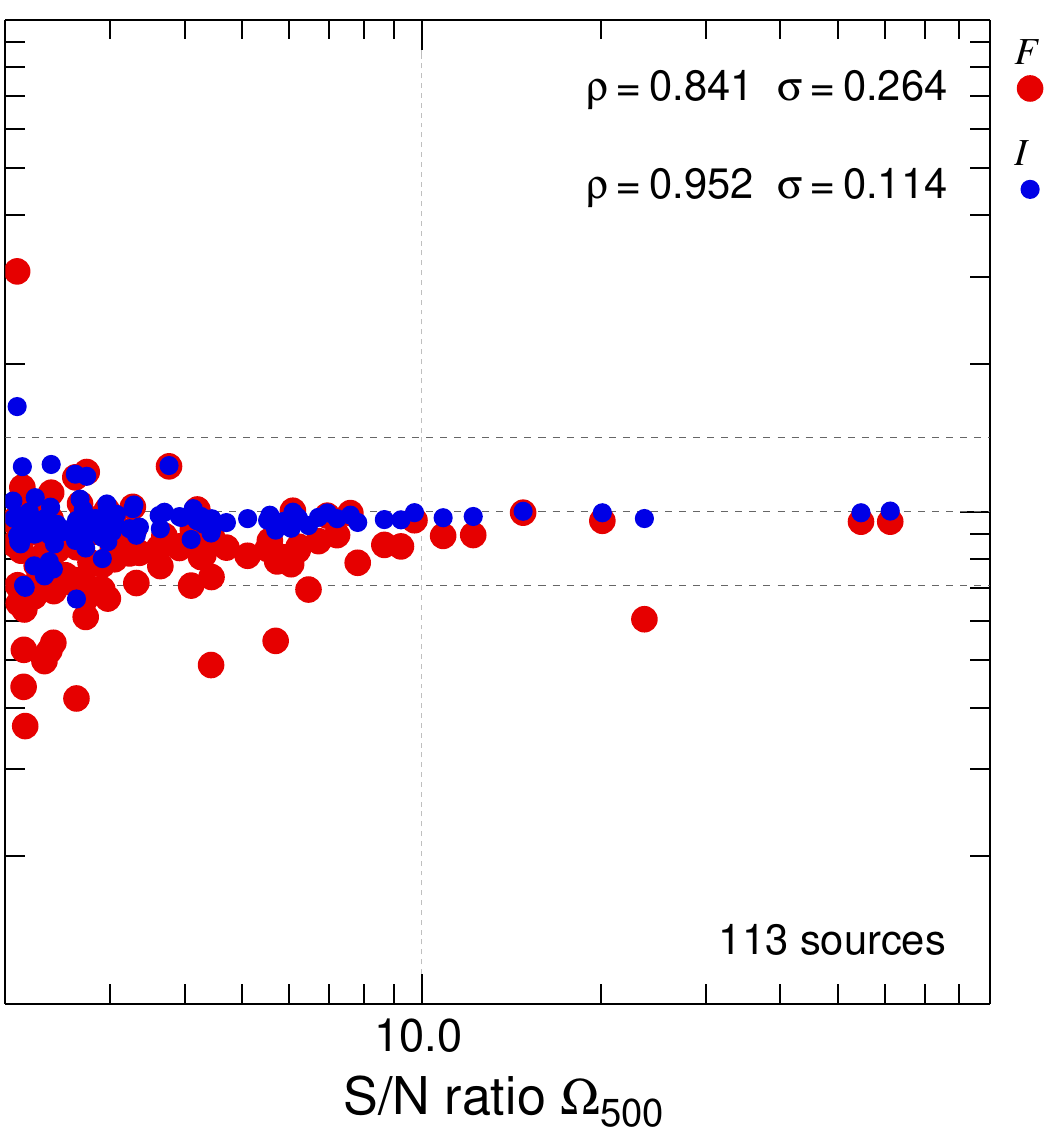}}}
\vspace{1.0mm}
\centerline{\resizebox{0.2695\hsize}{!}{\includegraphics{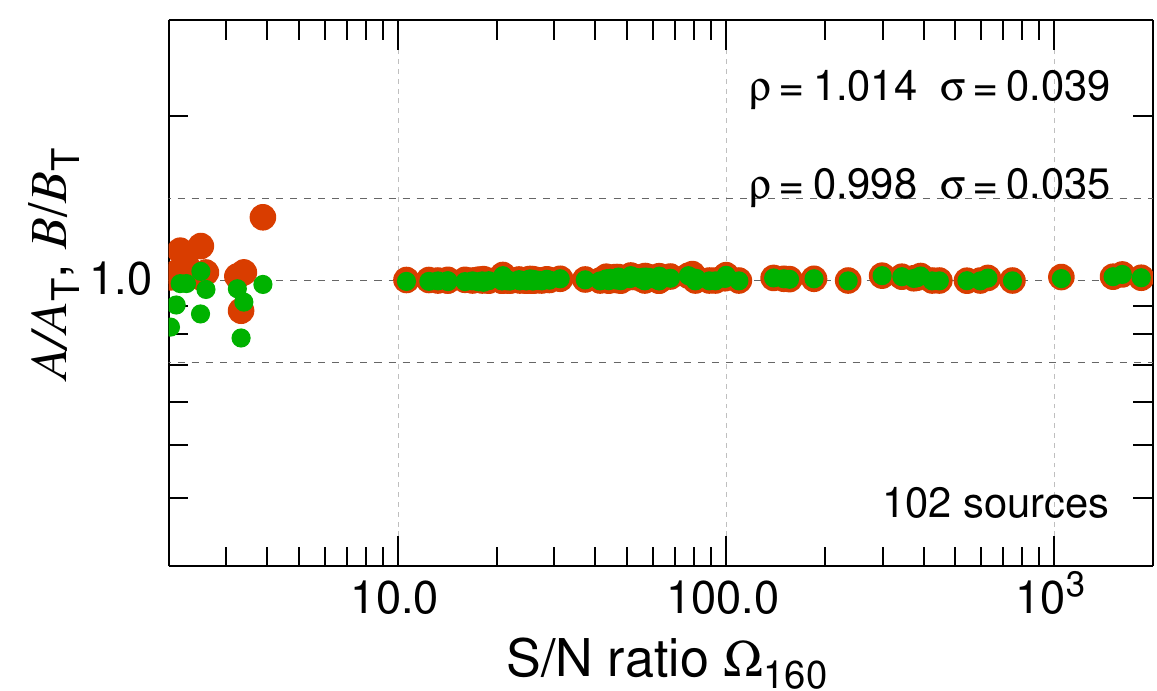}}
            \resizebox{0.2315\hsize}{!}{\includegraphics{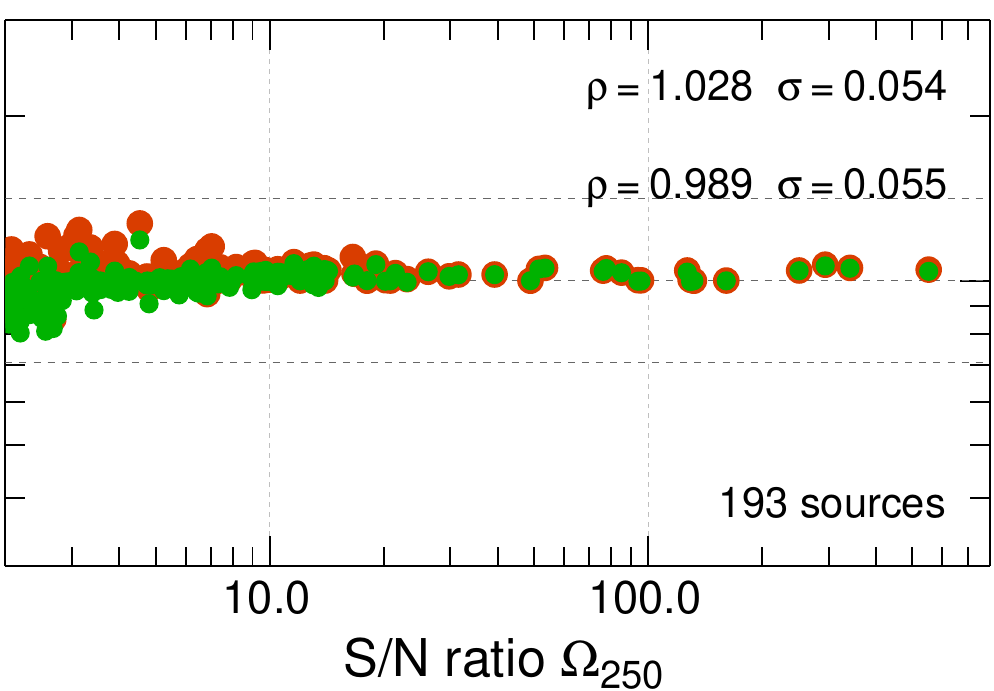}}
            \resizebox{0.2315\hsize}{!}{\includegraphics{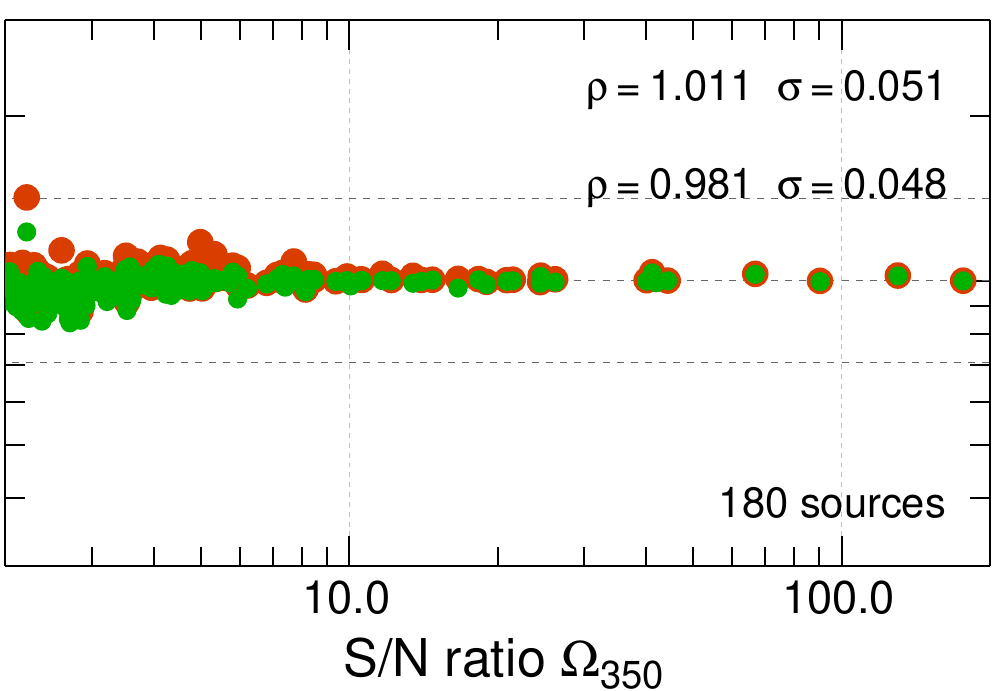}}
            \resizebox{0.2435\hsize}{!}{\includegraphics{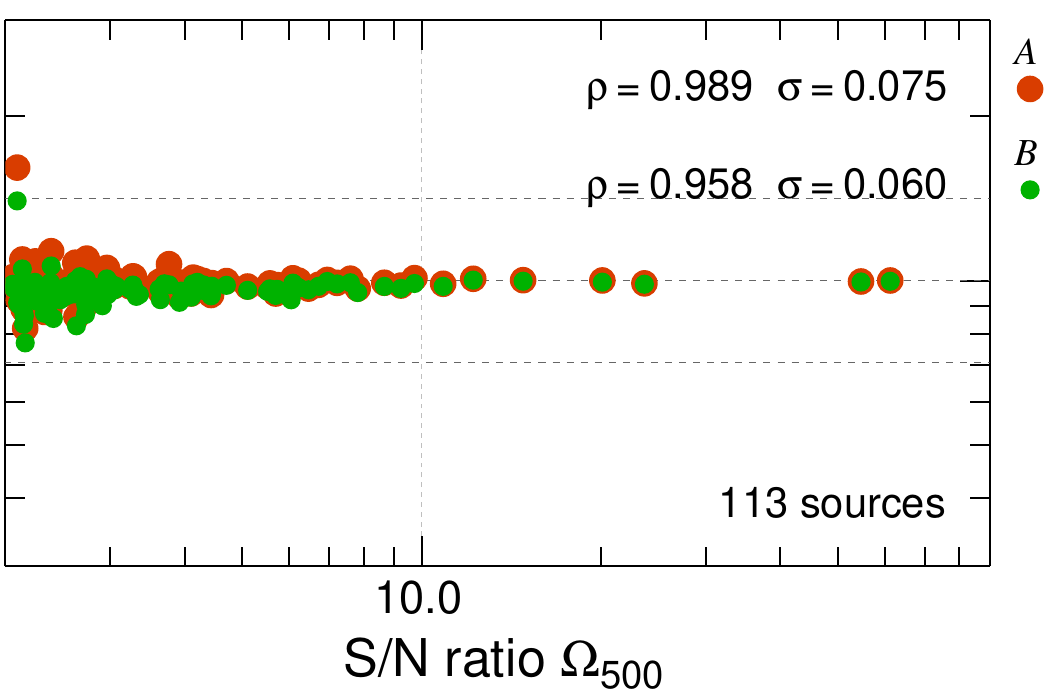}}}
\vspace{1.0mm}
\centerline{\resizebox{0.2695\hsize}{!}{\includegraphics{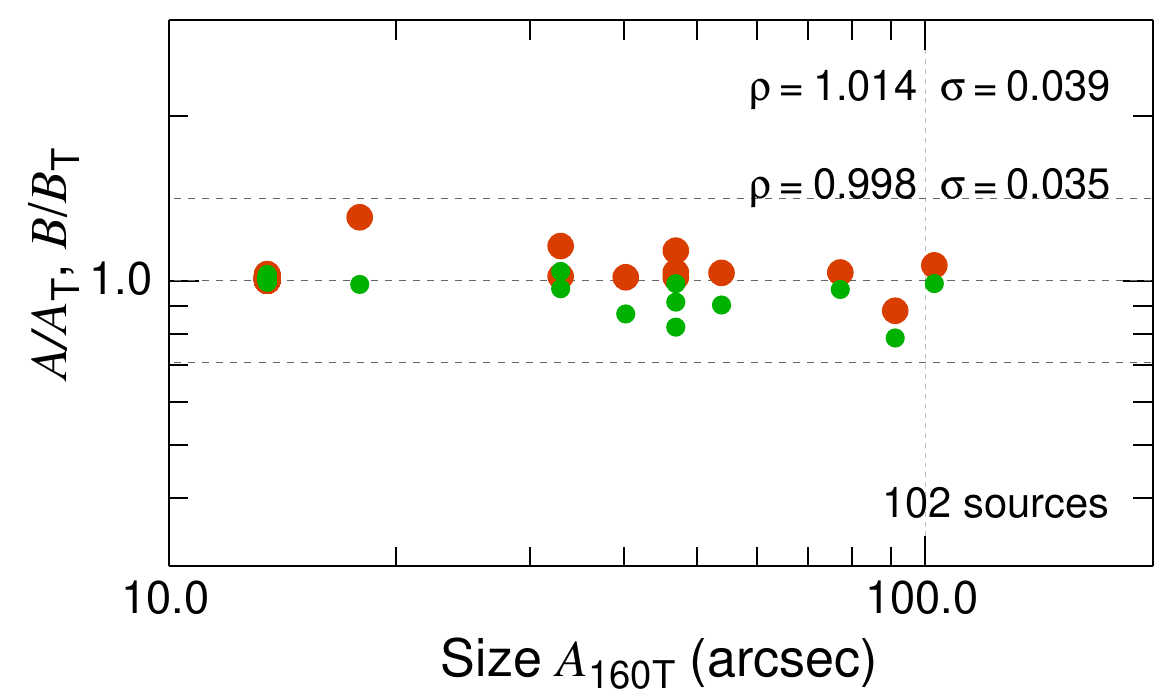}}
            \resizebox{0.2315\hsize}{!}{\includegraphics{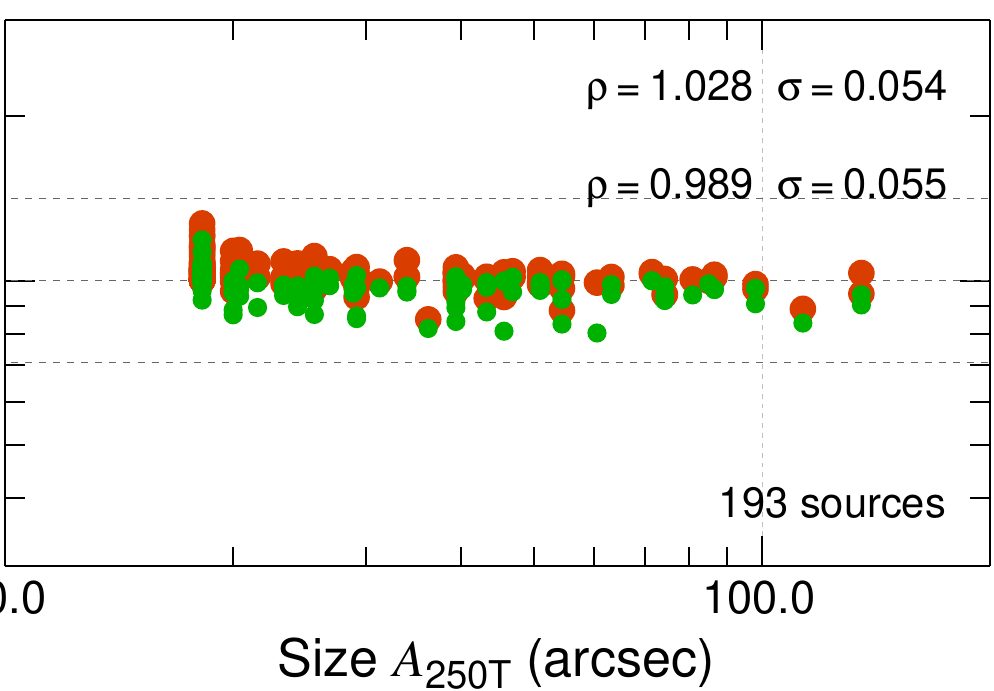}}
            \resizebox{0.2315\hsize}{!}{\includegraphics{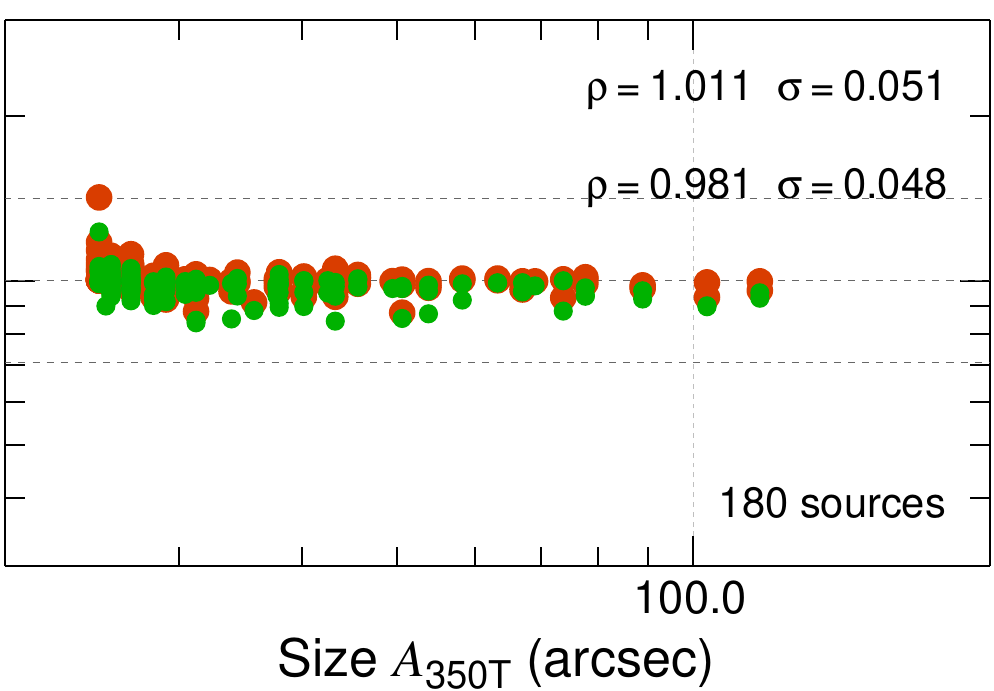}}
            \resizebox{0.2435\hsize}{!}{\includegraphics{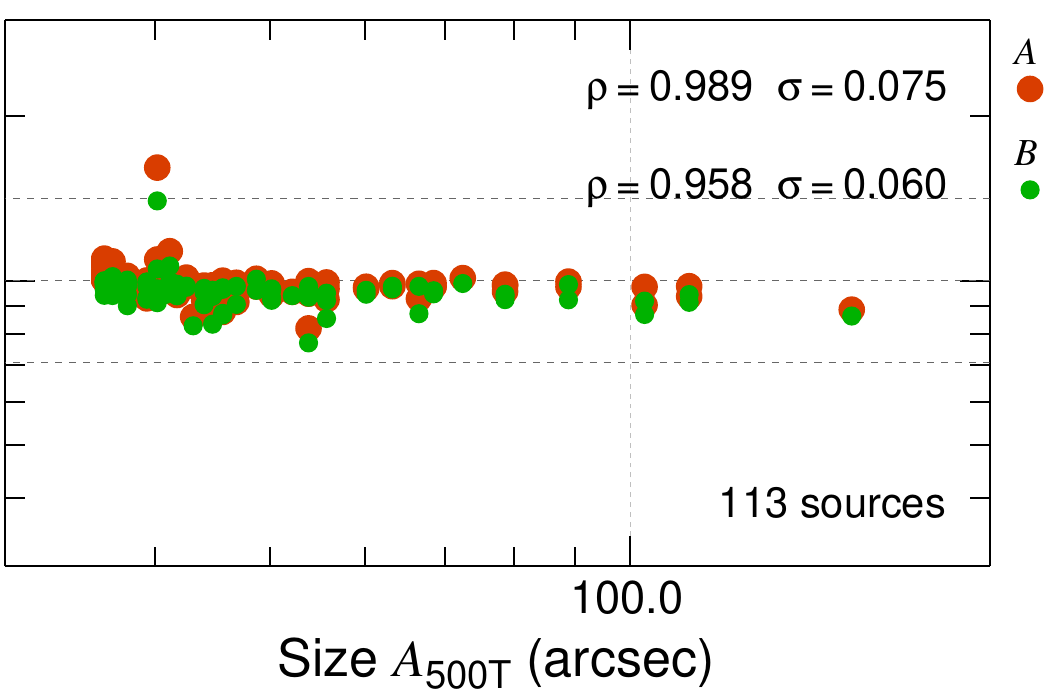}}}
\vspace{2.3mm}
\centerline{\resizebox{0.2695\hsize}{!}{\includegraphics{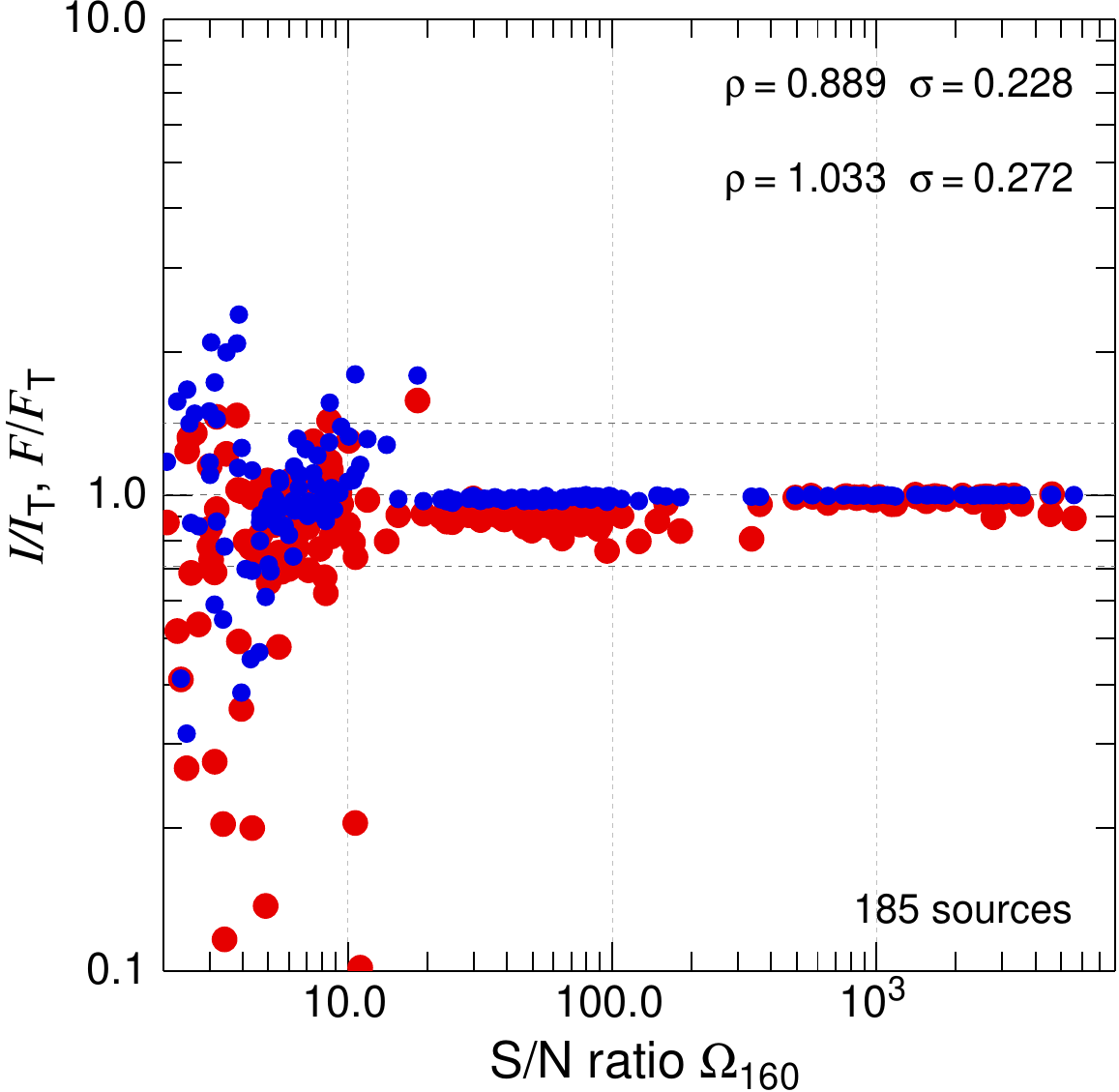}}
            \resizebox{0.2315\hsize}{!}{\includegraphics{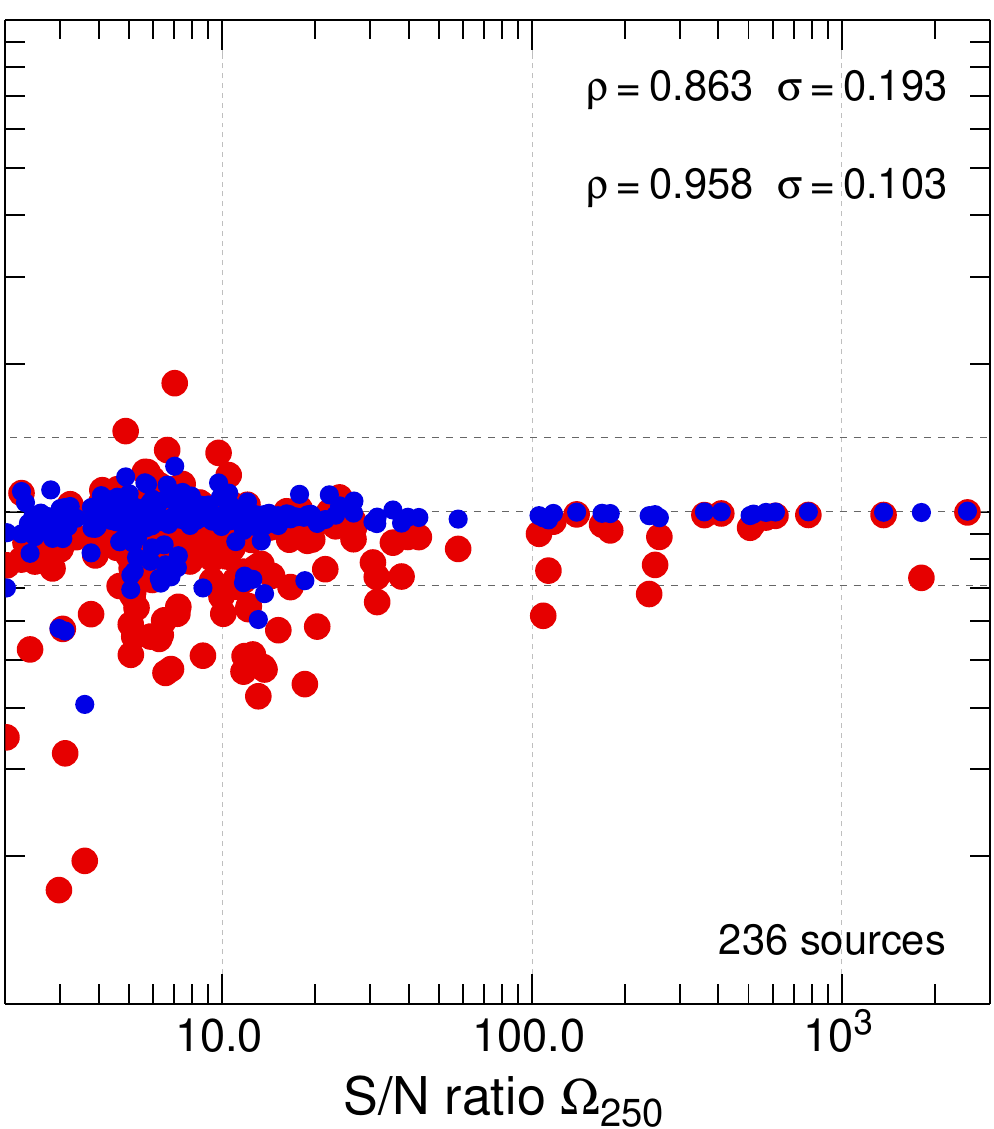}}
            \resizebox{0.2315\hsize}{!}{\includegraphics{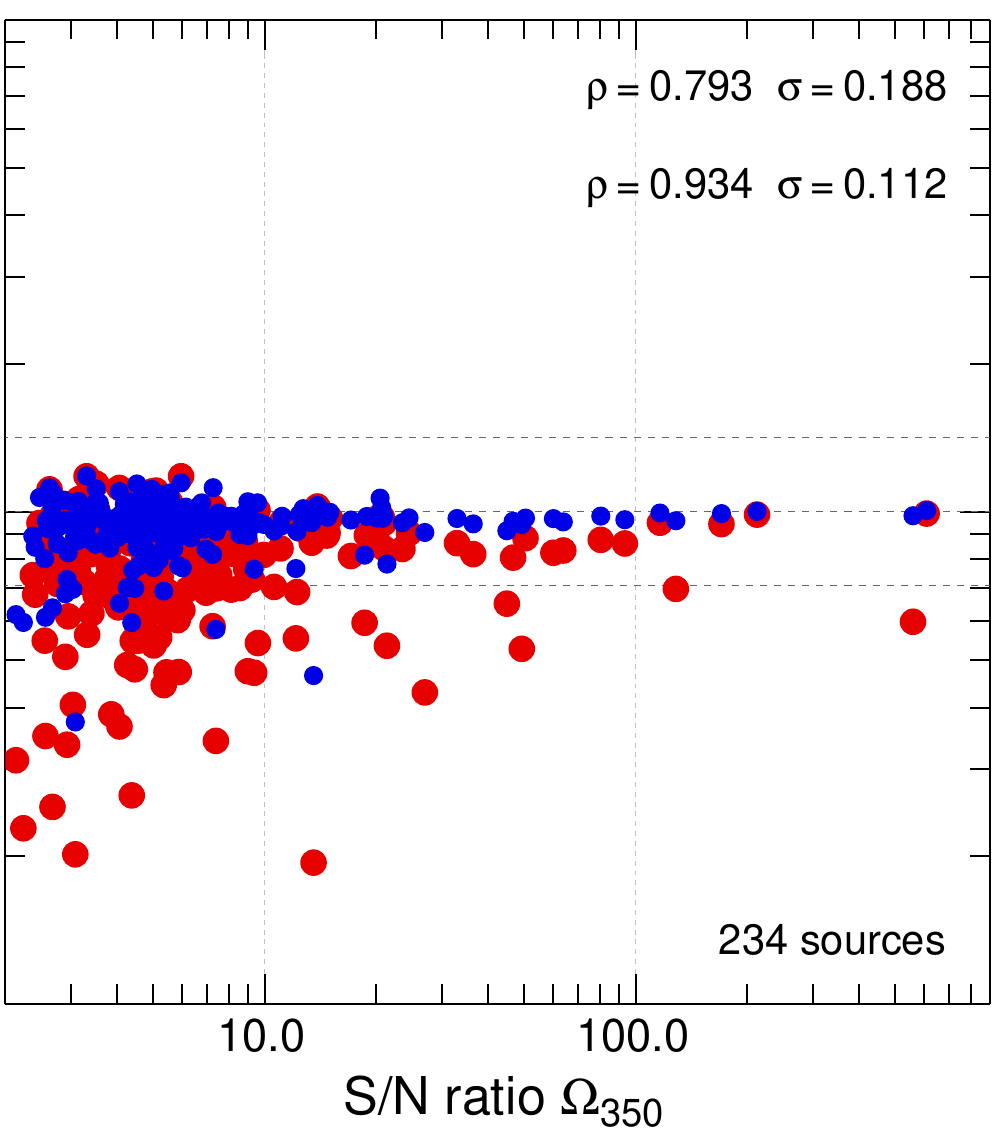}}
            \resizebox{0.2435\hsize}{!}{\includegraphics{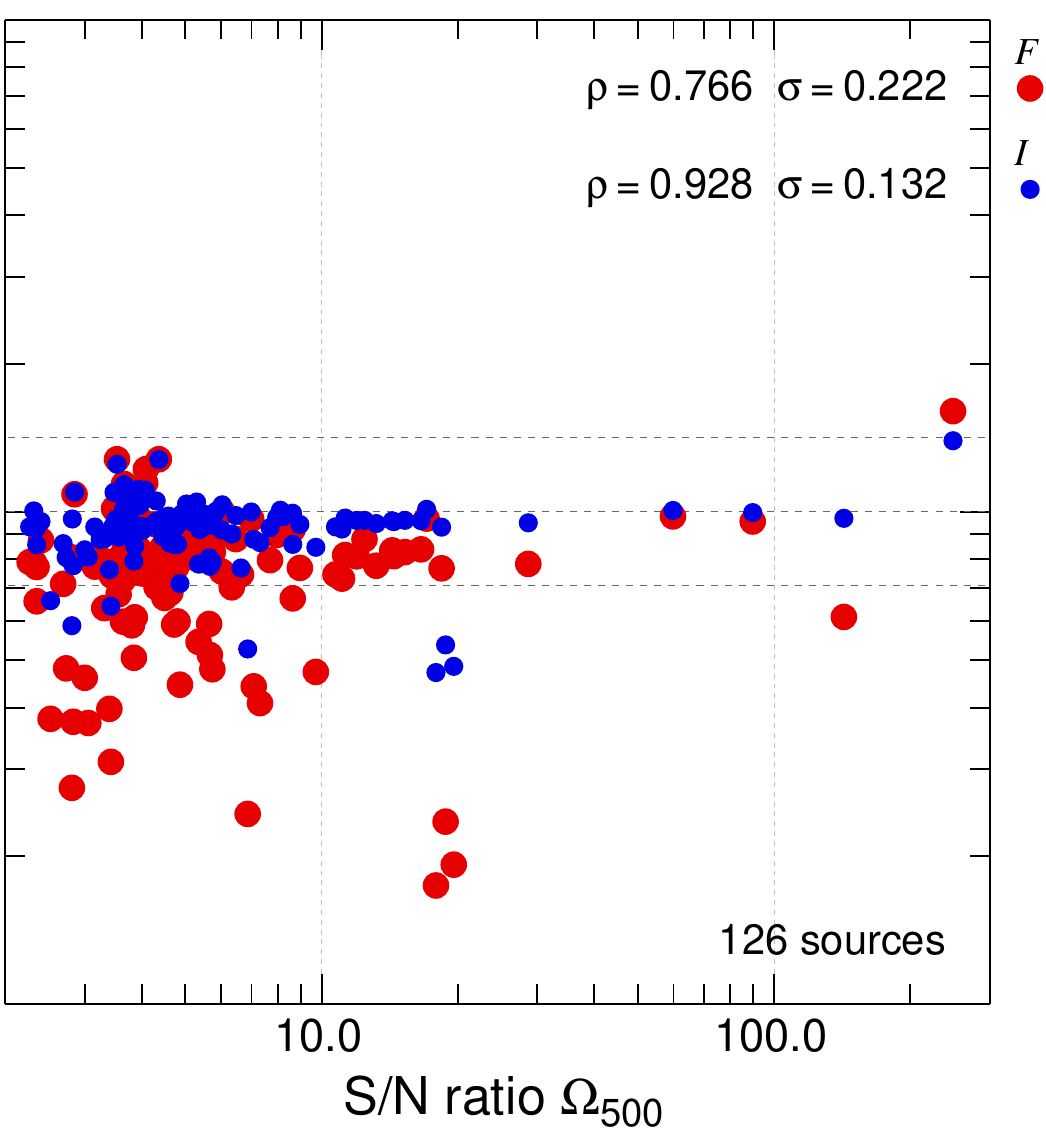}}}
\vspace{1.0mm}
\centerline{\resizebox{0.2695\hsize}{!}{\includegraphics{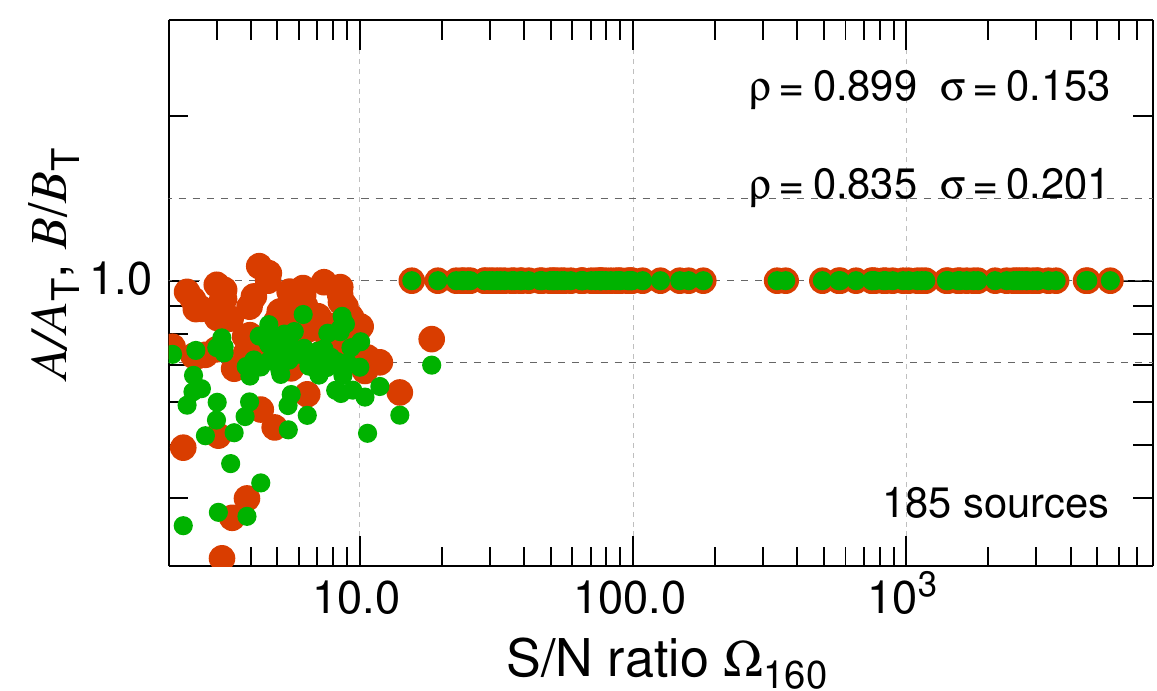}}
            \resizebox{0.2315\hsize}{!}{\includegraphics{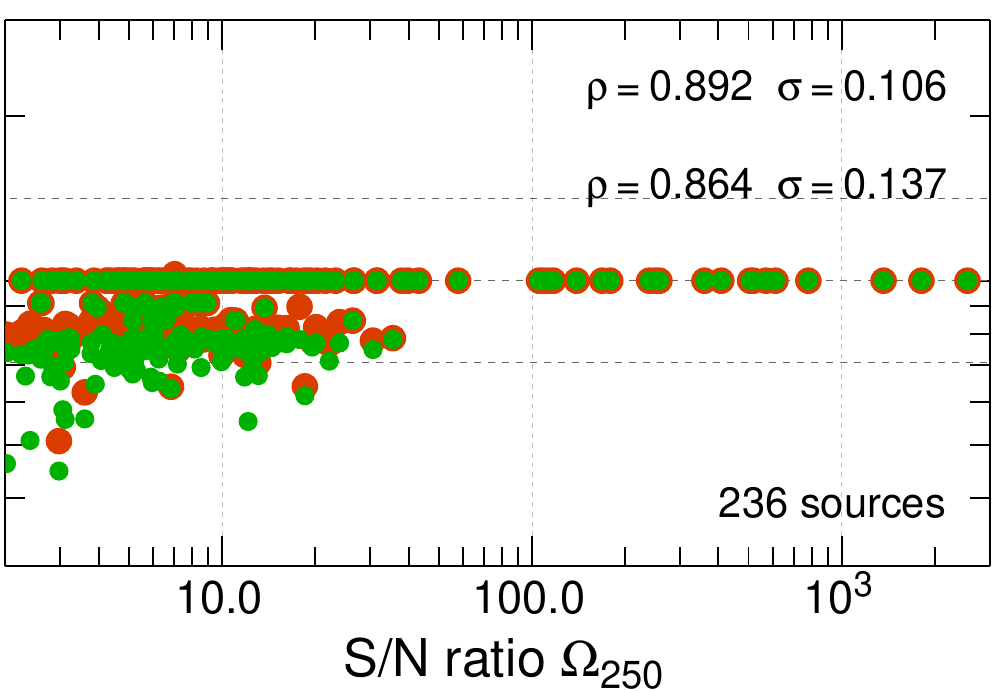}}
            \resizebox{0.2315\hsize}{!}{\includegraphics{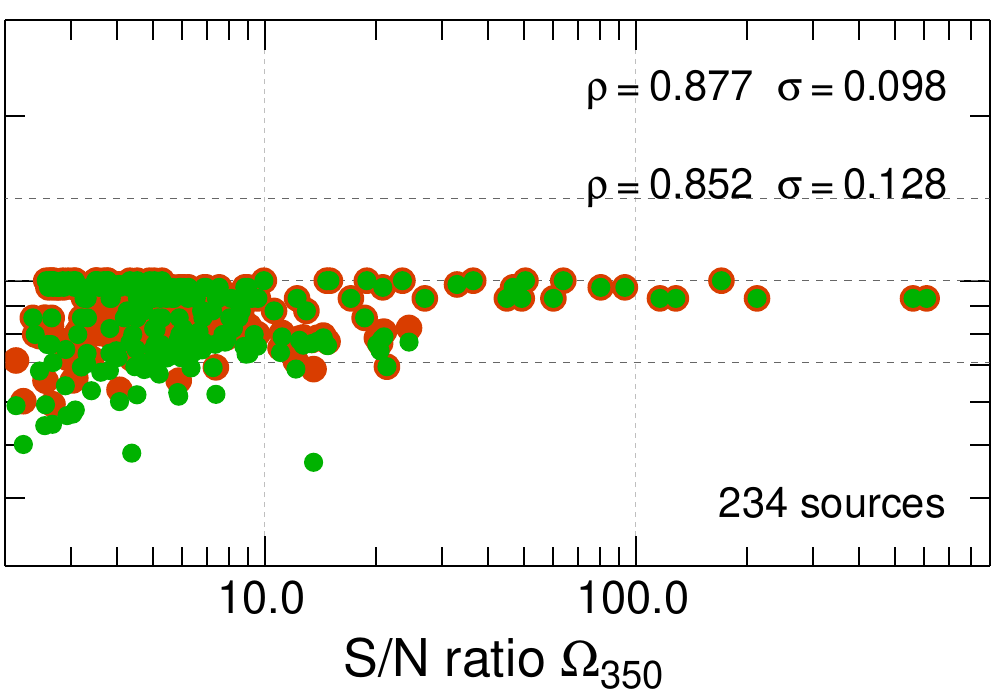}}
            \resizebox{0.2435\hsize}{!}{\includegraphics{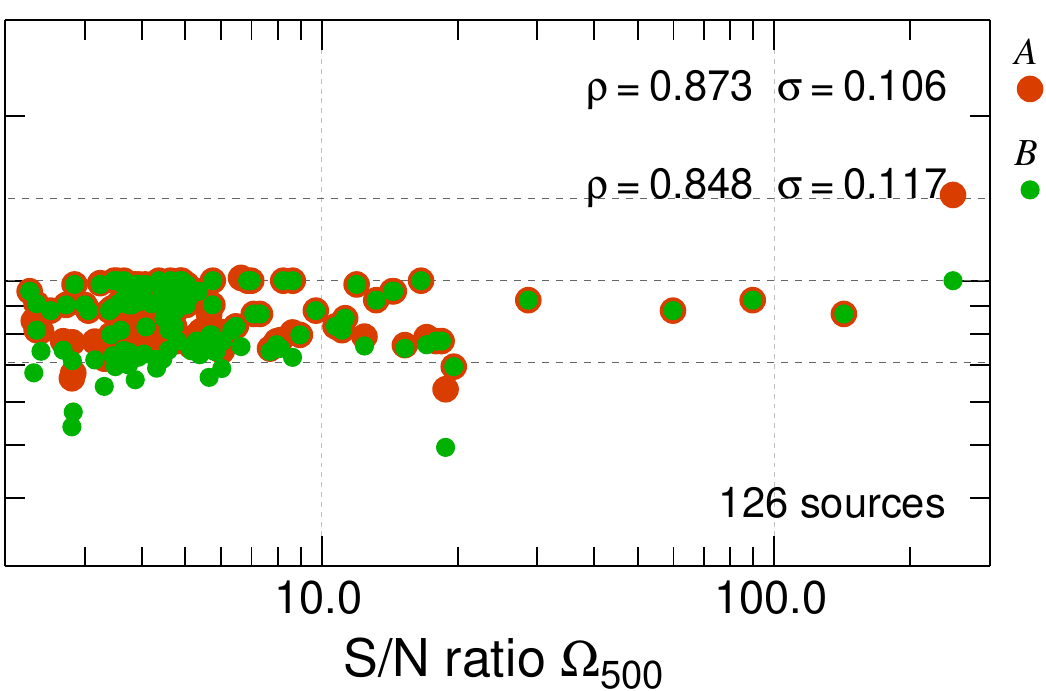}}}
\vspace{1.0mm}
\centerline{\resizebox{0.2695\hsize}{!}{\includegraphics{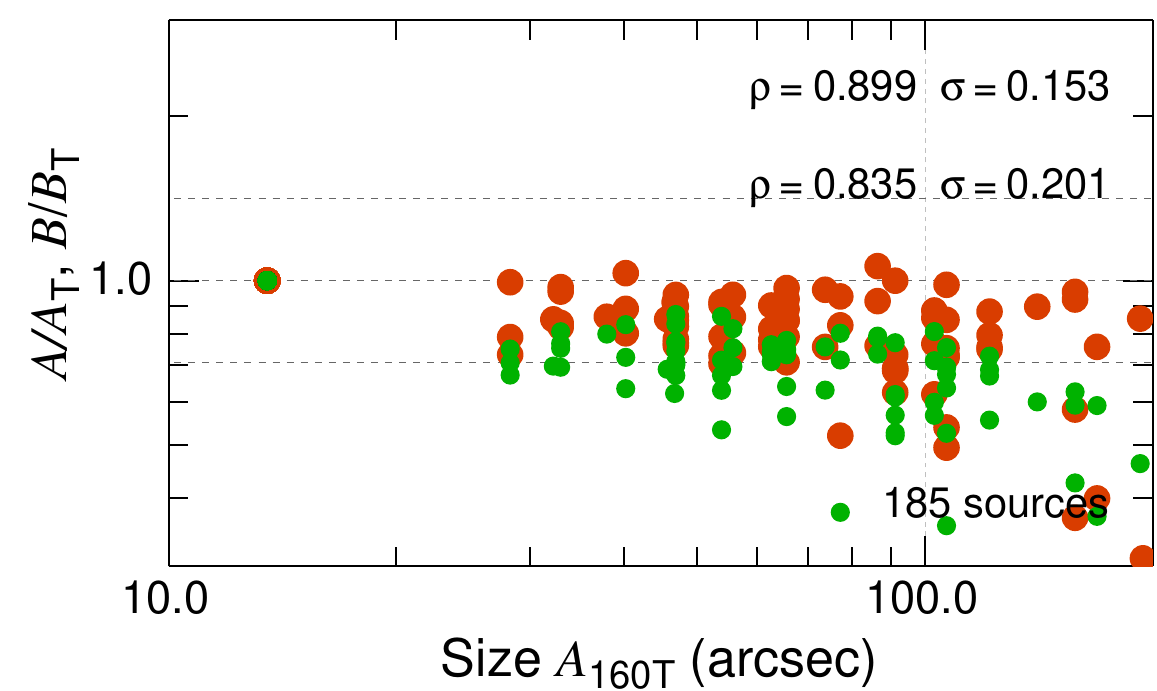}}
            \resizebox{0.2315\hsize}{!}{\includegraphics{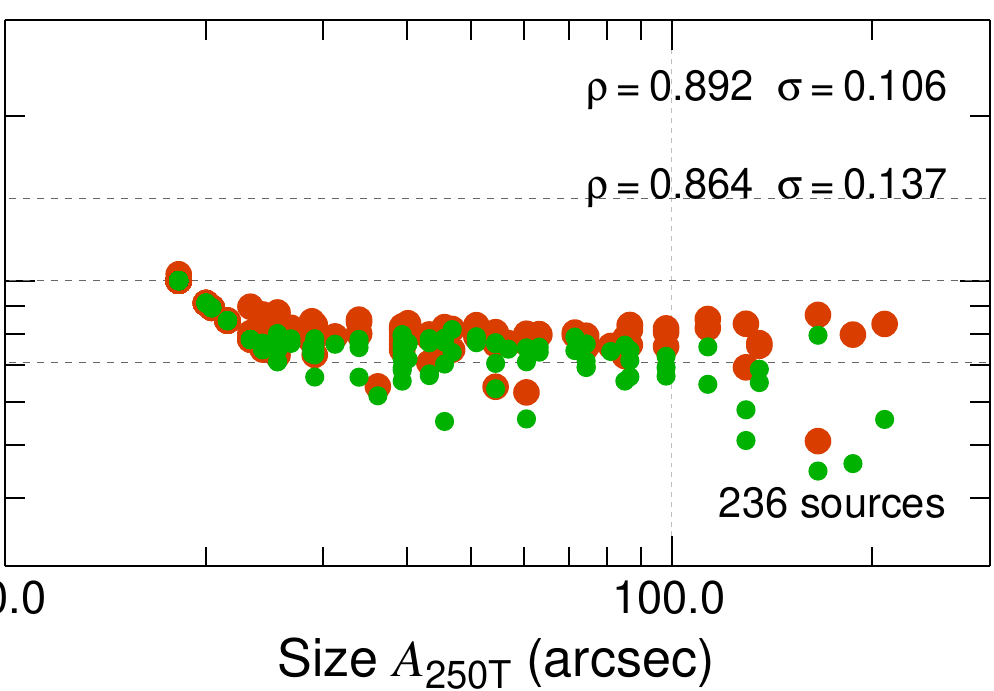}}
            \resizebox{0.2315\hsize}{!}{\includegraphics{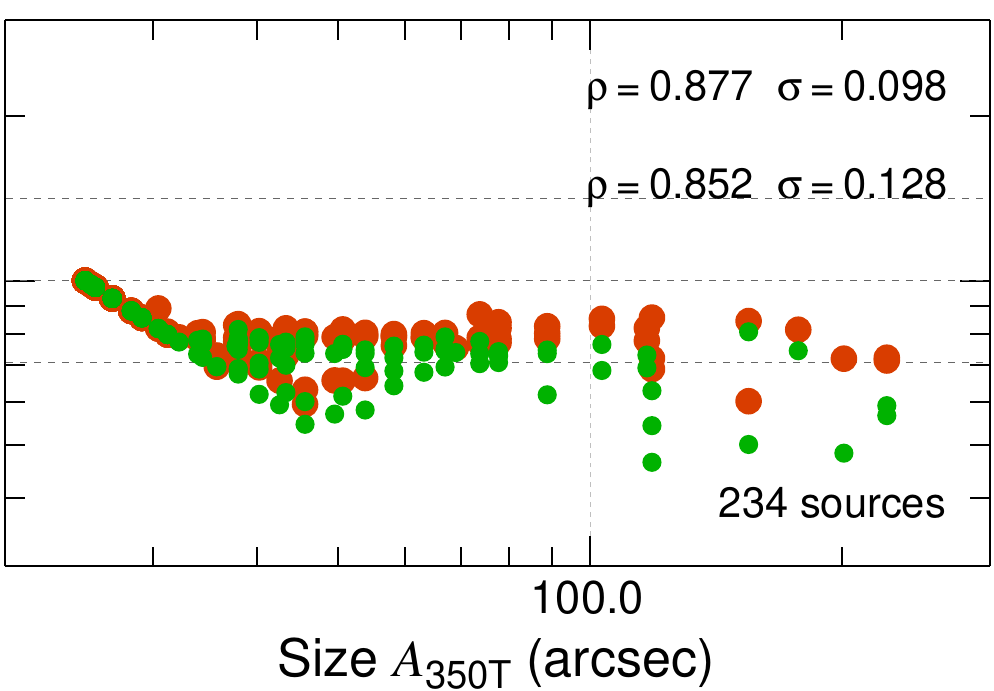}}
            \resizebox{0.2435\hsize}{!}{\includegraphics{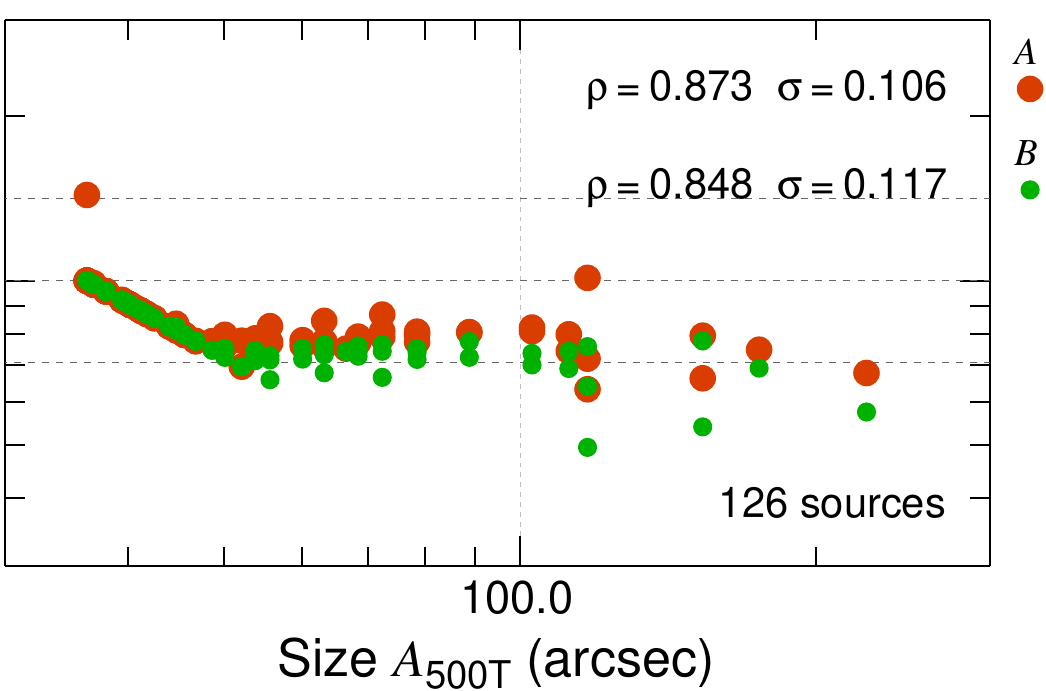}}}
\caption
{ 
Benchmark B$_4$ extraction with \textsl{getsf} (three \emph{top} rows) and \textsl{getold} (three \emph{bottom} rows). Ratios of
the measured fluxes $F_{{\rm T}{\lambda}{n}}$, peak intensities $F_{{\rm P}{\lambda}{n}}$, and sizes $\{A,B\}_{{\lambda}{n}}$ to
their true values ($F/F_{\rm T}$, $I/I_{\rm T}$, $A/A_{\rm T}$, and $B/B_{\rm T}$) are shown as a function of the S/N ratio
$\Omega_{{\lambda}{n}}$. The size ratios $A/A_{\rm T}$ and $B/B_{\rm T}$ are also shown as a function of the true sizes
$\{A,B\}_{{\lambda}{n}{\rm T}}$. The mean $\varrho_{{\rm \{P|T|A|B\}}{\lambda}}$ and standard deviation $\sigma_{{\rm
\{P|T|A|B\}}{\lambda}}$ of the ratios are displayed in the panels. Similar plots for $\lambda\le 100$\,$\mu$m with only bright
protostellar cores are not presented, because their measurements are quite accurate, with $\varrho_{\{{\rm
P|T|A|B}\}{\lambda}}\approx \{1.000|1.000|1.000|1.000\}$ and $\sigma_{\{{\rm P|T|A|B}\}{\lambda}}\approx
\{0.0004|0.0006|0.0002|0.0002\}$.
} 
\label{accuracyB4}
\end{figure*}

Figures \ref{accuracyA2}\,--\,\ref{accuracyB4} display the measurement accuracies of the peak intensity $F_{{\rm P}{\lambda}{n}}$,
integrated flux $F_{{\rm T}{\lambda}{n}}$, and sizes $\{A,B\}_{{\lambda}{n}}$ for each acceptable source $n$, represented by the
ratios of their measured and true values, as functions of their S/N ratios $\Omega_{{\lambda}{n}}$ and true FWHM sizes
$A_{{\lambda}{n}{\rm T}}$. The accuracy plots are not shown for $\lambda < 160$\,$\mu$m, because only the bright protostellar cores
are extractable in those images and their measurements are quite precise, with errors well below $1${\%}. The measurement results
in the derived $\mathcal{D}_{\{11|13\}{\arcsec}}$ are not shown either, because they are known to be inaccurate (e.g., Appendix~A
of Paper I). Some of the starless cores become measurable at $\{160|170\}$\,$\mu$m as faint sources with $\Omega_{{\lambda}{n}}\la
10$, the values well below the S/N of the bright protostellar cores. The faintness of the starless cores with respect to the
background and noise fluctuations makes measurements of some of them inaccurate, with a large spread of errors in total fluxes,
exceeding a factor of $2^{1/2}$. Toward the longer wavelengths ($250{-}500$\,$\mu$m), the starless cores become brighter, whereas
the protostellar cores become fainter, making their $\Omega_{{\lambda}{n}}$ ranges overlap for the two populations of sources.

Figures \ref{accuracyA2}\,--\,\ref{accuracyB4} reveal that \textsl{getold} systematically underestimates the FWHM sizes
$\{A,B\}_{{\lambda}{n}}$ of sources by ${\sim\,}20\%$. The problem is most clearly visible for the well-resolved sources, because
for the slightly resolved or unresolved sources \textsl{getold} adjusts the underestimated values $\{A,B\}_{{\lambda}{n}} <
O_{\lambda}$ by setting them to the angular resolution $O_{\lambda}$. The main reason for the systematic deficiency is the size
estimation algorithm that uses the source intensity moments, which can only be accurate for the Gaussian sources. In most practical
applications, however, there are no Gaussian-shaped sources, for several reasons. Firstly, the point-spread functions (PSFs, beams)
of telescopes are often non-Gaussian in their lower parts, which affects the shapes of mostly the unresolved sources. Secondly, the
radiative transfer models of the starless and protostellar cores (Sect.~\ref{skybench}) suggest that the real physical cores
produce non-Gaussian intensity profiles, thereby affecting the shapes of mostly the resolved sources. Finally, the backgrounds of
sources in bright fluctuating molecular clouds cannot be determined accurately, hence non-negligible over- or under-subtraction of
the background of even the Gaussian sources would create non-Gaussian shapes, in both resolved and unresolved cases. Background of
extracted sources is often overestimated, hence the intensity moments of the background-subtracted source would underestimate
$\{A,B\}_{{\lambda}{n}}$. For the protostellar cores that have power-law intensity profiles at large radii, the intensity moments
algorithm leads to strongly overestimated half-maximum sizes and for the starless cores with flat-topped shapes, the intensity
moments could significantly underestimate the half-maximum sizes (cf. Sect. 3.4.6 of Paper I).

Figures \ref{accuracyA2}\,--\,\ref{accuracyB4} demonstrate that \textsl{getsf} does not have such systematic problems with the FWHM
sizes $\{A,B\}_{{\lambda}{n}}$ of sources. This is because \textsl{getsf} evaluates them directly at the half-maximum intensity
(Sect. 3.4.6 of Paper I), unlike \textsl{getold} that employs the source intensity moments. Direct measurements are much less
affected by the background inaccuracies, but over-subtracted backgrounds of the (almost) unresolved sources could also lead to
unrealistically small $\{A,B\}_{{\lambda}{n}} < O_{\lambda}$ and underestimated fluxes $F_{{\rm P}{\lambda}{n}}$ and $F_{{\rm
T}{\lambda}{n}}$. The sizes and fluxes of such unresolved or slightly resolved sources are rectified by \textsl{getsf} using the
correction factors (Appendix \ref{corrections}) derived for an unresolved Gaussian source, assuming that it is the background
over-subtraction that makes the source have the sub-resolution sizes $\{A,B\}_{{\lambda}{n}} < O_{\lambda}$. The Gaussian model is
used to obtain the correction factors, not the measurements themselves. Unfortunately, similar corrections cannot be derived for
the well-resolved sources, nor for the sources with underestimated backgrounds and overestimated sizes and fluxes.

Figures \ref{accuracyA2}\,--\,\ref{accuracyB4} show a behavior that is qualitatively similar for the different variants of
Benchmarks A and B. There is an expected general trend that the numbers of acceptable sources in the accuracy plots become lower
for the backgrounds with increasing complexity, in the sequence from $\{\mathrm{A},\mathrm{B}\}_2$ to $\{\mathrm{A},\mathrm{B}\}_3$
and to B$_4$. This is caused by the much stronger variations in the immediate surroundings of the sources, especially those located
on the densest parts of the background cloud, which strongly reduce the S/N ratio $\Omega_{{\lambda}{n}}$ of the extracted sources.
As a result, some of those sources that were acceptable in the simpler variants of the benchmarks, are pushed off the acceptability
domain by their lower values $\Omega_{{\lambda}{n}} < 2$. In all benchmarks, the measurement errors significantly increase for the
faint sources, because their estimated individual backgrounds become more strongly affected by the fluctuations of the filamentary
cloud and noise. The resulting over- or underestimation of the backgrounds depends on whether the sources happen to be located on
the hollow- or hill-like fluctuation, correspondingly, as well as on the other types of background inaccuracies (cf. Appendices
\ref{backgrounds} and \ref{corrections}).


\begin{figure*}
\centering
\centerline{
  \resizebox{0.3612\hsize}{!}{\includegraphics{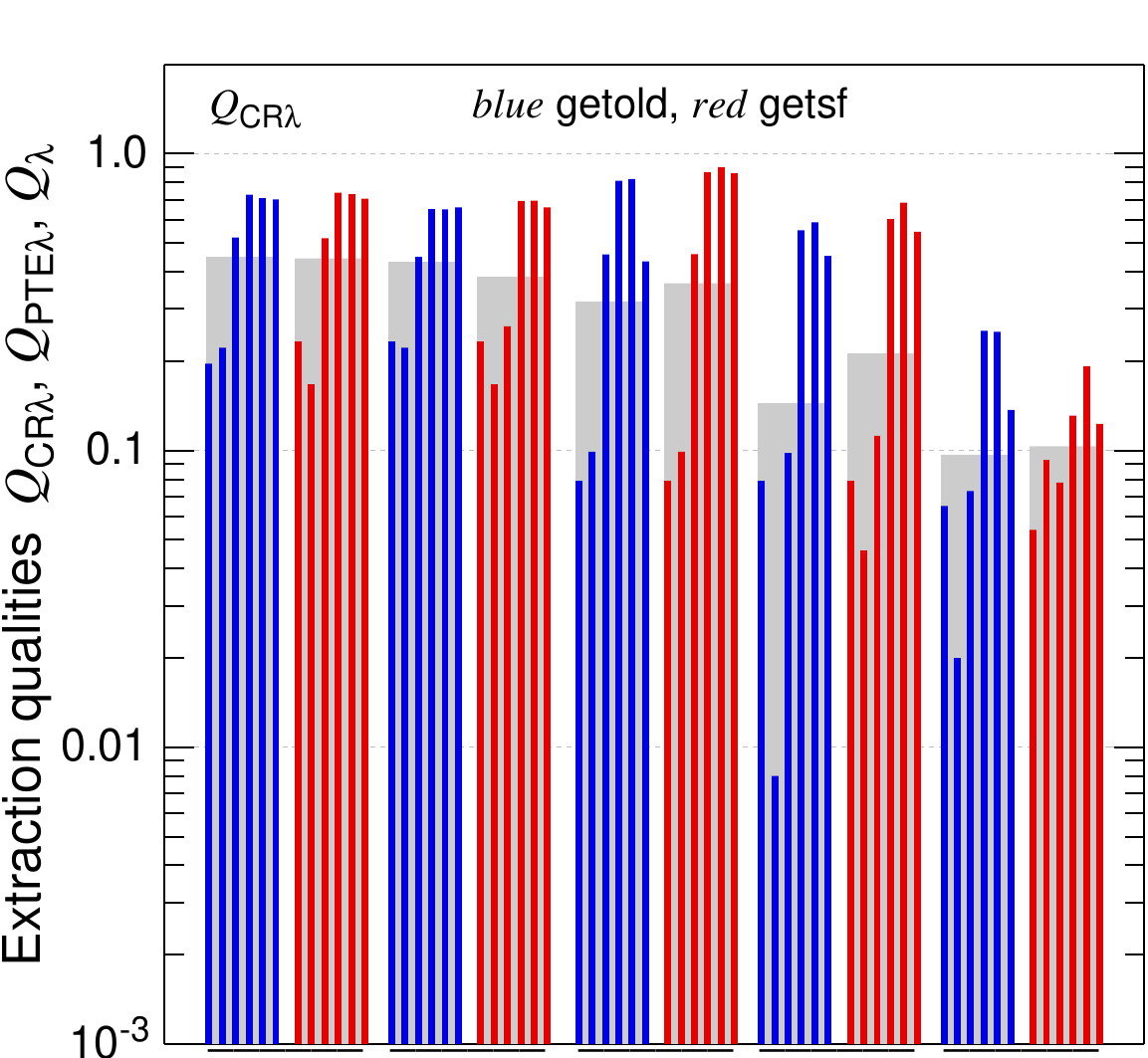}}
  \resizebox{0.3100\hsize}{!}{\includegraphics{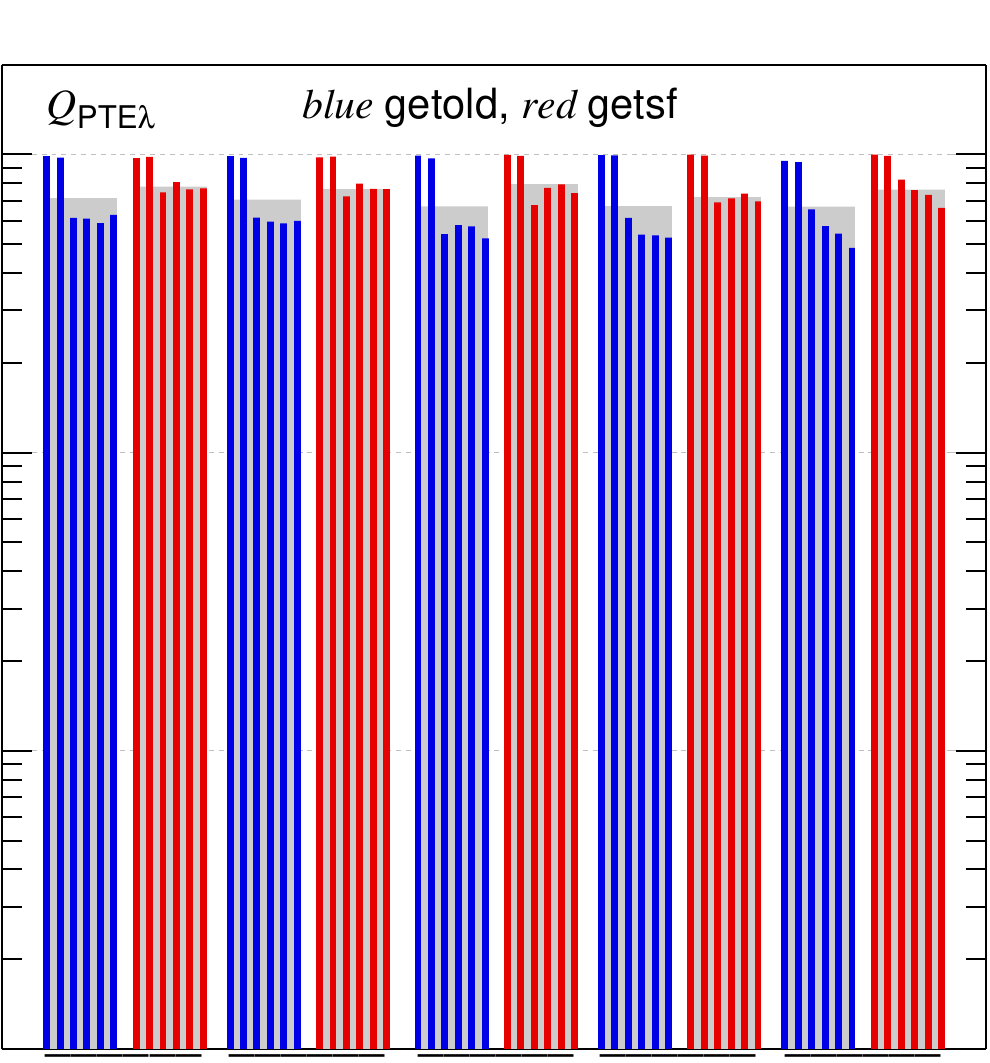}}
  \resizebox{0.3100\hsize}{!}{\includegraphics{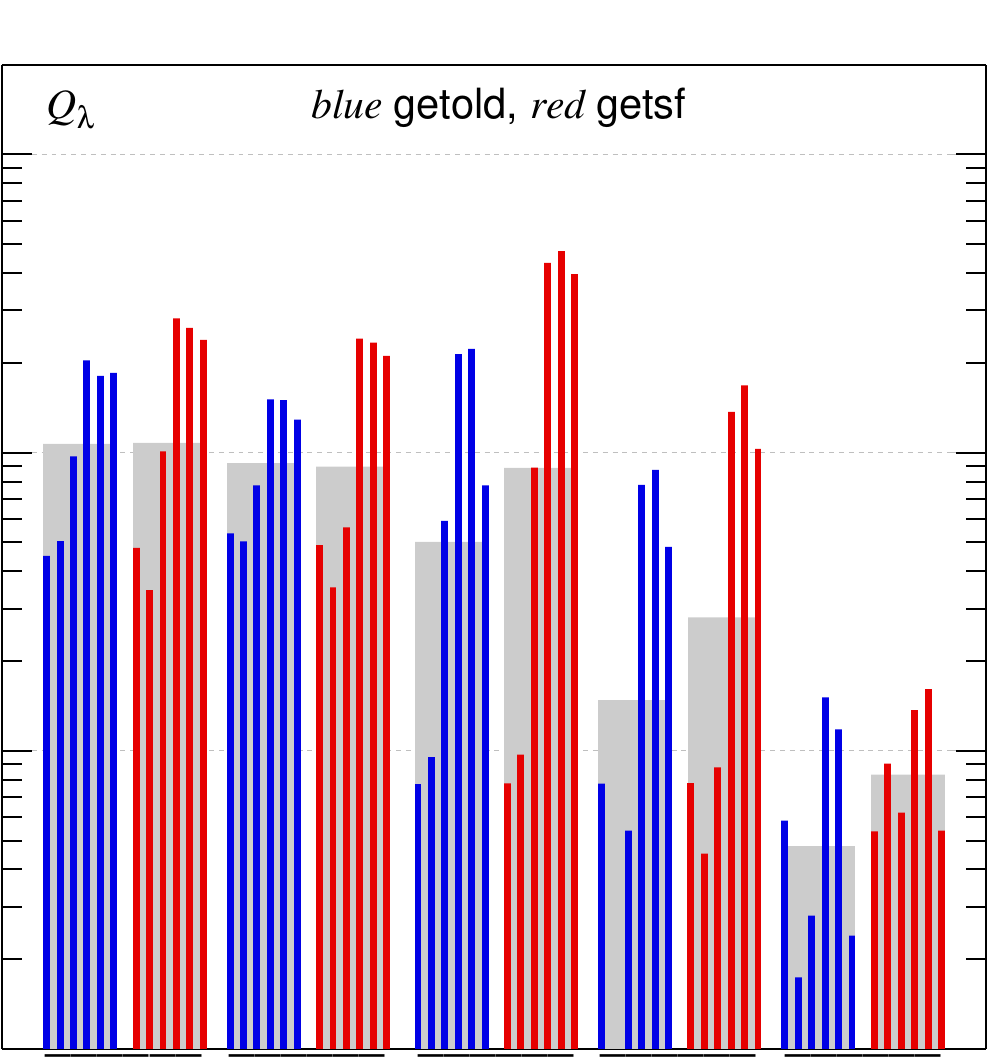}}}
\centerline{
  \resizebox{0.3612\hsize}{!}{\includegraphics{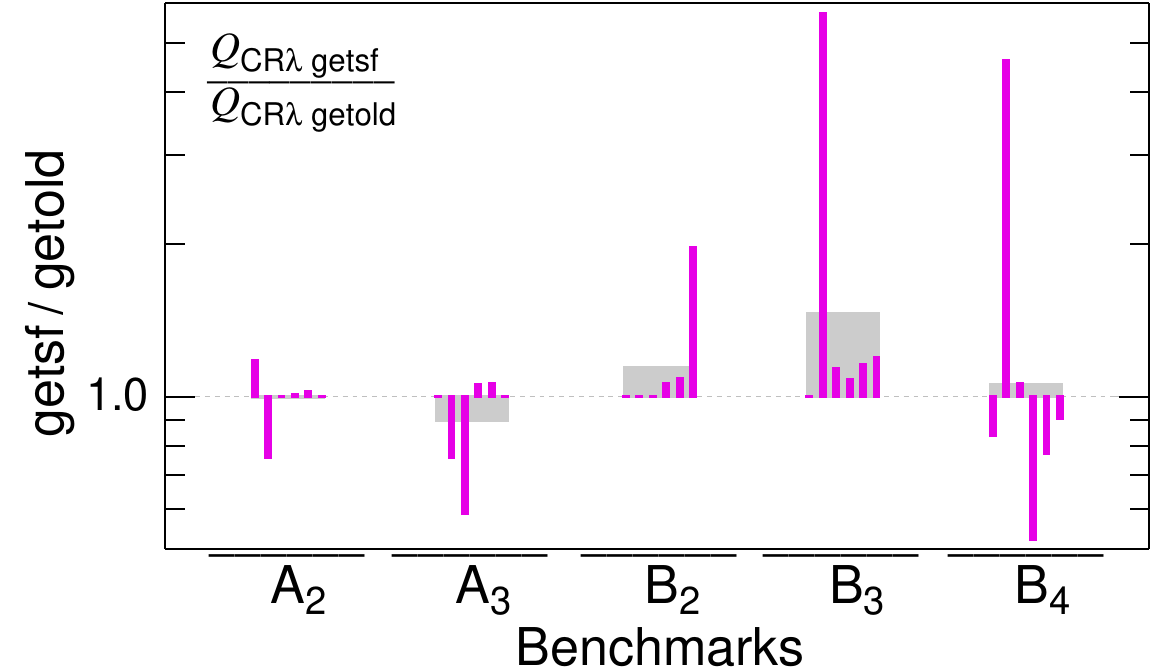}}
  \resizebox{0.3100\hsize}{!}{\includegraphics{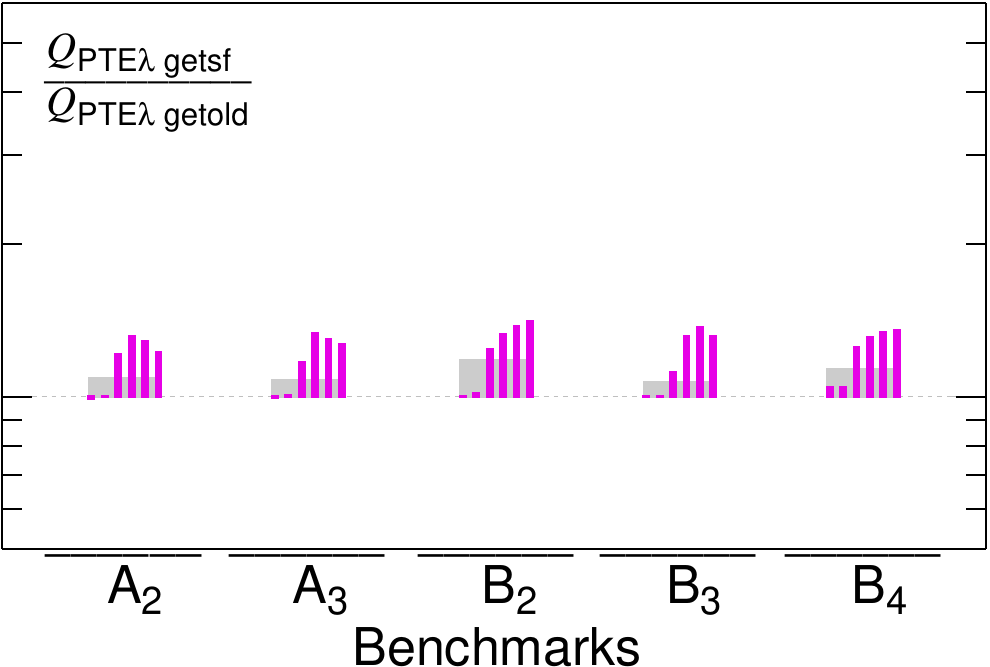}}
  \resizebox{0.3100\hsize}{!}{\includegraphics{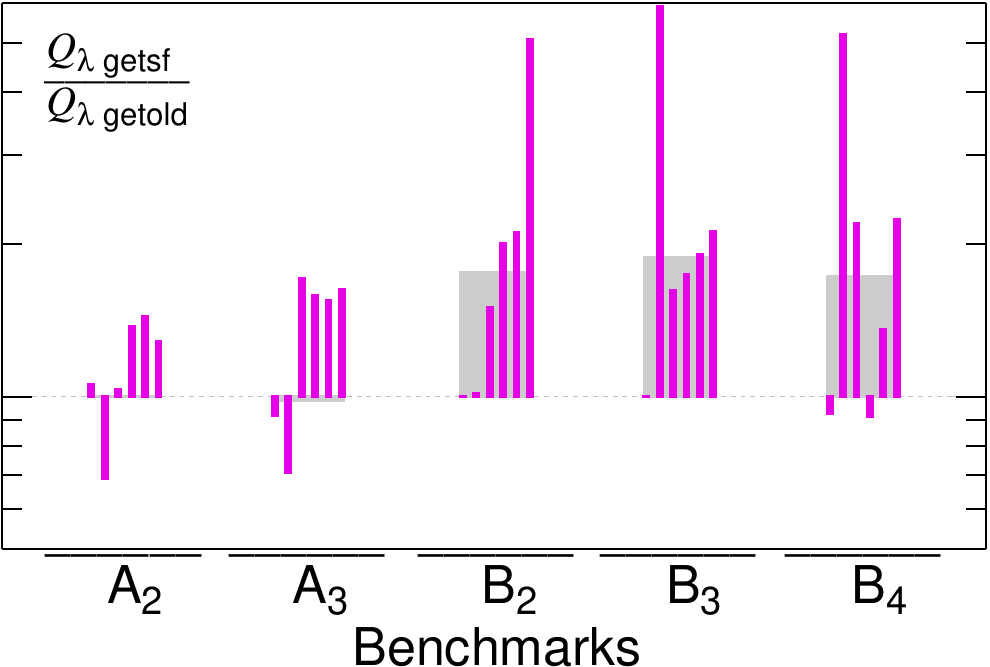}}}
\caption
{ 
Overview of the benchmarking results for the source extractions with \textsl{getold} and \textsl{getsf} (Tables \ref{qualAB23} and
\ref{qualB4}). The \emph{top} panels show the detection quality $Q_{{\rm CR}{\lambda}}$, the measurement quality $Q_{{\rm
PTE}{\lambda}}$, and the overall quality $Q_{{\lambda}}$ from Eq. (\ref{finalqualities}), represented by vertical bars for each
wavelength (3 PACS and 3 SPIRE bands, from left to right), with an exception of the fictitious $\lambdabar$ of the surface density
$\mathcal{D}_{\{11|13\}\arcsec}$. The global qualities of the methods, $Q_{{\rm CR}}$, $Q_{{\rm PTE}}$, and $Q$, defined as the
geometric means over the wavelengths, are represented by the wide gray bars. The \emph{bottom} panels help visualize the ratios of
the qualities for the \textsl{getsf} and \textsl{getold} extractions for each benchmark and wavelength.
} 
\label{extrqual}
\end{figure*}

\begin{figure*}
\centering
\centerline{
  \resizebox{0.3612\hsize}{!}{\includegraphics{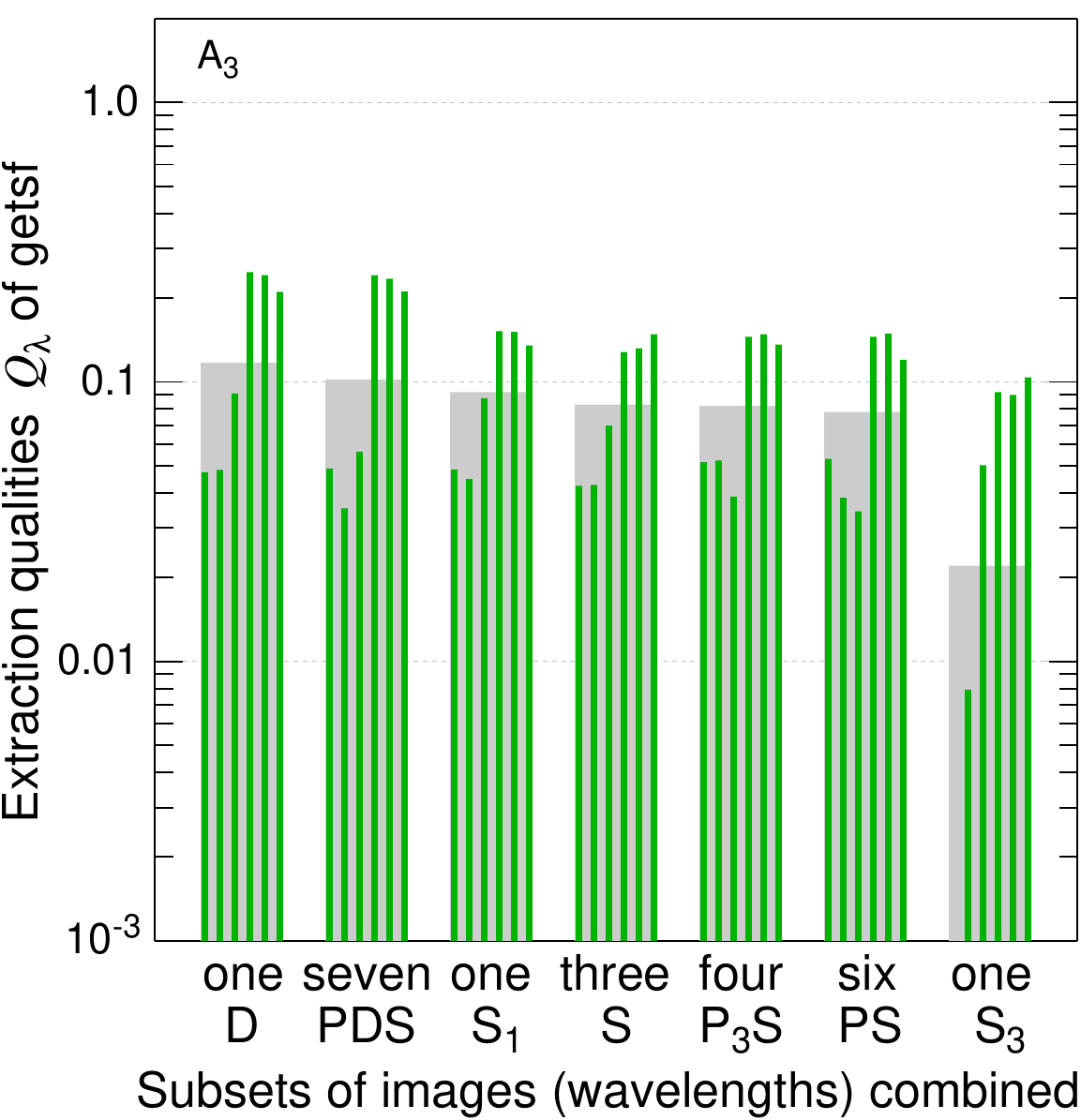}}
  \resizebox{0.3100\hsize}{!}{\includegraphics{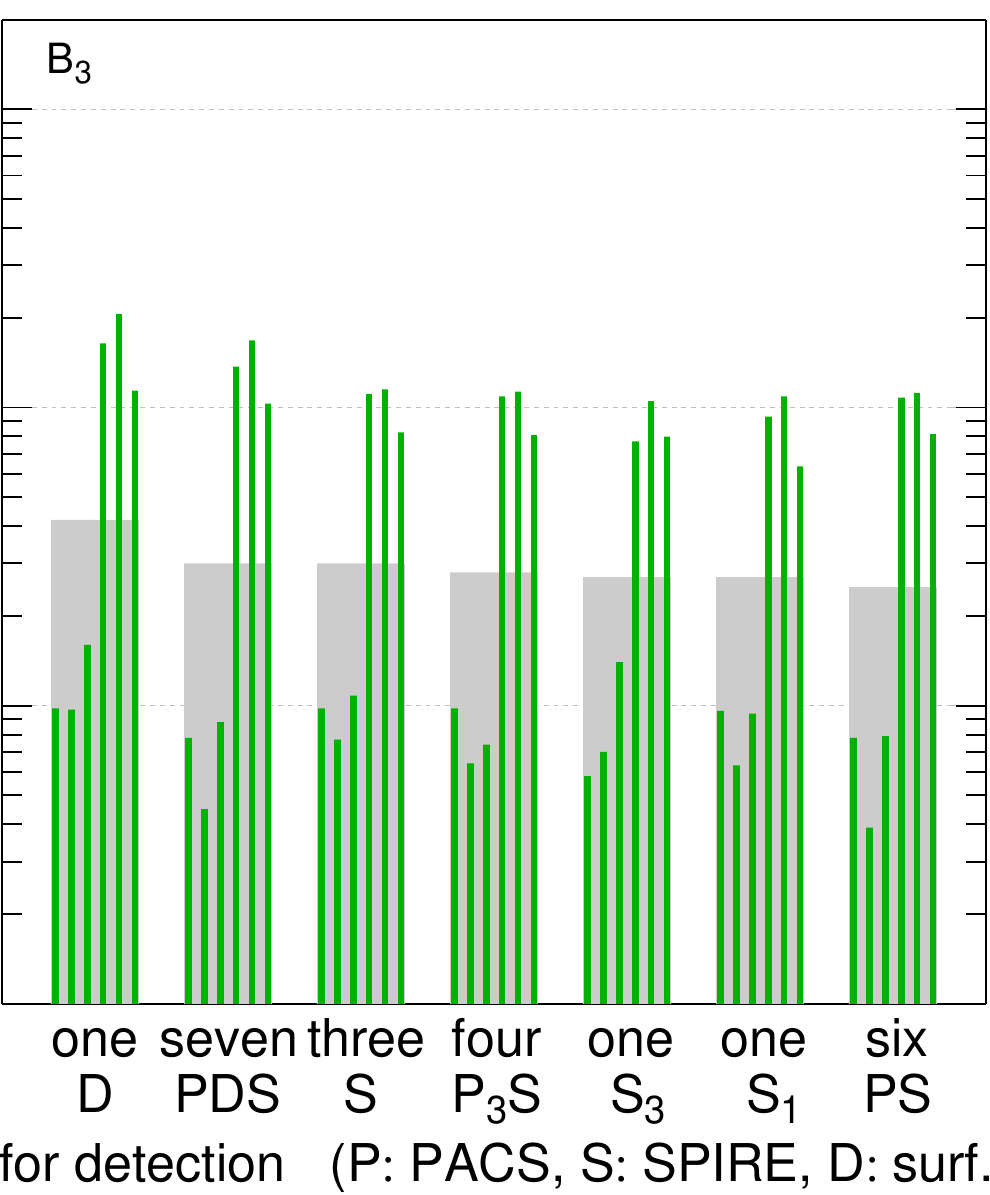}}
  \resizebox{0.3100\hsize}{!}{\includegraphics{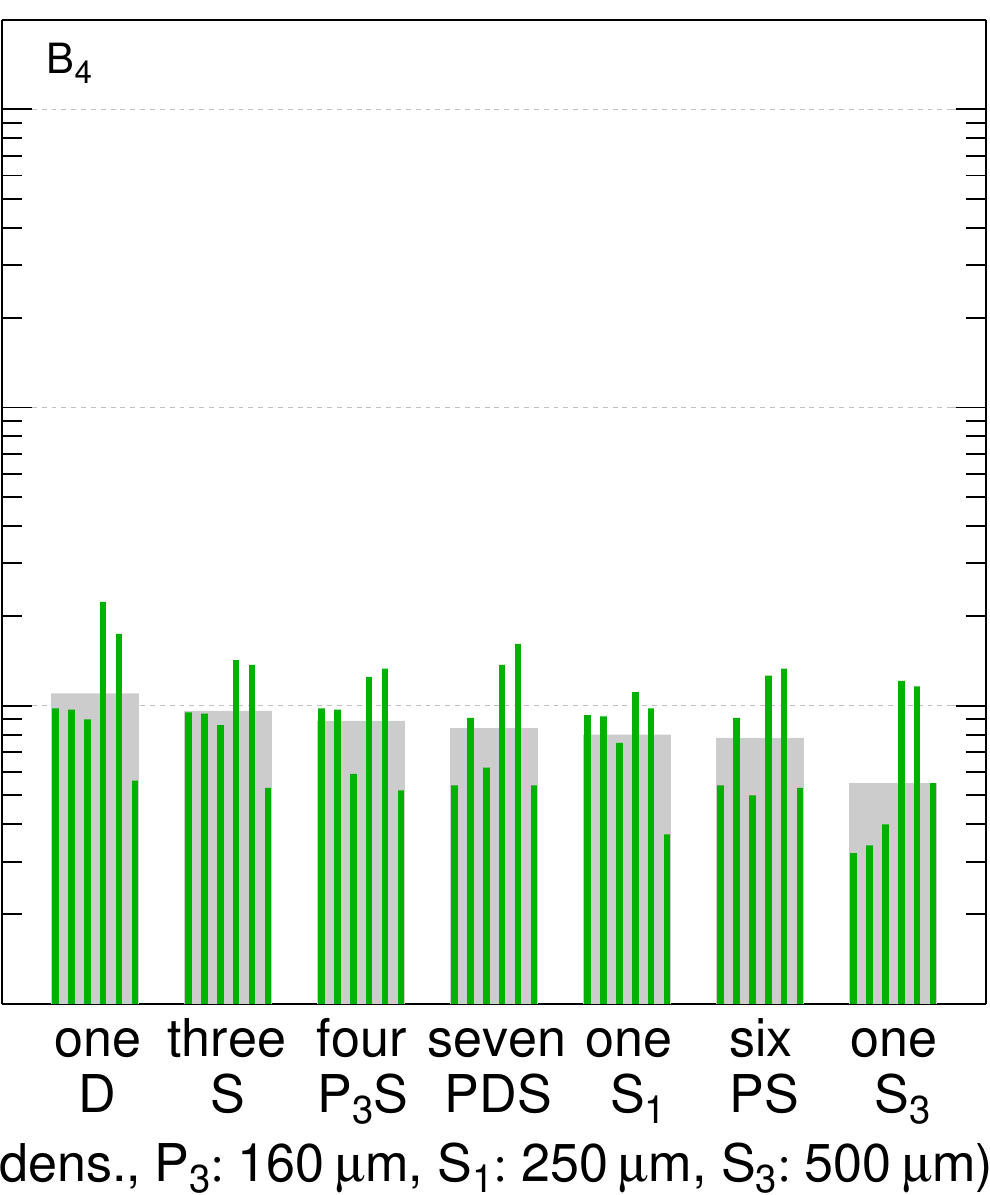}}}
\caption
{ 
Overview of the dependence of the \textsl{getsf} extraction qualities (Tables \ref{qualA3sets}\,--\,\ref{qualB4sets}) on various
subsets of wavelengths (images) used in detecting sources (cf. Sects. 3.4.3 and 3.4.4 of Paper I) in benchmarks A$_3$, B$_3$, and
B$_4$. The overall qualities $Q_{\lambda}$ from Eq. (\ref{finalqualities}) are represented by the green vertical bars for each
wavelength (3 PACS and 3 SPIRE bands), with an exception of the fictitious $\lambdabar$ of the surface density
$\mathcal{D}_{\{11|13\}\arcsec}$. The results for the subsets are annotated on the horizontal axis with the number of images used
for detection. In each panel, the \textsl{getsf} extractions are sorted in the order of their decreasing global quality $Q$,
defined as the geometric mean over the wavelengths and represented by the wide gray bars.
} 
\label{qualA3B3B4}
\end{figure*}


\subsubsection{Extraction qualities}
\label{qualities}

Figure \ref{extrqual} presents an overview of the extraction qualities of \textsl{getold} and \textsl{getsf}, displaying $Q_{{\rm
CR}{\lambda}}$, $Q_{{\rm PTE}{\lambda}}$, and $Q_{{\lambda}}$ from Tables \ref{qualAB23} and \ref{qualB4}. The first two qualities
conveniently evaluate the extraction methods at their independent detection and measurement steps, whereas the third one combines
the two in the overall extraction quality. To facilitate their analysis, the plots display also the global qualities $Q_{{\rm
CR}}$, $Q_{{\rm PTE}}$, and $Q$, the geometric mean values over the wavelengths, for each benchmark. All features of the plots in
Fig.~\ref{extrqual} can be readily understood by comparisons of the tabulated qualities (Tables \ref{qualAB23} and \ref{qualB4}).

The source-detection quality $Q_{{\rm CR}{\lambda}}$ is the product of the extraction completeness $C_{\lambda}$ and reliability
$R_{\lambda}$. As expected, the global detection quality $Q_{{\rm CR}}$ of both methods decreases from A$_2$ to B$_4$, toward the
more complex benchmarks (Fig.~\ref{extrqual}), demonstrating better results for \textsl{getsf} in all benchmarks, except A$_3$. In
A$_3$, \textsl{getsf} has a $13${\%} lower quality, because of several spurious (very noisy) sources extracted at $110$ and
$170$\,$\mu$m with very low significance levels, within just a few percent above the cleaning threshold $\varpi_{{\lambda}{\rm
S}{j}} = 5\sigma_{{\!\lambda}{\rm S}{j}}$ (Sect.~3.4.2 of Paper I). At $\{70|75\}$ and $\{100|110\}$\,$\mu$m, $Q_{{\rm
CR}{\lambda}}$ shows lower values, because only the protostellar cores are detectable, whereas at $\{160|170\}$\,$\mu$m, some of
the starless cores appear as faint detectable sources, hence the quality gets higher. For some benchmarks, $Q_{{\rm CR}{\lambda}}$
becomes significantly lower, which usually indicates that more spurious sources were extracted, hence the lower reliability
$R_{\lambda}$.

The measurement quality $Q_{{\rm PTE}{\lambda}}$ is a product of the respective qualities $Q_{{\rm P}{\lambda}}$, $Q_{{\rm
T}{\lambda}}$, and $Q_{{\rm E}{\lambda}}$ of peak intensity, integrated flux, and source area. The global measurement quality
$Q_{{\rm PTE}}$ for \textsl{getsf} is better by $20${\%} than that for \textsl{getold}, across all benchmarks
(Fig.~\ref{extrqual}). At $\{70|75\}$ and $\{100|110\}$\,$\mu$m, $Q_{{\rm PTE}{\lambda}}$ is within $2${\%} of unity, because the
protostellar cores are bright, hence they can be accurately measured (cf. Figs.~\ref{accuracyA2}\,--\,\ref{accuracyB4}). The faint
starless cores at $\{160|170\}$\,$\mu$m are poorly measurable; therefore, $Q_{{\rm PTE}{\lambda}}$ becomes lower. The \textsl{getsf}
measurement quality is substantially higher at the SPIRE wavelengths, partly because \textsl{getold} systematically underestimates
source sizes (Sect. \ref{accuracies}).

The overall quality $Q_{{\lambda}}$ is the product of $Q_{{\rm CR}{\lambda}}$ and $Q_{{\rm PTE}{\lambda}}$, as well as of the
positional quality $Q_{{\rm D}{\lambda}}$ and goodness $G_{\lambda}$. In line with the expectations, the global quality $Q$ of both
methods decreases toward the more complex benchmarks (Fig.~\ref{extrqual}). For the simpler Benchmark A, \textsl{getsf} has a small
$10${\%} edge over \textsl{getold}, whereas for Benchmark B, the \textsl{getsf} quality reaches the values higher by a factor of
$2^{1/2}$. The quality evaluation system (Sect.~\ref{evalqual}) encapsulates all aspects of source extraction; therefore, the plots
in Fig.~\ref{extrqual}, based on Tables \ref{qualAB23} and \ref{qualB4}, justify the conclusion that the \textsl{getsf} is superior
to \textsl{getold} in both Benchmarks A and B.


\begin{figure*}                                                               
\centering
\centerline{
  \resizebox{0.328\hsize}{!}{\includegraphics{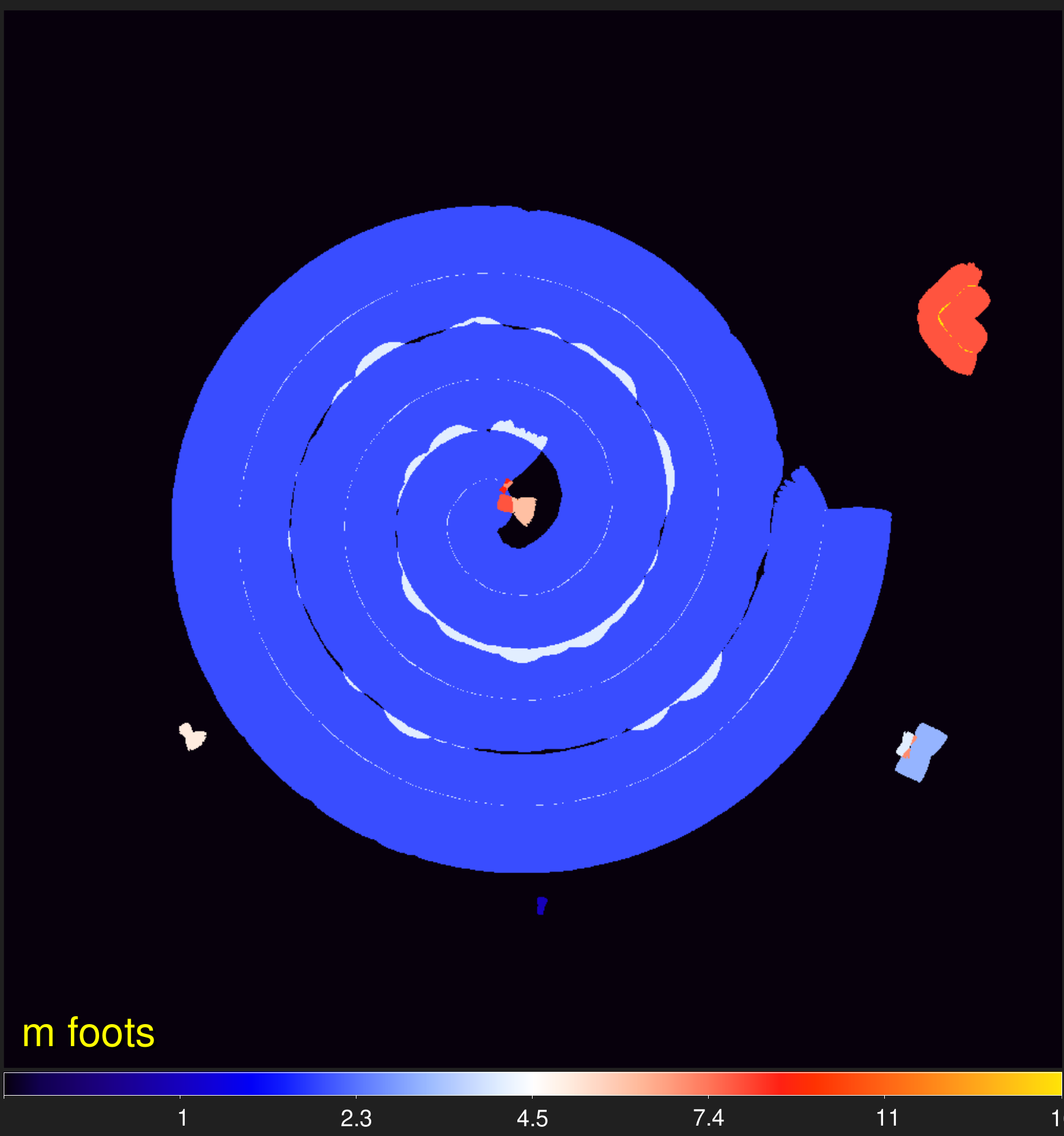}}
  \resizebox{0.328\hsize}{!}{\includegraphics{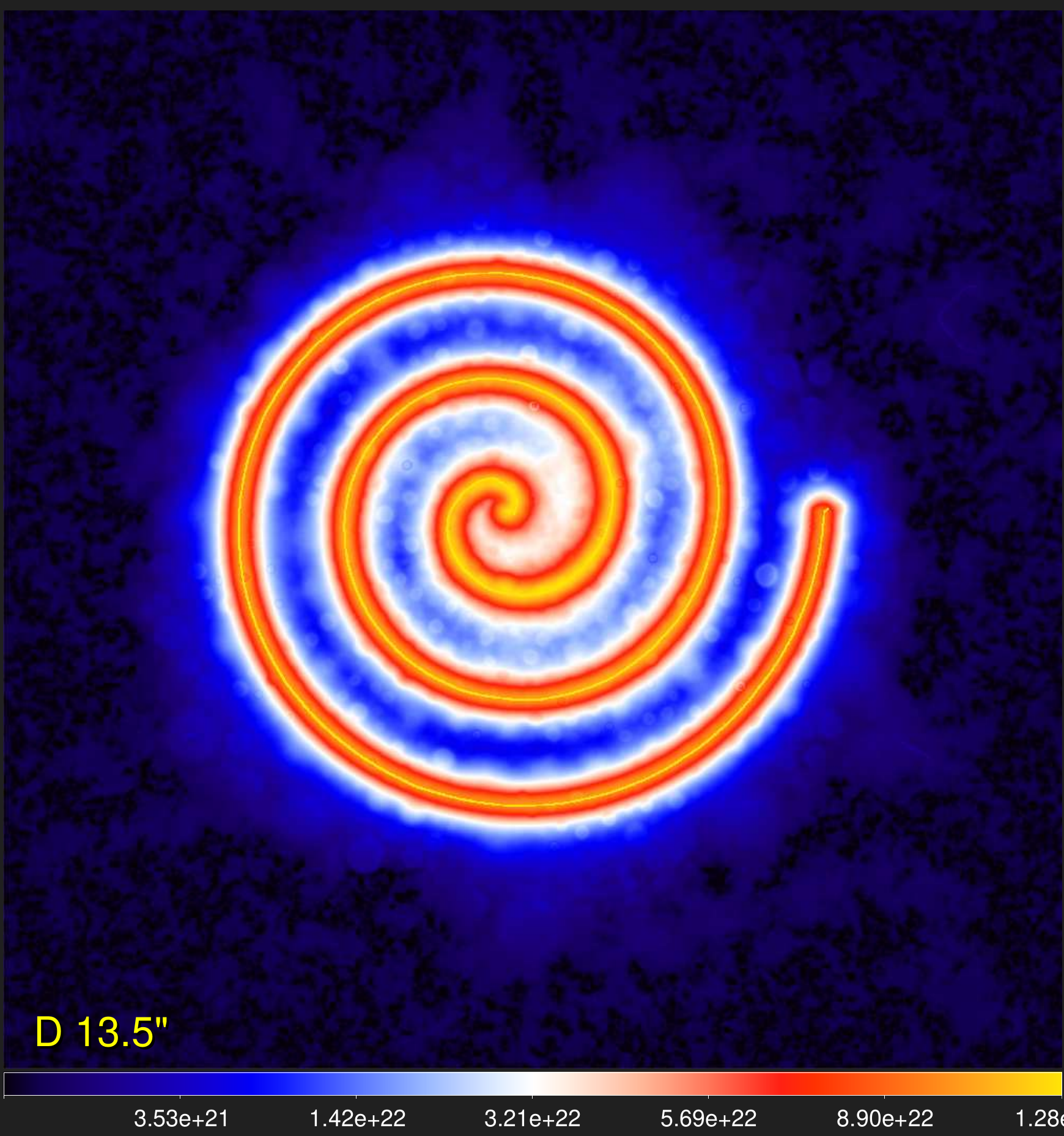}}
  \resizebox{0.328\hsize}{!}{\includegraphics{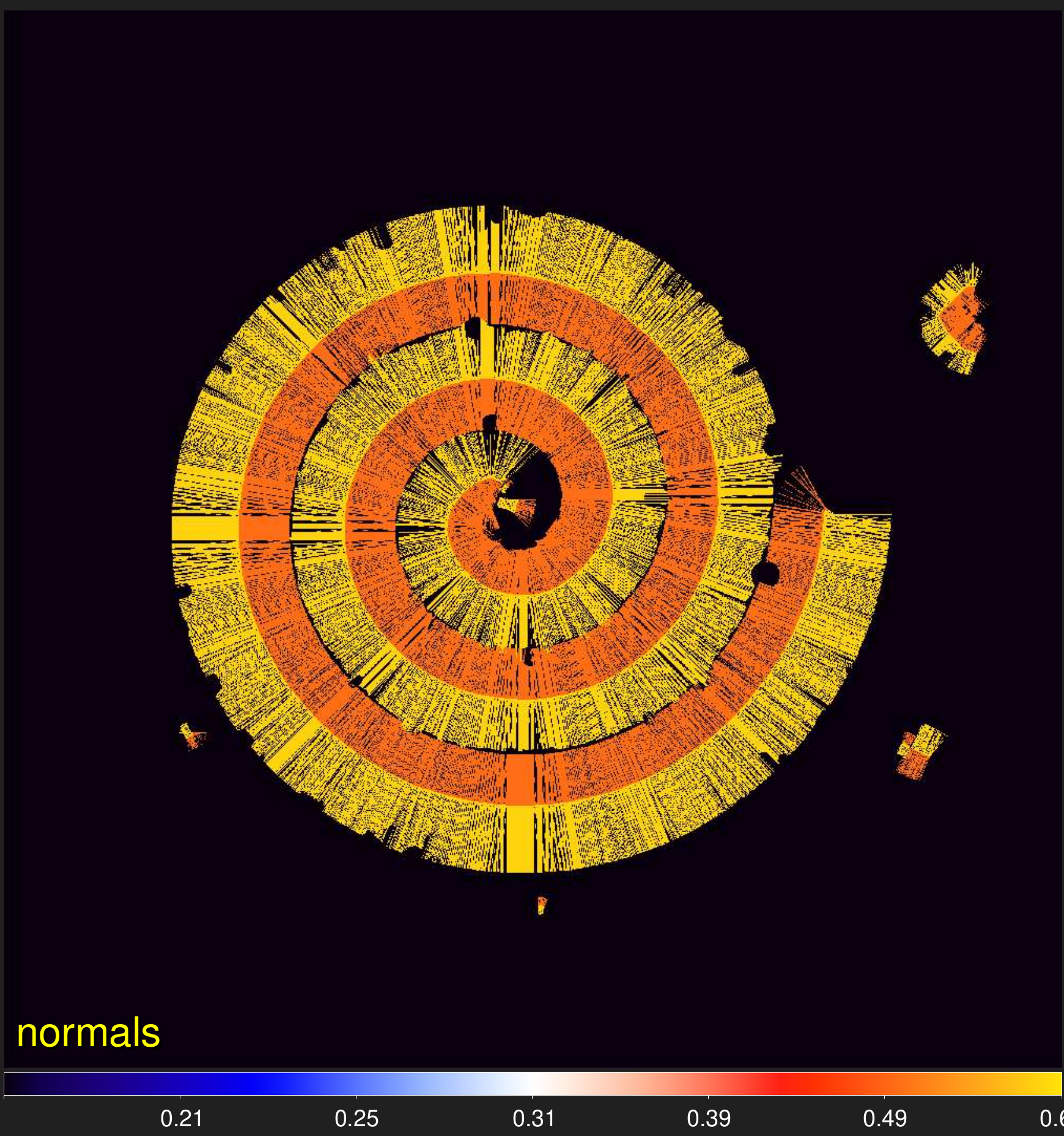}}}
\caption
{ 
Benchmark B$_4$ extraction of filaments with \textsl{getsf}. The component of filaments $\mathcal{F}_{{\lambdabar}{Y}}$ is shown in
the \emph{middle} panel for the surface density $\mathcal{D}_{13\arcsec}$, overlaid with the skeleton $\mathcal{K}_{{k}{4}}$
(detection significance $\xi = 4$) obtained from the combined $\mathcal{F}_{{\rm D}{j}{\rm C}}$ on the scales $S_{\!k}\approx
153${\arcsec}, corresponding to the adopted value $Y_{\lambdabar} = 150${\arcsec}. The \emph{left} panel displays the skeletons and
the footprints $\Upsilon_{{n}}$, with the pixel values equal to the skeleton (filament) number. The \emph{right} panel shows the
detected skeleton of the spiral filament, together with its one-sided normals. The first pixel of the skeleton is at the center,
hence the left normals point to the center (red) and the right normals point outward (orange). Square-root color mapping, except
the footprints with linear mapping.
} 
\label{filamentsB4}
\end{figure*}


\subsubsection{Dependence on the images used for detection}
\label{dependonimg}

In the present multiwavelength benchmarking, all seven images were combined in the wavelength-independent images for detecting
sources (Sect.~3.4.3 of Paper I). To some extent, however, source extraction results must depend on the images used for source
detection. Both \textsl{getsf} and \textsl{getold} combine images and detect sources with almost the same algorithms; therefore,
\textsl{getsf} alone may be used to evaluate the dependence of the extraction qualities on the subsets of images. Only the
realistic benchmark variants with backgrounds ($\{\mathrm{A},\mathrm{B}\}_3$ and B$_4$) may be used in these tests to keep the
amounts of results within reasonable limits.

Figure \ref{qualA3B3B4} presents an overview of the overall quality $Q_{\lambda}$ and its global counterpart $Q$ for source
extractions with \textsl{getsf} in A$_3$, B$_3$, and B$_4$ using 6 subsets of images combined for detection. The full set of seven
images (PDS) was discussed above (Sects. \ref{visuals}\,--\,\ref{qualities}) and is shown again for completeness. The subset of six
images (PS) tests the case when the surface density image $\mathcal{D}_{\{11|13\}\arcsec}$ is not used. The subset of four images
(P$_3$S) examines the absence of two PACS images (at $\{70|75\}$ and $\{100|110\}$\,$\mu$m). The subset of three images (S)
clarifies the effects of the source detection with only the SPIRE images. The three single-image subsets (S$_1$, S$_3$, and D)
explore the source extractions with the $250$\,$\mu$m image, the $500$\,$\mu$m image, and the surface density image
$\mathcal{D}_{\{11|13\}\arcsec}$, respectively. The subsets with only the PACS images are not considered, because no starless cores
appear in the images at $\lambda < 160$\,$\mu$m.

The results (Fig.~\ref{qualA3B3B4}) for the seven different cases are sorted from left to right in the order of decreasing global
quality $Q$. In all three benchmarks, the best extraction quality is found in the subset D, when the surface density
$\mathcal{D}_{\{11|13\}\arcsec}$ is the single image used to detect sources. It is obvious from the original images (e.g.,
Figs.~\ref{imagesA3}\,--\,\ref{imagesB4}) that the surface density image must be beneficial for source extractions, because the
sources are visible there most clearly. However, this result suggests that the high-resolution $\mathcal{D}_{\{11|13\}\arcsec}$ may
also be used alone to detect sources, with better results than in a combination with the \emph{Herschel} images. In the benchmarks
$\{\mathrm{A},\mathrm{B}\}_3$, the second best global quality $Q$ is shown by the complete set (PDS), when all seven images are
used to detect sources. In B$_4$, however, the extraction quality with this subset of images is only the fourth, which was caused
by a few more spurious sources extracted at $70$ and $160$\,$\mu$m. The spurious peaks are clearly identifiable with the background
and noise fluctuations in those images that happened to be slightly brighter than the cleaning threshold $\varpi_{{\lambda}{\rm
S}{j}} = 5\sigma_{{\!\lambda}{\rm S}{j}}$ (Sect.~3.4.2 of Paper I). Without the spurious sources, the PDS set would have the second
best $Q$ value in all three benchmarks.

When the subset S of only the three SPIRE images is used for source detection, the global quality $Q$ becomes the fourth, the
third, and the second best in the benchmarks A$_3$, B$_3$, and B$_4$, correspondingly (Fig.~\ref{qualA3B3B4}). An addition of the
PACS $160$\,$\mu$m image to the SPIRE images in P$_3$S leads to the fifth, the fourth, and the third best $Q$ values among all
subsets, always just below the global quality for the subset S. The slightly lower (by $5{-}10${\%}) qualities in P$_3$S can be
traced to the chance extraction of a few spurious sources at the longest PACS wavelength. Using the subset S$_1$ with the single
$250$\,$\mu$m image for source detection makes the global quality in A$_3$ the third best, whereas in B$_3$ and B$_4$ it becomes
only the sixth and fifth, respectively. The absence of the high-resolution $\mathcal{D}_{\{11|13\}\arcsec}$ in the subset PS of the
six \emph{Herschel} images makes the extraction one of the two worst ones. However, the differences between the $Q$ values outside
the top three best-quality results is very small, at the levels of a few percent. An exception is the worst extraction for S$_3$,
whose quality is well below all others in A$_3$ and B$_4$, because of the lowest angular resolution of the detection image.

Formally taking all the benchmarking results, it is possible to rank the \textsl{getsf} source extraction qualities by summing up
their places in the three benchmarks shown in Fig.~\ref{qualA3B3B4}. The two best subsets of the \emph{Herschel} images to be used
for source detection are D ($\mathcal{D}_{\{11|13\}\arcsec}$) and PDS ($\mathcal{D}_{\{11|13\}\arcsec}$ together with all PACS and
SPIRE images), and the next good subset is S (SPIRE images at $250{-}500$\,$\mu$m). The three worst subsets appear to be S1
($250$\,$\mu$m image), PS (all \emph{Herschel} images), and S3 ($500$\,$\mu$m image). It must be emphasized that the actual choices
in real-life applications depend on the research interests. For example, if the goal is to study the protostellar cores, then the
shortest PACS wavelength, where they are the brightest and with the highest resolution, is the best choice for their detection.
However, if the aim is to study the starless cores that are the strongest at the SPIRE wavelengths, then the high-resolution
surface density $\mathcal{D}_{\{11|13\}\arcsec}$ (possibly together with the $250{-}500$\,$\mu$m images) is likely the best choice
for the source detection with \textsl{getsf}. This is an important decision to make when preparing for source extractions.



\begin{figure*}
\centering
\centerline{
  \resizebox{0.3440\hsize}{!}{\includegraphics{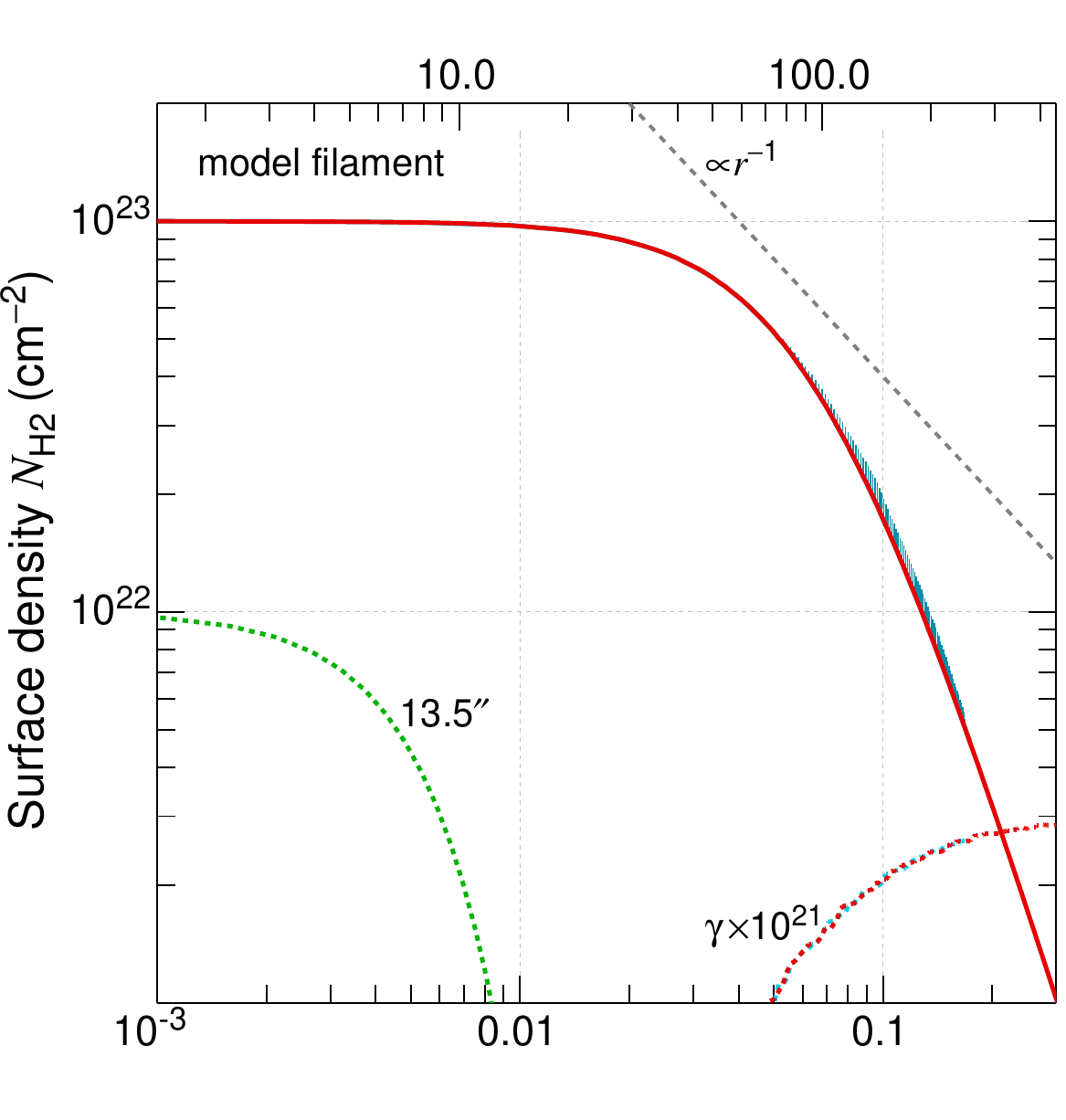}}
  \resizebox{0.3060\hsize}{!}{\includegraphics{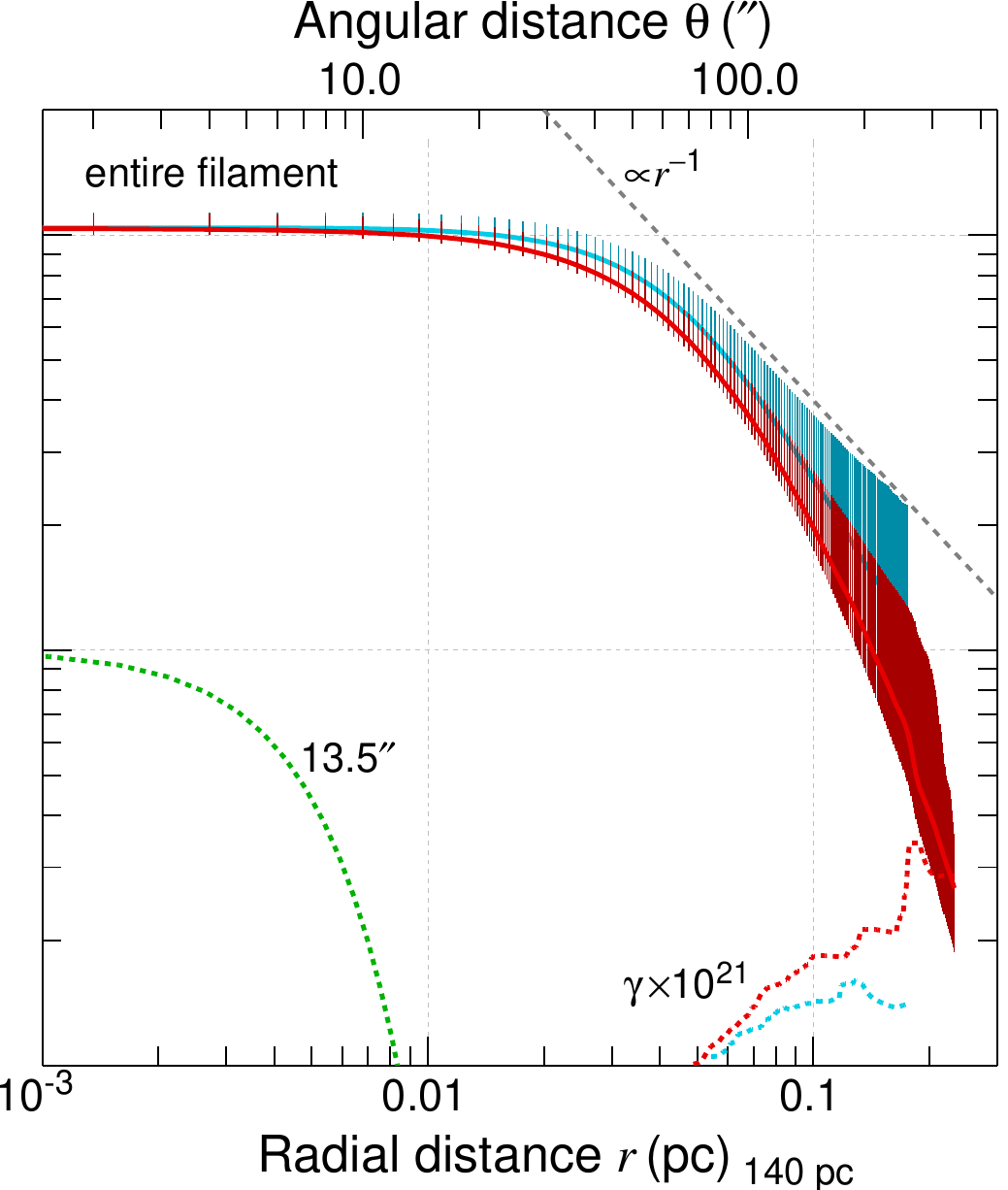}}
  \resizebox{0.3366\hsize}{!}{\includegraphics{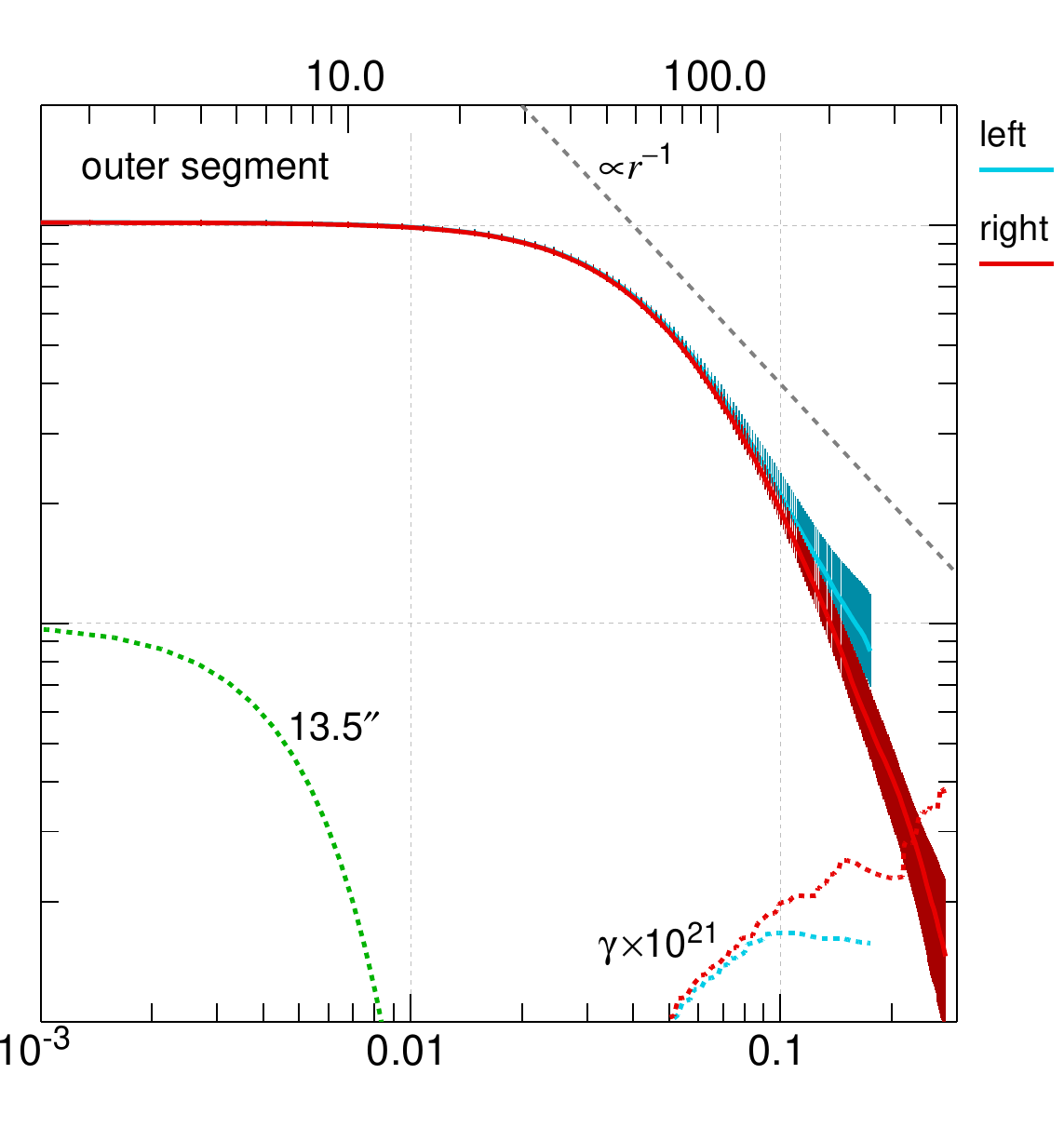}}}
\caption
{ 
Benchmark B$_4$ extraction of filaments with \textsl{getsf}. Plotted are the one-sided (left and right) profiles
$D_{\{{\alpha|\beta}\}}(r)$ of the filament surface densities $\mathcal{F}_{{\lambdabar}{Y}}$, their slopes
$\gamma_{\{\alpha|\beta\}}(r)$, multiplied by $10^{21}$ for convenience, the reference line with a slope of $-1$, and the Gaussian
beam with the half-maximum width $O_{\rm H} = 13.5{\arcsec}$. The profiles represent the median densities along the skeleton and
the vertical bars are their positive and negative deviations $\varsigma_{\{{\alpha|\beta}\}\pm}(r)$ about the median profile. The
\emph{left} panel shows the true profiles of the simulated filament $\mathcal{D}_{\rm F}$ across its crest. The \emph{middle} panel
presents the profiles, measured in the $\mathcal{F}_{{\lambdabar}{Y}}$ component of $\mathcal{D}_{13{\arcsec}}$ along the entire
detected skeleton length (Fig.~\ref{filamentsB4}). The \emph{right} panel displays the more accurate profiles, measured across the
outermost skeleton loop, where the filament is least affected by the inaccuracies of its background.
} 
\label{filB4profs}
\end{figure*}


\subsection{Filament extraction in Benchmark B$_4$}
\label{benchB4fil}

Filaments are separated from both backgrounds $\mathcal{B}_{{\lambda}{Y}}$ and sources $\mathcal{S}_{{\lambda}}$ and detected as
skeletons in their own flattened, wavelength-combined component $\mathcal{F}_{{\rm D}{j}{\rm C}}$ (Sects. 3.2\,--\,3.4 in Paper I).
The separation allows the filament crests to be traced more precisely, reducing the interference from the sources that could
significantly affect the results. In the standard approach to the multiwavelength benchmarking adopted in this paper, the filament
detection image is combined from six wavelengths, excluding the $70$ and $100$\,$\mu$m images, because the simulated filament is
very faint and noisy at those wavelength.

Figure \ref{filamentsB4} shows the skeletons and footprints of eight filaments detected in B$_4$ at the significance level $\xi =
4$ (Sect.~3.4.5 in Paper I). All but one of them are spurious, the short fluctuations of background that happened to be elongated
and slightly denser than the filament detection threshold $\varpi_{{\lambda}{\rm F}{j}} = 2\sigma_{{\!\lambda}{\rm F}{j}}$
(Sect.~3.4.2 in Paper I). The rate of spurious filaments can be reduced using a higher value of the skeleton significance $\xi$
when detecting filaments. Spurious filaments usually have their lengths shorter than their widths, hence they can be discarded from
further analysis after a visual inspection. This is not done by \textsl{getsf} automatically, because also the real filaments are
often split into relatively short segments by the sources, other intersecting filaments, or background fluctuations.

The sides of a filament are known to \textsl{getsf} as left ($\alpha$) or right ($\beta$) with respect to the path from the first
pixel of the skeleton to its last pixel. The first pixel of the spiral skeleton is in the center, hence the left normals to the
skeleton are pointing inside the loops and the right normals are pointing outward (Fig.~\ref{filamentsB4}). Although the one-sided
footprints and normals touch each other, which indicates overlapping of the two sides, the model filament is not affected by
self-blending (Sect.~\ref{skybenchB}), hence the orthogonal profiles of each loop must follow the true model profile, unaltered by
the blending that would complicate the measurements of the observed filaments. The central loops of the filament are blended,
however, with the dense background cloud (Fig.~\ref{filamentsB4}), which makes the separated background of the filament less
accurate, underestimated in the central area (Fig. 8 in Paper I). The outermost loop of the spiral filament has a more accurate
background and is filament-free along the right normals. Therefore, the right-sided measurements of the filament along the
outermost loop may be expected to produce more accurate results than those over the inner parts of the spiral filament that have a
contribution from the strongly fluctuating background.

Figure \ref{filB4profs} presents the filament radial profiles $D_{\{{\alpha|\beta}\}}(r)$ along the skeleton normals,
median-averaged over the filament length. The standard deviations $\varsigma_{\{{\alpha|\beta}\}\pm}(r)$ about the median profiles
are computed separately for the positive and negative differences. The true filament profiles correspond to the model surface
densities (Eq. (2) in Paper I) and display practically no differences between the filament sides. The slopes
$\gamma_{\{\alpha|\beta\}}(r)$ accurately represent the true model values, increasing from $\gamma(r)\approx 1$ at the half-maximum
radius of $0.05$\,pc to $\gamma(r)\approx 3$ (at much larger distances $r\ga 0.3$\,pc). For the filament extracted with
\textsl{getsf} in B$_4$, the radial profiles obtained from the entire filament are less accurate, with significantly larger
dispersions (Fig.~\ref{filB4profs}). This is caused by the underestimated fluctuating background in the central area (Fig. 8 in
Paper I) that in effect makes a substantial contribution to the background-subtracted filament $\mathcal{F}_{{\lambdabar}{Y}}$. For
a comparison, the profiles $D_{\{{\alpha|\beta}\}}(r)$ obtained over only the outer filament loop, where its background is more
accurate (Fig.~\ref{filB4profs}), much better reproduce the true model surface density distribution, with much smaller dispersions
of their values along the segment.

From the entire filament component $\mathcal{F}_{{\lambdabar}{Y}}$, \textsl{getsf} integrated the one-sided masses
$M_{\{{\alpha|\beta}\}} = \{3.48|3.78\}\times 10^{3}$\,$M_{\sun}$, overestimated by $\{14|24\}${\%} with respect to the true model
value (Sect.~\ref{skybenchB}). The one-sided linear densities $\Lambda_{\{{\alpha|\beta}\}} = \{343|306\}$\,$M_{\sun}$\,pc$^{-1}$
(from Eq.~(49) of Paper I) are overestimated by $\{18|6\}${\%} with respect to the true model value (Sect.~\ref{skybenchB}). The
discrepancies are caused by the residual contribution of the incompletely subtracted background, underestimated by up to
${\sim\,}50${\%} in the center of the filamentary cloud (Fig. 8 in Paper I). The filament footprint covers the entire cloud (Fig.
\ref{filamentsB4}), which makes it especially difficult to separate the filament from its blended background.

Benchmark B$_4$ provides a good test for filament extraction methods. The source extraction in B$_4$ with \textsl{getold},
described in Sect.~\ref{benchAB}, executed also \textsl{getfilaments} (Paper II), an integral part of \textsl{getsources}. Although
the method passed simpler filament extraction tests (Sect. 3 in Paper II), \textsl{getfilaments} was unable to properly reconstruct
the filament in B$_4$. The crest values of the filament were underestimated by a factor of ${\sim\,}5$ in the central area of the
dense background cloud, whereas the values were either correct or overestimated within ${\sim\,}40${\%} in some segments of the
outermost loop of the filament. Even though the filament one-sided widths were determined fairly accurately (within $10{-}20${\%}),
the distant fainter areas of the filament profile (beyond a radius of $0.1$\,pc) were completely missing. Therefore, the mass and
linear density of the filament were also strongly underestimated (by factors of ${\sim\,}3$).

The dense spiral filament in B$_4$ represents just the simplest benchmark. The filament crest must not create problems for any
skeletonization algorithm; its detection is not the main goal of this benchmark. The simulated filament was created primarily to
test the accuracy of various methods in measuring the filament profile and physical properties. Observed filaments display have
various masses, densities, lengths, widths, curvatures, and signal-to-noise ratios. The filaments imaged with \emph{Herschel} are
embedded in strongly fluctuating backgrounds and arranged in complex networks with hundreds of interconnected segments. A proper
benchmarking would require the simulated images that resemble the observations, as well as a quality evaluation system, similar to
that used in this paper for testing the source-extraction methods. Realistic and rigorous benchmarking of filament-extraction
methods are the subject of a future work.


\section{Discussion}
\label{discussion}

Astronomical images are known to be very dissimilar across the electromagnetic spectrum (e.g., Figs. 16\,--\,23 in Paper I).
Therefore, the source- or filament-extraction methods, developed for different research areas and types of observed images, have
heterogeneous properties and qualities. Benchmarking of the extraction tools must also depend on the research project, and the
simulated images must resemble the complexity and structural components of the typical observed images. For example, if the sources
of interest are all unresolved and there is no strong fluctuating background in the observed images, then the benchmark images must
also contain just the unresolved sources (with a similar spatial distribution) and faint background. In this simple case, it may
well be that a simple source-extraction tool employing a PSF-fitting algorithm could give more accurate results than a more general
method designed to work for both unresolved and resolved sources on complex filamentary backgrounds.

The benchmarks described and applied in this study were designed to resemble the mid- to far-infrared (submm) imaging observations
obtained with \emph{Herschel} for the nearby star-forming regions. The simulated images contain a bright, fluctuating filamentary
background cloud and starless and protostellar cores with a wide range of sizes, from unresolved to strongly resolved. By
construction, these benchmark images are most suitable for testing the source- and filament-extraction methods to be applied in the
studies of star formation. If observed images are significantly different, the benchmarks explored in this paper may not be
directly applicable for testing extraction methods. For example, the substantial differences between the ALMA interferometric
images of distant star-forming regions (e.g., Fig. 23 in Paper I) and the \emph{Herschel} images of the nearby star-forming clouds
required creation of dedicated benchmarks with unresolved sources and background from MHD simulations (Pouteau et al., in prep.).
To make the benchmark images better resemble the real interferometric observations, they were also processed with the ALMA
observations simulator.

For testing the source-extraction methods, it is the model sources that are the most important (primary) component of the
benchmarks, and it must resemble the sources in real observations as closely as possible. Similarly, for testing the
filament-extraction methods, it is the model filaments that are the main component, with all their parameters tabulated in a truth
catalog. The other components of the benchmark images (e.g., fluctuating background, instrumental noise) just complicate the
extraction of the primary component. They may be scaled up or down to create variants of the same benchmark with diverse
contributions of the secondary components. For example, this paper employed several benchmark variants (\{A|B\}$_{2}$,
\{A|B\}$_{3}$, and B$_{4}$) of different complexity, expanding the applicability of the two benchmarks to other types of images of
the nearby star-forming regions.

Benchmarking source- or filament-extraction methods, it is important to make ensure that the simulated images contain realistic
enough models of the sources or filaments that are expected to be extracted in the real-life observations. This may be a potential
problem, because that requires an advance knowledge of the physical reality being observed. In practice, there usually exists a
good deal of previous studies that would allow the creation of the suitable primary and secondary components of the benchmarks.
However, if an application of the extraction tools to the observed images shows the component properties that are significantly
different from the ones simulated for the benchmarks, the latter may need to be adjusted and the testing of the methods to be
repeated.


\section{Conclusions}
\label{conclusions}

This paper described detailed benchmarking of two multiwavelength source and filament extraction methods, \textsl{getsf} and
\textsl{getold}, to quantitatively evaluate their performance in Benchmarks A and B. In total, the two methods of source extraction
were tested and compared using five variants of the simulated multiwavelength images of different complexity. Although the
benchmarks were designed to resemble the \emph{Herschel} observations of star-forming regions, the images are suitable for
evaluating extraction methods for various astronomical projects and applications.

Benchmark B includes the complex fluctuating background cloud, the long dense filament, and the multitude of sources (starless and
protostellar cores) with wide ranges of sizes, masses, and intensity profiles, computed with a radiative transfer code. In
Benchmark A with similar properties of the structural components (no filaments), the sources are allowed to arbitrarily overlap
with each other. The benchmarks enable conclusive comparisons between different methods and allow a quantitative comparison of
their qualities, using the formalism given in this paper, in terms of the extraction completeness, reliability, and goodness, as
well as the detection and measurement accuracies and the overall quality. All benchmark images, the truth catalogs containing the
model parameters, and the reference extraction catalogs produced by the author are available for download on the \textsl{getsf}
website\footnote{\url{http://irfu.cea.fr/Pisp/alexander.menshchikov/\#bench}}.


The quantitative analysis of the benchmark source extractions showed that the \textsl{getsf} method has superior qualities in
comparison with \textsl{getold}. The benchmark filament extraction with \textsl{getsf} recovered parameters of the model filament,
in contrast to the extraction with \textsl{getold} that was unable to properly reconstruct the filament to an acceptable accuracy.
An investigation of the dependence of the source extraction results on different sets of images used to detect sources suggested
that the best choice for source detection with \textsl{getsf} is the high-resolution surface density, either alone or together with
other \emph{Herschel} images. The worst choice for source detection would be the lowest-resolution observed images.

The benchmarks explored in this paper are proposed as the standard benchmarks for calibrating existing and future source and
filament extraction methods before any astrophysical applications of the methods. It is critically important to use only the best
calibrated tools with known properties that are fully understood on the basis of the standard benchmarking. Applications of various
uncalibrated extraction tools with unknown qualities that have never been quantitatively compared, could lead to a proliferation
of incompatible results and severe long-term problems in understanding of the astrophysical reality.


\begin{acknowledgements} 
This study used the \textsl{cfitsio} library \citep{Pence1999}, developed at HEASARC NASA (USA), \textsl{saoimage ds9}
\citep{JoyeMandel2003} and \textsl{wcstools} \citep{Mink2002}, developed at the Smithsonian Astrophysical Observatory (USA), and
the \textsl{stilts} software \citep{Taylor2006}, developed at Bristol University (UK). The \textsl{plot} utility and \textsl{ps12d}
library, used in this work to draw figures directly in the PostScript language, were written by the author using the
\textsl{psplot} library (by Kevin E. Kohler), developed at Nova Southeastern University Oceanographic Center (USA), and the
plotting subroutines from the MHD code \textsl{azeus} \citep{Ramsey2012}, developed by David Clarke and the author at Saint Mary's
University (Canada). HGBS and HOBYS are the \emph{Herschel} Key Projects jointly carried out by SPIRE Specialist Astronomy Group 3
(SAG3), scientists of several institutes in the PACS Consortium (e.g., CEA Saclay, INAF-IAPS Rome, LAM/OAMP Marseille), and
scientists of the \emph{Herschel} Science Center (HSC).
\end{acknowledgements} 


\bibliographystyle{aa}
\bibliography{aamnem99,bench}


\begin{appendix}

\section{Fluctuating backgrounds and the measurement accuracy for faint sources}
\label{backgrounds}

Exact shapes of the molecular clouds under faint sources are practically impossible to separate from the observed emission peaks
with any acceptable accuracy. The observed backgrounds of sources fluctuate on all spatial scales. Instrumental noise further
complicates the source backgrounds by adding random fluctuations on scales of the angular resolution $O_{\lambda}$. The background
and noise fluctuations are totally blended with the sources and no source extraction method is able to precisely deblend the
components. This makes the measured sizes and fluxes of faint sources uncertain, often significantly over- or underestimated,
depending on the unknown shapes of the fluctuations within the source footprints. Naturally, the background inaccuracies become
relatively less important for increasingly stronger sources.

\begin{figure}
\centering
\centerline{
  \resizebox{0.905\hsize}{!}{\includegraphics{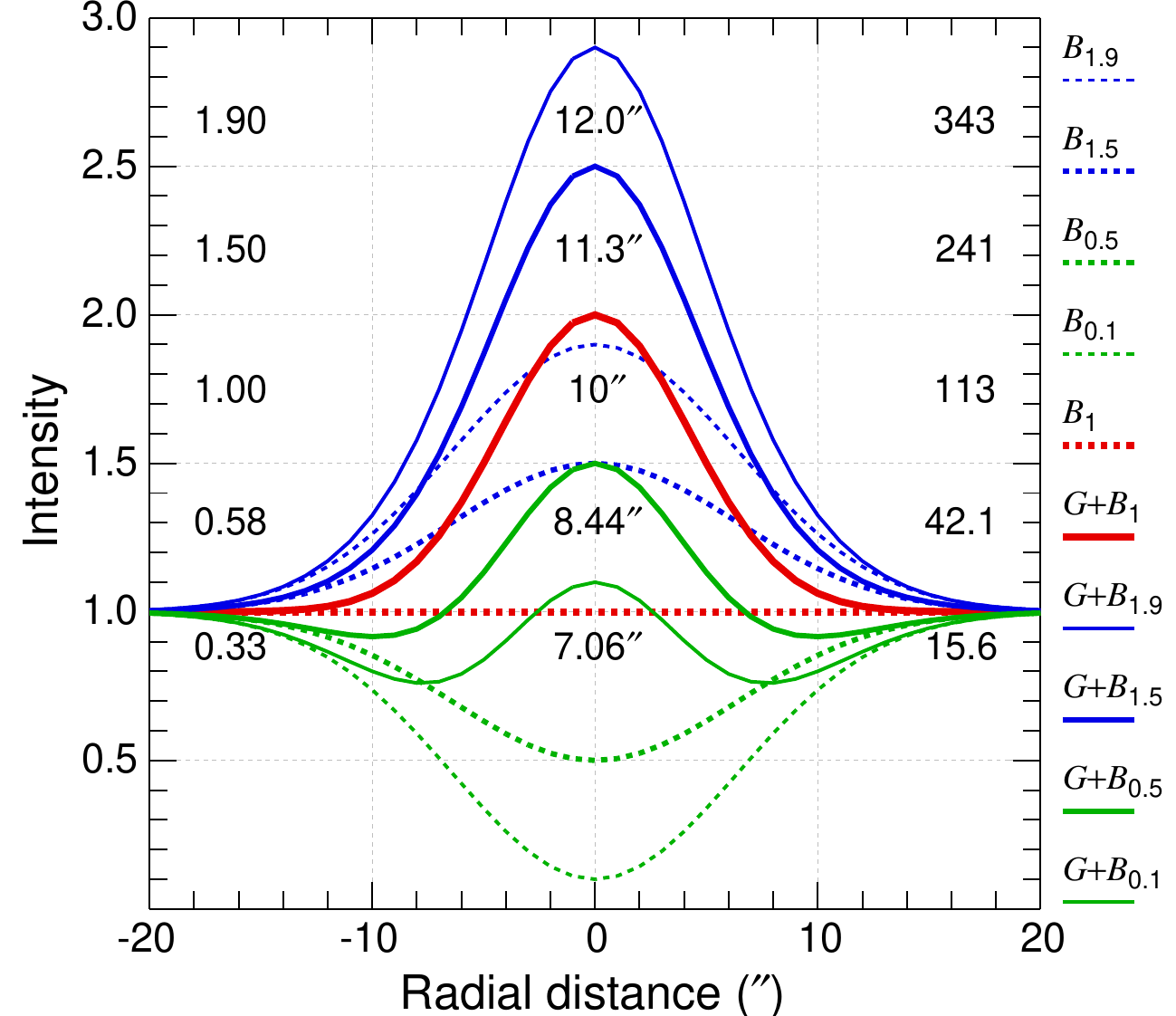}}}
\caption
{ 
Fluctuating backgrounds and measurements of faint sources. The model Gaussian source $\mathcal{G}$ with a size of $10${\arcsec}
(FWHM) and a peak intensity of $1$ corresponds to the flat background $\mathcal{B}_{1} = 1$ (red lines). The same source
$\mathcal{G}$ is also added to the nonuniform (hill- and hollow-like) backgrounds $\mathcal{B}_{1.9}$, $\mathcal{B}_{1.5}$,
$\mathcal{B}_{0.5}$, and $\mathcal{B}_{0.1}$ (blue and green lines). The fluctuating backgrounds were obtained from the flat
background $\mathcal{B}_{1}$ by adding or subtracting the Gaussians with a size of $15${\arcsec} (FWHM) and the peak values of
$0.5$ and $0.9$. Extraction methods would not be able to recognize that the real backgrounds are hill- or hollow-like, hence they
would instead subtract the flat backgrounds, based on the intensities just outside the apparent source footprints. Therefore, the
source $\mathcal{G}$ would be extracted with over- or underestimated FWHM sizes $A$, peak intensities $F_{{\rm P}}$, and total
fluxes $F_{{\rm T}}$ (the middle, left, and right columns of numbers, respectively).
} 
\label{faintback}
\end{figure}

Figure \ref{faintback} illustrates the problem using a simple Gaussian source $\mathcal{G}$ of a FWHM size of $10${\arcsec} and
several differently shaped backgrounds (flat and hill- or hollow-like). To simplify the matters, the source may be considered as
unresolved, although the extended sources are also affected by the same problem. In the simplest (unrealistic) case, the source
could be observed against constant background ($\mathcal{B}_{1} = 1$). The fluctuating backgrounds were modeled by adding the
positive or negative $15${\arcsec} (FWHM) Gaussians with peak values of $0.5$ and $0.9$ to the flat background. For simplicity, the
source position is assumed to be aligned with the background extrema, which is sufficient to illustrate the roots of the problem.

The flat background would normally present no difficulties for accurate source measurements. However, the strongly fluctuating
backgrounds pose severe problems for source extraction methods. Measurements of the faint sources could be quite different from the
true values, depending on the sign and magnitude of the background fluctuation within their footprints (Fig. \ref{faintback}). When
the source is blended with a hill-like background, its shape remains very similar to a Gaussian source and contains no information
that the background is not flat. The source footprint usually widens and the hill-like background contributes to the overestimated
width and fluxes of the source. On the other hand, when the source is blended with a hollow-like background, its apparent footprint
shrinks to the area limited by the intensity minimum that appears around the peak. As a consequence, the sizes and fluxes of such
sources become underestimated, sometimes quite strongly (Fig. \ref{faintback}).

It is clear that the backgrounds of sources in the benchmark simulations and real observations are much more complex than the above
simple model. However, the model illustrates the fundamental reasons behind the increasingly larger inaccuracies for the sources
with low S/N ratios in Benchmarks A and B (Figs. \ref{accuracyA2}\,--\,\ref{accuracyB4}). In general, measurement accuracy for such
sources is impossible to improve, because the necessary information is practically lost, when the source peak is blended with the
background and noise fluctuations. Fortunately, the unresolved or slightly resolved sources are the exception, for which it is
possible to (approximately) correct the underestimated sizes and fluxes.

\section{Corrections for the measurements of unresolved or slightly resolved sources}
\label{corrections}

The PSFs (the telescope beams) set a natural lower limit to the source sizes $\{A,B\}_{{\lambda}{n}}$, their values must be larger
or at least equal to the angular resolution $O_{\lambda}$. However, the benchmarking discussed in this paper has revealed numerous
examples of sources with sizes $\{A,B\}_{{\lambda}{n}} < O_{\lambda}$ and underestimated peak intensities $F_{{\rm P}{\lambda}{n}}$
and integrated fluxes $F_{{\rm T}{\lambda}{n}}$. An analysis of the results showed that the underestimated parameters are related
to the overestimated backgrounds of the faint sources. When the FWHM sizes are directly measured at half-maximum intensities, like
in \textsl{getsf} (cf. Sect. 3.4.6 of Paper I), the measurements can be improved, as shown below. However, such corrections are not
feasible for \textsl{getold}, because the sizes obtained with intensity moments often correspond to uncertain levels, significantly
deviating from the half-maximum intensity.

\begin{figure}
\centering
\centerline{
  \resizebox{0.935\hsize}{!}{\includegraphics{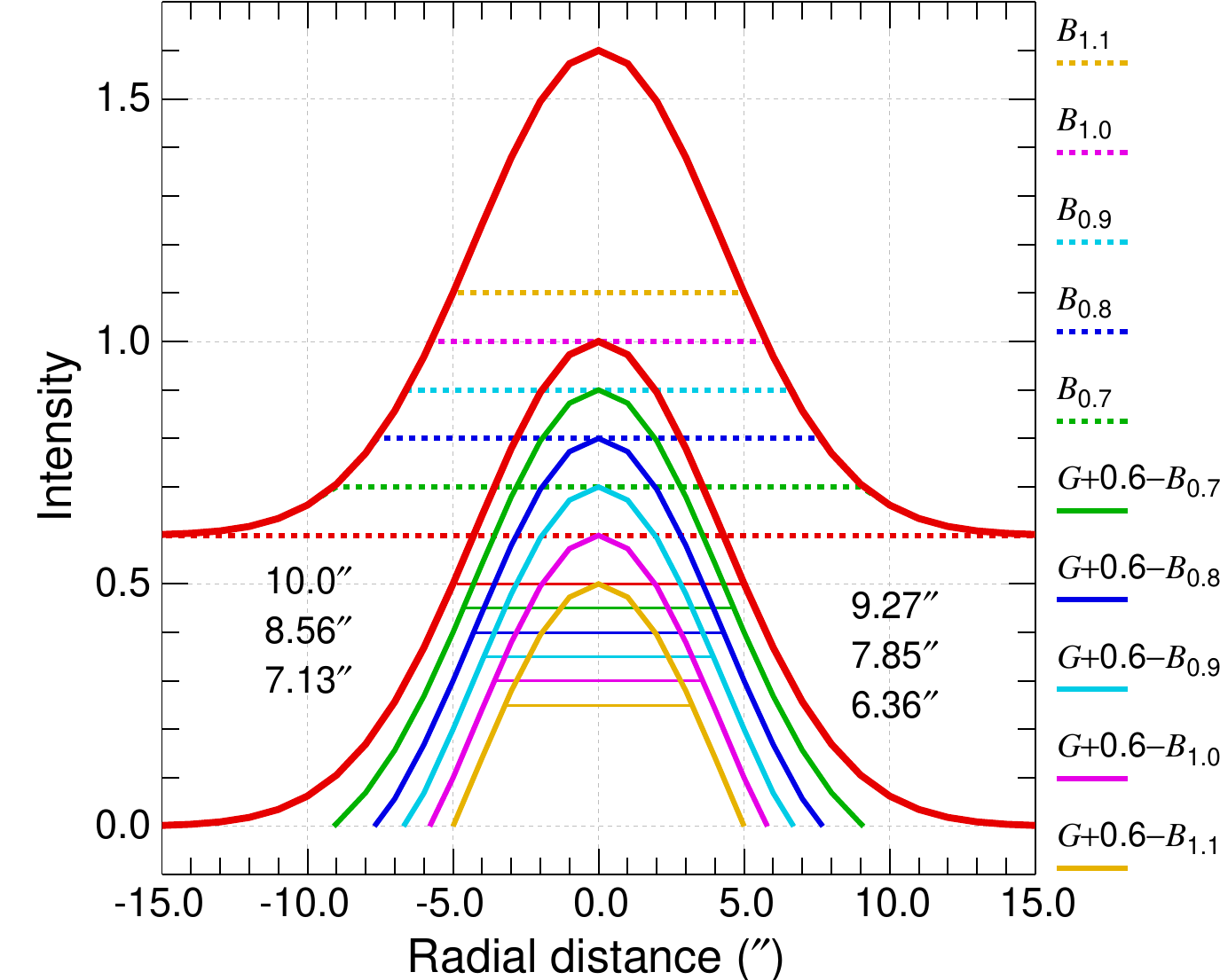}}}
\caption
{ 
Approximate corrections for unresolved or slightly resolved sources with overestimated backgrounds. The model Gaussian source
$\mathcal{G}$ with a size of $10${\arcsec} (FWHM) and a peak intensity of $1$ is superposed on a flat background with a constant
intensity of $\mathcal{B}_{0.9} = 0.6$ (upper red lines). The increasingly overestimated backgrounds $\mathcal{B}_{0.7}$,
$\mathcal{B}_{0.8}$, $\mathcal{B}_{0.9}$, $\mathcal{B}_{1.0}$, and $\mathcal{B}_{1.1}$ are shown with the dotted lines of different
color. When such backgrounds are subtracted (lower colored curves), the source FWHM sizes $A$, peak intensities $F_{{\rm P}}$, and
total fluxes $F_{{\rm T}}$ become increasingly underestimated. The measured $A$ values (given in the plot) get progressively
smaller than the angular resolution of $10${\arcsec}, which clearly indicates an increasing inaccuracy of the background. Requiring
that a source cannot be narrower than the telescope beam, it is possible to improve the measurements, substantially reducing their
errors.
} 
\label{correct}
\end{figure}

Figure \ref{correct} illustrates the Gaussian model, adopted by \textsl{getsf} to correct the underestimated sizes and fluxes of
faint sources, when their measured sizes are smaller than the beam size. The model assumes that the unresolved or slightly resolved
faint sources have Gaussian shapes, which is an appropriate assumption, because most telescopes have Gaussian beams in their
central (upper) parts. Various deviations and artifacts that often appear in the PSFs at larger angular distances from their peak
are invisible for the faint sources. The model also supposes that the actual source background is flat, which is the only
reasonable assumption that could be made. Although there are many possible shapes of the fluctuating background within a source
footprint, they cannot be accurately recovered from the blended source intensity distribution.

With the above two assumptions, Fig. \ref{correct} demonstrates how the measured properties of a Gaussian source $\mathcal{G}$
would be affected by the increasingly overestimated backgrounds $\mathcal{B}_{0.7}$, $\mathcal{B}_{0.8}$, $\mathcal{B}_{0.9}$,
$\mathcal{B}_{1.0}$, and $\mathcal{B}_{1.1}$. In the simplest case of an intrinsically flat background, the background could be
progressively overestimated for stronger instrumental noise, which would effectively represent a fluctuating background of the
source. For the blended sources or those in crowded areas, absence of the source-free pixels in their immediate environments is
often the reason for the background to be overestimated. Whatever the actual cause, an over-subtraction of the increasingly
inaccurate background leads to the progressively underestimated FWHM sizes, peak intensities, and total fluxes. For the Gaussian
$\mathcal{G}$ in Fig. \ref{correct}, it is possible to determine the correction factors that would recover the true properties of
the source from their underestimated values.

The multiplicative correction factors $f_{{\rm S}{\lambda}{n}}$, $f_{{\rm P}{\lambda}{n}}$, and $f_{{\rm T}{\lambda}{n}}$ for the
sizes $\{A,B\}_{{\lambda}{n}}$, peak intensity $F_{{\rm P}{\lambda}{n}}$, and total flux $F_{{\rm T}{\lambda}{n}}$, respectively,
are obtained empirically by approximating the results for the Gaussian model (Fig. \ref{correct}) using different overestimated
backgrounds,
\begin{eqnarray} 
\left.\begin{aligned}
f_{{\rm S}{\lambda}{n}} &=\max\left(O_{\lambda}\left(A_{{\lambda}{n}}B_{{\lambda}{n}}\right)^{-1/2},1\right), \\
f_{{\rm P}{\lambda}{n}} &=f^{1.75}_{{\rm S}{\lambda}{n}\!}+0.35\left(1-f_{{\rm S}{\lambda}{n}}\right), \\
f_{{\rm T}{\lambda}{n}} &=1.15f^{3.87}_{{\rm S}{\lambda}{n}}-0.0813\left(f_{{\rm S}{\lambda}{n}}-0.04\right)^{-15},
\end{aligned}\right.
\label{correcfacs}
\end{eqnarray} 
where the factors differ from unity only when $(A_{{\lambda}{n}}B_{{\lambda}{n}})^{1/2} < O_{\lambda}$. They are applied when
creating the final catalog at the end of the measurement iterations. The factors are implemented in \textsl{getsf}, hence the
benchmark extraction catalogs discussed in this paper contain improved measurements for the faint unresolved or slightly resolved
sources.

By their definition, the factors from Eq.~(\ref{correcfacs}) provide precise results for only Gaussian sources on flat backgrounds.
In most cases, however, the real backgrounds of sources have more complex shapes, in which case the formulas from
Eq.~(\ref{correcfacs}) provide less accurate corrections to the measured quantities. Despite being approximate, the corrections are
nevertheless very useful, because they significantly improve the measurements. For example, the hollow-like backgrounds
$\mathcal{B}_{0.5}$ and $\mathcal{B}_{0.1}$ of a Gaussian source $\mathcal{G}$ from Fig. \ref{faintback} lead to substantially
overestimated derived backgrounds. The corrections $f_{{\rm S}{\lambda}{n}}$, $f_{{\rm P}{\lambda}{n}}$, and $f_{{\rm
T}{\lambda}{n}}$, obtained for the $\mathcal{G} + \mathcal{B}_{0.5}$ model improve the measurements by the factors of $1.18$,
$1.28$, and $2.21$, whereas the improvement factors for the $\mathcal{G} + \mathcal{B}_{0.1}$ model are $1.42$, $1.69$, and
$4.42$, respectively.



\section{Tabulated qualities of source extractions in Benchmarks A and B}
\label{tabresults}

Tables \ref{qualAB23} and \ref{qualB4} collect all qualities of the source extractions with \textsl{getsf} and \textsl{getold} in
the benchmark variants $\{\rm{A},\rm{B}\}_2$, $\{\rm{A},\rm{B}\}_3$, and B$_4$, discussed in Sects.
\ref{visuals}\,--\,\ref{qualities}. Tables \ref{qualA3sets}, \ref{qualB3sets}, and \ref{qualB4sets} present the qualities of source
extractions with \textsl{getsf} in the benchmarks $\{\rm{A},\rm{B}\}_3$ and B$_4$ for different subsets of images used to detect
sources, discussed in Sect. \ref{dependonimg}.

\begin{table*}  
\caption
{ 
Benchmarks $\{\rm{A},\rm{B}\}_2$ and $\{\rm{A},\rm{B}\}_3$ with \textsl{getsf} and \textsl{getold} in the standard approach, with
all seven images combined for source detection. The extraction qualities, defined in
Eqs.~(\ref{qualities0})\,--\,(\ref{finalqualities}), are evaluated for only acceptable sources, cf. Eq.~(\ref{acceptable}), with
errors in measurements within a factor of $2^{1/2}$. The numbers of model sources are $N_{\rm T} = 459$ in Benchmark A and $N_{\rm
T} = 919$ in Benchmark B. Source measurements in the image of derived surface densities are known to be inaccurate (e.g.,
Appendix~A of Paper I), hence the data are not presented.
} 
\begin{tabular}{rrrrrrlllllllllll}
\hline
\noalign{\smallskip}
\!\!\!Bench A$_2$&$\lambda$\,\,&\!\!\!$N_{{\rm D}{{\lambda}}}$&\!\!\!\!$N_{{\rm G}{{\lambda}}}$&\!\!\!$N_{{\rm B}{{\lambda}}}$&
\!$N_{\rm S}$&\,\,\,$C_{\lambda}$&\,\,\,$R_{\lambda}$&\,\,$G_{\lambda}$&$Q_{{\rm P}{\lambda}}$&$Q_{{\rm T}{\lambda}}$&
$Q_{{\rm E}{\lambda}}$&$Q_{{\rm D}{\lambda}}$&$Q_{{\rm CR}{\lambda}}$&$Q_{{\rm PTE}{\lambda}}$&\,\,$Q_{{\lambda}}$\\
\noalign{\smallskip}
\hline
\noalign{\smallskip}
\textsl{getsf}&
                  75 & 107 & 107 &  0 & 0 & 0.233 & 1.000 & 0.233 & 0.997 & 0.992 & 0.984 & 0.906 & 0.233 & 0.973 & 0.0479 \\
\textsl{getsf}&
                 110 & 107 & 107 &  0 & 3 & 0.233 & 0.717 & 0.233 & 0.997 & 0.990 & 0.993 & 0.906 & 0.167 & 0.981 & 0.0346 \\
\textsl{getsf}&
                 170 & 238 & 209 & 29 & 0 & 0.519 & 1.000 & 0.455 & 0.903 & 0.893 & 0.925 & 0.574 & 0.519 & 0.746 & 0.101  \\
\textsl{getsf}&
                 250 & 341 & 313 & 28 & 1 & 0.743 & 0.995 & 0.682 & 0.948 & 0.895 & 0.950 & 0.694 & 0.739 & 0.807 & 0.282  \\
\textsl{getsf}&
                 350 & 337 & 310 & 27 & 1 & 0.734 & 0.995 & 0.675 & 0.936 & 0.867 & 0.940 & 0.697 & 0.730 & 0.763 & 0.262  \\
\textsl{getsf}&
                 500 & 326 & 288 & 38 & 1 & 0.710 & 0.994 & 0.627 & 0.944 & 0.871 & 0.937 & 0.702 & 0.706 & 0.770 & 0.239  \\
\noalign{\smallskip}
\hline
\noalign{\smallskip}
\textsl{getold}&
                   75 & 107 & 107 &  0 & 2 & 0.233 & 0.839 & 0.233 & 0.997 & 0.992 & 0.998 & 1.000 & 0.196 & 0.987 & 0.0450 \\
\textsl{getold}&
                  110 & 107 & 107 &  0 & 1 & 0.233 & 0.951 & 0.233 & 0.996 & 0.983 & 0.996 & 1.000 & 0.222 & 0.976 & 0.0505 \\
\textsl{getold}&
                  170 & 242 & 197 & 45 & 1 & 0.527 & 0.990 & 0.429 & 0.913 & 0.912 & 0.736 & 0.708 & 0.522 & 0.613 & 0.0971 \\
\textsl{getold}&
                  250 & 336 & 313 & 23 & 1 & 0.732 & 0.995 & 0.682 & 0.948 & 0.897 & 0.717 & 0.673 & 0.728 & 0.609 & 0.204  \\
\textsl{getold}&
                  350 & 332 & 309 & 23 & 2 & 0.723 & 0.979 & 0.673 & 0.934 & 0.855 & 0.738 & 0.645 & 0.708 & 0.589 & 0.181  \\
\textsl{getold}&
                  500 & 329 & 301 & 28 & 2 & 0.717 & 0.979 & 0.656 & 0.940 & 0.859 & 0.776 & 0.643 & 0.701 & 0.627 & 0.185  \\
\noalign{\smallskip}
\hline
\noalign{\smallskip}
\!\!\!Bench B$_2$&$\lambda$\,\,&\!\!\!$N_{{\rm D}{{\lambda}}}$&\!\!\!\!$N_{{\rm G}{{\lambda}}}$&\!\!\!$N_{{\rm B}{{\lambda}}}$&
\!$N_{\rm S}$&\,\,\,$C_{\lambda}$&\,\,\,$R_{\lambda}$&\,\,$G_{\lambda}$&$Q_{{\rm P}{\lambda}}$&$Q_{{\rm T}{\lambda}}$&
$Q_{{\rm E}{\lambda}}$&$Q_{{\rm D}{\lambda}}$&$Q_{{\rm CR}{\lambda}}$&$Q_{{\rm PTE}{\lambda}}$&\,\,$Q_{{\lambda}}$\\
\noalign{\smallskip}
\hline
\noalign{\smallskip}
\textsl{getsf}&
                  70 &  91 &  91 &  0 & 2 & 0.099 & 0.796 & 0.099 & 1.000 & 0.998 & 1.000 & 1.000 & 0.079 & 0.997 & 0.0078 \\
\textsl{getsf}&
                 100 &  91 &  91 &  0 & 0 & 0.099 & 1.000 & 0.099 & 1.000 & 0.994 & 0.997 & 1.000 & 0.099 & 0.990 & 0.0097 \\
\textsl{getsf}&
                 160 & 422 & 342 & 80 & 0 & 0.459 & 1.000 & 0.372 & 0.865 & 0.873 & 0.894 & 0.770 & 0.459 & 0.676 & 0.0890 \\
\textsl{getsf}&
                 250 & 796 & 768 & 28 & 0 & 0.866 & 1.000 & 0.836 & 0.951 & 0.906 & 0.896 & 0.774 & 0.866 & 0.772 & 0.433  \\
\textsl{getsf}&
                 350 & 827 & 822 &  5 & 0 & 0.900 & 1.000 & 0.894 & 0.945 & 0.874 & 0.962 & 0.742 & 0.900 & 0.794 & 0.474  \\
\textsl{getsf}&
                 500 & 789 & 782 &  7 & 0 & 0.859 & 1.000 & 0.851 & 0.938 & 0.836 & 0.946 & 0.733 & 0.859 & 0.742 & 0.397  \\
\noalign{\smallskip}
\hline
\noalign{\smallskip}
\textsl{getold}&
                   70 &  91 &  91 &  0 & 2 & 0.099 & 0.796 & 0.099 & 0.999 & 0.992 & 1.000 & 1.000 & 0.079 & 0.991 & 0.0077 \\
\textsl{getold}&
                  100 &  91 &  91 &  0 & 0 & 0.099 & 1.000 & 0.099 & 0.996 & 0.975 & 1.000 & 1.000 & 0.099 & 0.971 & 0.0095 \\
\textsl{getold}&
                  160 & 421 & 309 &112 & 0 & 0.458 & 1.000 & 0.336 & 0.884 & 0.889 & 0.688 & 0.708 & 0.458 & 0.541 & 0.0590 \\
\textsl{getold}&
                  250 & 745 & 737 &  8 & 1 & 0.811 & 0.999 & 0.802 & 0.951 & 0.905 & 0.674 & 0.569 & 0.810 & 0.579 & 0.214  \\
\textsl{getold}&
                  350 & 762 & 759 &  3 & 3 & 0.829 & 0.991 & 0.826 & 0.947 & 0.851 & 0.711 & 0.572 & 0.822 & 0.574 & 0.223  \\
\textsl{getold}&
                  500 & 549 & 525 & 24 &15 & 0.597 & 0.726 & 0.571 & 0.935 & 0.799 & 0.700 & 0.598 & 0.434 & 0.523 & 0.0776 \\
\noalign{\smallskip}
\hline
\noalign{\smallskip}
\!\!\!Bench A$_3$&$\lambda$\,\,&\!\!\!$N_{{\rm D}{{\lambda}}}$&\!\!\!\!$N_{{\rm G}{{\lambda}}}$&\!\!\!$N_{{\rm B}{{\lambda}}}$&
\!$N_{\rm S}$&\,\,\,$C_{\lambda}$&\,\,\,$R_{\lambda}$&\,\,$G_{\lambda}$&$Q_{{\rm P}{\lambda}}$&$Q_{{\rm T}{\lambda}}$&
$Q_{{\rm E}{\lambda}}$&$Q_{{\rm D}{\lambda}}$&$Q_{{\rm CR}{\lambda}}$&$Q_{{\rm PTE}{\lambda}}$&\,\,$Q_{{\lambda}}$\\
\noalign{\smallskip}
\hline
\noalign{\smallskip}
\textsl{getsf}&
                  75 & 107 & 107 &  0 & 0 & 0.233 & 1.000 & 0.233 & 0.998 & 0.992 & 0.988 & 0.921 & 0.233 & 0.978 & 0.0490 \\
\textsl{getsf}&
                 110 & 107 & 107 &  0 & 3 & 0.233 & 0.717 & 0.233 & 0.998 & 0.991 & 0.995 & 0.921 & 0.167 & 0.984 & 0.0353 \\
\textsl{getsf}&
                 170 & 213 & 186 & 27 & 9 & 0.464 & 0.564 & 0.405 & 0.882 & 0.880 & 0.931 & 0.733 & 0.262 & 0.723 & 0.0562 \\
\textsl{getsf}&
                 250 & 320 & 287 & 33 & 1 & 0.697 & 0.994 & 0.625 & 0.944 & 0.895 & 0.946 & 0.697 & 0.693 & 0.799 & 0.241  \\
\textsl{getsf}&
                 350 & 319 & 294 & 25 & 0 & 0.695 & 1.000 & 0.641 & 0.938 & 0.868 & 0.941 & 0.686 & 0.695 & 0.766 & 0.234  \\
\textsl{getsf}&
                 500 & 303 & 273 & 30 & 0 & 0.660 & 1.000 & 0.595 & 0.933 & 0.870 & 0.942 & 0.704 & 0.660 & 0.765 & 0.211  \\
\noalign{\smallskip}
\hline
\noalign{\smallskip}
\textsl{getold}&
                   75 & 107 & 107 &  0 & 0 & 0.233 & 1.000 & 0.233 & 0.997 & 0.992 & 0.998 & 1.000 & 0.233 & 0.987 & 0.0537 \\
\textsl{getold}&
                  110 & 107 & 107 &  0 & 1 & 0.233 & 0.951 & 0.233 & 0.996 & 0.981 & 0.996 & 1.000 & 0.222 & 0.974 & 0.0504 \\
\textsl{getold}&
                  170 & 209 & 171 & 38 & 1 & 0.455 & 0.987 & 0.373 & 0.894 & 0.902 & 0.761 & 0.756 & 0.449 & 0.614 & 0.0776 \\
\textsl{getold}&
                  250 & 301 & 274 & 27 & 1 & 0.656 & 0.993 & 0.597 & 0.942 & 0.887 & 0.712 & 0.651 & 0.651 & 0.595 & 0.151  \\
\textsl{getold}&
                  350 & 300 & 269 & 31 & 1 & 0.654 & 0.993 & 0.586 & 0.940 & 0.849 & 0.735 & 0.673 & 0.649 & 0.587 & 0.150  \\
\textsl{getold}&
                  500 & 305 & 270 & 35 & 1 & 0.664 & 0.994 & 0.588 & 0.929 & 0.830 & 0.776 & 0.557 & 0.660 & 0.599 & 0.129  \\
\noalign{\smallskip}
\hline
\noalign{\smallskip}
\!\!\!Bench B$_3$&$\lambda$\,\,&\!\!\!$N_{{\rm D}{{\lambda}}}$&\!\!\!\!$N_{{\rm G}{{\lambda}}}$&\!\!\!$N_{{\rm B}{{\lambda}}}$&
\!$N_{\rm S}$&\,\,\,$C_{\lambda}$&\,\,\,$R_{\lambda}$&\,\,$G_{\lambda}$&$Q_{{\rm P}{\lambda}}$&$Q_{{\rm T}{\lambda}}$&
$Q_{{\rm E}{\lambda}}$&$Q_{{\rm D}{\lambda}}$&$Q_{{\rm CR}{\lambda}}$&$Q_{{\rm PTE}{\lambda}}$&\,\,$Q_{{\lambda}}$\\
\noalign{\smallskip}
\hline
\noalign{\smallskip}
\textsl{getsf}&
                  70 &  91 &  91 &  0 & 2 & 0.099 & 0.796 & 0.099 & 1.000 & 0.999 & 1.000 & 1.000 & 0.079 & 0.998 & 0.0078 \\
\textsl{getsf}&
                 100 &  91 &  91 &  0 & 5 & 0.099 & 0.465 & 0.099 & 1.000 & 0.995 & 0.997 & 1.000 & 0.046 & 0.991 & 0.0045 \\
\textsl{getsf}&
                 160 & 185 & 156 & 29 & 8 & 0.201 & 0.555 & 0.170 & 0.851 & 0.887 & 0.914 & 0.669 & 0.112 & 0.690 & 0.0088 \\
\textsl{getsf}&
                 250 & 578 & 506 & 72 & 5 & 0.629 & 0.958 & 0.551 & 0.914 & 0.877 & 0.890 & 0.578 & 0.602 & 0.713 & 0.137  \\
\textsl{getsf}&
                 350 & 629 & 570 & 59 & 0 & 0.684 & 1.000 & 0.620 & 0.931 & 0.849 & 0.934 & 0.537 & 0.684 & 0.739 & 0.168  \\
\textsl{getsf}&
                 500 & 502 & 450 & 52 & 0 & 0.546 & 1.000 & 0.490 & 0.924 & 0.813 & 0.927 & 0.555 & 0.546 & 0.696 & 0.103  \\
\noalign{\smallskip}
\hline
\noalign{\smallskip}
\textsl{getold}&
                   70 &  91 &  91 &  0 & 2 & 0.099 & 0.796 & 0.099 & 0.999 & 0.995 & 1.000 & 1.000 & 0.079 & 0.995 & 0.0078 \\
\textsl{getold}&
                  100 &  91 &  91 &  2 &35 & 0.101 & 0.076 & 0.099 & 1.000 & 0.993 & 1.000 & 1.000 & 0.008 & 0.993 & 0.0008 \\
\textsl{getold}&
                  160 & 219 & 144 & 89 &16 & 0.254 & 0.388 & 0.157 & 0.893 & 0.896 & 0.767 & 0.577 & 0.098 & 0.613 & 0.0054 \\
\textsl{getold}&
                  250 & 507 & 440 & 67 & 0 & 0.552 & 1.000 & 0.479 & 0.922 & 0.875 & 0.668 & 0.548 & 0.552 & 0.539 & 0.0780 \\
\textsl{getold}&
                  350 & 539 & 490 & 49 & 0 & 0.587 & 1.000 & 0.533 & 0.929 & 0.831 & 0.693 & 0.523 & 0.587 & 0.535 & 0.0876 \\
\textsl{getold}&
                  500 & 416 & 355 & 61 & 0 & 0.453 & 1.000 & 0.386 & 0.925 & 0.799 & 0.711 & 0.526 & 0.453 & 0.526 & 0.0483 \\
\noalign{\smallskip}
\hline
\end{tabular}
\label{qualAB23}
\end{table*}


\begin{table*}  
\caption
{ 
Benchmark B$_4$ with \textsl{getsf} and \textsl{getold} in the standard approach, with all seven images combined for source
detection. The extraction qualities, defined in Eqs.~(\ref{qualities0})\,--\,(\ref{finalqualities}), are evaluated for only
acceptable sources, cf. Eq.~(\ref{acceptable}), with errors in measurements within a factor of $2^{1/2}$. The number of model
sources is $N_{\rm T} = 919$. Source measurements in the image of derived surface densities are known to be inaccurate (e.g.,
Appendix~A of Paper I), hence the data are not presented.
} 
\begin{tabular}{rrrrrrlllllllllll}
\hline
\noalign{\smallskip}
\!\!\!Bench B$_4$&$\lambda$\,\,&\!\!\!$N_{{\rm D}{{\lambda}}}$&\!\!\!\!$N_{{\rm G}{{\lambda}}}$&\!\!\!$N_{{\rm B}{{\lambda}}}$&
\!$N_{\rm S}$&\,\,\,$C_{\lambda}$&\,\,\,$R_{\lambda}$&\,\,$G_{\lambda}$&$Q_{{\rm P}{\lambda}}$&$Q_{{\rm T}{\lambda}}$&
$Q_{{\rm E}{\lambda}}$&$Q_{{\rm D}{\lambda}}$&$Q_{{\rm CR}{\lambda}}$&$Q_{{\rm PTE}{\lambda}}$&\,\,$Q_{{\lambda}}$\\
\noalign{\smallskip}
\hline
\noalign{\smallskip}
\textsl{getsf}& 
                  70 &  91 &  91 &  0 &  4 & 0.099 & 0.549 & 0.099 & 1.000 & 0.999 & 1.000 & 1.000 & 0.054 & 0.998 & 0.0054 \\
\textsl{getsf}& 
                 100 &  91 &  91 &  0 &  1 & 0.099 & 0.935 & 0.099 & 0.999 & 0.993 & 0.997 & 1.000 & 0.093 & 0.989 & 0.0091 \\
\textsl{getsf}& 
                 160 & 102 & 100 &  2 &  3 & 0.111 & 0.700 & 0.109 & 0.939 & 0.926 & 0.947 & 0.888 & 0.078 & 0.824 & 0.0062 \\
\textsl{getsf}& 
                 250 & 193 & 173 & 20 &  7 & 0.210 & 0.623 & 0.188 & 0.951 & 0.875 & 0.911 & 0.734 & 0.131 & 0.759 & 0.0137 \\
\textsl{getsf}& 
                 350 & 180 & 151 & 29 &  1 & 0.196 & 0.982 & 0.164 & 0.932 & 0.838 & 0.936 & 0.697 & 0.192 & 0.732 & 0.0161 \\
\textsl{getsf}& 
                 500 & 113 &  91 & 22 &  0 & 0.123 & 1.000 & 0.099 & 0.909 & 0.796 & 0.913 & 0.677 & 0.123 & 0.662 & 0.0054 \\
\noalign{\smallskip}
\hline
\noalign{\smallskip}
\textsl{getold}&
                  70 &  91 &  91 &  0 &  3 & 0.099 & 0.659 & 0.099 & 0.992 & 0.958 & 1.000 & 0.950 & 0.065 & 0.951 & 0.0058 \\
\textsl{getold}&
                 100 &  91 &  91 &  0 & 13 & 0.099 & 0.198 & 0.099 & 0.991 & 0.950 & 1.000 & 0.950 & 0.020 & 0.942 & 0.0017 \\
\textsl{getold}&
                 160 & 174 & 120 & 54 & 12 & 0.189 & 0.386 & 0.131 & 0.925 & 0.870 & 0.813 & 0.456 & 0.073 & 0.655 & 0.0028 \\
\textsl{getold}&
                 250 & 235 & 179 & 56 &  1 & 0.256 & 0.989 & 0.195 & 0.944 & 0.845 & 0.721 & 0.533 & 0.253 & 0.575 & 0.0151 \\
\textsl{getold}&
                 350 & 233 & 160 & 73 &  1 & 0.254 & 0.989 & 0.174 & 0.932 & 0.819 & 0.711 & 0.498 & 0.251 & 0.543 & 0.0118 \\
\textsl{getold}&
                 500 & 126 &  84 & 42 &  0 & 0.137 & 1.000 & 0.091 & 0.912 & 0.787 & 0.677 & 0.387 & 0.137 & 0.486 & 0.0024 \\
\noalign{\smallskip}
\hline
\end{tabular}
\label{qualB4}
\end{table*}


\begin{table*}  
\caption
{ 
Benchmark A$_3$ with \textsl{getsf} using different subsets of images for the combination over wavelengths and detection. The
qualities, defined in Eqs.~(\ref{qualities0})\,--\,(\ref{finalqualities}), are evaluated for only acceptable sources, cf.
Eq.~(\ref{acceptable}), with errors in measurements within a factor of $2^{1/2}$. The number of model sources $N_{\rm T} = 459$.
Source measurements in the image of derived surface densities $\mathcal{D}_{11{\arcsec}}$ (at a fictitious wavelength $\lambdabar =
175\,\mu$m) are known to be inaccurate (e.g., Appendix~A of Paper I), hence the data are not presented. The extractions are sorted,
from top to bottom, by their global qualities $Q$ of $0.117$, $0.102$, $0.092$, $0.083$, $0.082$, $0.078$, and $0.022$.
} 
\begin{tabular}{crrrrrlllllllllll}
\hline
\noalign{\smallskip}
\!\!\!Bench A$_3$&$\lambda$\,\,&\!\!\!$N_{{\rm D}{{\lambda}}}$&\!\!\!\!$N_{{\rm G}{{\lambda}}}$&\!\!\!$N_{{\rm B}{{\lambda}}}$&
\!$N_{\rm S}$&\,\,\,$C_{\lambda}$&\,\,\,$R_{\lambda}$&\,\,$G_{\lambda}$&$Q_{{\rm P}{\lambda}}$&$Q_{{\rm T}{\lambda}}$&
$Q_{{\rm E}{\lambda}}$&$Q_{{\rm D}{\lambda}}$&$Q_{{\rm CR}{\lambda}}$&$Q_{{\rm PTE}{\lambda}}$&\,\,$Q_{{\lambda}}$\\
\noalign{\smallskip}
\hline
\noalign{\smallskip}
\!\!    --   &
                 75 & 107 & 107 &  0 & 0 & 0.233 & 1.000 & 0.233 & 0.998 & 0.979 & 0.992 & 0.902 & 0.233 & 0.969 & 0.0475 \\
\!\!    --   &
                110 & 107 & 107 &  0 & 0 & 0.233 & 1.000 & 0.233 & 0.998 & 0.994 & 0.995 & 0.902 & 0.233 & 0.987 & 0.0484 \\
\!\!    --   &
                170 & 207 & 184 & 23 & 1 & 0.451 & 0.986 & 0.401 & 0.878 & 0.878 & 0.928 & 0.712 & 0.445 & 0.716 & 0.0909 \\
\!\!detection&  175 & -- & -- & -- & -- & -- & -- & -- & -- & -- & -- & -- & -- & -- & -- \\
\!\!    --   &
                250 & 327 & 286 & 41 & 1 & 0.712 & 0.994 & 0.623 & 0.943 & 0.893 & 0.935 & 0.712 & 0.708 & 0.787 & 0.247 \\
\!\!    --   &
                350 & 324 & 292 & 32 & 0 & 0.706 & 1.000 & 0.636 & 0.935 & 0.867 & 0.941 & 0.703 & 0.706 & 0.764 & 0.241 \\
\!\!    --   &
                500 & 303 & 270 & 33 & 0 & 0.660 & 1.000 & 0.588 & 0.935 & 0.864 & 0.939 & 0.712 & 0.660 & 0.759 & 0.210 \\
\noalign{\smallskip}
\hline
\noalign{\smallskip}
\!\!detection&
                 75 & 107 & 107 &  0 & 0 & 0.233 & 1.000 & 0.233 & 0.998 & 0.992 & 0.988 & 0.921 & 0.233 & 0.978 & 0.0490 \\
\!\!detection&
                110 & 107 & 107 &  0 & 3 & 0.233 & 0.717 & 0.233 & 0.998 & 0.991 & 0.995 & 0.921 & 0.167 & 0.984 & 0.0353 \\
\!\!detection&
                170 & 213 & 186 & 27 & 9 & 0.464 & 0.564 & 0.405 & 0.882 & 0.880 & 0.931 & 0.733 & 0.262 & 0.723 & 0.0562 \\
\!\!detection&  175 & -- & -- & -- & -- & -- & -- & -- & -- & -- & -- & -- & -- & -- & -- \\
\!\!detection&
                250 & 320 & 287 & 33 & 1 & 0.697 & 0.994 & 0.625 & 0.944 & 0.895 & 0.946 & 0.697 & 0.693 & 0.799 & 0.241  \\
\!\!detection&
                350 & 319 & 294 & 25 & 0 & 0.695 & 1.000 & 0.641 & 0.938 & 0.868 & 0.941 & 0.686 & 0.695 & 0.766 & 0.234  \\
\!\!detection&
                500 & 303 & 273 & 30 & 0 & 0.660 & 1.000 & 0.595 & 0.933 & 0.870 & 0.942 & 0.704 & 0.660 & 0.765 & 0.211  \\
\noalign{\smallskip}
\hline
\noalign{\smallskip}
\!\!    --   &
                 75 & 106 & 105 &  1 & 0 & 0.231 & 1.000 & 0.229 & 0.998 & 0.956 & 0.985 & 0.978 & 0.231 & 0.939 & 0.0485 \\
\!\!    --   &
                110 & 107 & 107 &  0 & 0 & 0.233 & 1.000 & 0.233 & 0.979 & 0.990 & 0.961 & 0.888 & 0.233 & 0.931 & 0.0449 \\
\!\!    --   &
                170 & 217 & 191 & 26 & 1 & 0.473 & 0.987 & 0.416 & 0.885 & 0.878 & 0.932 & 0.621 & 0.467 & 0.724 & 0.0874 \\
\!\!    --   &  175 & -- & -- & -- & -- & -- & -- & -- & -- & -- & -- & -- & -- & -- & -- \\
\!\!detection&
                250 & 316 & 282 & 34 & 3 & 0.688 & 0.950 & 0.614 & 0.942 & 0.887 & 0.943 & 0.479 & 0.654 & 0.788 & 0.152 \\
\!\!    --   &
                350 & 308 & 282 & 26 & 0 & 0.671 & 1.000 & 0.614 & 0.935 & 0.862 & 0.937 & 0.484 & 0.671 & 0.755 & 0.151 \\
\!\!    --   &
                500 & 291 & 265 & 26 & 0 & 0.634 & 1.000 & 0.577 & 0.927 & 0.861 & 0.938 & 0.494 & 0.634 & 0.749 & 0.135 \\
\noalign{\smallskip}
\hline
\noalign{\smallskip}
\!\!    --   &
                 75 & 107 & 104 &  3 & 0 & 0.233 & 1.000 & 0.227 & 0.998 & 0.948 & 0.970 & 0.876 & 0.233 & 0.918 & 0.0425 \\
\!\!    --   &
                110 & 107 & 107 &  0 & 0 & 0.233 & 1.000 & 0.233 & 0.978 & 0.989 & 0.956 & 0.850 & 0.233 & 0.925 & 0.0428 \\
\!\!    --   &
                170 & 216 & 192 & 24 & 1 & 0.471 & 0.987 & 0.418 & 0.886 & 0.868 & 0.928 & 0.503 & 0.465 & 0.714 & 0.0698 \\
\!\!    --   &  175 & -- & -- & -- & -- & -- & -- & -- & -- & -- & -- & -- & -- & -- & -- \\
\!\!detection&
                250 & 286 & 260 & 26 & 3 & 0.623 & 0.940 & 0.566 & 0.945 & 0.899 & 0.937 & 0.484 & 0.586 & 0.796 & 0.128 \\
\!\!detection&
                350 & 284 & 264 & 20 & 1 & 0.619 & 0.993 & 0.575 & 0.941 & 0.870 & 0.929 & 0.491 & 0.614 & 0.760 & 0.132 \\
\!\!detection&
                500 & 278 & 255 & 23 & 0 & 0.606 & 1.000 & 0.556 & 0.934 & 0.863 & 0.944 & 0.579 & 0.606 & 0.761 & 0.148 \\
\noalign{\smallskip}
\hline
\noalign{\smallskip}
\!\!    --   &
                 75 & 107 & 107 &  0 & 0 & 0.233 & 1.000 & 0.233 & 0.998 & 0.980 & 0.997 & 0.973 & 0.233 & 0.975 & 0.0516 \\
\!\!    --   &
                110 & 107 & 107 &  0 & 0 & 0.233 & 1.000 & 0.233 & 0.998 & 0.993 & 0.998 & 0.973 & 0.233 & 0.989 & 0.0523 \\
\!\!detection&
                170 & 218 & 190 & 28 &12 & 0.475 & 0.464 & 0.414 & 0.885 & 0.880 & 0.936 & 0.584 & 0.221 & 0.729 & 0.0389 \\
\!\!    --   &  175 & -- & -- & -- & -- & -- & -- & -- & -- & -- & -- & -- & -- & -- & -- \\
\!\!detection&
                250 & 285 & 257 & 28 & 2 & 0.621 & 0.972 & 0.560 & 0.947 & 0.898 & 0.945 & 0.533 & 0.603 & 0.804 & 0.145 \\
\!\!detection&
                350 & 284 & 262 & 22 & 0 & 0.619 & 1.000 & 0.571 & 0.940 & 0.868 & 0.939 & 0.546 & 0.619 & 0.766 & 0.148 \\
\!\!detection&
                500 & 277 & 255 & 22 & 0 & 0.603 & 1.000 & 0.556 & 0.929 & 0.857 & 0.939 & 0.543 & 0.603 & 0.748 & 0.136 \\
\noalign{\smallskip}
\hline
\noalign{\smallskip}
\!\!detection&
                 75 & 107 & 107 &  0 & 0 & 0.233 & 1.000 & 0.233 & 0.997 & 0.981 & 0.997 & 1.000 & 0.233 & 0.975 & 0.0530 \\
\!\!detection&
                110 & 107 & 107 &  0 & 3 & 0.233 & 0.717 & 0.233 & 0.998 & 0.992 & 0.998 & 1.000 & 0.167 & 0.988 & 0.0385 \\
\!\!detection&
                170 & 215 & 191 & 24 & 9 & 0.468 & 0.568 & 0.416 & 0.878 & 0.874 & 0.934 & 0.555 & 0.266 & 0.717 & 0.0440 \\
\!\!    --   &  175 & -- & -- & -- & -- & -- & -- & -- & -- & -- & -- & -- & -- & -- & -- \\
\!\!detection&
                250 & 286 & 256 & 30 & 3 & 0.623 & 0.940 & 0.558 & 0.944 & 0.896 & 0.950 & 0.554 & 0.586 & 0.804 & 0.145 \\
\!\!detection&
                350 & 284 & 261 & 23 & 0 & 0.619 & 1.000 & 0.569 & 0.940 & 0.868 & 0.940 & 0.551 & 0.619 & 0.767 & 0.149 \\
\!\!detection&
                500 & 277 & 258 & 19 & 0 & 0.603 & 1.000 & 0.562 & 0.928 & 0.854 & 0.934 & 0.477 & 0.603 & 0.741 & 0.120 \\
\noalign{\smallskip}
\hline
\noalign{\smallskip}
\!\!    --   &
                 75 &  94 &   1 & 93 & 3 & 0.205 & 0.671 & 0.002 & 0.999 & 0.997 & 0.997 & 1.000 & 0.137 & 0.993 & 0.0003 \\
\!\!    --   &
                110 &  91 &  42 & 49 & 2 & 0.198 & 0.796 & 0.092 & 0.893 & 0.987 & 0.747 & 0.829 & 0.158 & 0.658 & 0.0079 \\
\!\!    --   &
                175 & 191 & 152 & 39 & 0 & 0.416 & 1.000 & 0.331 & 0.900 & 0.889 & 0.891 & 0.512 & 0.416 & 0.713 & 0.0503 \\
\!\!    --   &  175 & -- & -- & -- & -- & -- & -- & -- & -- & -- & -- & -- & -- & -- & -- \\
\!\!    --   &
                250 & 238 & 210 & 28 & 0 & 0.519 & 1.000 & 0.458 & 0.937 & 0.884 & 0.924 & 0.507 & 0.519 & 0.765 & 0.0920 \\
\!\!    --   &
                350 & 242 & 218 & 24 & 0 & 0.527 & 1.000 & 0.475 & 0.942 & 0.894 & 0.932 & 0.508 & 0.527 & 0.785 & 0.0999 \\
\!\!detection&
                500 & 248 & 226 & 22 & 0 & 0.540 & 1.000 & 0.492 & 0.936 & 0.868 & 0.946 & 0.508 & 0.540 & 0.768 & 0.1038 \\
\noalign{\smallskip}
\hline
\end{tabular}
\label{qualA3sets}
\end{table*}

\begin{table*}  
\caption
{ 
Benchmark B$_3$ with \textsl{getsf} using different subsets of images for the combination over wavelengths and detection. The
qualities, defined in Eqs.~(\ref{qualities0})\,--\,(\ref{finalqualities}), are evaluated for only acceptable sources, cf.
Eq.~(\ref{acceptable}), with errors in measurements within a factor of $2^{1/2}$. The number of model sources $N_{\rm T} = 919$.
Source measurements in the image of derived surface densities $\mathcal{D}_{13{\arcsec}}$ (at a fictitious wavelength $\lambdabar =
165\,\mu$m) are known to be inaccurate (e.g., Appendix~A of Paper I), hence the data are not presented. The extractions are sorted,
from top to bottom, by their global qualities $Q$ of $0.042$, $0.030$, $0.030$, $0.028$, $0.027$, $0.027$, $0.025$.
} 
\begin{tabular}{crrrrrlllllllllll}
\hline
\noalign{\smallskip}
\!\!\!Bench B$_3$&$\lambda$\,\,&\!\!\!$N_{{\rm D}{{\lambda}}}$&\!\!\!\!$N_{{\rm G}{{\lambda}}}$&\!\!\!$N_{{\rm B}{{\lambda}}}$&
\!$N_{\rm S}$&\,\,\,$C_{\lambda}$&\,\,\,$R_{\lambda}$&\,\,$G_{\lambda}$&$Q_{{\rm P}{\lambda}}$&$Q_{{\rm T}{\lambda}}$&
$Q_{{\rm E}{\lambda}}$&$Q_{{\rm D}{\lambda}}$&$Q_{{\rm CR}{\lambda}}$&$Q_{{\rm PTE}{\lambda}}$&\,\,$Q_{{\lambda}}$\\
\noalign{\smallskip}
\hline
\noalign{\smallskip}
\!\!    --   &
                 70 &  91 &  91 &  0 & 0 & 0.099 & 1.000 & 0.099 & 1.000 & 0.999 & 1.000 & 1.000 & 0.099 & 0.998 & 0.0098 \\
\!\!    --   &
                100 &  91 &  91 &  0 & 0 & 0.099 & 1.000 & 0.099 & 1.000 & 0.993 & 0.997 & 1.000 & 0.099 & 0.990 & 0.0097 \\
\!\!    --   &
                160 & 179 & 156 & 23 & 0 & 0.195 & 1.000 & 0.170 & 0.859 & 0.882 & 0.908 & 0.705 & 0.195 & 0.687 & 0.0160 \\
\!\!detection&  165 & -- & -- & -- & -- & -- & -- & -- & -- & -- & -- & -- & -- & -- & -- \\
\!\!    --   &
                250 & 584 & 507 & 77 & 3 & 0.635 & 0.985 & 0.552 & 0.918 & 0.871 & 0.892 & 0.667 & 0.626 & 0.713 & 0.164  \\
\!\!    --   &
                350 & 637 & 567 & 70 & 0 & 0.693 & 1.000 & 0.617 & 0.928 & 0.841 & 0.942 & 0.656 & 0.693 & 0.735 & 0.206  \\
\!\!    --   &
                500 & 495 & 420 & 75 & 0 & 0.539 & 1.000 & 0.457 & 0.922 & 0.890 & 0.928 & 0.671 & 0.539 & 0.693 & 0.114  \\
\noalign{\smallskip}
\hline
\noalign{\smallskip}
\!\!detection&
                 70 &  91 &  91 &  0 & 2 & 0.099 & 0.796 & 0.099 & 1.000 & 0.999 & 1.000 & 1.000 & 0.079 & 0.998 & 0.0078 \\
\!\!detection&
                100 &  91 &  91 &  0 & 5 & 0.099 & 0.465 & 0.099 & 1.000 & 0.995 & 0.997 & 1.000 & 0.046 & 0.991 & 0.0045 \\
\!\!detection&
                160 & 185 & 156 & 29 & 8 & 0.201 & 0.555 & 0.170 & 0.851 & 0.887 & 0.914 & 0.669 & 0.112 & 0.690 & 0.0088 \\
\!\!detection&  165 & -- & -- & -- & -- & -- & -- & -- & -- & -- & -- & -- & -- & -- & -- \\
\!\!detection&
                250 & 578 & 506 & 72 & 5 & 0.629 & 0.958 & 0.551 & 0.914 & 0.877 & 0.890 & 0.578 & 0.602 & 0.713 & 0.137  \\
\!\!detection&
                350 & 629 & 570 & 59 & 0 & 0.684 & 1.000 & 0.620 & 0.931 & 0.849 & 0.934 & 0.537 & 0.684 & 0.739 & 0.168  \\
\!\!detection&
                500 & 502 & 450 & 52 & 0 & 0.546 & 1.000 & 0.490 & 0.924 & 0.813 & 0.927 & 0.555 & 0.546 & 0.696 & 0.103  \\
\noalign{\smallskip}
\hline
\noalign{\smallskip}
\!\!    --   &
                 70 &  91 &  91 &  0 & 0 & 0.099 & 1.000 & 0.099 & 1.000 & 0.997 & 1.000 & 1.000 & 0.099 & 0.997 & 0.0098 \\
\!\!    --   &
                100 &  91 &  91 &  0 & 2 & 0.099 & 0.796 & 0.099 & 1.000 & 0.995 & 0.997 & 1.000 & 0.079 & 0.991 & 0.0077 \\
\!\!    --   &
                160 & 178 & 151 & 27 & 4 & 0.194 & 0.789 & 0.164 & 0.858 & 0.879 & 0.916 & 0.623 & 0.153 & 0.691 & 0.0108 \\
\!\!    --   &  165 & -- & -- & -- & -- & -- & -- & -- & -- & -- & -- & -- & -- & -- & -- \\
\!\!detection&
                250 & 524 & 483 & 41 & 6 & 0.570 & 0.930 & 0.526 & 0.927 & 0.877 & 0.898 & 0.547 & 0.530 & 0.730 & 0.111  \\
\!\!detection&
                350 & 561 & 519 & 42 & 1 & 0.610 & 0.998 & 0.565 & 0.930 & 0.845 & 0.931 & 0.457 & 0.609 & 0.732 & 0.115  \\
\!\!detection&
                500 & 486 & 435 & 51 & 0 & 0.529 & 1.000 & 0.473 & 0.923 & 0.816 & 0.930 & 0.470 & 0.529 & 0.701 & 0.0825 \\
\noalign{\smallskip}
\hline
\noalign{\smallskip}
\!\!    --   &
                 70 &  91 &  91 &  0 & 0 & 0.099 & 1.000 & 0.099 & 1.000 & 0.999 & 1.000 & 1.000 & 0.099 & 0.998 & 0.0098 \\
\!\!    --   &
                100 &  91 &  91 &  0 & 3 & 0.099 & 0.659 & 0.099 & 1.000 & 0.995 & 0.997 & 1.000 & 0.065 & 0.991 & 0.0064 \\
\!\!detection&
                160 & 178 & 154 & 24 & 7 & 0.194 & 0.592 & 0.168 & 0.851 & 0.868 & 0.910 & 0.573 & 0.115 & 0.672 & 0.0074 \\
\!\!    --   &  165 & -- & -- & -- & -- & -- & -- & -- & -- & -- & -- & -- & -- & -- & -- \\
\!\!detection&
                250 & 524 & 483 & 41 & 6 & 0.570 & 0.930 & 0.526 & 0.925 & 0.879 & 0.899 & 0.537 & 0.530 & 0.731 & 0.109  \\
\!\!detection&
                350 & 561 & 520 & 41 & 1 & 0.610 & 0.998 & 0.566 & 0.931 & 0.845 & 0.930 & 0.448 & 0.609 & 0.732 & 0.113  \\
\!\!detection&
                500 & 484 & 435 & 49 & 0 & 0.527 & 1.000 & 0.473 & 0.924 & 0.817 & 0.932 & 0.460 & 0.527 & 0.704 & 0.0808 \\
\noalign{\smallskip}
\hline
\noalign{\smallskip}
\!\!    --   &
                 70 &  86 &  80 &  6 & 0 & 0.094 & 1.000 & 0.087 & 0.957 & 0.992 & 0.881 & 0.848 & 0.094 & 0.836 & 0.0058 \\
\!\!    --   &
                100 &  91 &  91 &  0 & 0 & 0.099 & 1.000 & 0.099 & 0.955 & 0.994 & 0.909 & 0.825 & 0.099 & 0.863 & 0.0070 \\
\!\!    --   &
                160 & 179 & 146 & 33 & 0 & 0.195 & 1.000 & 0.159 & 0.854 & 0.885 & 0.922 & 0.647 & 0.195 & 0.697 & 0.0140 \\
\!\!    --   &  165 & -- & -- & -- & -- & -- & -- & -- & -- & -- & -- & -- & -- & -- & -- \\
\!\!    --   &
                250 & 443 & 382 & 61 & 0 & 0.482 & 1.000 & 0.416 & 0.930 & 0.869 & 0.897 & 0.529 & 0.482 & 0.726 & 0.0769 \\
\!\!    --   &
                350 & 490 & 458 & 32 & 0 & 0.533 & 1.000 & 0.498 & 0.935 & 0.878 & 0.903 & 0.534 & 0.533 & 0.742 & 0.105  \\
\!\!detection&
                500 & 459 & 423 & 36 & 0 & 0.499 & 1.000 & 0.460 & 0.926 & 0.820 & 0.932 & 0.491 & 0.499 & 0.708 & 0.0799 \\
\noalign{\smallskip}
\hline
\noalign{\smallskip}
\!\!    --   &
                 70 &  91 &  91 &  0 & 0 & 0.099 & 1.000 & 0.099 & 1.000 & 0.998 & 0.999 & 0.981 & 0.099 & 0.997 & 0.0096 \\
\!\!    --   &
                100 &  91 &  91 &  0 & 3 & 0.099 & 0.659 & 0.099 & 1.000 & 0.995 & 0.997 & 0.981 & 0.065 & 0.991 & 0.0063 \\
\!\!    --   &
                160 & 186 & 159 & 27 & 4 & 0.202 & 0.802 & 0.173 & 0.853 & 0.877 & 0.904 & 0.493 & 0.162 & 0.676 & 0.0094 \\
\!\!    --   &  165 & -- & -- & -- & -- & -- & -- & -- & -- & -- & -- & -- & -- & -- & -- \\
\!\!detection&
                250 & 555 & 501 & 54 & 8 & 0.604 & 0.895 & 0.545 & 0.927 & 0.857 & 0.897 & 0.444 & 0.540 & 0.713 & 0.0932 \\
\!\!    --   &
                350 & 577 & 521 & 56 & 2 & 0.628 & 0.993 & 0.567 & 0.926 & 0.833 & 0.931 & 0.428 & 0.623 & 0.718 & 0.109  \\
\!\!    --   &
                500 & 489 & 430 & 59 & 1 & 0.532 & 0.998 & 0.468 & 0.918 & 0.808 & 0.930 & 0.370 & 0.531 & 0.690 & 0.0634 \\
\noalign{\smallskip}
\hline
\noalign{\smallskip}
\!\!detection&
                 70 &  91 &  91 &  0 & 2 & 0.099 & 0.796 & 0.099 & 1.000 & 0.999 & 1.000 & 1.000 & 0.079 & 0.998 & 0.0078 \\
\!\!detection&
                100 &  91 &  91 &  0 & 6 & 0.099 & 0.401 & 0.099 & 1.000 & 0.995 & 0.997 & 1.000 & 0.040 & 0.991 & 0.0039 \\
\!\!detection&
                160 & 179 & 153 & 26 & 7 & 0.195 & 0.594 & 0.166 & 0.856 & 0.875 & 0.910 & 0.604 & 0.116 & 0.681 & 0.0079 \\
\!\!    --   &  165 & -- & -- & -- & -- & -- & -- & -- & -- & -- & -- & -- & -- & -- & -- \\
\!\!detection&
                250 & 526 & 477 & 49 & 6 & 0.572 & 0.930 & 0.519 & 0.925 & 0.879 & 0.897 & 0.538 & 0.532 & 0.730 & 0.108  \\
\!\!detection&
                350 & 562 & 518 & 44 & 1 & 0.612 & 0.998 & 0.564 & 0.930 & 0.845 & 0.928 & 0.448 & 0.610 & 0.729 & 0.112  \\
\!\!detection&
                500 & 488 & 436 & 52 & 0 & 0.531 & 1.000 & 0.474 & 0.925 & 0.818 & 0.932 & 0.458 & 0.531 & 0.705 & 0.0814 \\
\noalign{\smallskip}
\hline
\end{tabular}
\label{qualB3sets}
\end{table*}

\begin{table*}  
\caption
{ 
Benchmark B$_4$ with \textsl{getsf} using different subsets of images for the combination over wavelengths and detection. The
qualities, defined in Eqs.~(\ref{qualities0})\,--\,(\ref{finalqualities}), are evaluated for only acceptable sources, cf.
Eq.~(\ref{acceptable}), with errors in measurements within a factor of $2^{1/2}$. The number of model sources $N_{\rm T} = 919$.
Source measurements in the image of derived surface densities $\mathcal{D}_{13{\arcsec}}$ (at a fictitious wavelength $\lambdabar =
165\,\mu$m) are known to be inaccurate (e.g., Appendix~A of Paper I), hence the data are not presented. The extractions are sorted,
from top to bottom, by their global qualities $Q$ of $0.011$, $0.0096$, $0.0089$, $0.0084$, $0.0080$, $0.0078$, $0.0055$.
} 
\begin{tabular}{crrrrrlllllllllll}
\hline
\noalign{\smallskip}
\!\!\!Bench B$_4$&$\lambda$\,\,&\!\!\!$N_{{\rm D}{{\lambda}}}$&\!\!\!\!$N_{{\rm G}{{\lambda}}}$&\!\!\!$N_{{\rm B}{{\lambda}}}$&
\!$N_{\rm S}$&\,\,\,$C_{\lambda}$&\,\,\,$R_{\lambda}$&\,\,$G_{\lambda}$&$Q_{{\rm P}{\lambda}}$&$Q_{{\rm T}{\lambda}}$&
$Q_{{\rm E}{\lambda}}$&$Q_{{\rm D}{\lambda}}$&$Q_{{\rm CR}{\lambda}}$&$Q_{{\rm PTE}{\lambda}}$&\,\,$Q_{{\lambda}}$\\
\noalign{\smallskip}
\hline
\noalign{\smallskip}
\!\!    --   &
                 70 &  91 &  91 &  0 &  0 & 0.099 & 1.000 & 0.099 & 1.000 & 0.999 & 1.000 & 1.000 & 0.099 & 0.998 & 0.0098 \\
\!\!    --   &
                100 &  91 &  91 &  0 &  0 & 0.099 & 1.000 & 0.099 & 0.999 & 0.992 & 0.997 & 1.000 & 0.099 & 0.988 & 0.0097 \\
\!\!    --   &
                160 & 101 &  99 &  2 &  0 & 0.110 & 1.000 & 0.108 & 0.940 & 0.921 & 0.948 & 0.924 & 0.110 & 0.820 & 0.0090 \\
\!\!detection&  165 & -- & -- & -- & -- & -- & -- & -- & -- & -- & -- & -- & -- & -- & -- \\
\!\!    --   &
                250 & 194 & 168 & 26 &  1 & 0.211 & 0.984 & 0.183 & 0.955 & 0.876 & 0.914 & 0.768 & 0.208 & 0.765 & 0.0223 \\
\!\!    --   &
                350 & 179 & 148 & 31 &  0 & 0.195 & 1.000 & 0.161 & 0.931 & 0.833 & 0.944 & 0.760 & 0.195 & 0.732 & 0.0174 \\
\!\!    --   &
                500 & 113 &  84 & 29 &  0 & 0.123 & 1.000 & 0.091 & 0.912 & 0.793 & 0.913 & 0.749 & 0.123 & 0.660 & 0.0056 \\
\noalign{\smallskip}
\hline
\noalign{\smallskip}
\!\!    --   &
                 70 &  91 &  91 &  0 &  0 & 0.099 & 1.000 & 0.099 & 1.000 & 0.997 & 0.999 & 0.969 & 0.099 & 0.996 & 0.0095 \\
\!\!    --   &
                100 &  91 &  91 &  0 &  0 & 0.099 & 1.000 & 0.099 & 0.999 & 0.992 & 0.996 & 0.969 & 0.099 & 0.987 & 0.0094 \\
\!\!    --   &
                160 & 101 &  99 &  2 &  0 & 0.110 & 1.000 & 0.108 & 0.942 & 0.934 & 0.944 & 0.870 & 0.110 & 0.831 & 0.0086 \\
\!\!    --   &  165 & -- & -- & -- & -- & -- & -- & -- & -- & -- & -- & -- & -- & -- & -- \\
\!\!detection&
                250 & 187 & 172 & 15 &  5 & 0.203 & 0.734 & 0.187 & 0.952 & 0.883 & 0.911 & 0.664 & 0.149 & 0.766 & 0.0142 \\
\!\!detection&
                350 & 177 & 153 & 24 &  2 & 0.193 & 0.931 & 0.166 & 0.935 & 0.838 & 0.934 & 0.627 & 0.179 & 0.732 & 0.0137 \\
\!\!detection&
                500 & 120 &  93 & 27 &  0 & 0.131 & 1.000 & 0.101 & 0.910 & 0.798 & 0.918 & 0.603 & 0.131 & 0.667 & 0.0053 \\
\noalign{\smallskip}
\hline
\noalign{\smallskip}
\!\!    --   &
                 70 &  91 &  91 &  0 &  0 & 0.099 & 1.000 & 0.099 & 1.000 & 0.999 & 1.000 & 1.000 & 0.099 & 0.998 & 0.0098 \\
\!\!    --   &
                100 &  91 &  91 &  0 &  0 & 0.099 & 1.000 & 0.099 & 0.999 & 0.993 & 0.997 & 1.000 & 0.099 & 0.989 & 0.0097 \\
\!\!detection&
                160 & 102 & 100 &  2 &  3 & 0.111 & 0.700 & 0.109 & 0.942 & 0.929 & 0.944 & 0.847 & 0.078 & 0.827 & 0.0059 \\
\!\!    --   &  165 & -- & -- & -- & -- & -- & -- & -- & -- & -- & -- & -- & -- & -- & -- \\
\!\!detection&
                250 & 186 & 171 & 15 &  6 & 0.202 & 0.667 & 0.186 & 0.952 & 0.882 & 0.911 & 0.651 & 0.135 & 0.765 & 0.0125 \\
\!\!detection&
                350 & 176 & 152 & 24 &  2 & 0.192 & 0.931 & 0.165 & 0.936 & 0.839 & 0.933 & 0.617 & 0.178 & 0.733 & 0.0133 \\
\!\!detection&
                500 & 120 &  93 & 27 &  0 & 0.131 & 1.000 & 0.101 & 0.910 & 0.798 & 0.917 & 0.590 & 0.131 & 0.666 & 0.0052 \\
\noalign{\smallskip}
\hline
\noalign{\smallskip}
\!\!detection& 
                 70 &  91 &  91 &  0 &  4 & 0.099 & 0.549 & 0.099 & 1.000 & 0.999 & 1.000 & 1.000 & 0.054 & 0.998 & 0.0054 \\
\!\!detection& 
                100 &  91 &  91 &  0 &  1 & 0.099 & 0.935 & 0.099 & 0.999 & 0.993 & 0.997 & 1.000 & 0.093 & 0.989 & 0.0091 \\
\!\!detection& 
                160 & 102 & 100 &  2 &  3 & 0.111 & 0.700 & 0.109 & 0.939 & 0.926 & 0.947 & 0.888 & 0.078 & 0.824 & 0.0062 \\
\!\!detection&  165 & -- & -- & -- & -- & -- & -- & -- & -- & -- & -- & -- & -- & -- & -- \\
\!\!detection& 
                250 & 193 & 173 & 20 &  7 & 0.210 & 0.623 & 0.188 & 0.951 & 0.875 & 0.911 & 0.734 & 0.131 & 0.759 & 0.0137 \\
\!\!detection& 
                350 & 180 & 151 & 29 &  1 & 0.196 & 0.982 & 0.164 & 0.932 & 0.838 & 0.936 & 0.697 & 0.192 & 0.732 & 0.0161 \\
\!\!detection& 
                500 & 113 &  91 & 22 &  0 & 0.123 & 1.000 & 0.099 & 0.909 & 0.796 & 0.913 & 0.677 & 0.123 & 0.662 & 0.0054 \\
\noalign{\smallskip}
\hline
\noalign{\smallskip}
\!\!    --   &
                 70 &  91 &  91 &  0 &  0 & 0.099 & 1.000 & 0.099 & 1.000 & 0.998 & 0.999 & 0.950 & 0.099 & 0.996 & 0.0093 \\
\!\!    --   &
                100 &  91 &  91 &  0 &  0 & 0.099 & 1.000 & 0.099 & 0.999 & 0.993 & 0.996 & 0.950 & 0.099 & 0.988 & 0.0092 \\
\!\!    --   &
                160 & 101 & 100 &  1 &  0 & 0.110 & 1.000 & 0.109 & 0.935 & 0.927 & 0.950 & 0.762 & 0.110 & 0.824 & 0.0075 \\
\!\!    --   &  165 & -- & -- & -- & -- & -- & -- & -- & -- & -- & -- & -- & -- & -- & -- \\
\!\!detection&
                250 & 193 & 173 & 20 &  6 & 0.210 & 0.680 & 0.188 & 0.948 & 0.868 & 0.915 & 0.545 & 0.143 & 0.754 & 0.0111 \\
\!\!    --   &
                350 & 175 & 146 & 29 &  3 & 0.190 & 0.860 & 0.159 & 0.930 & 0.826 & 0.936 & 0.525 & 0.164 & 0.718 & 0.0098 \\
\!\!    --   &
                500 & 117 &  86 & 31 &  1 & 0.127 & 0.959 & 0.094 & 0.909 & 0.792 & 0.916 & 0.487 & 0.122 & 0.659 & 0.0037 \\
\noalign{\smallskip}
\hline
\noalign{\smallskip}
\!\!detection&
                 70 &  91 &  91 &  0 &  4 & 0.099 & 0.549 & 0.099 & 1.000 & 0.999 & 1.000 & 1.000 & 0.054 & 0.998 & 0.0054 \\
\!\!detection&
                100 &  91 &  91 &  0 &  1 & 0.099 & 0.935 & 0.099 & 0.999 & 0.993 & 0.997 & 1.000 & 0.093 & 0.989 & 0.0091 \\
\!\!detection&
                160 & 102 &  99 &  3 &  4 & 0.111 & 0.593 & 0.108 & 0.943 & 0.925 & 0.944 & 0.858 & 0.066 & 0.823 & 0.0050 \\
\!\!    --   &  165 & -- & -- & -- & -- & -- & -- & -- & -- & -- & -- & -- & -- & -- & -- \\
\!\!detection&
                250 & 187 & 171 & 16 &  6 & 0.203 & 0.669 & 0.186 & 0.951 & 0.880 & 0.911 & 0.655 & 0.136 & 0.762 & 0.0126 \\
\!\!detection&
                250 & 176 & 151 & 25 &  2 & 0.192 & 0.931 & 0.164 & 0.934 & 0.838 & 0.936 & 0.618 & 0.178 & 0.733 & 0.0133 \\
\!\!detection&
                500 & 121 &  93 & 28 &  0 & 0.132 & 1.000 & 0.101 & 0.911 & 0.800 & 0.919 & 0.591 & 0.132 & 0.669 & 0.0053 \\
\noalign{\smallskip}
\hline
\noalign{\smallskip}
\!\!    --   &
                 70 &  84 &  65 & 19 &  0 & 0.091 & 1.000 & 0.071 & 0.883 & 0.993 & 0.776 & 0.736 & 0.091 & 0.681 & 0.0032 \\
\!\!    --   &
                100 &  85 &  68 & 17 &  0 & 0.092 & 1.000 & 0.074 & 0.893 & 0.976 & 0.789 & 0.712 & 0.092 & 0.687 & 0.0034 \\
\!\!    --   &
                160 &  95 &  85 & 10 &  0 & 0.103 & 1.000 & 0.092 & 0.894 & 0.904 & 0.794 & 0.649 & 0.103 & 0.641 & 0.0040 \\
\!\!    --   &  165 & -- & -- & -- & -- & -- & -- & -- & -- & -- & -- & -- & -- & -- & -- \\
\!\!    --   &
                250 & 167 & 153 & 14 &  0 & 0.182 & 1.000 & 0.166 & 0.926 & 0.863 & 0.819 & 0.613 & 0.182 & 0.654 & 0.0121 \\
\!\!    --   &
                350 & 165 & 137 & 28 &  0 & 0.180 & 1.000 & 0.149 & 0.934 & 0.856 & 0.894 & 0.606 & 0.180 & 0.715 & 0.0116 \\
\!\!detection&
                500 & 118 &  96 & 22 &  0 & 0.128 & 1.000 & 0.104 & 0.911 & 0.800 & 0.919 & 0.609 & 0.128 & 0.670 & 0.0055 \\
\noalign{\smallskip}
\hline
\end{tabular}
\label{qualB4sets}
\end{table*}

\end{appendix}

\end{document}